\DocumentMetadata{}
\documentclass[%
reprint,
superscriptaddress,
nofootinbib,
amsmath,amssymb,
aps,
prb,
]{revtex4-2}

\usepackage[T1]{fontenc}
\usepackage{graphicx}% Include figure files
\usepackage{dcolumn}% Align table columns on decimal point
\usepackage{bm}% bold math
\usepackage{mathtools}
\usepackage{mathrsfs}
\usepackage{braket}
\usepackage{color}
\usepackage[table]{xcolor}
\usepackage{comment}
\usepackage{subcaption}
\usepackage{ragged2e}
\usepackage{multirow}
\usepackage[version=3]{mhchem}
\usepackage{dashbox}
\usepackage{tikz}
\usetikzlibrary{arrows.meta,shapes.geometric,calc}
\usepackage{pdfpages}
\usepackage{pgffor}
\usepackage{newpax}

\usepackage{hyperref}% add hypertext capabilities
\hypersetup{
     colorlinks   = true,
     linkcolor    = blue,
     citecolor    = blue,
     urlcolor     = blue
}

\DeclareCaptionFormat{justified}{%
  \justifying#1#2#3
}

\newcounter{subsubsubsection}[subsubsection]
\renewcommand{\thesubsubsubsection}{\roman{subsubsubsection}.}

\newcommand{\subsubsubsection}[1]{%
    \par\addvspace{2ex}%
    \refstepcounter{subsubsubsection}%
    \noindent
    \begin{minipage}{\linewidth}
        \centering\normalfont
        \textit{\thesubsubsubsection\quad #1}\par
    \end{minipage}%
    \par\nobreak\vspace{1.2ex}%
}

\makeatletter
\AtBeginDocument{\let\LS@rot\@undefined}
\makeatother

\def\supplementfilename{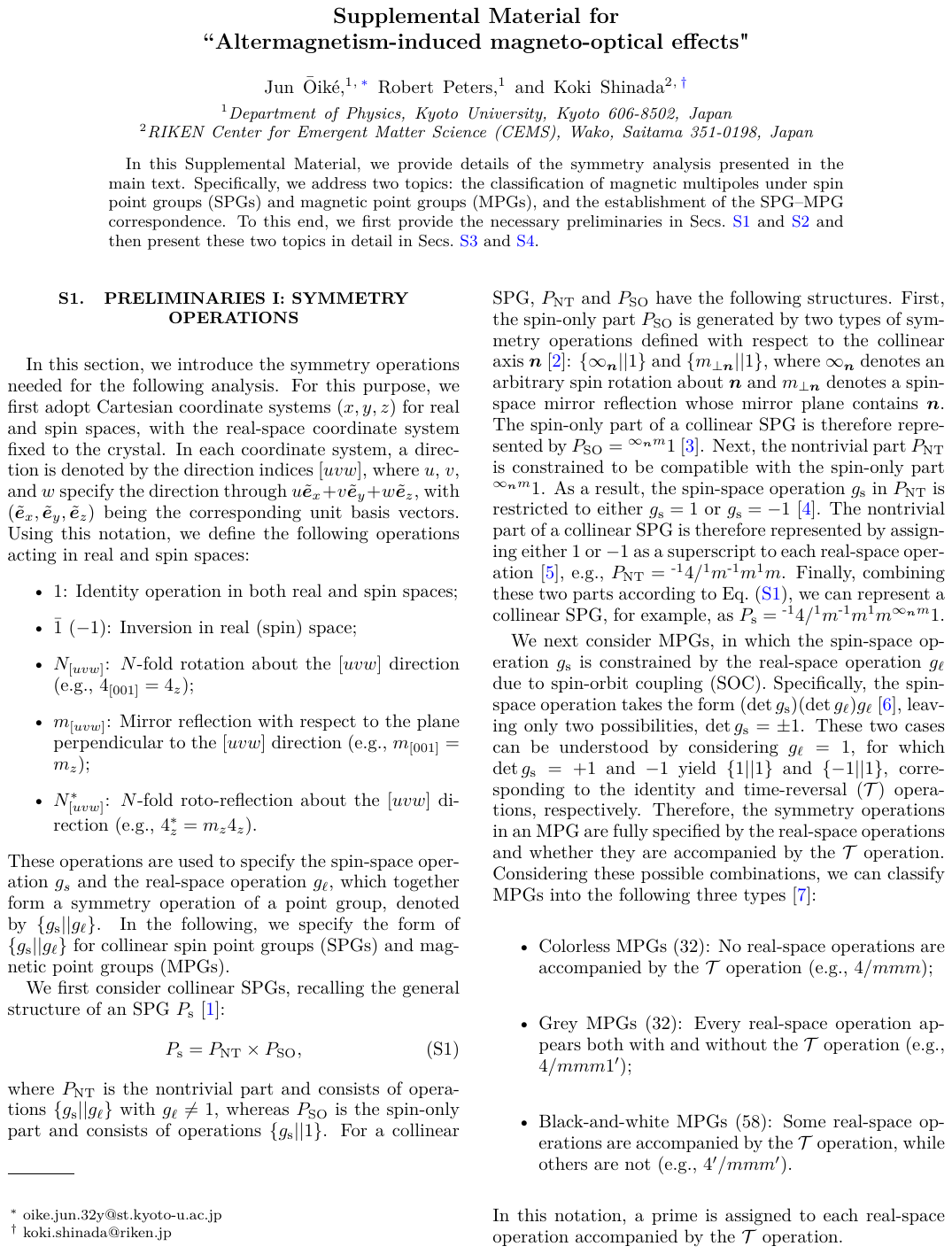}

\newcount\numbersupplementpages

\pdfximage{\supplementfilename}
\numbersupplementpages=\pdflastximagepages

\ExplSyntaxOn
\cs_generate_variant:Nn \tl_gset:Nn { Ne }
\cs_generate_variant:Nn \tl_if_empty:nTF { e }
\ExplSyntaxOff

\begin{document}

\preprint{APS/123-QED}

\title{Altermagnetism-induced magneto-optical effects}

\author{Jun {\=O}ik\'{e}}
\altaffiliation[oike.jun.32y@st.kyoto-u.ac.jp]{}%Lines break automatically or can be forced with \\
\affiliation{Department of Physics, Kyoto University, Kyoto 606-8502, Japan}
\author{Robert Peters}
\affiliation{Department of Physics, Kyoto University, Kyoto 606-8502, Japan}
\author{Koki Shinada}
\altaffiliation[koki.shinada@riken.jp]{}%Lines break automatically or can be forced with \\
\affiliation{RIKEN Center for Emergent Matter Science (CEMS), Wako, Saitama 351-0198, Japan}

\date{\today}
             
\begin{abstract}
Altermagnets break time-reversal symmetry despite their zero net magnetization,
enabling time-reversal-odd responses while retaining the advantages of antiferromagnets.
However, magneto-optical (MO) effects are not universal among altermagnets because MO effects typically arise from magnetic dipole order (ferromagnetism),
which is generally absent in altermagnets.
In this work, we address this limitation by establishing octupolar MO effects, which arise from magnetic octupole order through spatial dispersion.
Owing to this distinct origin, octupolar MO effects do not necessarily require spin-orbit coupling even in collinear magnets
and can arise without an external magnetic field in antiferromagnets where conventional MO effects are symmetry-forbidden.
Such antiferromagnets include $d$-wave altermagnets such as \ce{RuO2} and \ce{KV2Se2O}, certain $g$- and $i$-wave altermagnets, including \ce{CrSb}, 
as well as noncoplanar antiferromagnets such as \ce{MnTe2} and pyrochlore iridates with all-in-all-out magnetic order.
From an experimental perspective, octupolar MO effects can reach detectable magnitudes and offer potential for magnetic-domain imaging,
as supported by numerical calculations showing $\mu$rad-scale Faraday rotation for a 100-nm-thick sample in the visible-light range and a sign reversal between opposite domains.
Furthermore, octupolar MO effects are linked to an $f$-sum rule, enabling a quantitative estimate of the spin magnetic octupole moment from optical measurements.
These results extend the scope of MO effects to a broad class of antiferromagnets, including altermagnets, and open a field-free optical route to exploring magnetic octupole order.
\end{abstract}

\maketitle

%%%%%%%%%%%%%%%%%%%%%%%%%%%%%%%%%%%%%%%%%%%%%%%%%%%%%%%%%%%%%%%%%%%%%%%%%%%%%%%%%%%%%%%%%%%%%%%%%%%%%%%%%%%%%%%%%%%%%%%%%%%%%%%%%%%%

\section{Introduction}

Altermagnets exhibit momentum-dependent spin splitting despite their antiferromagnetic order with zero net magnetization \cite{Smejkal2022-xr,Smejkal2022-xx}.
This spin splitting persists in the nonrelativistic limit without spin-orbit coupling (SOC) \cite{Noda2016-wt,Okugawa2018-ws,Ahn2019-cy,Naka2019-uk,Hayami2019-zk,Smejkal2020-zu,Yuan2020-al,Naka2020-pm,Hayami2020-tg,Yuan2021-xz,Egorov2021-uo,Mazin2021-ea}, 
enabling the use of spin as an information carrier while mitigating SOC-induced spin relaxation.
In addition, their antiferromagnetic dynamics offer the potential for terahertz-speed operation, while 
their zero net magnetization suppresses stray magnetic fields, allowing high-density device integration.
Therefore, altermagnets have attracted considerable attention as promising materials for antiferromagnetic spintronics \cite{Jungwirth2016-td,Baltz2018-or,Dal_Din2024-kd}.
To date, several materials have been proposed as candidate altermagnets, 
including \ce{RuO2} \cite{Feng2022-jm,Fedchenko2024-lw,Zhang_2025-lb} and \ce{KV2Se2O} \cite{Jiang2025-au,Wang2025-wm,Guofei2026-tf} with $d$-wave spin splitting 
and $\alpha$-\ce{MnTe} \cite{Lee2024-hl,Krempasky2024-tz,Osumi2024-kd,Hajlaoui2024-ol,Aoyama2024-nz} and \ce{CrSb} \cite{Reimers2024-gw,Zeng2024-ea,Ding2024-qn,Yang2025-ko} with $g$-wave spin splitting.

Another distinctive feature of altermagnets is that their magnetic order breaks time-reversal ($\mathcal{T}$) symmetry.
This broken $\mathcal{T}$ symmetry provides a means to probe their magnetic structures through $\mathcal{T}$-odd responses.
Among such responses, magnetic optical activity (MOA) 
\footnote{Throughout this paper, we use the term ``magnetic optical activity'' to refer to magneto-optical effects.
} \cite{Faraday1846-bb,Kerr1877-ap,Cotton1896-je} is characterized by polarization rotation and ellipticity in linearly polarized light,
providing a spatially resolved probe of magnetic domains.
Indeed, MOA has been experimentally observed in $\alpha$-\ce{MnTe} and successfully employed to image its altermagnetic domains \cite{Gray2024-us,Amin2024-ay,Yamamoto2025-kp,Yang2026-xr,Watanabe2026-ek,Yang2026-rw}.
However, in other candidate altermagnets, such as \ce{RuO2}, \ce{KV2Se2O}, and \ce{CrSb}, MOA has not yet been observed, 
suggesting that MOA is not a universal response of altermagnets.

\begin{figure*}[t]
    \centering
        \begin{tikzpicture}[>=Stealth]
            \node[font=\fontsize{12pt}{17pt}\selectfont] (eq) at (0,3) {%
            $\varepsilon_{ij}(\bm{q},\omega)=\varepsilon_{ij}(\omega)+\varepsilon_{ij,k}(\omega)q_k+\varepsilon_{ij,kl}(\omega)q_kq_l+\cdots$
            };

            \node[inner sep=0] (fig1) at (-3.4,-0.4) {
                \includegraphics[width=4.4cm, height=3.0cm, trim=15 25 40 0, clip]{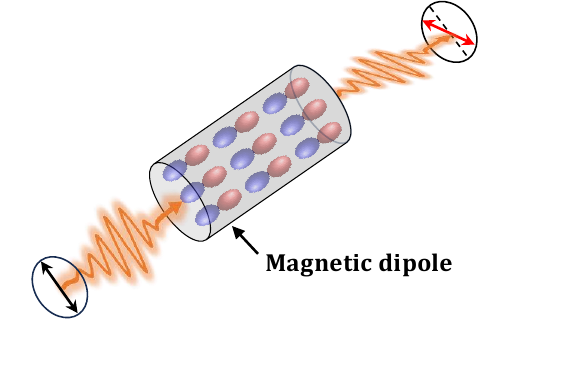} 
            };

            \node[above=1mm, font=\fontsize{10.5pt}{17pt}\selectfont] (fig1_caption) at (fig1.north) {
                (a) Dipolar MOA
            };

            \node[inner sep=0] (fig2) at (3.6,-0.4) {
                \includegraphics[width=4.4cm, height=3.0cm, trim=15 25 40 0, clip]{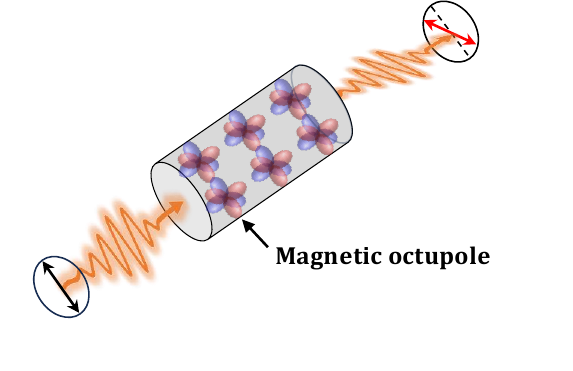} 
            };

            \node[above=1mm, font=\fontsize{10.5pt}{17pt}\selectfont] (fig2_caption) at (fig2.north) {
                (b) Octupolar MOA
            };

            \draw[->, thick] ([xshift=-2cm,yshift=-2pt]eq.south) -- ([yshift=2pt]fig1_caption.north);
            \draw[->, thick] ([xshift=2.3cm,yshift=-2pt]eq.south) -- ([yshift=2pt]fig2_caption.north);

            \node[below=2.5mm, font=\fontsize{10pt}{17pt}\selectfont] at (fig1.south) {$\alpha$-\ce{MnTe}, \ce{Mn3Sn}};
            \node[below=2.5mm, align=center, font=\fontsize{10pt}{14pt}\selectfont] at (fig2.south) {\ce{RuO2}, \ce{KV2Se2O}, \ce{CrSb}, \\ \ce{MnTe2}, \ce{\textit{R}2Ir2O7} (\textit{R}=rare earth) \\ \textcolor{red}{\textbf{(This work)}}};
        
            \phantomsubcaption
            \label{fig:General_concept1}
            \phantomsubcaption
            \label{fig:General_concept2}
        \end{tikzpicture}
    \caption{Schematic illustrations of polarization rotation of transmitted light for (a) dipolar MOA and (b) octupolar MOA.
    These two types of MOA arise from magnetic dipole and octupole orders, respectively, 
    and are described by different terms in the wave-vector expansion of the permittivity tensor $\varepsilon_{ij}(\bm{q},\omega)$, where $\bm{q}$ is the wave vector and $\omega$ is the angular frequency.
    Specifically, dipolar MOA and octupolar MOA correspond to the zeroth-order term $\varepsilon_{ij}(\omega)$ and the second-order term $\varepsilon_{ij,kl}(\omega)q_kq_l$, respectively.
    We also list examples of antiferromagnets that can exhibit each type of MOA below the corresponding illustrations.
    Notably, the antiferromagnets listed in (b) can exhibit octupolar MOA even though dipolar MOA is symmetry-forbidden.
    Note that the MOA observed in the noncollinear antiferromagnet \ce{Mn3Sn} \cite{Higo2018}, although often discussed in connection with magnetic octupole order, is classified as dipolar rather than octupolar MOA.}
    \label{fig:General_concept}
\end{figure*}
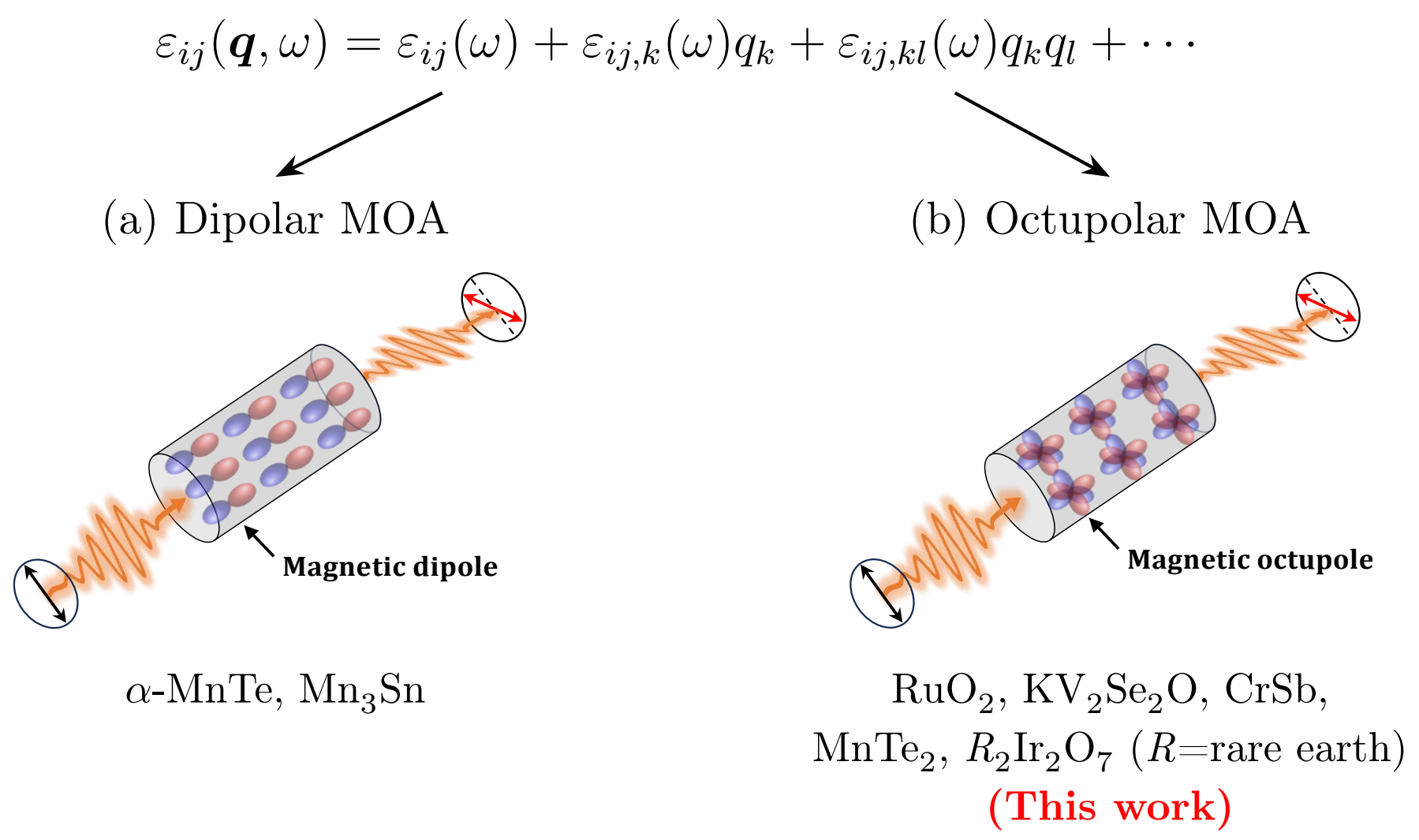

The nonuniversality of MOA in altermagnets can be understood in terms of their magnetic multipole order.
Altermagnetism is generally characterized by higher-rank magnetic multipole order because zero net magnetization precludes magnetic dipole order (ferromagnetism).
For example, $d$-wave altermagnetism is characterized by magnetic octupole order \cite{Fernandes2023-cw,Bhowal2024-vj,McClarty2024-eo}.
Given this higher-rank nature of altermagnetism and the fact that MOA typically originates from magnetic dipole order,
the presence of MOA in altermagnets can be regarded as rather exceptional.
A notable exception is $\alpha$-\ce{MnTe}, where SOC induces weak magnetic dipole order in the ground state \cite{Kluczyk2024-ec,Mazin2024-yc,Chen_Ye2026-jt}, thereby enabling MOA.
In most other altermagnets, however, magnetic dipole order remains symmetry-forbidden even in the presence of SOC, calling for an alternative mechanism of MOA beyond magnetic dipole order.

A possible route to MOA beyond magnetic dipole order was already explored in the 1990s by Graham and Raab \cite{Graham1991-xf}.
They showed that MOA can arise through spatial dispersion in cubic antiferromagnets without magnetic dipole order.
However, this MOA has received little attention since then, partly because its magnetic origin has remained unclear.
We now interpret this MOA as originating from magnetic octupole order in light of modern multipole theory \cite{Hayami2018-ip,Watanabe2018-xc,Yatsushiro2021-af},
in which magnetic octupole order is the next higher-rank even-parity magnetic multipole order after magnetic dipole order.
Based on this interpretation, MOA may also arise in materials with magnetic octupole order, including $d$-wave altermagnets, even in the absence of magnetic dipole order.
We therefore refer to MOA arising from magnetic dipole and octupole orders as dipolar and octupolar MOA, respectively,
and focus on octupolar MOA as an alternative to dipolar MOA (Fig.~\ref{fig:General_concept}).

Specifically, we develop a general and comprehensive theoretical framework for octupolar MOA.
This framework naturally encompasses the earlier results of Graham and Raab \cite{Graham1991-xf} 
while revealing several key features of octupolar MOA in addition to its ability to arise without an external magnetic field.
First, octupolar MOA can arise not only in $d$-wave altermagnets, such as \ce{RuO2} and \ce{KV2Se2O}, but also in certain $g$- and $i$-wave altermagnets, including \ce{CrSb}.
Second, octupolar MOA extends beyond collinear magnets to coplanar and noncoplanar magnets with magnetic octupole order, 
including \ce{MnTe2} \cite{Zhu2024-oc,Wulferding2026-kn} 
and pyrochlore iridates \ce{\textit{R}2Ir2O7} (\textit{R}=rare earth) with all-in-all-out (AIAO) magnetic order \cite{Tomiyasu2012-cy,Sagayama2013-xe,Donnerer2016-xq,Guo2016-is}.
Third, we derive the complex Faraday angle for octupolar MOA, and 
numerical calculations show that the resulting response is of the order of $10^{-6}$ for a 100-nm-thick sample in the visible-light range and changes sign upon reversal of the magnetic octupole order.
These results establish octupolar MOA as a broadly applicable and experimentally accessible route to field-free MOA beyond dipolar MOA,
with potential applications to magnetic-domain imaging.

Beyond these experimental implications, our framework also reveals several fundamental aspects of octupolar MOA.
First, octupolar MOA is described by the octupolar MOA vector $\tilde{\bm{G}}^{(\mathrm{o})}$, 
constructed from the second-order term in the wave-vector expansion of the permittivity tensor.
The form of this vector determines the light-propagation directions along which octupolar MOA is allowed.
Second, octupolar MOA does not necessarily require SOC in collinear magnets, in sharp contrast to dipolar MOA in collinear magnets, which requires SOC \cite{Watanabe2024-us}.
Third, we derive a microscopic expression for $\tilde{\bm{G}}^{(\mathrm{o})}$ and show that it contains a component describing a magnetoelectric response that occurs under magnetic octupole order 
and extends the conventional magnetoelectric response activated by magnetic quadrupole order.
This connection to the magnetoelectric response provides a physical interpretation of octupolar MOA, paralleling the relation between dipolar MOA and the optical Hall conductivity.
Furthermore, this magnetoelectric response obeys an $f$-sum rule that establishes a direct relation to the spin magnetic octupole moment, 
thereby enabling a quantitative estimate of the octupole moment.

The rest of this paper is organized into three main parts.
The first part develops the general theory of octupolar MOA.
Specifically, we establish a macroscopic theory of octupolar MOA in Sec.~\ref{Sec:Macroscopic_theory_MO} and 
discuss the conditions for observing it in Sec.~\ref{Sec:MO_conditions}.
The second part investigates the characteristic properties of octupolar MOA.
Specifically, we examine its SOC dependence in Sec.~\ref{Sec:SOC_dependence_MOA} and 
derive analytical expressions for its complex Faraday angle in Sec.~\ref{Sec:Faraday_Rotation}.
The third part provides a microscopic and quantitative analysis of octupolar MOA.
Specifically, we derive a microscopic expression for $\tilde{\bm{G}}^{(\mathrm{o})}$, clarify its physical interpretation, and establish the associated $f$-sum rule in Sec.~\ref{Sec:Microscopic_theory_MO}.
We then numerically evaluate the complex Faraday angle for octupolar MOA using models of \ce{\textit{R}2Ir2O7}, \ce{RuO2}, and \ce{CrSb} in Sec.~\ref{Sec:Numerical_results}.
Finally, we summarize this work in Sec.~\ref{Sec:Conclusion}.

Before proceeding, we introduce a convention for the Cartesian coordinate systems used throughout this paper.
This convention is to use the same notation $(x,y,z)$ for two different coordinate systems: one fixed to the crystal and the other with one of its axes along the light-propagation direction.
Which coordinate system $(x,y,z)$ refers to depends on the section.
The former is used in Secs.~\ref{Sec:MO_conditions}, \ref{Sec:SOC_dependence_MOA}, and \ref{Sec:Numerical_results}, 
whereas the latter is used in Secs.~\ref{Sec:Macroscopic_theory_MO} and \ref{Sec:Faraday_Rotation}.
Note that sections requiring no specific choice of coordinate system are omitted from this list, and 
that in Sec.~\ref{Sec:MO_conditions}, where both coordinate systems are used, the latter is denoted by $(X,Y,Z)$.

%%%%%%%%%%%%%%%%%%%%%%%%%%%%%%%%%%%%%%%%%%%%%%%%%%%%%%%%%%%%%%%%%%%%%%%%%%%%%%%%%%%%%%%%%%%%%%%%%%%%%%%%%%%%%%%%%%%%%%%%%%%%%%%%%%%%

\section{Macroscopic theory of octupolar MOA} \label{Sec:Macroscopic_theory_MO}

In this section, we establish a macroscopic theory of octupolar MOA by starting from a general description of optical activity.
Here, optical activity refers to the polarization rotation and ellipticity of light arising from its interaction with a medium.
We first derive the eigenvalue equation describing optical activity in Sec.~\ref{Subsec:optical_activity_general}.
The resulting equation is expressed in terms of the permittivity tensor, 
whose components corresponding to octupolar MOA are identified in Sec.~\ref{Subsec:symmetry of optical conductivity tensor}.
We then review established mechanisms of optical activity, including dipolar MOA, in Sec.~\ref{Subsec:theory of MD-induced magneto-optical effects}.
Building on these established mechanisms, we formulate octupolar MOA in Sec.~\ref{Subsec:theory of MO-induced magneto-optical effects}.
Finally, Sec.~\ref{Subsec:short summary of theory} summarizes the main results of this section.

%%%%%%%%%%%%%%%%%%%%%%%%%%%%%%%%%%%%%%%%%%%%%%%%%%%%%%%%%%%%%%%%%%

\subsection{Eigenvalue equation for optical activity} \label{Subsec:optical_activity_general}

To derive the eigenvalue equation, we first consider Maxwell's equations and the linear constitutive relations
for a monochromatic plane wave with spatial and temporal dependence $e^{i(\bm{q}\cdot\bm{r}-\omega t)}$.
Here, $\bm{r}$ is the position vector, $t$ is time, $\bm{q}$ is the wave vector, and $\omega$ is the angular frequency.
For this plane wave, Maxwell's equations in SI units reduce to
\begin{subequations} \label{eq:Maxwell} 
    \begin{align}
        \bm{q} \times \bm{E}_0&=\omega \bm{B}_0, \label{eq:Maxwell1} \\ 
        \bm{q} \times \bm{H}_0&=-\omega \bm{D}_0, \label{eq:Maxwell2}
    \end{align}
\end{subequations}
while the linear constitutive relations are given by
\begin{subequations} \label{eq:constitutive relation}
\begin{align}
    D_{0i}&=\varepsilon_0\varepsilon_{ij}(\bm{q},\omega)E_{0j}, \label{eq:constitutive relation1} \\
    \bm{H}_0&=\frac{1}{\mu_0}\bm{B}_0. \label{eq:constitutive relation2} 
\end{align}
\end{subequations}
The notation used in these equations is as follows.
The quantities $\bm{E}_0$, $\bm{B}_0$, $\bm{D}_0$, and $\bm{H}_0$ are the complex amplitudes of 
the electric field $\bm{E}$, magnetic flux density $\bm{B}$, electric displacement field $\bm{D}$, and magnetic field intensity $\bm{H}$, respectively.
The constants $\varepsilon_0$ and $\mu_0$ are the vacuum permittivity and permeability, respectively, while
$\varepsilon_{ij}(\bm{q},\omega)$ is the complex relative permittivity tensor characterizing linear optical responses of the medium.
Latin indices label Cartesian components, and the Einstein summation convention is used for repeated indices.

We next combine Maxwell's equations with the linear constitutive relations 
to derive the eigenvalue equation describing optical activity.
Specifically, eliminating $\bm{B}_0$ and $\bm{H}_0$ from Eq.~\eqref{eq:Maxwell} using Eq.~\eqref{eq:constitutive relation2}, we obtain
\begin{align}
    \bm{D}_0=\frac{1}{\mu_0 \omega^2} \bigl[(\bm{q} \cdot \bm{q})\bm{E}_0-(\bm{q}\cdot \bm{E}_0) \bm{q} \bigr], \label{eq:D0andH0toD0}
\end{align}
which, together with Eq.~\eqref{eq:constitutive relation1}, yields the eigenvalue equation for $\bm{E}_0$.
However, we instead formulate the eigenvalue equation for $\bm{D}_0$ because $\bm{D}_0$ always satisfies the transversality condition $\bm{q} \cdot \bm{D}_0=0$,
eliminating the need to consider its component along $\bm{q}$.
This advantage can be fully exploited by aligning one of the coordinate axes with $\bm{q}$.
We therefore choose a Cartesian coordinate system ($x,y,z$) with the $z$ axis along $\bm{q}$,
so that $\bm{D}_0$ lies entirely in the $x$-$y$ plane.
In this coordinate system, the eigenvalue equation for $\bm{D}_0$ is obtained by eliminating $\bm{E}_0$ from Eq.~\eqref{eq:D0andH0toD0} using Eq.~\eqref{eq:constitutive relation1}:
\begin{align}
    \Bigl[\frac{1}{n^2}\delta_{\alpha \beta}-{\varepsilon^{-1}}_{\alpha \beta}(\bm{q},\omega)\Bigr]D_{0\beta}=0. \label{eq:eigenequation_general}
\end{align}
Here, $\alpha,\beta$ run over $x$ and $y$, $n$ is the refractive index satisfying $n^2=(\bm{q} \cdot \bm{q})c^2/\omega^2$, where $c=1/\sqrt{\varepsilon_0 \mu_0}$ is the speed of light, and 
$\varepsilon^{-1}$ is the inverse matrix of $\varepsilon$, satisfying
\begin{align}
    \varepsilon_0 E_{0i}= {\varepsilon^{-1}}_{ij}(\bm{q},\omega)D_{0j}. \label{eq:constitutive relation3}
\end{align}

The eigenvalue equation in Eq.~\eqref{eq:eigenequation_general} shows that optical activity is governed by the permittivity tensor $\varepsilon_{ij}(\bm{q},\omega)$,
which determines the refractive indices $n$ and eigenvectors $(D_{0x},D_{0y})$ of the optical eigenmodes.
Consequently, different symmetry-allowed components of the permittivity tensor can give rise to different contributions to optical activity.
We therefore examine the symmetry of the permittivity tensor and identify the components corresponding to octupolar MOA.

%%%%%%%%%%%%%%%%%%%%%%%%%%%%%%%%%%%%%%%%%%%%%%%%%%%%%%%%%%%%%%%%%%

\subsection{Symmetry of the permittivity tensor} \label{Subsec:symmetry of optical conductivity tensor}

The components of the permittivity tensor $\varepsilon_{ij}(\bm{q},\omega)$ corresponding to octupolar MOA are identified through the following three-step procedure:
(i) $\bm{q}$ expansion, (ii) $\mathcal{T}$ decomposition, and (iii) tensor-symmetry analysis.
Before describing each step in detail, we briefly review the definition of magnetic multipoles.
Magnetic multipoles are obtained from the spatial gradient expansion of the magnetic field $\bm{B}(\bm{r})$ 
in the interaction energy with the magnetization density $\bm{m}(\bm{r})$, $E_{\mathrm{int}}=-\int m_\mu(\bm{r})B_\mu(\bm{r}) d\bm{r}$ \cite{Bhowal2024-vj,Urru2022-ww}:
\begin{align}
    E_{\mathrm{int}}&=-\int m_\mu(\bm{r}) B_\mu(\bm{0}) d\bm{r} -\int m_\mu (\bm{r}) r_k \partial_{r_k}B_\mu(\bm{0}) d\bm{r} \notag \\
    &\quad - \frac{1}{2}\int m_\mu(\bm{r})r_kr_l \partial_{r_k}\partial_{r_l}B_\mu(\bm{0}) d\bm{r}+\cdots,
\end{align}
where the Greek index $\mu$ also labels a Cartesian component and $\partial_{r_k}=\partial/\partial r_k$.
The first, second, and third terms correspond to the magnetic dipole $M_\mu=\int m_\mu d\bm{r}$, magnetic quadrupole $M_{\mu,k}=\int m_\mu r_kd\bm{r}$, and magnetic octupole $M_{\mu,kl}=\int m_\mu r_kr_l d\bm{r}$ ($M_{\mu,kl}=M_{\mu,lk}$), respectively,
with higher-order terms representing higher-rank magnetic multipoles $M_{\mu,kl\ldots}=\int m_\mu r_kr_l \cdots d\bm{r}$.
The ferroic ordering of these magnetic multipoles is referred to as magnetic multipole order, and we hereafter assume such ordering and use the terms multipole and multipole order interchangeably.
Having reviewed magnetic multipoles, we now describe the three steps and explain why each is required.

The first step is the $\bm{q}$ expansion of $\varepsilon_{ij}(\bm{q},\omega)$, which is required because the magnetic octupole $M_{\mu,kl}$ is a rank-3 tensor, whereas $\varepsilon_{ij}(\bm{0},\omega)$ is a rank-2 tensor.
This expansion is justified in the long-wavelength limit, where the wavelength $\lambda$ is much larger than the lattice constant $a$, i.e., $qa \sim a/\lambda \ll 1$.
In practice, this condition is well satisfied for most materials in the visible-light range; for example, taking $\lambda=500~$nm and $a=5~\text{\AA}$ gives $a/\lambda \sim 10^{-3}$.
Accordingly, $\varepsilon_{ij}(\bm{q},\omega)$ can be expanded as 
\begin{align}
    \varepsilon_{ij}(\bm{q},\omega)=\varepsilon_{ij}(\omega)+\varepsilon_{ij,k}(\omega)q_k+\varepsilon_{ij,kl}(\omega)q_kq_l+\cdots, \label{eq:qexpansion_permittivity}
\end{align}
where the expansion coefficients are defined by
\begin{subequations} \label{eq:qexpansion_coefficients}
    \begin{align}
        \varepsilon_{ij}(\omega)&=\varepsilon_{ij}(\bm{0},\omega), \\
        \varepsilon_{ij,k}(\omega)&=\Bigl(\frac{\partial \varepsilon_{ij}(\bm{q},\omega)}{\partial q_k}\Bigr)_{\bm{q} \to 0}, \\
        \varepsilon_{ij,kl}(\omega)&=\frac{1}{2}\Bigl(\frac{\partial^2 \varepsilon_{ij}(\bm{q},\omega)}{\partial q_k \partial q_l}\Bigr)_{\bm{q} \to 0}.
    \end{align} 
\end{subequations}
Here, $\varepsilon_{ij,kl}=\varepsilon_{ij,lk}$, and this symmetry is used implicitly hereafter.

The second step is the decomposition of $\varepsilon_{ij}(\bm{q},\omega)$ into the $\mathcal{T}$-even and $\mathcal{T}$-odd components, which is required 
because $M_{\mu,kl}$ is $\mathcal{T}$-odd, whereas $\varepsilon_{ij}(\bm{q},\omega)$ generally contains both $\mathcal{T}$-even and $\mathcal{T}$-odd components.
This decomposition is achieved using the Onsager reciprocal relation~\cite{Onsager1931-dc,Onsager1931-wc,Casimir1945-ve}:
\begin{align}
    \varepsilon_{ij}(\bm{q},\omega,M)=\varepsilon_{ji}(-\bm{q},\omega,-M), \label{eq:Onsager_general}
\end{align}
where $M$ is a $\mathcal{T}$-odd quantity, such as magnetic multipoles.
Applying this Onsager reciprocal relation to the $\bm{q}$ expansion in Eq.~\eqref{eq:qexpansion_permittivity} yields
\begin{subequations}
    \begin{align}
        \varepsilon_{ij}(\omega,M)&= \varepsilon_{ji}(\omega,-M), \label{eq:Onsager_0th} \\
        \varepsilon_{ij,k}(\omega,M)&=-\varepsilon_{ji,k}(\omega,-M), \\
        \varepsilon_{ij,kl}(\omega,M)&=\varepsilon_{ji,kl}(\omega,-M), 
    \end{align}
\end{subequations}
which lead to the decomposition
\begin{subequations} \label{eq:qexpansion_even_odd} 
    \begin{align}
        \varepsilon^{(\mathrm{even})}_{ij}(\bm{q},\omega)&=\varepsilon^{(\mathrm{S})}_{ij}(\omega)+\varepsilon^{(\mathrm{A})}_{ij,k}(\omega)q_k+\varepsilon^{(\mathrm{S})}_{ij,kl}(\omega)q_kq_l+\cdots, \label{eq:qexpansion_even} \\
        \varepsilon^{(\mathrm{odd})}_{ij}(\bm{q},\omega)&=\varepsilon^{(\mathrm{A})}_{ij}(\omega)+\varepsilon^{(\mathrm{S})}_{ij,k}(\omega)q_k+\varepsilon^{(\mathrm{A})}_{ij,kl}(\omega)q_kq_l+\cdots. \label{eq:qexpansion_odd}
    \end{align}
\end{subequations}
Here, the superscripts (even) and (odd) denote the $\mathcal{T}$-even and $\mathcal{T}$-odd components, respectively, 
while (S) and (A) denote the symmetric and antisymmetric parts under the interchange of the indices $i$ and $j$, respectively.

The last step is the tensor-symmetry analysis of $\varepsilon_{ij}(\bm{q},\omega)$, which is required because $M_{\mu,kl}$ is an axial tensor, whereas $\varepsilon_{ij}(\bm{q},\omega)$ is a polar tensor.
To establish the correspondence between them, $\varepsilon_{ij}(\bm{q},\omega)$ must be contracted with the Levi-Civita symbol $e_{ijk}$.
Applying this contraction to Eq.~\eqref{eq:qexpansion_odd} establishes the following correspondence:
\begin{subequations} \label{eq:Multipole tensor_conductivity tensor}
    \begin{align} 
        M_\mu &\leftrightarrow e_{\mu ij}\varepsilon^{(\mathrm{A})}_{ij}, \label{eq:Multipole tensor_conductivity tensor1} \\
        M_{\mu,k} &\leftrightarrow e_{\mu ij}\varepsilon^{(\mathrm{S})}_{ki,j}, \label{eq:Multipole tensor_conductivity tensor2} \\
        M_{\mu,kl} &\leftrightarrow e_{\mu ij}\varepsilon^{(\mathrm{A})}_{ij,kl}, \label{eq:Multipole tensor_conductivity tensor3}
    \end{align}
\end{subequations}
where the quantities on both sides of each correspondence share the same tensor rank, $\mathcal{T}$-odd character, and axial character. 
This correspondence identifies $\varepsilon^{(\mathrm{A})}_{ij}$, $\varepsilon^{(\mathrm{S})}_{ij,k}$, and $\varepsilon^{(\mathrm{A})}_{ij,kl}$
as the coefficients describing MOA induced by magnetic dipoles, quadrupoles,\footnote{MOA induced by magnetic quadrupoles is related to optical magnetoelectric effects in noncentrosymmetric magnets, such as gyrotropic birefringence and directional dichroism \cite{Arima2008-xq,Malashevich2010-il}. } 
and octupoles, respectively.
Accordingly, we focus on $\varepsilon^{(\mathrm{A})}_{ij,kl}$ to discuss octupolar MOA.

However, we also need to consider the $\mathcal{T}$-even component in Eq.~\eqref{eq:qexpansion_even}
because it can induce optical activity, which may coexist with MOA.
It is therefore essential to establish a clear distinction between these two types of optical activity.
To this end, we first consider the zeroth-order term $\varepsilon_{ij}(\omega)$, which provides the simplest setting for illustrating this distinction.
Having established this distinction, we then proceed to the second-order term $\varepsilon_{ij,kl}(\omega)q_kq_l$ and analyze octupolar MOA.

Before proceeding, we briefly discuss the $\mathcal{T}$-even component.
This component can include contributions that depend on magnetic order and those that do not.
Of these contributions, those independent of magnetic order are typically dominant, and we therefore focus below on the optical activity arising from them,
which we refer to as nonmagnetic optical activity (NMOA).

%%%%%%%%%%%%%%%%%%%%%%%%%%%%%%%%%%%%%%%%%%%%%%%%%%%%%%%%%%%%%%%%%%

\subsection{Optical activity induced by the zeroth-order term $\varepsilon_{ij}(\omega)$} \label{Subsec:theory of MD-induced magneto-optical effects}

To analyze the optical activity induced by the zeroth-order term $\varepsilon_{ij}(\omega)$,
we solve the eigenvalue equation in Eq.~\eqref{eq:eigenequation_general} using the inverse permittivity tensor ${\varepsilon^{-1}}_{ij}(\omega)$.
To this end, we decompose ${\varepsilon^{-1}}_{ij}(\omega)$ into its symmetric and antisymmetric parts as
\begin{align}
    {\varepsilon^{-1}}_{ij}(\omega)&=\eta^{(\mathrm{S})}_{ij}(\omega)+e_{\mu ij}G^{(\mathrm{d})}_\mu(\omega). \label{eq:epsilon_inverse_MD}
\end{align}
Here, the symmetric tensor $\eta^{(\mathrm{S})}_{ij}$ is $\mathcal{T}$-even and can give rise to NMOA,
whereas the axial vector $\bm{G}^{(\mathrm{d})}$ representing the antisymmetric part is $\mathcal{T}$-odd and describes dipolar MOA, with the superscript (d) denoting its magnetic-dipole origin.
With this decomposition, the eigenvalue equation is written as 
\begin{align}
    \begin{pmatrix}
    1/n^2-\eta^{(\mathrm{S})}_{xx} & -(\eta^{(\mathrm{S})}_{xy}+G^{(\mathrm{d})}_z) \\
    -(\eta^{(\mathrm{S})}_{xy}-G^{(\mathrm{d})}_z) & 1/n^2-\eta^{(\mathrm{S})}_{yy}
    \end{pmatrix}
    \begin{pmatrix}
    D_{0x} \\ D_{0y}
    \end{pmatrix}=\bm{0}, \label{eq:eigenequation_MD}
\end{align}
from which the eigenvalues and eigenvectors are obtained as
\begin{subequations} \label{eq:eigenvalue_eigenvector_MD}
    \begin{align}
        \frac{1}{n^2}&=\frac{1}{2} \bigl[\eta^{(\mathrm{S})}_{x+y} \pm F(\eta^{(\mathrm{S})}_{x-y},\eta^{(\mathrm{S})}_{xy},G^{(\mathrm{d})}_z) \bigr], \label{eq:eigenvalue_MD} \\
        \frac{D_{0y}}{D_{0x}}&=\frac{-1}{2(\eta^{(\mathrm{S})}_{xy}+G^{(\mathrm{d})}_z)}\bigl[\eta^{(\mathrm{S})}_{x-y} \mp F(\eta^{(\mathrm{S})}_{x-y},\eta^{(\mathrm{S})}_{xy},G^{(\mathrm{d})}_z) \bigr], \label{eq:eigenvector_MD} 
    \end{align}
\end{subequations}
where $\eta^{(\mathrm{S})}_{x \pm y}=\eta^{(\mathrm{S})}_{xx}\pm\eta^{(\mathrm{S})}_{yy}$, $F(A,B,C)=\sqrt{A^2+4(B^2-C^2)}$, and 
the argument $\omega$ is omitted hereafter unless otherwise noted.
Equation~\eqref{eq:eigenvalue_eigenvector_MD} enables us to analyze the optical activity,
for which we first examine $\eta^{(\mathrm{S})}_{ij}$ and $\bm{G}^{(\mathrm{d})}$ in detail.

Here, we express $\eta^{(\mathrm{S})}_{ij}$ and $\bm{G}^{(\mathrm{d})}$ in terms of the permittivity tensor $\varepsilon_{ij}(\omega)$
by decomposing $\varepsilon_{ij}(\omega)$ into its symmetric and antisymmetric parts as
\begin{align}
    \varepsilon_{ij}(\omega)&=\varepsilon^{(\mathrm{S})}_{ij}(\omega)+e_{\mu ij}g^{(\mathrm{d})}_\mu(\omega), \label{eq:epsilon_normal_MD} 
\end{align}
where $\bm g^{(\mathrm{d})}$ is the axial vector representing the antisymmetric part.
With this decomposition, $\eta^{(\mathrm{S})}_{ij}$ and $\bm{G}^{(\mathrm{d})}$ are expressed as \cite{Landau-Lifshitz}
\begin{subequations} \label{eq:eta_G_relation_MD}
    \begin{align}
        \eta^{(\mathrm{S})}_{ij}&=\frac{1}{|\varepsilon^{(\mathrm{S})}|+\varepsilon^{(\mathrm{S})}_{kl}g^{(\mathrm{d})}_kg^{(\mathrm{d})}_l} 
        \bigl[ {(\varepsilon^{(\mathrm{S})})^{-1}}_{ij}|\varepsilon^{(\mathrm{S})}|+g^{(\mathrm{d})}_ig^{(\mathrm{d})}_j\bigr], \label{eq:eta_relation_MD} \\
        G^{(\mathrm{d})}_\mu&=\frac{-1}{|\varepsilon^{(\mathrm{S})}|+\varepsilon^{(\mathrm{S})}_{kl}g^{(\mathrm{d})}_kg^{(\mathrm{d})}_l}\varepsilon^{(\mathrm{S})}_{\mu \nu}g^{(\mathrm{d})}_\nu, \label{eq:G_relation_MD}
    \end{align}
\end{subequations}
where $|\cdot|$ denotes the determinant of its argument.
These expressions can be further simplified by assuming that the magnetic contribution $\bm{g}^{(\mathrm{d})}$ is sufficiently small, i.e., $\bm{g}^{(\mathrm{d})} \ll 1$, and 
neglecting terms beyond first order in $\bm{g}^{(\mathrm{d})}$:
\begin{align}
    \eta^{(\mathrm{S})}_{ij}&\simeq {(\varepsilon^{(\mathrm{S})})^{-1}}_{ij}, \quad G^{(\mathrm{d})}_\mu \simeq \frac{-1}{|\varepsilon^{(\mathrm{S})}|}\varepsilon^{(\mathrm{S})}_{\mu \nu}g^{(\mathrm{d})}_\nu. \label{eq:eta_G_relation_MD_approx}
\end{align}

Equation~\eqref{eq:eta_G_relation_MD_approx} shows that $\eta^{(\mathrm{S})}_{ij}$ is determined by $\varepsilon^{(\mathrm{S})}_{ij}$,
allowing us to classify the possible forms of $\eta^{(\mathrm{S})}_{ij}$ into the following three cases:
\begin{subequations} \label{eq:Symmetric_eta_three}
\begin{alignat}{2} 
    \mkern-20mu\mathrm{(i)}\ & \text{Isotropic case:} \notag \\
    &\varepsilon^{(\mathrm{S})}_{xx}=\varepsilon^{(\mathrm{S})}_{yy},\ \varepsilon^{(\mathrm{S})}_{i \neq j}=0 &\ \Rightarrow 
    \eta^{(\mathrm{S})}_{xx}=\eta^{(\mathrm{S})}_{yy},\ \eta^{(\mathrm{S})}_{xy}=0, \label{eq:Symmetric_eta_1} \\
    \mkern-20mu\mathrm{(ii)}\ & \rlap{\text{Anisotropic case (without off-diagonal components):}} \notag \\ 
    &\varepsilon^{(\mathrm{S})}_{xx} \neq \varepsilon^{(\mathrm{S})}_{yy},\ \varepsilon^{(\mathrm{S})}_{i \neq j}=0 &\ \Rightarrow 
    \eta^{(\mathrm{S})}_{xx} \neq \eta^{(\mathrm{S})}_{yy},\ \eta^{(\mathrm{S})}_{xy}=0, \label{eq:Symmetric_eta_2} \\
    \mkern-20mu\mathrm{(iii)}\ & \rlap{\text{Anisotropic case (with off-diagonal components):}} \notag \\
    &\varepsilon^{(\mathrm{S})}_{xx} \neq \varepsilon^{(\mathrm{S})}_{yy},\ \varepsilon^{(\mathrm{S})}_{xy} \neq 0 &\ \Rightarrow 
    \eta^{(\mathrm{S})}_{xx} \neq \eta^{(\mathrm{S})}_{yy},\ \eta^{(\mathrm{S})}_{xy} \neq 0, \label{eq:Symmetric_eta_3}
\end{alignat}
\end{subequations}
where $\varepsilon^{(\mathrm{S})}_{i \neq j}$ denotes the off-diagonal components of $\varepsilon^{(\mathrm{S})}_{ij}$.
These three cases correspond to the symmetry-allowed forms of $\varepsilon^{(\mathrm{S})}_{ij}$ for different crystal systems, with the conventional choice of principal rotation axis.
Specifically, case (i) applies to tetragonal, trigonal, hexagonal, and cubic crystal systems; case (ii) to orthorhombic crystal systems; and case (iii) to triclinic and monoclinic crystal systems \cite{Agranovich1984}. 
Based on this classification, we now analyze the optical activity for each case.

%%%%%%%%%%%%%%%%%%%%%%%%%%%%%%%%%

\subsubsection{Isotropic case}

First, we consider case (i) $\eta^{(\mathrm{S})}_{xx}=\eta^{(\mathrm{S})}_{yy}=1/\varepsilon^{(0)}$ and $\eta^{(\mathrm{S})}_{xy}=0$,
where Eq.~\eqref{eq:eigenvalue_eigenvector_MD} reduces to
\begin{align}
    n_{\pm} \simeq \sqrt{\varepsilon^{(0)}} \Bigl(1 \mp \frac{i \varepsilon^{(0)} G^{(\mathrm{d})}_z}{2} \Bigr), \quad \frac{D_{0y}}{D_{0x}}=\pm i. \label{eq:eigenvalue_MD_MD}
\end{align}
The eigenmodes corresponding to $n_+$ and $n_-$ are left- and right-circularly polarized modes, respectively.
These two eigenmodes are degenerate when $G^{(\mathrm{d})}_z=0$, 
whereas a nonzero $G^{(\mathrm{d})}_z$ lifts this degeneracy ($n_+ \neq n_-$), leading to dipolar MOA without NMOA.
Here, we illustrate how $n_+ \neq n_-$ gives rise to dipolar MOA, using transmitted light as an example.
First, we consider linearly polarized light incident on the medium.
This light is decomposed into the two eigenmodes in the medium.
During propagation, these eigenmodes acquire a relative phase difference and experience different attenuation because $n_+ \neq n_-$,
leading to polarization rotation and ellipticity in transmitted light.
The resulting optical activity is known as the Faraday effect \cite{Faraday1846-bb}, while its counterpart in reflected light is known as the Kerr effect \cite{Kerr1877-ap}.

%%%%%%%%%%%%%%%%%%%%%%%%%%%%%%%%%

\subsubsection{Anisotropic case}

Second, we consider case (ii) $\eta^{(\mathrm{S})}_{xx} \neq \eta^{(\mathrm{S})}_{yy}$ and $\eta^{(\mathrm{S})}_{xy}=0$, 
and case (iii) $\eta^{(\mathrm{S})}_{xx} \neq \eta^{(\mathrm{S})}_{yy}$ and $\eta^{(\mathrm{S})}_{xy}\neq0$ together.
Here, we set $G^{(\mathrm{d})}_z=0$ to eliminate dipolar MOA and examine whether these cases give rise to NMOA.
Under this condition, Eq.~\eqref{eq:eigenvalue_eigenvector_MD} reduces, for case (ii), to 
\begin{subequations} \label{eq:eigenvalue_MD_LB_diag}
    \begin{align}
        n_x&=\sqrt{1/\eta^{(\mathrm{S})}_{xx}}, \quad D_{0y}=0, \label{eq:eigenvalue_MD_LB1_diag} \\
        n_y&=\sqrt{1/\eta^{(\mathrm{S})}_{yy}}, \quad D_{0x}=0, \label{eq:eigenvalue_MD_LB2_diag}
    \end{align}
\end{subequations}
and, for case (iii), to 
\begin{subequations} \label{eq:eigenvalue_MD_LB_offdiag}
    \begin{align}
        \frac{1}{n_{xy}^2}&=\frac{1}{2} \bigl[\eta^{(\mathrm{S})}_{x+y} \pm F(\eta^{(\mathrm{S})}_{x-y},\eta^{(\mathrm{S})}_{xy},0)\bigr], \label{eq:eigenvalue_MD_LB1_offdiag} \\
        \frac{D_{0y}}{D_{0x}}&=\frac{-1}{2\eta^{(\mathrm{S})}_{xy}}\bigl[\eta^{(\mathrm{S})}_{x-y} \mp F(\eta^{(\mathrm{S})}_{x-y},\eta^{(\mathrm{S})}_{xy},0) \bigr]. \label{eq:eigenvalue_MD_LB2_offdiag}
    \end{align}
\end{subequations}
In both cases, the eigenmodes corresponding to $n_{x/y}$ and $n_{xy}$ are linearly polarized along the eigenmode principal axes.
The two eigenmodes in each case generally have different refractive indices, giving rise to NMOA through the same mechanism as dipolar MOA.
Since this NMOA originates from the uniform permittivity tensor $\varepsilon^{(\mathrm{S})}_{ij}(\omega)$, we refer to it as uniform NMOA.
In case (ii), for example, uniform NMOA is realized through the refractive-index anisotropy ($n_x \neq n_y$) arising from $\eta^{(\mathrm{S})}_{xx} \neq \eta^{(\mathrm{S})}_{yy}$.

However, unlike dipolar MOA, uniform NMOA does not necessarily arise for every incident linear polarization direction.
Specifically, uniform NMOA vanishes when the linear polarization direction is aligned with an eigenmode principal axis, because the light excites only one of the two eigenmodes.
Thus, the linear polarization directions that allow uniform NMOA depend on the eigenmode principal axes, whose orientations differ between cases (ii) and (iii).
In case (ii), the eigenmode principal axes coincide with the coordinate axes, whereas in case (iii), they are rotated away from the coordinate axes due to a nonzero $\eta^{(\mathrm{S})}_{xy}$.
As a result, for certain linear polarization directions, uniform NMOA can emerge in one case but not in the other.
This difference becomes crucial in the discussion of strategies for avoiding NMOA in Sec.~\ref{Subsec:short summary of theory}.

Finally, we remark that refractive-index anisotropy can also be induced by magnetic order.
This anisotropy arises from the magnetic-order-dependent contribution to the $\mathcal{T}$-even component $\varepsilon^{(\mathrm{S})}_{ij}(\omega)$
and gives rise to optical activity, such as the Voigt (or Cotton--Mouton) effect \cite{GDZPPN002499460,CottonMouton1905,CottonMouton1905combined}.
This type of optical activity has recently been proposed as a powerful probe of altermagnetism with the aid of external magnetic fields \cite{Sunko2026-ty}; 
a brief overview is presented in Appendix~\ref{Sec:AppA}.
By contrast, octupolar MOA discussed below does not require external magnetic fields, offering a field-free optical route to probing altermagnetism
and, more broadly, magnetic octupole order.

%%%%%%%%%%%%%%%%%%%%%%%%%%%%%%%%%%%%%%%%%%%%%%%%%%%%%%%%%%%%%%%%%%

\subsection{Optical activity induced by the second-order term $\varepsilon_{ij,kl}(\omega)q_kq_l$} \label{Subsec:theory of MO-induced magneto-optical effects}

We next extend the discussion to the second-order term $\varepsilon_{ij,kl}(\omega)q_kq_l$ and analyze octupolar MOA.
Here, we provide an overview of the analysis that follows.
We first express the inverse permittivity tensor ${\varepsilon^{-1}}_{ij}(\bm{q},\omega)$ in terms of the permittivity tensor $\varepsilon_{ij}(\bm{q},\omega)$ 
and then solve the eigenvalue equation using the resulting expressions.
In this process, however, the lower-order terms $\varepsilon_{ij}(\omega)$ and $\varepsilon_{ij,k}(\omega)q_k$ also need to be included.
We therefore make the following two assumptions to simplify the analysis and focus on octupolar MOA:
(i) magnetic octupole order is present while magnetic dipole order is symmetry-forbidden, and (ii) inversion symmetry is preserved.
These assumptions are naturally satisfied in antiferromagnets such as those shown in Fig.~\ref{fig:General_concept2}
and lead to $\varepsilon^{(\mathrm{A})}_{ij}(\omega)=0$ and $\varepsilon_{ij,k}(\omega)=0$ under assumptions (i) and (ii), respectively.
However, $\varepsilon^{(\mathrm{S})}_{ij}(\omega)$ remains, so that the analysis proceeds along the lines of the three cases summarized in Eq.~\eqref{eq:Symmetric_eta_three}.
Within this framework, we clarify how octupolar MOA described by $\varepsilon^{(\mathrm{A})}_{ij,kl}(\omega)q_kq_l$ manifests itself in each case.

To express ${\varepsilon^{-1}}_{ij}(\bm{q},\omega)$ in terms of $\varepsilon_{ij}(\bm{q},\omega)$,
we first decompose ${\varepsilon^{-1}}_{ij}(\bm{q},\omega)$ into its symmetric and antisymmetric parts as 
\begin{align}
    {\varepsilon^{-1}}_{ij}(\bm{q},\omega)&=K^{(\mathrm{S})}_{ij}(\bm{q},\omega)+e_{\mu ij}G^{(\mathrm{o})}_\mu(\bm{q},\omega), \label{eq:epsilon_inverse_MO}
\end{align}
where the symmetric tensor $K^{(\mathrm{S})}_{ij}$ is $\mathcal{T}$-even and can give rise to NMOA,
whereas the axial vector $\bm{G}^{(\mathrm{o})}$ representing the antisymmetric part is $\mathcal{T}$-odd and describes octupolar MOA, with the superscript (o) denoting its magnetic-octupole origin.
Similarly, $\varepsilon_{ij}(\bm{q},\omega)$ can be decomposed in the same form as 
\begin{align}
    \varepsilon_{ij}(\bm{q},\omega)&=\kappa^{(\mathrm{S})}_{ij}(\bm{q},\omega)+n^2e_{\mu ij}\tilde{g}^{(\mathrm{o})}_\mu(\bm{q},\omega). \label{eq:epsilon_normal_MO} 
\end{align}
Here, the symmetric tensor $\kappa^{(\mathrm{S})}_{ij}$ and the axial vector $\tilde{\bm{g}}^{(\mathrm{o})}$ representing the antisymmetric part are given by
\begin{subequations} \label{eq:epsilon_normal_MO_coefficients_normalized}
    \begin{align}
        \kappa^{(\mathrm{S})}_{ij}(\bm{q},\omega) &=\varepsilon^{(\mathrm{S})}_{ij}(\omega)+n^2\tilde{g}^{(\mathrm{S})}_{ij}(\bm{q},\omega), \\
        e_{\mu ij}\tilde{g}^{(\mathrm{o})}_\mu(\bm{q},\omega) &=(\omega/c)^2\varepsilon^{(\mathrm{A})}_{ij,kl}(\omega)\tilde{q}_k\tilde{q}_l, \label{eq:epsilon_normal_MO_coefficients_normalized_MO}
    \end{align}
\end{subequations}
where we introduce the normalized vector $\tilde{\bm{q}}=\bm{q}/\sqrt{\bm{q} \cdot \bm{q}}$, with quantities expressed in terms of $\tilde{\bm{q}}$ denoted by a tilde.
With this notation, $\tilde{g}^{(\mathrm{S})}_{ij}$ is similarly expressed as 
\begin{align}
    \tilde{g}^{(\mathrm{S})}_{ij}(\bm{q},\omega)=(\omega/c)^2\varepsilon^{(\mathrm{S})}_{ij,kl}(\omega)\tilde{q}_k\tilde{q}_l. \label{eq:epsilon_normal_MO_coefficients_normalized_OA}
\end{align}

We next express $K^{(\mathrm{S})}_{ij}$ and $\bm{G}^{(\mathrm{o})}$ in terms of $\varepsilon_{ij}(\bm{q},\omega)$
by assuming $\tilde{g}^{(\mathrm{S})}_{ij} \ll 1$ and $\tilde{\bm{g}}^{(\mathrm{o})} \ll 1$.
These approximations are justified by the smallness of the second-order term $\varepsilon_{ij,kl}(\omega)q_kq_l$ and 
are applied sequentially, retaining terms up to first order in $\tilde{g}^{(\mathrm{S})}_{ij}$ and $\tilde{\bm{g}}^{(\mathrm{o})}$.
First, applying $\tilde{\bm{g}}^{(\mathrm{o})} \ll 1$ and following the same procedure as for Eq.~\eqref{eq:eta_G_relation_MD_approx}, we obtain
\begin{align}
    K^{(\mathrm{S})}_{ij}& \simeq {(\kappa^{(\mathrm{S})})^{-1}}_{ij}, \quad 
    \tilde{G}^{(\mathrm{o})}_\mu \simeq \frac{-1}{|\kappa^{(\mathrm{S})}|}\kappa^{(\mathrm{S})}_{\mu \nu} \tilde{g}^{(\mathrm{o})}_\nu, \label{eq:eta_G_relation_MO_approx}
\end{align}
where $\bm{G}^{(\mathrm{o})}=n^2 \tilde{\bm{G}}^{(\mathrm{o})}$, and the arguments $\bm{q}$ and $\omega$ are omitted hereafter unless otherwise noted.
Next, applying $\tilde{g}^{(\mathrm{S})}_{ij} \ll 1$, we further simplify Eq.~\eqref{eq:eta_G_relation_MO_approx} to 
\begin{subequations} \label{eq:epsilon_inverse_MO_coefficients}
    \begin{align}
        K^{(\mathrm{S})}_{ii} &\simeq \eta^{(\mathrm{S})}_{ii}(1-n^2\tilde{\eta}^{(\mathrm{S})}_{ii}), \label{eq:epsilon_inverse_MO_coefficients_diag} \\
        K^{(\mathrm{S})}_{ij} &\simeq \eta^{(\mathrm{S})}_{ij}(1-n^2\tilde{\eta}_0)+n^2\tilde{G}^{(\mathrm{S})}_{ij}, \quad (i \neq j) \label{eq:epsilon_inverse_MO_coefficients_offdiag} \\
        \tilde{G}^{(\mathrm{o})}_\mu & \simeq \frac{-1}{|\varepsilon^{(\mathrm{S})}|} \varepsilon^{(\mathrm{S})}_{\mu \nu}\tilde{g}^{(\mathrm{o})}_\nu, \label{eq:epsilon_inverse_MO_coefficients_MO}
    \end{align}
\end{subequations}
where the first and second expressions correspond to the diagonal and off-diagonal components, respectively.
Here, $\tilde{\eta}_0$, $\tilde{\eta}^{(\mathrm{S})}_{ii}$, and $\tilde{G}^{(\mathrm{S})}_{ij}$ are first-order quantities in $\tilde{g}^{(\mathrm{S})}_{ij}$ and are explicitly given by
\begin{subequations}
    \begin{align}
        \tilde{\eta}_0 &=\eta^{(\mathrm{S})}_{xx}\tilde{g}^{(\mathrm{S})}_{xx}+\eta^{(\mathrm{S})}_{yy}\tilde{g}^{(\mathrm{S})}_{yy}+\eta^{(\mathrm{S})}_{zz}\tilde{g}^{(\mathrm{S})}_{zz} \notag \\
        & \quad +2(\eta^{(\mathrm{S})}_{yz}\tilde{g}^{(\mathrm{S})}_{yz}+\eta^{(\mathrm{S})}_{zx}\tilde{g}^{(\mathrm{S})}_{zx}+\eta^{(\mathrm{S})}_{xy}\tilde{g}^{(\mathrm{S})}_{xy}), \label{eq:tilde_eta0_definition} \\
        \tilde{\eta}^{(\mathrm{S})}_{ii} &= \tilde{\eta}_0-\frac{1}{\bar{\varepsilon}^{(\mathrm{S})}_{ii}}
        (\varepsilon^{(\mathrm{S})}_{jj}\tilde{g}^{(\mathrm{S})}_{kk}+\varepsilon^{(\mathrm{S})}_{kk}\tilde{g}^{(\mathrm{S})}_{jj}-2\varepsilon^{(\mathrm{S})}_{jk}\tilde{g}^{(\mathrm{S})}_{jk}), \label{eq:tilde_etaii_definition}  \\
        \tilde{G}^{(\mathrm{S})}_{ij} &= \frac{-1}{|\varepsilon^{(\mathrm{S})}|}(\varepsilon^{(\mathrm{S})}_{kk}\tilde{g}^{(\mathrm{S})}_{ij}+\varepsilon^{(\mathrm{S})}_{ij}\tilde{g}^{(\mathrm{S})}_{kk} \notag \\
        & \qquad \qquad -\varepsilon^{(\mathrm{S})}_{jk}\tilde{g}^{(\mathrm{S})}_{ki}-\varepsilon^{(\mathrm{S})}_{ki}\tilde{g}^{(\mathrm{S})}_{jk}), \quad (i \neq j) \label{eq:Gij_definition} 
    \end{align}
\end{subequations}
where $\bar{\varepsilon}^{(\mathrm{S})}$ denotes the cofactor matrix of $\varepsilon^{(\mathrm{S})}$, 
and $(i,j,k)$ runs over the cyclic permutations of $(x,y,z)$, with no summation over $i$, $j$, or $k$.

We now solve the eigenvalue equation in Eq.~\eqref{eq:eigenequation_general} using the expression for ${\varepsilon^{-1}}_{ij}(\bm{q},\omega)$ [Eq.~\eqref{eq:epsilon_inverse_MO}]
to analyze the optical activity arising from $\varepsilon_{ij,kl}(\omega)q_kq_l$.
The eigenvalue equation then takes the form 
\begin{align}
    \begin{pmatrix}
    1/n^2-K^{(\mathrm{S})}_{xx} & -(K^{(\mathrm{S})}_{xy}+G^{(\mathrm{o})}_z) \\
    -(K^{(\mathrm{S})}_{xy}-G^{(\mathrm{o})}_z) & 1/n^2-K^{(\mathrm{S})}_{yy}
    \end{pmatrix}
    \begin{pmatrix}
    D_{0x} \\ D_{0y}
    \end{pmatrix}=\bm{0}. \label{eq:eigenequation_MO_original}
\end{align}
However, since an exact solution would be cumbersome, we solve this equation perturbatively to first order in $\tilde{g}^{(\mathrm{S})}_{ij}$ and $\tilde{\bm{g}}^{(\mathrm{o})}$.
The results are presented below separately for the three cases summarized in Eq.~\eqref{eq:Symmetric_eta_three}, 
while the detailed derivation is provided in Appendix~\ref{Sec:AppB}.

%%%%%%%%%%%%%%%%%%%%%%%%%%%%%%%%%

\subsubsection{Isotropic case}

First, we consider case (i) $\eta^{(\mathrm{S})}_{xx}=\eta^{(\mathrm{S})}_{yy}=1/\varepsilon^{(0)}$ and $\eta^{(\mathrm{S})}_{xy}=0$,
for which Eq.~\eqref{eq:eigenequation_MO_original} yields the following eigenvalues and eigenvectors:
\begin{subequations} \label{eq:eigenvalue_MO1}
    \begin{align}
        n^2 & \simeq \varepsilon^{(0)} \Bigl\{ 1+\frac{\varepsilon^{(0)}}{2} \notag \\
        & \quad \times \bigl[\tilde{\eta}^{(\mathrm{S})}_{x+y} \pm F(\tilde{\eta}^{(\mathrm{S})}_{x-y},\varepsilon^{(0)}\tilde{G}^{(\mathrm{S})}_{xy},\varepsilon^{(0)}\tilde{G}^{(\mathrm{o})}_z)\bigr] \Bigr\}, \\
        \frac{D_{0y}}{D_{0x}} &\simeq \frac{1}{2\varepsilon^{(0)}(\tilde{G}^{(\mathrm{S})}_{xy}+\tilde{G}^{(\mathrm{o})}_z)} \notag \\
        & \quad \times \bigl[ \tilde{\eta}^{(\mathrm{S})}_{x-y}\mp F(\tilde{\eta}^{(\mathrm{S})}_{x-y},\varepsilon^{(0)}\tilde{G}^{(\mathrm{S})}_{xy},\varepsilon^{(0)}\tilde{G}^{(\mathrm{o})}_z)\bigr], \label{eq:eigenvalue_MO1_vector}
    \end{align}
\end{subequations}
where $\tilde{\eta}^{(\mathrm{S})}_{x \pm y}=\tilde{\eta}^{(\mathrm{S})}_{xx}\pm\tilde{\eta}^{(\mathrm{S})}_{yy}$.
These expressions have the same structure as Eq.~\eqref{eq:eigenvalue_eigenvector_MD}.
Moreover, $\tilde{\eta}^{(\mathrm{S})}_{ij}$ and $\tilde{G}^{(\mathrm{S})}_{xy}$ can be classified into the following three cases, in analogy with Eq.~\eqref{eq:Symmetric_eta_three}:
\begin{subequations} \label{eq:Symmetric_tilde_eta1_three}
\begin{alignat}{2} 
    \mathrm{(i\text{-}a)}\ &\tilde{\eta}^{(\mathrm{S})}_{xx}=\tilde{\eta}^{(\mathrm{S})}_{yy}, &\  \tilde{G}^{(\mathrm{S})}_{xy}=0, \label{eq:Symmetric_tilde_eta1_1} \\
    \mathrm{(i\text{-}b)}\ &\tilde{\eta}^{(\mathrm{S})}_{xx} \neq \tilde{\eta}^{(\mathrm{S})}_{yy}, &\ \tilde{G}^{(\mathrm{S})}_{xy}=0, \label{eq:Symmetric_tilde_eta1_2} \\
    \mathrm{(i\text{-}c)}\ &\tilde{\eta}^{(\mathrm{S})}_{xx} \neq \tilde{\eta}^{(\mathrm{S})}_{yy}, &\ \tilde{G}^{(\mathrm{S})}_{xy}\neq 0. \label{eq:Symmetric_tilde_eta1_3} 
\end{alignat}
\end{subequations}
Taken together, these observations imply that the discussion of optical activity developed in Sec.~\ref{Subsec:theory of MD-induced magneto-optical effects} can be applied directly to Eq.~\eqref{eq:eigenvalue_MO1},
from which it follows that case (i-a) exhibits only MOA, whereas cases (i-b) and (i-c) can exhibit NMOA.
In the following, we explicitly demonstrate these conclusions.

We first consider case (i-a), for which Eq.~\eqref{eq:eigenvalue_MO1} reduces to 
\begin{align} \label{eq:eigenvalue_MO1_MO}
    n_{\pm} \simeq \sqrt{\varepsilon^{(0)}} \Bigl(1+\frac{\varepsilon^{(0)}\tilde{\eta}^{(0)}}{2} \mp \frac{i(\varepsilon^{(0)})^2\tilde{G}^{(\mathrm{o})}_z}{2} \Bigr), \quad \frac{D_{0y}}{D_{0x}} = \pm i,
\end{align}
where we set $\tilde{\eta}^{(\mathrm{S})}_{xx}=\tilde{\eta}^{(\mathrm{S})}_{yy}=\tilde{\eta}^{(0)}$.
This expression has the same structure as Eq.~\eqref{eq:eigenvalue_MD_MD}, 
showing that $\tilde{G}^{(\mathrm{o})}_z$ gives rise to octupolar MOA through the same mechanism as that of dipolar MOA driven by $G^{(\mathrm{d})}_z$ in case (i).
This type of MOA was first pointed out by Graham and Raab in cubic antiferromagnets \cite{Graham1991-xf}, 
although its magnetic origin and generality beyond cubic antiferromagnets have remained unclear.
In this work, we identify its origin as magnetic octupole order and extend it to a broader class of antiferromagnets (see Sec.~\ref{Subsec:MO_effects_compedition}).

We next consider cases (i-b) and (i-c) together, focusing on case (i-b) to demonstrate the presence of NMOA.
To this end, we set $\tilde{G}^{(\mathrm{o})}_z=0$, whereby Eq.~\eqref{eq:eigenvalue_MO1} reduces to 
\begin{subequations} \label{eq:eigenvalue_MO1_LB}
    \begin{align}
        n_x &\simeq \sqrt{\varepsilon^{(0)}}\Bigl(1+\frac{\varepsilon^{(0)}\tilde{\eta}^{(\mathrm{S})}_{xx}}{2}\Bigr), \quad D_{0y}=0, \\
        n_y &\simeq \sqrt{\varepsilon^{(0)}}\Bigl(1+\frac{\varepsilon^{(0)}\tilde{\eta}^{(\mathrm{S})}_{yy}}{2}\Bigr), \quad D_{0x}=0. 
    \end{align}
\end{subequations}
These expressions reveal that NMOA is absent without the $\bm{q}$-dependent term, 
but can arise from $\tilde{\eta}^{(\mathrm{S})}_{xx} \neq \tilde{\eta}^{(\mathrm{S})}_{yy}$ once the $\bm{q}$ dependence is taken into account.
This type of NMOA was first pointed out by Lorentz \cite{Lorentz1878Voortplantingssnelheid}; accordingly, we refer to it as Lorentz NMOA.
Lorentz NMOA was later theoretically attributed to the second-order term $\varepsilon^{(\mathrm{S})}_{ij,kl}(\omega)q_kq_l$ \cite{Condon1932-ay,Graham1990-sc},
and experiments on Si confirmed that its magnitude is of the order of $10^{-6}$ \cite{Pastrnak1971-iv}.
Lorentz NMOA also arises in case (i-c), although the eigenmode principal axes differ from those in case (i-b) owing to a nonzero $\tilde{G}^{(\mathrm{S})}_{xy}$.

Finally, we comment on cases (i-b) and (i-c) in the presence of a finite $\tilde{G}^{(\mathrm{o})}_z$.
In this situation, Lorentz NMOA coexists with octupolar MOA and provides a background to its observation.
However, neither case is realized among the light-propagation directions considered in this work for which $\tilde{G}^{(\mathrm{o})}_z$ is finite (see Sec.~\ref{Subsec:MO_effects_compedition}).
We therefore restrict the following discussion of octupolar MOA within case (i) to case (i-a).

%%%%%%%%%%%%%%%%%%%%%%%%%%%%%%%%%

\subsubsection{Anisotropic case (without off-diagonal components)}

Second, we turn to case (ii) $\eta^{(\mathrm{S})}_{xx} \neq \eta^{(\mathrm{S})}_{yy}$ and $\eta^{(\mathrm{S})}_{xy}=0$, 
for which Eq.~\eqref{eq:eigenequation_MO_original} yields the following eigenvalues and eigenvectors:
\begin{subequations} \label{eq:eigenvalue_MO2_MO}
    \begin{align}
        n_x & \simeq \sqrt{\frac{1}{\eta^{(\mathrm{S})}_{xx}}}\Bigl(1+\frac{\tilde{\eta}^{(\mathrm{S})}_{xx}}{2\eta^{(\mathrm{S})}_{xx}} \Bigr), 
        \quad \frac{D_{0y}}{D_{0x}} \simeq \frac{\tilde{G}^{(\mathrm{S})}_{xy}-\tilde{G}^{(\mathrm{o})}_z}{\eta^{(\mathrm{S})}_{xx}\eta^{(\mathrm{S})}_{x-y}}, \label{eq:eigenvalue_MO2_MO_x} \\
        n_y & \simeq \sqrt{\frac{1}{\eta^{(\mathrm{S})}_{yy}}}\Bigl(1+\frac{\tilde{\eta}^{(\mathrm{S})}_{yy}}{2\eta^{(\mathrm{S})}_{yy}} \Bigr),
        \quad \frac{D_{0x}}{D_{0y}} \simeq -\frac{\tilde{G}^{(\mathrm{S})}_{xy}+\tilde{G}^{(\mathrm{o})}_z}{\eta^{(\mathrm{S})}_{yy}\eta^{(\mathrm{S})}_{x-y}}. \label{eq:eigenvalue_MO2_MO_y}
    \end{align}
\end{subequations}
The eigenvalues exhibit refractive-index anisotropy underlying both uniform NMOA and Lorentz NMOA.
The eigenvectors indicate that the eigenmode principal axes are rotated away from the coordinate axes by $\tilde{G}^{(\mathrm{S})}_{xy}$ and $\tilde{G}^{(\mathrm{o})}_z$.
In particular, $\tilde{G}^{(\mathrm{o})}_z$ enters the eigenvectors but not the eigenvalues at first order, 
suggesting that octupolar MOA arises through a different mechanism from that in case (i-a).

We note, however, that Eq.~\eqref{eq:eigenvalue_MO2_MO} is valid only when 
\begin{align}
    |\eta^{(\mathrm{S})}_{x-y}| \gg \left| \frac{\tilde{G}^{(\mathrm{S})}_{xy}-\tilde{G}^{(\mathrm{o})}_z}{\eta^{(\mathrm{S})}_{xx}} \right|,\ 
    \left|\frac{\tilde{G}^{(\mathrm{S})}_{xy}+\tilde{G}^{(\mathrm{o})}_z}{\eta^{(\mathrm{S})}_{yy}}\right|. \label{eq:eigenequation_MO2_validity}
\end{align}
Otherwise, at least one of the conditions $|D_{0y}/D_{0x}| \ll 1$ and $|D_{0x}/D_{0y}| \ll 1$ is no longer satisfied, so that the corresponding eigenvector cannot be treated perturbatively.
In such cases, a different treatment is required, as discussed in Appendix~\ref{Sec:AppB2}.
In what follows, however, we restrict our discussion to the regime where the condition in Eq.~\eqref{eq:eigenequation_MO2_validity} is satisfied.
In particular, even when considering the regime $|\eta^{(\mathrm{S})}_{x-y}| \ll 1$, we assume that this condition remains satisfied.

%%%%%%%%%%%%%%%%%%%%%%%%%%%%%%%%%

\subsubsection{Anisotropic case (with off-diagonal components)}

Finally, we address case (iii) $\eta^{(\mathrm{S})}_{xx} \neq \eta^{(\mathrm{S})}_{yy}$ and $\eta^{(\mathrm{S})}_{xy} \neq 0$,
for which Eq.~\eqref{eq:eigenequation_MO_original} yields the following eigenvalues and eigenvectors: 
\begin{subequations} \label{eq:eigenvalue_MO3}
    \begin{align}
        n &\simeq n_{xy} \biggl\{1+\frac{n_{xy}^4}{4}\Bigl[(\eta^{(\mathrm{S})}_{xx}\tilde{\eta}^{(\mathrm{S})}_{xx}+\eta^{(\mathrm{S})}_{yy}\tilde{\eta}^{(\mathrm{S})}_{yy}) \notag \\
        & \quad \pm \frac{\eta^{(\mathrm{S})}_{x-y}(\eta^{(\mathrm{S})}_{xx}\tilde{\eta}^{(\mathrm{S})}_{xx}-\eta^{(\mathrm{S})}_{yy}\tilde{\eta}^{(\mathrm{S})}_{yy})
        +4(\eta^{(\mathrm{S})}_{xy})^2\tilde{\eta}^{(\mathrm{S})}_{xy}}{F^{(\mathrm{S})}_{xy}}\Bigr]\biggr\}, \\
        \frac{D_{0y}}{D_{0x}} &\simeq \frac{-1}{2\eta^{(\mathrm{S})}_{xy}}(\eta^{(\mathrm{S})}_{x-y}\mp F^{(\mathrm{S})}_{xy}) \biggl[1 - n^2_{xy} \notag \\
        & \quad \times \Bigl( \frac{\tilde{G}^{(\mathrm{o})}_z}{\eta^{(\mathrm{S})}_{xy}}
        \mp \frac{\eta^{(\mathrm{S})}_{xx}\tilde{\eta}^{(\mathrm{S})}_{xx}-\eta^{(\mathrm{S})}_{yy}\tilde{\eta}^{(\mathrm{S})}_{yy}-\eta^{(\mathrm{S})}_{x-y}\tilde{\eta}^{(\mathrm{S})}_{xy}}{F^{(\mathrm{S})}_{xy}} \Bigr) \biggr],
    \end{align}
\end{subequations}
where $F^{(\mathrm{S})}_{xy}=F(\eta^{(\mathrm{S})}_{x-y},\eta^{(\mathrm{S})}_{xy},0)$, $\tilde{\eta}^{(\mathrm{S})}_{xy}=\tilde{\eta}_0-\tilde{G}^{(\mathrm{S})}_{xy}/\eta^{(\mathrm{S})}_{xy}$,
and $n_{xy}$ denotes the eigenvalues given in Eq.~\eqref{eq:eigenvalue_MD_LB1_offdiag}.
As in Eq.~\eqref{eq:eigenvalue_MO2_MO}, $\tilde{G}^{(\mathrm{o})}_z$ enters the eigenvectors but not the eigenvalues at first order.
The difference, however, is that $\eta^{(\mathrm{S})}_{xy}$ also contributes to the rotation of the eigenmode principal axes, 
leading to more involved expressions.

%%%%%%%%%%%%%%%%%%%%%%%%%%%%%%%%%%%%%%%%%%%%%%%%%%%%%%%%%%%%%%%%%%

\subsection{Short summary} \label{Subsec:short summary of theory}

\begin{table}[t]
    \caption{Classification of optical activity induced by the zeroth-order term $\varepsilon_{ij}(\omega)$ and the second-order term $\varepsilon_{ij,kl}(\omega)q_kq_l$.
    The components responsible for each type are listed beneath the corresponding entry.}
    %\vspace{1mm}
    \label{tab:classification_OA}
    \centering
    \tabcolsep = 1.2mm
    \renewcommand\arraystretch{1.2}
    \begin{tabular}{c||c|c} \hline \hline
        & MOA ($\mathcal{T}$-odd) & NMOA ($\mathcal{T}$-even)  \\ \hline 
        \multirow{2}{*}{$\varepsilon_{ij}(\omega)$} & Dipolar MOA & Uniform NMOA \\ 
        & ($G^{(\mathrm{d})}_z$) & ($\eta^{(\mathrm{S})}_{xx} \neq \eta^{(\mathrm{S})}_{yy}$, $\eta^{(\mathrm{S})}_{xy}$) \\ \hline 
        \multirow{2}{*}{$\varepsilon_{ij,kl}(\omega)q_kq_l$} & Octupolar MOA & Lorentz NMOA  \\
        & ($\tilde{G}^{(\mathrm{o})}_z$) & ($\tilde{\eta}^{(\mathrm{S})}_{xx} \neq \tilde{\eta}^{(\mathrm{S})}_{yy}$, $\tilde{G}^{(\mathrm{S})}_{xy}$)  \\ \hline \hline
    \end{tabular}
\end{table}

In summary, we identified four types of optical activity from the zeroth-order term $\varepsilon_{ij}(\omega)$ and the second-order term $\varepsilon_{ij,kl}(\omega)q_kq_l$:
dipolar MOA, octupolar MOA, uniform NMOA, and Lorentz NMOA, as listed in Table~\ref{tab:classification_OA}.
These four types have distinct origins with respect to $\mathcal{T}$ symmetry and the expansion order.
Specifically, dipolar MOA and octupolar MOA arise from the $\mathcal{T}$-odd components of the zeroth- and second-order terms, respectively, 
and we refer to the corresponding axial vectors $\bm{G}^{(\mathrm{d})}$ and $\tilde{\bm{G}}^{(\mathrm{o})}$ as the dipolar and octupolar MOA vectors.
In contrast, uniform NMOA and Lorentz NMOA arise from the $\mathcal{T}$-even components of the zeroth- and second-order terms, respectively.

Having identified the four types of optical activity,
we next consider how to separate octupolar MOA from the other contributions that may coexist with it.
As a starting point, we consider centrosymmetric systems where magnetic dipole order is symmetry-forbidden.
These conditions prohibit the optical activity induced by the first-order term $\varepsilon_{ij,k}(\omega)q_k$ and dipolar MOA, respectively.
The remaining task is therefore to separate octupolar MOA from uniform NMOA and Lorentz NMOA.
To this end, we propose two strategies.

The first strategy is to align the incident linear polarization direction with an eigenmode principal axis, 
thereby precluding NMOA (see Sec.~\ref{Subsec:theory of MD-induced magneto-optical effects}).
However, its experimental implementation is not always straightforward because the eigenmode principal axes can be rotated away from the coordinate axes.
To simplify the situation, we therefore consider an experimentally practical configuration in which the linear polarization direction is aligned with a coordinate axis, 
hereafter referred to as coordinate-axis alignment (CAA).
Under CAA, the presence or absence of NMOA is determined by the presence or absence of the off-diagonal components $\eta^{(\mathrm{S})}_{xy}$ and $\tilde{G}^{(\mathrm{S})}_{xy}$, which rotate the eigenmode principal axes.
Taking their presence or absence into account, we can identify four distinct types of optical activity involving octupolar MOA.

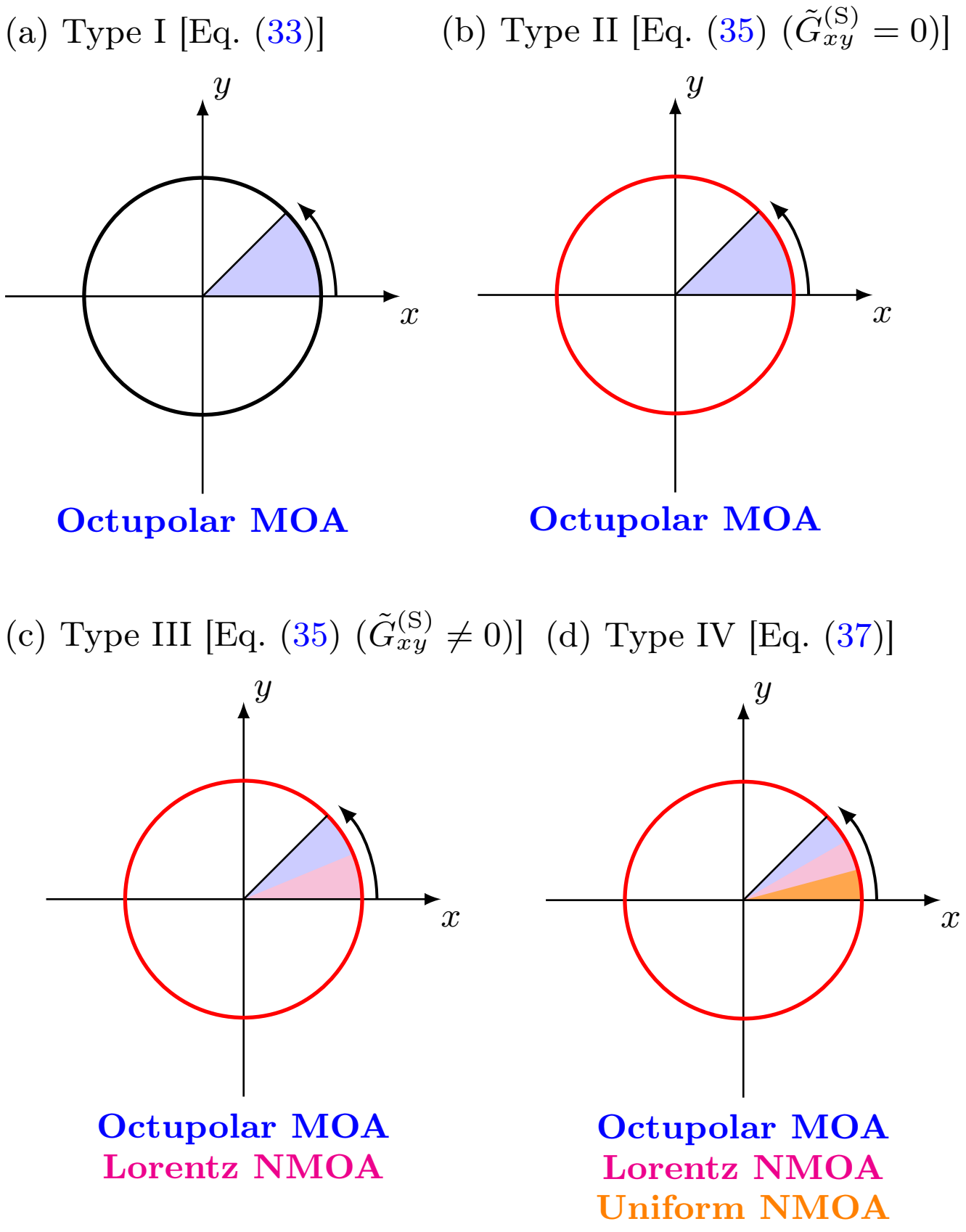
\begin{figure}[t]
    \centering
    \begin{subfigure}[t]{0.44\linewidth}
        \caption{Type I [Eq.~\eqref{eq:eigenvalue_MO1_MO}]}
        \begin{tikzpicture}[scale=0.9]
            \fill[blue!20] (0,0) -- (1.2, 0) arc[start angle=0, end angle=45, radius=1.2] -- cycle; % 塗りつぶし
            \draw[-latex, line width=0.5pt] (-2,0) -- (2,0);      % x軸
            \node[] at (2.1, -0.2) {$x$};
            \draw[-latex, line width=0.5pt] (0,-2) -- (0,2);      % y軸
            \node[] at (0.2, 2.1) {$y$};
            \draw[line width=1pt] (0,0) circle(1.2);              % 円
            \draw[line width=0.5pt, rotate=45] (0,0) -- (1.2,0);  % 30度回転
            \draw[-latex, line width=0.75pt] (1.35,0) arc[start angle=0, end angle=45, radius=1.35]; % 90度の弧で矢印を描く
            \node[text=blue] at (0.0, -2.3) {\textbf{Octupolar MOA}};
        \end{tikzpicture}
        \label{fig:MO_Type1}
    \end{subfigure}
    \hfill
    \begin{subfigure}[t]{0.54\linewidth}
        \vspace{-2.1pt}
        \caption{Type II [Eq.~\eqref{eq:eigenvalue_MO2_MO} ($\tilde{G}^{(\mathrm{S})}_{xy}=0$)]}
        \begin{tikzpicture}[scale=0.9]
            \fill[blue!20] (0,0) -- (1.2, 0) arc[start angle=0, end angle=45, radius=1.2] -- cycle; % 塗りつぶし
            \draw[-latex, line width=0.5pt] (-2,0) -- (2,0);      % x軸
            \node[] at (2.1, -0.2) {$x$};
            \draw[-latex, line width=0.5pt] (0,-2) -- (0,2);      % y軸
            \node[] at (0.2, 2.1) {$y$};
            \draw[line width=1pt, draw=red] (0,0) circle(1.2);              % 円
            \draw[line width=0.5pt, rotate=45] (0,0) -- (1.2,0);  % 30度回転
            \draw[-latex, line width=0.75pt] (1.35,0) arc[start angle=0, end angle=45, radius=1.35]; % 90度の弧で矢印を描く
            \node[text=blue] at (0.0, -2.3) {\textbf{Octupolar MOA}};
        \end{tikzpicture}
        \label{fig:MO_Type2}
    \end{subfigure}
    %\vfill
    \begin{subfigure}[t]{0.55\linewidth}
        \vspace{-2.0pt}
        \caption{Type III [Eq.~\eqref{eq:eigenvalue_MO2_MO} ($\tilde{G}^{(\mathrm{S})}_{xy}\neq0$)]}
        \begin{tikzpicture}[scale=0.9]
            \fill[blue!20] (0,0) -- (1.2, 0) arc[start angle=0, end angle=45, radius=1.2] -- cycle; % 塗りつぶし
            \fill[magenta!30] (0,0) -- (1.2, 0) arc[start angle=0, end angle=22.5, radius=1.2] -- cycle; % 塗りつぶし
            \draw[-latex, line width=0.5pt] (-2,0) -- (2,0);      % x軸
            \node[] at (2.1, -0.2) {$x$};
            \draw[-latex, line width=0.5pt] (0,-2) -- (0,2);      % y軸
            \node[] at (0.2, 2.1) {$y$};
            \draw[line width=1pt, draw=red] (0,0) circle(1.2);              % 円
            \draw[line width=0.5pt, rotate=45] (0,0) -- (1.2,0);  % 30度回転
            \draw[-latex, line width=0.75pt] (1.35,0) arc[start angle=0, end angle=45, radius=1.35]; % 90度の弧で矢印を描く
            \node[align=center] at (0.0, -2.5) {\textcolor{blue}{\textbf{Octupolar MOA}} \\ \textcolor{magenta}{\textbf{Lorentz NMOA}}};
        \end{tikzpicture}
        \label{fig:MO_Type3}
    \end{subfigure}
    \hfill
    \begin{subfigure}[t]{0.43\linewidth}
        \caption{Type IV [Eq.~\eqref{eq:eigenvalue_MO3}]}
        \begin{tikzpicture}[scale=0.9]
            \fill[blue!20] (0,0) -- (1.2, 0) arc[start angle=0, end angle=45, radius=1.2] -- cycle; % 塗りつぶし
            \fill[magenta!30] (0,0) -- (1.2, 0) arc[start angle=0, end angle=30, radius=1.2] -- cycle; % 塗りつぶし
            \fill[orange!70] (0,0) -- (1.2, 0) arc[start angle=0, end angle=15, radius=1.2] -- cycle; % 塗りつぶし
            \draw[-latex, line width=0.5pt] (-2,0) -- (2,0);      % x軸
            \node[] at (2.1, -0.2) {$x$};
            \draw[-latex, line width=0.5pt] (0,-2) -- (0,2);      % y軸
            \node[] at (0.2, 2.1) {$y$};
            \draw[line width=1pt, draw=red] (0,0) circle(1.2);              % 円
            \draw[line width=0.5pt, rotate=45] (0,0) -- (1.2,0);  % 30度回転
            \draw[-latex, line width=0.75pt] (1.35,0) arc[start angle=0, end angle=45, radius=1.35]; % 90度の弧で矢印を描く
            \node[align=center] at (0.0, -2.7) {\textcolor{blue}{\textbf{Octupolar MOA}} \\ \textcolor{magenta}{\textbf{Lorentz NMOA}} \\ \textcolor{orange}{\textbf{Uniform NMOA}}};
        \end{tikzpicture}
        \label{fig:MO_Type4}
    \end{subfigure}
    \caption{Schematic illustration of four types of optical activity involving octupolar MOA under CAA.
             Each panel illustrates which optical-activity contributions rotate the polarization plane initially aligned along the $x$ axis.
             These contributions are indicated below each panel, and the corresponding equations are given in the subcaption.
             Note that the red circles schematically indicate that NMOA can emerge away from CAA.
             }
    \label{fig:MO_Type}
\end{figure}

These four types are schematically illustrated in Fig.~\ref{fig:MO_Type}.
Types I and II involve only octupolar MOA under CAA due to the absence of the off-diagonal components.
The difference between these two types emerges away from CAA: Type I does not involve NMOA, whereas Type II does.
In contrast, Types III and IV also involve Lorentz NMOA, with Type IV additionally involving uniform NMOA, due to the presence of the corresponding off-diagonal components.
Thus, Types I and II are advantageous for isolating octupolar MOA, 
whereas Types III and IV require the second strategy to eliminate the NMOA contributions.

The second strategy is to use Sagnac interferometry \cite{Kapitulnik1994-jz,Xia2006-rm}, which selectively isolates the $\mathcal{T}$-odd contribution to optical activity.
This technique allows octupolar MOA to be separated from NMOA regardless of the linear polarization direction, unlike CAA.
For the remainder of this work, however, we focus on CAA as a simple means of avoiding NMOA.

Finally, we note that observing octupolar MOA does not necessarily require strong spatial dispersion.
Spatial dispersion describes the nonlocal electromagnetic response of a material and manifests itself as $\bm{q}$-dependent terms in the permittivity tensor \cite{Agranovich1984}.
Since $\bm{q}$-dependent terms are usually regarded as corrections to the local, $\bm{q}$-independent response,
one might expect their observation to require strong spatial dispersion.
However, this expectation does not necessarily apply when a $\bm{q}$-dependent term constitutes the leading nonvanishing response.
In this case, the response may be small in magnitude but can remain observable due to the absence of lower-order responses.
For octupolar MOA, such a case is realized when magnetic dipole order is symmetry-forbidden and inversion symmetry is preserved, which together eliminate the lower-order $\mathcal{T}$-odd responses.
Under these conditions, octupolar MOA can therefore be observed even without strong spatial dispersion, with its magnitude lying well within an experimentally accessible range (see Sec.~\ref{Sec:Numerical_results}).

%%%%%%%%%%%%%%%%%%%%%%%%%%%%%%%%%%%%%%%%%%%%%%%%%%%%%%%%%%%%%%%%%%%%%%%%%%%%%%%%%%%%%%%%%%%%%%%%%%%%%%%%%%%%%%%%%%%%%%%%%%%%%%%%%%%%

\section{Conditions for observing octupolar MOA} \label{Sec:MO_conditions}

In this section, we discuss the conditions for observing octupolar MOA.
Here, we provide an overview of the analysis by recalling the macroscopic theory established in the previous section.
In this theory, we assumed that the octupolar MOA vector $\tilde{\bm{G}}^{(\mathrm{o})}$ has a finite component and showed that octupolar MOA emerges when light propagates along the direction of this component.
In the present analysis, we examine the conditions under which $\tilde{\bm{G}}^{(\mathrm{o})}$ has a finite component and identify the light-propagation directions along which octupolar MOA is allowed.
The resulting directions and the corresponding expressions for $\tilde{\bm{G}}^{(\mathrm{o})}$ are determined by the magnetic-octupole components allowed by crystal symmetry.
For each of these directions, we further examine the coexistence of octupolar MOA with NMOA.
These two analyses are carried out in Secs.~\ref{Subsec:MO_effects_conditions} and \ref{Subsec:MO_effects_compedition}, respectively,
and their results are summarized in Table~\ref{tab:Mijk_list}.

Before proceeding, we introduce two distinct Cartesian coordinate systems: (i) $(x,y,z)$ fixed to the crystal and (ii) $(X,Y,Z)$ with one of its axes along the light-propagation direction.
Coordinate system (i) is used to describe crystal symmetry, allowing us to identify the symmetry-allowed magnetic-octupole components.
In addition, it is used as a common reference frame for specifying light-propagation directions.
Specifically, a light-propagation direction is denoted by the direction indices [$uvw$] defined with respect to coordinate system (i),
where $u$, $v$, and $w$ specify the direction of the light wave vector $\tilde{\bm{q}}$ through
$\tilde{\bm{q}} \propto u\tilde{\bm{e}}_x+v\tilde{\bm{e}}_y+w\tilde{\bm{e}}_z$,
with $(\tilde{\bm{e}}_x,\tilde{\bm{e}}_y,\tilde{\bm{e}}_z)$ being the corresponding unit basis vectors and an overbar indicating a negative index, e.g., $\bar{u}=-u$.
In contrast, coordinate system (ii) is used to describe the polarization state of light.
This polarization state, however, remains constrained by the crystal symmetry described in coordinate system (i).
Therefore, transforming from coordinate system (i) to (ii) carries these symmetry constraints into the description in coordinate system (ii),
allowing us to analyze the coexistence of octupolar MOA with NMOA.

%%%%%%%%%%%%%%%%%%%%%%%%%%%%%%%%%%%%%%%%%%%%%%%%%%%%%%%%%%%%%%%%%%

\subsection{Light-propagation directions for octupolar MOA} \label{Subsec:MO_effects_conditions}

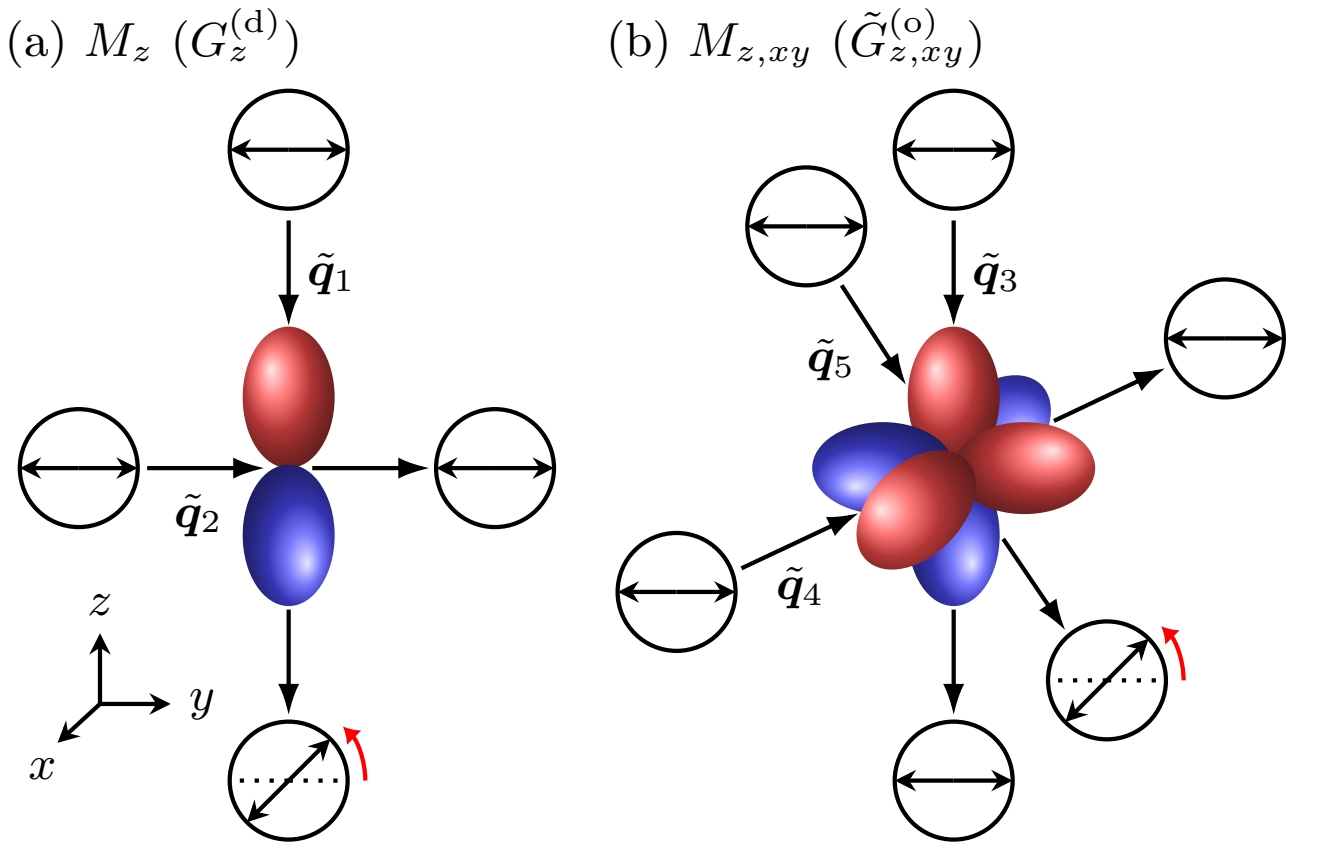
\begin{figure}[t]
    \centering
    \begin{subfigure}[t]{0.44\linewidth}
        \caption{$M_z\ (G^{(\mathrm{d})}_z)$}
        \begin{tikzpicture}[scale=0.78]
            % Dipole
            \begin{scope}[scale=0.3]
                \shade[ball color=red!70!white,shading angle=0] (0,0) ellipse (1.3 and 2.0);
                \shade[ball color=blue!70!white,shading angle=180] (0,-3.9) ellipse (1.3 and 2.0);
            \end{scope}
            % \node[yshift=-0.5mm] at (1.0,-1.7) {$M_z$};

            % Light direction1
            \draw[-{Latex[length=6, width=4]}, line width=0.8pt] (0,1.5) -- (0,0.6) node[midway, right] {$\tilde{\bm{q}}_1$};
            \draw[-{Latex[length=6, width=4]}, line width=0.8pt] (0,-1.8) -- (0,-2.7);
            % \node at (0.0,2.9) {$\bm{q}=(0,0,-1)$};

            % Light plane1 (in)
            \draw[line width=0.8pt, draw=black] (0, 2.1) circle (0.5);
            \draw[-{Stealth[length=4, width=4]}, line width=0.8pt, draw=black] (0, 2.1) -- ++(0:0.5);
            \draw[-{Stealth[length=4, width=4]}, line width=0.8pt, draw=black] (0, 2.1) -- ++(0+180:0.5);

            % Light plane1 (out)
            \draw[line width=0.8pt, draw=black] (0, -3.25) circle (0.5);
            \draw[-{Stealth[length=4, width=4]}, line width=0.8pt, draw=black] (0, -3.25) -- ++(45:0.5);
            \draw[-{Stealth[length=4, width=4]}, line width=0.8pt, draw=black] (0, -3.25) -- ++(45+180:0.5);
            \draw[dotted, line width=0.8pt, draw=black] (0, -3.25) -- ++(0:0.5);
            \draw[dotted, line width=0.8pt, draw=black] (0, -3.25) -- ++(0+180:0.5);
            \draw[-{Latex[length=3, width=4]}, line width=0.8pt, draw=red] (0, -3.25) ++ (0:0.65) arc (0:45:0.65);

            % Light direction2
            \draw[-{Latex[length=6, width=4]}, line width=0.8pt] (-1.2,-0.6) -- (-0.2,-0.6) node[midway, below, xshift=-0.5mm, yshift=-0.5mm] {$\tilde{\bm{q}}_2$};
            \draw[-{Latex[length=6, width=4]}, line width=0.8pt] (0.2,-0.6) -- (1.2,-0.6);
            % \node at (-1.75,0.2) {$\bm{q}=(0,1,0)$};

            % Light plane2 (in)
            \draw[line width=0.8pt, draw=black] (-1.78, -0.6) circle (0.5);
            \draw[-{Stealth[length=4, width=4]}, line width=0.8pt, draw=black] (-1.78, -0.6) -- ++(0:0.5);
            \draw[-{Stealth[length=4, width=4]}, line width=0.8pt, draw=black] (-1.78, -0.6) -- ++(0+180:0.5);

            % Light plane2 (out)
            \draw[line width=0.8pt, draw=black] (1.75, -0.6) circle (0.5);
            \draw[-{Stealth[length=4, width=4]}, line width=0.8pt, draw=black] (1.75, -0.6) -- ++(0:0.5);
            \draw[-{Stealth[length=4, width=4]}, line width=0.8pt, draw=black] (1.75, -0.6) -- ++(0+180:0.5);

            \begin{scope}[shift={(-1.6,-2.6)}, scale=0.6]
                \draw[-{Stealth[length=4, width=4]}, line width=0.8pt] (0,0) -- (-0.6,-0.55) node[below, xshift=-1.0mm, yshift=0.2mm] {$x$};
                \draw[-{Stealth[length=4, width=4]}, line width=0.8pt] (0,0) -- (1.0,0) node[right] {$y$};
                \draw[-{Stealth[length=4, width=4]}, line width=0.8pt] (0,0) -- (0,1.0) node[above] {$z$};
            \end{scope}
        \end{tikzpicture}
        \label{fig:MO_condition1}
    \end{subfigure}
    \hfill
    \begin{subfigure}[t]{0.54\linewidth}
        \caption{$M_{z,xy}\ (\tilde{G}^{(\mathrm{o})}_{z,xy})$}
        \begin{tikzpicture}[scale=0.78]
            % Octupole
            \begin{scope}[scale=0.3]
                \shade[ball color=blue!70!white,shading angle=225, rotate=135] (-1.5,0.0) ellipse (1.3 and 2.0);

                \shade[ball color=red!70!white,shading angle=0] (0,0) ellipse (1.3 and 2.0);
                \shade[ball color=blue!70!white,shading angle=180] (0,-3.9) ellipse (1.3 and 2.0);

                \shade[ball color=red!70!white,shading angle=270] (2.0,-2.0) ellipse (2.0 and 1.3);
                \shade[ball color=blue!70!white,shading angle=90] (-2.0,-2.0) ellipse (2.0 and 1.3);

                \shade[ball color=red!70!white,shading angle=45, rotate=135] (-1.5,3.0) ellipse (1.3 and 2.0);
            \end{scope}
            % \node[xshift=-0.8mm] at (1.8,-1.2) {$M_{zxy}$};

            % Light direction1
            \draw[-{Latex[length=6, width=4]}, line width=0.8pt] (0,1.5) -- (0,0.6) node[midway, right] {$\tilde{\bm{q}}_3$};
            \draw[-{Latex[length=6, width=4]}, line width=0.8pt] (0,-1.8) -- (0,-2.7);
            % \node at (0.0,2.9) {$\bm{q}=(0,0,-1)$};

            % Light plane1 (in)
            \draw[line width=0.8pt, draw=black] (0, 2.1) circle (0.5);
            \draw[-{Stealth[length=4, width=4]}, line width=0.8pt, draw=black] (0, 2.1) -- ++(0:0.5);
            \draw[-{Stealth[length=4, width=4]}, line width=0.8pt, draw=black] (0, 2.1) -- ++(0+180:0.5);

            % Light plane1 (out)
            \draw[line width=0.8pt, draw=black] (0, -3.25) circle (0.5);
            \draw[-{Stealth[length=4, width=4]}, line width=0.8pt, draw=black] (0, -3.25) -- ++(0:0.5);
            \draw[-{Stealth[length=4, width=4]}, line width=0.8pt, draw=black] (0, -3.25) -- ++(0+180:0.5);

            % Light direction2
            \draw[-{Latex[length=6, width=4]}, line width=0.8pt] (-1.8,-1.45) -- (-0.8,-0.98) node[midway, below, yshift=-0.5mm] {$\tilde{\bm{q}}_4$};
            \draw[-{Latex[length=6, width=4]}, line width=0.8pt] (0.85,-0.2) -- (1.8,0.25);
            % \node at (-2.3,-0.8) {$\bm{q}=(-1,1,0)$};

            % Light plane2 (in)
            \draw[line width=0.8pt, draw=black] (-2.35, -1.65) circle (0.5);
            \draw[-{Stealth[length=4, width=4]}, line width=0.8pt, draw=black] (-2.35, -1.65) -- ++(0:0.5);
            \draw[-{Stealth[length=4, width=4]}, line width=0.8pt, draw=black] (-2.35, -1.65) -- ++(0+180:0.5);

            % Light plane1 (out)
            \draw[line width=0.8pt, draw=black] (2.3, 0.5) circle (0.5);
            \draw[-{Stealth[length=4, width=4]}, line width=0.8pt, draw=black] (2.3, 0.5) -- ++(0:0.5);
            \draw[-{Stealth[length=4, width=4]}, line width=0.8pt, draw=black] (2.3, 0.5) -- ++(0+180:0.5);

            % Light direction3
            \draw[-{Latex[length=6, width=4]}, line width=0.8pt] (-0.95,0.95) -- (-0.4,0.1) node[midway, below left, yshift=1.0mm] {$\tilde{\bm{q}}_5$};
            \draw[-{Latex[length=6, width=4]}, line width=0.8pt] (0.42,-1.2) -- (0.95,-1.98);

            % Light plane3 (in)
            \draw[line width=0.8pt, draw=black] (-1.25, 1.45) circle (0.5);
            \draw[-{Stealth[length=4, width=4]}, line width=0.8pt, draw=black] (-1.25, 1.45) -- ++(0:0.5);
            \draw[-{Stealth[length=4, width=4]}, line width=0.8pt, draw=black] (-1.25, 1.45) -- ++(0+180:0.5);

            % Light plane3 (out)
            \draw[line width=0.8pt, draw=black] (1.3, -2.4) circle (0.5);
            \draw[-{Stealth[length=4, width=4]}, line width=0.8pt, draw=black] (1.3, -2.4) -- ++(45:0.5);
            \draw[-{Stealth[length=4, width=4]}, line width=0.8pt, draw=black] (1.3, -2.4) -- ++(45+180:0.5);
            \draw[dotted, line width=0.8pt, draw=black] (1.3, -2.4) -- ++(0:0.5);
            \draw[dotted, line width=0.8pt, draw=black] (1.3, -2.4) -- ++(0+180:0.5);
            \draw[-{Latex[length=3, width=4]}, line width=0.8pt, draw=red] (1.3, -2.4) ++ (0:0.65) arc (0:45:0.65);
        \end{tikzpicture}
        \label{fig:MO_condition2}
    \end{subfigure}
    \caption{Schematic illustration of light-propagation directions that allow or forbid MOA for (a) a magnetic dipole $M_z$ and (b) a magnetic octupole $M_{z,xy}$.
             Each panel shows whether the polarization plane perpendicular to the wave vector $\tilde{\bm{q}}$ rotates after passing through the multipole.
             In panel~(a), dipolar MOA is allowed for $\tilde{\bm{q}}_1=(0,0,-1)$ (Faraday configuration), but forbidden for $\tilde{\bm{q}}_2=(0,1,0)$ (Voigt configuration).
             In panel~(b), octupolar MOA is forbidden for $\tilde{\bm{q}}_3=(0,0,-1)$ and $\tilde{\bm{q}}_4=(-1,1,0)/\sqrt{2}$, but allowed for $\tilde{\bm{q}}_5=(1,1,-1)/\sqrt{3}$.
             Note that all multipole components and light-propagation directions are expressed in coordinate system (i).
             }
    \label{fig:MO_condition}
\end{figure}

Here, we examine the conditions under which $\tilde{\bm{G}}^{(\mathrm{o})}$ has a finite component.
In particular, we focus on the component along a light-propagation direction, $\tilde{\bm{G}}^{(\mathrm{o})} \cdot \tilde{\bm{q}}$, and 
examine for which directions of $\tilde{\bm{q}}$ this component is nonzero: $\tilde{\bm{G}}^{(\mathrm{o})} \cdot \tilde{\bm{q}} \neq 0$.
This condition can be readily understood in the case of the dipolar MOA, for which the corresponding condition is $\bm{G}^{(\mathrm{d})} \cdot \tilde{\bm{q}} \neq 0$.
This is because the dipolar MOA vector $\bm{G}^{(\mathrm{d})}$ is independent of $\tilde{\bm{q}}$ [Eq.~\eqref{eq:epsilon_inverse_MD}] and is related to the magnetic dipole $\bm{M}$ [Eq.~\eqref{eq:Multipole tensor_conductivity tensor1}].
In other words, $\bm{G}^{(\mathrm{d})} \cdot \tilde{\bm{q}} \neq 0$ can be satisfied when $\tilde{\bm{q}}$ has a finite component along $\bm{M}$,
whereas $\tilde{\bm{q}} \perp \bm{M}$ does not satisfy $\bm{G}^{(\mathrm{d})}\cdot\tilde{\bm{q}}\neq 0$, thereby precluding dipolar MOA.
This distinction is commonly represented by the Faraday ($\tilde{\bm{q}} \parallel \bm{M}$) and Voigt ($\tilde{\bm{q}} \perp \bm{M}$) configurations [Fig.~\ref{fig:MO_condition1}], 
providing a practical guideline for experimentally observing dipolar MOA.
It is therefore natural to seek an analogous guideline for octupolar MOA.
Unlike $\bm{G}^{(\mathrm{d})}$, however, $\tilde{\bm{G}}^{(\mathrm{o})}$ explicitly depends on $\tilde{\bm{q}}$ [Eq.~\eqref{eq:epsilon_inverse_MO}], 
preventing a direct application of the same reasoning.

To overcome this difficulty, we separate the $\tilde{\bm{q}}$ dependence from $\tilde{\bm{G}}^{(\mathrm{o})}(\bm{q},\omega)$.
To this end, following Eq.~\eqref{eq:epsilon_normal_MO_coefficients_normalized_MO}, we express $\tilde{\bm{g}}^{(\mathrm{o})}(\bm{q},\omega)$ as a quadratic form in $\tilde{\bm{q}}$ 
by introducing a $\tilde{\bm{q}}$-independent tensor $\tilde{g}^{(\mathrm{o})}_{\mu,kl}(\omega)$ as
\begin{align}
    \tilde{g}^{(\mathrm{o})}_\mu(\bm{q},\omega)=\tilde{g}^{(\mathrm{o})}_{\mu,kl}(\omega)\tilde{q}_k\tilde{q}_l. \label{eq:tildegijk_definition}
\end{align}
This tensor is related to the second-order expansion coefficient $\varepsilon^{(\mathrm{A})}_{ij,kl}(\omega)$ through
\begin{align}
    e_{\mu ij}\tilde{g}^{(\mathrm{o})}_{\mu, kl}(\omega)= (\omega/c)^2\varepsilon^{(\mathrm{A})}_{ij,kl}(\omega). \label{eq:tildegijk_eijkl}
\end{align} 
Substituting Eq.~\eqref{eq:tildegijk_definition} into $\tilde{\bm{G}}^{(\mathrm{o})}(\bm{q},\omega)$ defined in Eq.~\eqref{eq:epsilon_inverse_MO_coefficients_MO}, 
we can similarly express it as a quadratic form in $\tilde{\bm{q}}$:
\begin{subequations} \label{eq:tildeGijk_definition}
\begin{align}
    \tilde{G}^{(\mathrm{o})}_\mu(\bm{q},\omega)=\tilde{G}^{(\mathrm{o})}_{\mu,kl}(\omega)\tilde{q}_k\tilde{q}_l, \label{eq:tildeGijk_definition_original}
\end{align}
where $\tilde{G}^{(\mathrm{o})}_{\mu,kl}(\omega)$ is given by
\begin{align}
    \tilde{G}^{(\mathrm{o})}_{\mu,kl}(\omega)=\frac{-1}{|\varepsilon^{(\mathrm{S})}|}\varepsilon^{(\mathrm{S})}_{\mu \nu}\tilde{g}^{(\mathrm{o})}_{\nu,kl}(\omega). \label{eq:tildeGijk_expression}
\end{align}
\end{subequations}
Equation~\eqref{eq:tildeGijk_definition} has two important features.
First, it separates the $\tilde{\bm{q}}$ dependence from $\tilde{\bm{G}}^{(\mathrm{o})}(\bm{q},\omega)$, leaving the $\tilde{\bm{q}}$-independent coefficient $\tilde{G}^{(\mathrm{o})}_{\mu,kl}(\omega)$.
Second, this coefficient is related to the magnetic octupole $M_{\mu,kl}$ 
because $\tilde{g}^{(\mathrm{o})}_{\mu, kl}(\omega)$ is related to $\varepsilon^{(\mathrm{A})}_{ij,kl}(\omega)$ [Eq.~\eqref{eq:tildegijk_eijkl}], 
which in turn corresponds to $M_{\mu,kl}$ [Eq.~\eqref{eq:Multipole tensor_conductivity tensor3}].
Therefore, Eq.~\eqref{eq:tildeGijk_definition} allows us to discuss the condition $\tilde{\bm{G}}^{(\mathrm{o})} \cdot \tilde{\bm{q}} \neq 0$ in terms of magnetic octupoles.

\begin{table*}[t]
    \caption{Summary of the conditions for observing octupolar MOA in centrosymmetric MPGs with magnetic octupole order but without magnetic dipole order.
    For each MPG, we list the symmetry-allowed magnetic-octupole components, light-propagation directions $[uvw]$ along which octupolar MOA is allowed, optical-activity type under CAA, and candidate materials.
    In the light-propagation-direction column, each direction is followed by the equation number in the main text that gives the corresponding expression for $\tilde{G}^{(\mathrm{o})}_{[uvw]}$.
    We also list, for Type-II optical activity, the linear polarization directions that satisfy CAA for the light-propagation direction given in the same row.
    Note that all magnetic-octupole components, light-propagation directions, and linear polarization directions are expressed in coordinate system (i), and that $4/mmm$ and $6/mmm$ do not exhibit octupolar MOA.
    }
    %\vspace{1mm}
    \label{tab:Mijk_list}
    \centering
    \tabcolsep = 1.3mm
    \renewcommand\arraystretch{1.2}
    \begin{tabular}{c||c|c|c|c|c} \hline \hline
        \multirow{2}{*}{MPG} & \multirow{2}{*}{Magnetic-octupole component $M_{\mu,kl}$} & Light-propagation & \multirow{2}{*}{Type} & Linear polarization & \multirow{2}{*}{Candidate materials} \\ 
        & & directions & & directions & \\ \hline 
        $mmm$ & $M_{x,yz},M_{y,zx},M_{z,xy}$ & $[111]$ \eqref{eq:OAV_111_orthorombic} & IV & \\ \hline
        \multirow{3}{*}{$4'/m$} & \multirow{2}{*}{$M_{x,zx}=-M_{y,yz}, M_{z,xx}=-M_{z,yy},$} & $[111]$ \eqref{eq:OAV_111_tetragonal} & \multirow{3}{*}{III} & \\
        & \multirow{2}{*}{$M_{x,yz}=M_{y,zx}\neq M_{z,xy}$} & $[101]$ \eqref{eq:OAV_101} & & \\ 
        & & $[011]$ \eqref{eq:OAV_011} & & \\ \hline
        $4/mmm$ & $M_{x,yz}=-M_{y,zx}$ & $-$ & \multicolumn{3}{>{\columncolor{gray!20}[\tabcolsep][\tabcolsep]}c}{} \\ \hline
        \multirow{2}{*}{$4'/mmm'$} & \multirow{2}{*}{$M_{x,yz}=M_{y,zx}\neq M_{z,xy}$} & \multirow{2}{*}{$[111]$ \eqref{eq:OAV_111_tetragonal}} & \multirow{2}{*}{II} & \multirow{2}{*}{$[1\bar{1}0],[11\bar{2}]$} & \ce{RuO2} \cite{Feng2022-jm,Fedchenko2024-lw,Zhang_2025-lb}, \\
        & & & & & \ce{MnF2} \cite{Quentin2025-zx} \\ \hline
        \multirow{3}{*}{$4'/mm'm$} & \multirow{3}{*}{$M_{x,zx}=-M_{y,yz}, M_{z,xx}=-M_{z,yy}$} & \multirow{2}{*}{$[101]$ \eqref{eq:OAV_101}} & \multirow{3}{*}{II} & \multirow{2}{*}{$[10\bar{1}],[010]$} & \ce{KV2Se2O} \cite{Jiang2025-au,Wang2025-wm,Guofei2026-tf}, \\ 
        & & \multirow{2}{*}{$[011]$ \eqref{eq:OAV_011}} & & \multirow{2}{*}{$[100],[01\bar{1}]$} & \ce{RbV2Te2O} \cite{Zhang2025-iv}, \\ 
        & & & & & \ce{La2O3Mn2Se2} \cite{Wei2025-nn,Zhang2026-rf} \\ \hline
        \multirow{2}{*}{$\bar{3}m$} & $M_{x,xx}=-M_{x,yy}=-M_{y,xy},$ & $[100]$ \eqref{eq:OAV_100} & \multirow{2}{*}{III} & & \multirow{2}{*}{$\alpha$-\ce{Fe2O3}\footnote{$\bar{3}m$ is realized below the Morin temperature.} \cite{Galindez-Ruales2025-qc,Sun2025-hw,Yamamoto2026-zn}}\\ 
        & $M_{x,yz}=-M_{y,zx}$ & $[110]$ (\ref{eq:OAV_110}, $\tilde{g}^{(\mathrm{o})}_{x,xy}=0$) & & & \\ \hline
        \multirow{3}{*}{$6'/m'$} & \multirow{2}{*}{$M_{x,xx}=-M_{x,yy}=-M_{y,xy},$} & $[100]$ \eqref{eq:OAV_100} & \multirow{3}{*}{II} & $[010],[001]$ & \\
        & \multirow{2}{*}{$M_{y,xx}=-M_{y,yy}=M_{x,xy}$} & $[010]$ \eqref{eq:OAV_010} & & $[001],[100] $& \\ 
        & & $[110]$ \eqref{eq:OAV_110} & & $[\bar{1}10],[001]$ & \\ \hline
        $6/mmm$ & $M_{x,yz}=-M_{y,zx}$ & $-$ & \multicolumn{3}{>{\columncolor{gray!20}[\tabcolsep][\tabcolsep]}c}{} \\ \hline
        \multirow{2}{*}{$6'/m'mm'$} & \multirow{2}{*}{$M_{x,xx}=-M_{x,yy}=-M_{y,xy}$} & $[100]$ \eqref{eq:OAV_100} & \multirow{2}{*}{II} & $[010],[001]$ \\
        & & $[110]$ (\ref{eq:OAV_110}, $\tilde{g}^{(\mathrm{o})}_{x,xy}=0$) & & $[\bar{1}10],[001]$  \\ \hline
        \multirow{2}{*}{$6'/m'm'm$} & \multirow{2}{*}{$M_{y,xx}=-M_{y,yy}=M_{x,xy}$} & $[010]$ \eqref{eq:OAV_010} & \multirow{2}{*}{II} & $[001],[100]$ & \ce{CrSb} \cite{Reimers2024-gw,Zeng2024-ea,Ding2024-qn,Yang2025-ko}, \\ 
        & & $[110]$ (\ref{eq:OAV_110}, $\tilde{g}^{(\mathrm{o})}_{y,xy}=0$) & & $[\bar{1}10],[001]$ & \ce{CoNb4Se8} \cite{Regmi2025-as,Nicholas2024-fm,De-Vita2025-pj,Sakhya2025-qy} \\ \hline
        $m\bar{3}$ & $M_{x,yz}=M_{y,zx}=M_{z,xy}$ & $[111]$ \eqref{eq:OAV_111_cubic} & I & & \ce{MnTe2} \cite{Zhu2024-oc,Wulferding2026-kn} \\ \hline
        $m\bar{3}m'$ & $M_{x,yz}=M_{y,zx}=M_{z,xy}$ & $[111]$ \eqref{eq:OAV_111_cubic} & I & & \ce{\textit{R}2Ir2O7} \cite{Tomiyasu2012-cy,Sagayama2013-xe,Donnerer2016-xq,Guo2016-is} \\ \hline \hline
    \end{tabular}
\end{table*}

Using Eq.~\eqref{eq:tildeGijk_definition}, we now examine the condition $\tilde{\bm{G}}^{(\mathrm{o})} \cdot \tilde{\bm{q}} \neq 0$, which can be written as
\begin{align}
    \tilde{\bm{G}}^{(\mathrm{o})} \cdot \tilde{\bm{q}}=\tilde{G}^{(\mathrm{o})}_{\mu,kl}\tilde{q}_\mu\tilde{q}_k\tilde{q}_l \neq 0. \label{eq:MO_condition_MO}
\end{align}
This expression indicates that octupolar MOA is allowed only when the light-propagation direction has simultaneously nonzero $\tilde{q}_\mu$, $\tilde{q}_k$, and $\tilde{q}_l$ components 
corresponding to the magnetic octupole $M_{\mu,kl}$.
For example, in the case of the magnetic octupole $M_{z,xy}$, octupolar MOA is allowed along a body diagonal 
but forbidden along an edge or a face diagonal, as shown in Fig.~\ref{fig:MO_condition2}.
This example illustrates that magnetic-octupole components determine the light-propagation directions along which octupolar MOA is allowed.
Therefore, identifying these directions requires knowledge of the magnetic-octupole components allowed by crystal symmetry.

Motivated by this requirement, we systematically investigate magnetic point groups (MPGs) with magnetic octupole order 
and identify the symmetry-allowed magnetic-octupole components in coordinate system (i).
Here, we restrict our analysis to centrosymmetric MPGs without magnetic dipole order to isolate octupolar MOA from lower-order MOA,
leaving the following ten MPGs \cite{Yatsushiro2021-af,Oike2024-bl}: $mmm$, $4'/m$, $4/mmm$, $4'/mmm'$ ($4'/mm'm$), $\bar{3}m$, $6'/m'$, $6/mmm$, $6'/m'mm'$ ($6'/m'm'm$), $m\bar{3}$, and $m\bar{3}m'$.
For these ten MPGs, the symmetry-allowed magnetic-octupole components are listed in Table~\ref{tab:Mijk_list}, with the detailed derivations presented in the Supplemental Material (SM) \cite{Supplemental}.

However, not all of these symmetry-allowed magnetic-octupole components allow octupolar MOA.
This is because Eq.~\eqref{eq:MO_condition_MO} imposes an additional constraint on the magnetic octupole $M_{\mu,kl}$.
Specifically, $M_{\mu,kl}$ must have a nonzero totally symmetric part $M_{(\mu, kl)}$ with respect to permutations of the indices $\mu$, $k$, and $l$,
\begin{align}
    M_{(\mu, kl)}=\frac{1}{3}(M_{\mu,kl}+M_{k,l \mu}+M_{l,\mu k}). \label{eq:Additional_constraints}
\end{align} 
This constraint excludes $4/mmm$ and $6/mmm$,
leaving the eight MPGs that allow octupolar MOA, which we refer to as octupolar MPGs.
For each octupolar MPG, light-propagation directions along which octupolar MOA is allowed are determined by the corresponding nonzero totally symmetric parts $M_{(\mu,kl)}$.
Among the allowed directions, we select representative directions and list them in Table~\ref{tab:Mijk_list},
providing a practical guideline for experimentally observing octupolar MOA.

Finally, we derive the expression for $\tilde{\bm{G}}^{(\mathrm{o})} \cdot \tilde{\bm{q}} \eqqcolon \tilde{G}^{(\mathrm{o})}_{[uvw]}$ 
for each representative light-propagation direction $[uvw]$ listed in Table~\ref{tab:Mijk_list}. 
Here, we present only the final expressions written in terms of $\tilde{g}^{(\mathrm{o})}_{\mu,kl}$ (see Eq.~\eqref{eq:tildeGijk_definition}),
leaving the detailed derivation to Appendix~\ref{Sec:AppC}.
For the [111] direction, the expressions take one of the following three forms depending on the MPGs:
\begin{subequations} \label{eq:OAV_111}
    \begin{align}
        \tilde{G}^{(\mathrm{o})}_{[111]}&=-\frac{2}{3\sqrt{3}}\biggl(\frac{\tilde{g}^{(\mathrm{o})}_{x,yz}}{\varepsilon^{(\mathrm{S})}_{yy}\varepsilon^{(\mathrm{S})}_{zz}}
        +\frac{\tilde{g}^{(\mathrm{o})}_{y,zx}}{\varepsilon^{(\mathrm{S})}_{zz}\varepsilon^{(\mathrm{S})}_{xx}}+\frac{\tilde{g}^{(\mathrm{o})}_{z,xy}}{\varepsilon^{(\mathrm{S})}_{xx}\varepsilon^{(\mathrm{S})}_{yy}} \biggr), \label{eq:OAV_111_orthorombic} \\
        \tilde{G}^{(\mathrm{o})}_{[111]}&=-\frac{2}{3\sqrt{3}}\frac{1}{\varepsilon^{(\mathrm{S})}_{xx}}\biggl(\frac{2\tilde{g}^{(\mathrm{o})}_{x,yz}}{\varepsilon^{(\mathrm{S})}_{zz}}
        +\frac{\tilde{g}^{(\mathrm{o})}_{z,xy}}{\varepsilon^{(\mathrm{S})}_{xx}} \biggr), \label{eq:OAV_111_tetragonal} \\
        \tilde{G}^{(\mathrm{o})}_{[111]}&=-\frac{2}{\sqrt{3}}\frac{\tilde{g}^{(\mathrm{o})}_{z,xy}}{(\varepsilon^{(\mathrm{S})}_{xx})^2}, \label{eq:OAV_111_cubic}
    \end{align}
\end{subequations}
where the first line applies to $mmm$, the second line to $4'/m$ and $4'/mmm'$, and the third line to $m\bar{3}$ and $m\bar{3}m'$.
Note that $4'/m$ and $4'/mmm'$ impose $\varepsilon^{(\mathrm{S})}_{xx}=\varepsilon^{(\mathrm{S})}_{yy}$ and $\tilde{g}^{(\mathrm{o})}_{x,yz}=\tilde{g}^{(\mathrm{o})}_{y,zx}$, while 
$m\bar{3}$ and $m\bar{3}m'$ impose $\varepsilon^{(\mathrm{S})}_{xx}=\varepsilon^{(\mathrm{S})}_{yy}=\varepsilon^{(\mathrm{S})}_{zz}$ and $\tilde{g}^{(\mathrm{o})}_{x,yz}=\tilde{g}^{(\mathrm{o})}_{y,zx}=\tilde{g}^{(\mathrm{o})}_{z,xy}$,
with representative tensor components used in Eqs.~\eqref{eq:OAV_111_tetragonal} and \eqref{eq:OAV_111_cubic}, respectively.
For the remaining directions, the expressions are given by
\begin{subequations} \label{eq:OAV_other}
    \begin{align}
        \tilde{G}^{(\mathrm{o})}_{[101]}&=-\frac{1}{2\sqrt{2}} \frac{1}{\varepsilon^{(\mathrm{S})}_{xx}} \biggl( \frac{2\tilde{g}^{(\mathrm{o})}_{x,zx}}{\varepsilon^{(\mathrm{S})}_{zz}} + \frac{\tilde{g}^{(\mathrm{o})}_{z,xx}}{\varepsilon^{(\mathrm{S})}_{xx}} \biggr), \label{eq:OAV_101} \\
        \tilde{G}^{(\mathrm{o})}_{[011]}&=-\frac{1}{2\sqrt{2}} \frac{1}{\varepsilon^{(\mathrm{S})}_{xx}} \biggl( \frac{2\tilde{g}^{(\mathrm{o})}_{y,yz}}{\varepsilon^{(\mathrm{S})}_{zz}} + \frac{\tilde{g}^{(\mathrm{o})}_{z,yy}}{\varepsilon^{(\mathrm{S})}_{xx}} \biggr), \label{eq:OAV_011} \\
        \tilde{G}^{(\mathrm{o})}_{[100]}&=-\frac{\tilde{g}^{(\mathrm{o})}_{x,xx}}{\varepsilon^{(\mathrm{S})}_{xx} \varepsilon^{(\mathrm{S})}_{zz}}, \label{eq:OAV_100} \\
        \tilde{G}^{(\mathrm{o})}_{[010]}&=-\frac{\tilde{g}^{(\mathrm{o})}_{y,yy}}{\varepsilon^{(\mathrm{S})}_{xx} \varepsilon^{(\mathrm{S})}_{zz}}, \label{eq:OAV_010} \\
        \tilde{G}^{(\mathrm{o})}_{[110]}&=-\frac{1}{\sqrt{2}} \frac{\tilde{g}^{(\mathrm{o})}_{x,xy}+\tilde{g}^{(\mathrm{o})}_{y,xy}}{\varepsilon^{(\mathrm{S})}_{xx}\varepsilon^{(\mathrm{S})}_{zz}}, \label{eq:OAV_110}
    \end{align}
\end{subequations}
where the MPGs corresponding to each light-propagation direction can be identified from Table~\ref{tab:Mijk_list}.
Note that, as in Eqs.~\eqref{eq:OAV_111_tetragonal} and \eqref{eq:OAV_111_cubic}, representative tensor components are chosen according to the symmetry of each MPG,
with either $\tilde{g}^{(\mathrm{o})}_{x,xy}$ or $\tilde{g}^{(\mathrm{o})}_{y,xy}$ in Eq.~\eqref{eq:OAV_110} additionally vanishing for $\bar{3}m$, $6'/m'mm'$, and $6'/m'm'm$, as shown in Table~\ref{tab:Mijk_list}.

%%%%%%%%%%%%%%%%%%%%%%%%%%%%%%%%%%%%%%%%%%%%%%%%%%%%%%%%%%%%%%%%%%

\subsection{Coexistence of octupolar MOA with NMOA} \label{Subsec:MO_effects_compedition}

We next examine the coexistence of octupolar MOA with NMOA for each representative light-propagation direction identified in the previous subsection.
Specifically, we consider CAA for each direction and determine which of the optical-activity types shown in Fig.~\ref{fig:MO_Type} is realized.
To this end, we construct coordinate system (ii) from coordinate system (i) through a coordinate transformation and apply the macroscopic theory established in Sec.~\ref{Sec:Macroscopic_theory_MO}.\footnote{
For the $[100]$ and $[010]$ directions, coordinate system (i) can be used directly as coordinate system (ii).
}
The resulting classification, together with candidate materials, is summarized in Table~\ref{tab:Mijk_list}, while the details of the analysis are presented in Appendix~\ref{Sec:AppC}.

A notable outcome of this classification is that many antiferromagnets, including those shown in Fig.~\ref{fig:General_concept2}, exhibit Type-I or Type-II optical activity.
For example, noncoplanar antiferromagnets such as \ce{MnTe2} and AIAO-ordered pyrochlore iridates \ce{\textit{R}2Ir2O7} exhibit Type-I optical activity, while 
candidate altermagnets such as \ce{RuO2}, \ce{KV2Se2O}, and \ce{CrSb} exhibit Type-II optical activity.
In these materials, octupolar MOA can be observed without NMOA, but the conditions required for such observation differ between Type-I and Type-II optical activity.
Specifically, Type-I optical activity requires no specific choice of linear polarization direction,
whereas Type-II optical activity requires the linear polarization direction to satisfy CAA (see Sec.~\ref{Subsec:short summary of theory}).
Therefore, such linear polarization directions are also listed in Table~\ref{tab:Mijk_list}, where they are expressed in coordinate system (i).

Finally, we revisit the MOA pointed out by Graham and Raab \cite{Graham1991-xf} 
to identify where it falls within this classification.
Specifically, we point out that this MOA corresponds to Type-I optical activity based on the following three considerations.
First, they predicted this MOA for cubic antiferromagnets, consistent with Type-I optical activity in cubic MPGs ($m\bar{3}, m\bar{3}m'$).
Second, they showed that this MOA can be observed for light propagating along body-diagonal directions, consistent with the [111] direction for Type-I optical activity.
Third, the expression they derived for this MOA can be reproduced from our formulation of Type-I optical activity.
Specifically, substituting Eq.~\eqref{eq:OAV_111_cubic} into the refractive indices $n_{\pm}$ for Type-I optical activity given in Eq.~\eqref{eq:eigenvalue_MO1_MO},
we obtain the refractive-index difference
\begin{align}
    n_+-n_-=\frac{2i}{\sqrt{3}}\sqrt{\varepsilon^{(\mathrm{S})}_{xx}}\tilde{g}^{(\mathrm{o})}_{z,xy}, \label{eq:Deltan_TypeI}
\end{align}
which agrees with their result.
Thus, our framework accounts for the MOA pointed out by Graham and Raab while further identifying its origin as magnetic octupole order
and extending it to a broader class of antiferromagnets.

%%%%%%%%%%%%%%%%%%%%%%%%%%%%%%%%%%%%%%%%%%%%%%%%%%%%%%%%%%%%%%%%%%%%%%%%%%%%%%%%%%%%%%%%%%%%%%%%%%%%%%%%%%%%%%%%%%%%%%%%%%%%%%%%%%%%

\section{SOC dependence of octupolar MOA} \label{Sec:SOC_dependence_MOA}

\begin{table*}[t]
    \caption{Classification of even-parity SMMs $M^{(\mathrm{spin})}_{\mu,kl \ldots}$ for the 90 nontrivial collinear SPGs (the superscript (spin) is omitted below).
    The symbols $\checkmark$ and $\times$ denote whether each SMM is allowed ($\checkmark$) or forbidden ($\times$). }
    %\vspace{1mm}
    \label{tab:SPG_classification}
    \centering
    \tabcolsep = 1.1mm
    \renewcommand\arraystretch{1.2}
    \begin{tabular}{c|c||c|c|c|c} \hline \hline
        \multirow{2}{*}{Class} & \multirow{2}{*}{Nontrivial collinear SPGs} & $M_\mu$ & $M_{\mu,kl}$ & $M_{\mu,klmn}$ & $M_{\mu,klmnpq}$  \\
        & & ($s$) & ($d$) & ($g$) & ($i$)   \\ \hline
        \multirow{4}{*}{$s$-wave} & ${^1}1,{^1}\bar{1},{^1}2,{^1}m,{^1}2/{^1}m,{^1}2{^1}2{^1}2,{^1}m{^1}m{^1}2,{^1}m{^1}m{^1}m,$ & \multirow{4}{*}{$\checkmark$} & \multirow{4}{*}{$\checkmark$} & \multirow{4}{*}{$\checkmark$} & \multirow{4}{*}{$\checkmark$} \\  
        & ${^1}4,{^1}\bar{4},{^1}4/{^1}m,{^1}4{^1}2{^1}2,{^1}4{^1}m{^1}m,{^1}\bar{4}{^1}2{^1}m,{^1}4/{^1}m{^1}m{^1}m,$ & & & & \\ 
        & ${^1}3,{^1}\bar{3},{^1}3{^1}2,{^1}3{^1}m,{^1}\bar{3}{^1}m,{^1}6,{^1}\bar{6},{^1}6/{^1}m,{^1}6{^1}2{^1}2,{^1}6{^1}m{^1}m,{^1}\bar{6}{^1}m{^1}2,{^1}6/{^1}m{^1}m{^1}m,$ & & & &  \\ 
        & ${^1}2{^1}3,{^1}2/{^1}m{^1}\bar{3},{^1}4{^1}3{^1}2,{^1}\bar{4}{^1}3{^1}m,{^1}4/{^1}m{^1}\bar{3}{^1}2/{^1}m$ (32 groups) & & & & \\ \hline 
        \multirow{2}{*}{$d$-wave} & ${^{\text{-}1}}2,{^{\text{-}1}}m,{^{\text{-}1}}2/{^{\text{-}1}}m,{^{\text{-}1}}2{^{\text{-}1}}2{^1}2,{^1}m{^{\text{-}1}}m{^{\text{-}1}}2,{^{\text{-}1}}m{^{\text{-}1}}m{^1}2,{^{\text{-}1}}m{^{\text{-}1}}m{^1}m$, & \multirow{2}{*}{$\times$} & \multirow{2}{*}{$\checkmark$} & \multirow{2}{*}{$\checkmark$} & \multirow{2}{*}{$\checkmark$}  \\  
        & ${^{\text{-}1}}4,{^{\text{-}1}}\bar{4},{^{\text{-}1}}4/{^1}m,{^{\text{-}1}}4{^{\text{-}1}}2{^1}2,{^{\text{-}1}}4{^{\text{-}1}}m{^1}m,{^{\text{-}1}}\bar{4}{^{\text{-}1}}2{^1}m,{^{\text{-}1}}\bar{4}{^1}2{^{\text{-}1}}m,{^{\text{-}1}}4/{^1}m{^{\text{-}1}}m{^1}m$ (15 groups) & & & & \\ \hline 
        \multirow{2}{*}{$g$-wave} & ${^1}4{^{\text{-}1}}2{^{\text{-}1}}2,{^1}4{^{\text{-}1}}m{^{\text{-}1}}m,{^1}\bar{4}{^{\text{-}1}}2{^{\text{-}1}}m,{^1}4/{^1}m{^{\text{-}1}}m{^{\text{-}1}}m,{^1}3{^{\text{-}1}}2,{^1}3{^{\text{-}1}}m,{^1}\bar{3}{^{\text{-}1}}m,$ & \multirow{2}{*}{$\times$} & \multirow{2}{*}{$\times$} & \multirow{2}{*}{$\checkmark$} & \multirow{2}{*}{$\checkmark$} \\  
        & ${^{\text{-}1}}6,{^{\text{-}1}}\bar{6},{^{\text{-}1}}6/{^{\text{-}1}}m,{^{\text{-}1}}6{^{\text{-}1}}2{^1}2,{^{\text{-}1}}6{^{\text{-}1}}m{^1}m,{^{\text{-}1}}\bar{6}{^1}m{^{\text{-}1}}2,{^{\text{-}1}}\bar{6}{^{\text{-}1}}m{^1}2,{^{\text{-}1}}6/{^{\text{-}1}}m{^{\text{-}1}}m{^1}m$ (15 groups) & & & & \\ \hline 
        \multirow{2}{*}{$i$-wave} & ${^1}6{^{\text{-}1}}2{^{\text{-}1}}2,{^1}6{^{\text{-}1}}m{^{\text{-}1}}m,{^1}\bar{6}{^{\text{-}1}}m{^{\text{-}1}}2,{^1}6/{^1}m{^{\text{-}1}}m{^{\text{-}1}}m,$ & \multirow{2}{*}{$\times$} & \multirow{2}{*}{$\times$} & \multirow{2}{*}{$\times$} & \multirow{2}{*}{$\checkmark$}  \\  
        & ${^{\text{-}1}}4{^1}3{^{\text{-}1}}2,{^{\text{-}1}}\bar{4}{^1}3{^{\text{-}1}}m,{^{\text{-}1}}4/{^1}m{^1}\bar{3}{^{\text{-}1}}2/{^{\text{-}1}}m$ (7 groups) & & & &  \\ \hline 
        \multirow{4}{*}{PT} & ${^{\text{-}1}}\bar{1},{^1}2/{^{\text{-}1}}m,{^{\text{-}1}}2/{^1}m,{^1}m{^1}m{^{\text{-}1}}m,{^{\text{-}1}}m{^{\text{-}1}}m{^{\text{-}1}}m,$  & \multirow{4}{*}{$\times$} & \multirow{4}{*}{$\times$} & \multirow{4}{*}{$\times$} & \multirow{4}{*}{$\times$}  \\ 
        & ${^1}4/{^{\text{-}1}}m,{^{\text{-}1}}4/{^{\text{-}1}}m,{^1}4/{^{\text{-}1}}m{^1}m{^1}m,{^{\text{-}1}}4/{^{\text{-}1}}m{^{\text{-}1}}m{^1}m,{^1}4/{^{\text{-}1}}m{^{\text{-}1}}m{^{\text{-}1}}m,$  & & & & \\ 
        & ${^{\text{-}1}}\bar{3},{^{\text{-}1}}\bar{3}{^1}m,{^{\text{-}1}}\bar{3}{^{\text{-}1}}m,{^{1}}6/{^{\text{-}1}}m,{^{\text{-}1}}6/{^1}m,{^1}6/{^{\text{-}1}}m{^1}m{^1}m,{^{\text{-}1}}6/{^1}m{^{\text{-}1}}m{^1}m,{^1}6/{^{\text{-}1}}m{^{\text{-}1}}m{^{\text{-}1}}m,$ & & & &  \\
        & ${^1}2/{^{\text{-}1}}m{^{\text{-}1}}\bar{3},{^{\text{-}1}}4/{^{\text{-}1}}m{^{\text{-}1}}\bar{3}{^{\text{-}1}}2/{^1}m,{^1}4/{^{\text{-}1}}m{^{\text{-}1}}\bar{3}{^1}2/{^{\text{-}1}}m$ (21 groups) & & & & \\ \hline \hline
    \end{tabular}
\end{table*}

In this section, we examine the SOC dependence of octupolar MOA by analyzing magnetic-octupole components responsible for it.
Our approach combines MPGs and spin point groups (SPGs) \cite{Litvin1974-hd,Litvin1977-kw,Opechowski1986}, which describe symmetry with and without SOC, respectively.
Specifically, we first consider the symmetry-allowed magnetic-octupole components for a given MPG while also identifying those for SPGs.
We then identify an SPG that reduces to the MPG when SOC is introduced and compare their allowed components.
If they share at least one magnetic-octupole component, octupolar MOA does not require SOC owing to the shared component; otherwise, it requires SOC.
On this basis, octupolar MOA in $d$-wave altermagnets does not require SOC, while SOC makes octupolar MOA symmetry-allowed in certain $g$- and $i$-wave altermagnets.

Following this approach requires two steps.
The first is to identify the symmetry-allowed magnetic-octupole components for SPGs; 
those for MPGs, particularly octupolar MPGs, have already been identified in the previous section (see Table~\ref{tab:Mijk_list}).
In this analysis, however, we consider magnetic multipoles in general, rather than restricting ourselves to magnetic octupoles.
The second is to determine which SPGs reduce to each MPG when SOC is introduced, and we refer to the resulting relation as the SPG--MPG correspondence.
These two analyses are carried out in Secs.~\ref{Subsec:MO_effects_SPG} and \ref{Subsec:SPG_MPG correspondence}, respectively.
Based on the results of these analyses, we finally introduce the strong/weak $X$-wave classification ($X=s,d,g,i$) of collinear magnets in Sec.~\ref{Subsec:strong_weak_altermagnets}.

%%%%%%%%%%%%%%%%%%%%%%%%%%%%%%%%%%%%%%%%%%%%%%%%%%%%%%%%%%%%%%%%%%

\subsection{SPG-based classification of magnetic multipoles} \label{Subsec:MO_effects_SPG}

We first briefly describe SPGs, in which the spin-space operation $g_\mathrm{s}$ and the real-space operation $g_\ell$ are treated independently.
An SPG $P_\mathrm{s}$ consists of symmetry operations of the form $\{g_\mathrm{s}||g_\ell\}$ and can be decomposed into the direct product \cite{Opechowski1986}
\begin{align}
    P_\mathrm{s}=P_{\mathrm{NT}}\times P_{\mathrm{SO}}. \label{eq:SPG_decomposition}
\end{align}
Here, $P_{\mathrm{NT}}$ is the nontrivial part and consists of operations $\{g_\mathrm{s}||g_\ell\}$ with $g_\ell \neq 1$, 
whereas $P_{\mathrm{SO}}$ is the spin-only part and consists of operations $\{g_\mathrm{s}||1\}$, with $1$ denoting the identity operation in real space.
In particular, $P_{\mathrm{SO}}$ reflects the magnetic structure, taking different forms for collinear, coplanar, and noncoplanar magnetic structures \cite{Opechowski1986}.
Among these three cases, we focus on collinear magnetic structures, which are relevant to altermagnetism.
In this case, the spin-only part is represented by $^{\infty_{\bm{n}}m}1$ in the Chen--Liu notation \cite{Chen2024-qg}, where $\bm{n}$ denotes the collinear axis.

The collinear nature of magnetic structures simplifies the symmetry analysis of magnetic multipoles (see the SM \cite{Supplemental} for details).
To see this, we recall the definition of magnetic multipoles: $M_{\mu,kl\ldots}=\int m_\mu r_kr_l \cdots d\bm{r}$.
Based on this definition, magnetic multipoles can be divided into spin magnetic multipoles (SMMs) and orbital magnetic multipoles 
according to whether they originate from the spin or orbital contribution to the magnetization density $\bm{m}$.
With this distinction in mind, only SMMs need to be considered because orbital magnetic multipoles are forbidden by symmetry \cite{Watanabe2024-us,Etxebarria2025-bh}.
Furthermore, only the nontrivial part of the SPG needs to be considered because the spin-only part only constrains the spin component of an SMM
and does not affect whether an SMM is allowed or forbidden by symmetry \cite{Etxebarria2025-bh}.
Thus, the analysis reduces to identifying the symmetry-allowed SMMs for the nontrivial collinear SPGs, 
which number 90 \cite{Liu2022-mk} out of the 598 nontrivial SPGs \cite{Litvin1977-kw}.

Based on the above considerations, we now identify the symmetry-allowed SMMs for the 90 nontrivial collinear SPGs.
The results are summarized in Table~\ref{tab:SPG_classification}, which lists even-parity SMMs up to spin magnetic triakontahectapoles (rank-7 SMMs).
The full classification, including odd-parity SMMs and the SMM components of both parities, is provided in the SM \cite{Supplemental}.
In Table~\ref{tab:SPG_classification}, the SPGs can be divided into two cases: those that allow at least one even-parity SMM and those that allow none.
In particular, the former can be further classified according to the lowest-rank even-parity SMM, with the even-parity SMMs labeled $s$, $d$, $g$, and $i$ in order of increasing rank.
Based on this notation, we refer to each class as the $X$-wave class ($X=s,d,g,i$).

A notable feature of the $X$-wave classes ($X=d,g,i$) is that the SPGs in each class coincide with those that allow $X$-wave spin splitting
according to the SPG-based definition of altermagnetism \cite{Smejkal2022-xr,Smejkal2022-xx}.
This result not only supports the previously proposed connection between altermagnetism and higher-rank magnetic multipoles \cite{Bhowal2024-vj,McClarty2024-eo},
but also establishes a comprehensive correspondence between altermagnetism and even-parity SMMs. 
Therefore, probing even-parity SMMs provides an alternative route to detecting altermagnetism without directly measuring spin splitting.
Moreover, even-parity SMMs can serve as order parameters, providing a thermodynamic characterization of altermagnetism across the phase diagram.

By contrast, the remaining SPGs possess $\mathcal{P}\mathcal{T}$ symmetry, which forbids even-parity SMMs and prevents spin splitting.
We therefore refer to this class as the PT class, which is instead characterized by odd-parity SMMs (see the SM \cite{Supplemental}).
In this sense, the PT class complements the $X$-wave classes, and together they complete the classification of the 90 nontrivial collinear SPGs.

%%%%%%%%%%%%%%%%%%%%%%%%%%%%%%%%%%%%%%%%%%%%%%%%%%%%%%%%%%%%%%%%%%

\subsection{SPG--MPG correspondence} \label{Subsec:SPG_MPG correspondence}

Here, we establish the SPG--MPG correspondence by starting from a collinear SPG and identifying the MPG to which it reduces when SOC is introduced.
However, a collinear SPG by itself does not uniquely determine the corresponding MPG.
This is because the orientation of the collinear axis $\bm{n}$ relative to the lattice is fixed in MPGs, 
whereas it remains unspecified in SPGs due to the independence of spin space and real space.
We therefore use the oriented-SPG framework \cite{Liu2026-ossg,Yu2026-zs}, in which the spin-space coordinate system is aligned with the real-space coordinate system.
Within this framework, specifying the orientation of $\bm{n}$ in spin space also determines its orientation relative to the lattice,
thereby uniquely identifying the corresponding MPG, which is a subgroup of the SPG.
In this setting, we express the orientation of $\bm{n}$ as $[uvw]$ in coordinate system (i), following the same notation introduced for $\tilde{\bm{q}}$ (see Sec.~\ref{Sec:MO_conditions}).

Once the SPG--MPG correspondence is established in this way, we can directly compare the symmetry-allowed magnetic-octupole components of the corresponding SPG and MPG.
Here, we focus on octupolar MPGs and perform this comparison to examine the SOC dependence of octupolar MOA.
Before proceeding, we extend the $X$-wave classification for SPGs to MPGs by assigning each MPG to an $X$-wave class 
according to its lowest-rank symmetry-allowed even-parity magnetic multipole.
Under this classification, all octupolar MPGs belong to the $d$-wave class.
We now turn to the analysis, first illustrating the procedure using two collinear SPGs as examples and then presenting the results for octupolar MPGs.
The detailed analysis is presented in the SM \cite{Supplemental}.

As examples, we consider two collinear SPGs ${^{\text{-}1}}4/{^1}m{^{\text{-}1}}m{^1}m^{\infty_{\bm{n}}m}1$ and ${^{\text{-}1}}6/{^{\text{-}1}}m{^1}m{^{\text{-}1}}m^{\infty_{\bm{n}}m}1$ 
for orientations of the collinear axis $\bm{n}=[001]$, $[100]$, and $[010]$.
The corresponding MPGs for these orientations are listed in Table~\ref{tab:SPG_MPG_relation}, together with representative materials.
Among these materials, \ce{RuO2} and \ce{CrSb} exhibit $d$-wave $\to$ $d$-wave and $g$-wave $\to$ $d$-wave transitions, respectively, between the $X$-wave classes of their SPGs and MPGs.
These correspondences show that the SPG and MPG of \ce{RuO2} share at least one magnetic-octupole component, whereas those of \ce{CrSb} do not.
Consequently, \ce{RuO2} can exhibit octupolar MOA without SOC, whereas \ce{CrSb} can exhibit octupolar MOA only in the presence of SOC.

We next consider octupolar MPGs and summarize their corresponding SPGs in Table~\ref{tab:SPG_MPG_relation_full}.
Among these correspondences, octupolar MOA is symmetry-allowed without SOC only when the SPGs also belong to the $d$-wave class, as illustrated by the above examples.
Accordingly, only $mmm$, $4'/m$, $4'/mmm'$, and $4'/mm'm$ can exhibit octupolar MOA without SOC.
For these MPGs, the shared magnetic-octupole components are also listed in Table~\ref{tab:SPG_MPG_relation_full}.

\begin{table}[t]
    \caption{SPG--MPG correspondences and representative materials for collinear SPGs 
    ${^{\text{-}1}}4/{^1}m{^{\text{-}1}}m{^1}m^{\infty_{\bm{n}}m}1$ and ${^{\text{-}1}}6/{^{\text{-}1}}m{^1}m{^{\text{-}1}}m^{\infty_{\bm{n}}m}1$ 
    with orientations of the collinear axis $\bm{n}=[001]$, $[100]$, and $[010]$.
    Each point group is assigned the corresponding $X$-wave class label.
    }
    %\vspace{1mm}
    \label{tab:SPG_MPG_relation}
    \centering
    \tabcolsep = 1.0mm
    \renewcommand\arraystretch{1.2}
    \begin{tabular}{c||c|c|c} \hline \hline
        Collinear SPG & $\bm{n}=[001]$ & $\bm{n}=[100]$ & $\bm{n}=[010]$ \\ \hline
        ${^{\text{-}1}}4/{^1}m{^{\text{-}1}}m{^1}m^{\infty_{\bm{n}}m}1$ & $4'/mmm'$ & $m'mm'$ & $mm'm'$ \\ 
        ($d$-wave)& ($d$-wave) & ($s$-wave) & ($s$-wave) \\ 
        & \ce{RuO2} & &  \\ \hline
        ${^{\text{-}1}}6/{^{\text{-}1}}m{^1}m{^{\text{-}1}}m^{\infty_{\bm{n}}m}1$  & $6'/m'm'm$ & $mmm$ & $m'm'm$ \\
        ($g$-wave) & ($d$-wave) & ($d$-wave) & ($s$-wave) \\ 
        & \ce{CrSb} & & $\alpha$-\ce{MnTe} \\ \hline \hline
    \end{tabular}
\end{table}

\begin{table}[t]
    \caption{SPG--MPG correspondences for octupolar MPGs. 
    The magnetic-octupole components $M_{\mu,kl}$ shared by the corresponding SPG and MPG are also listed, with $-$ indicating the absence of shared components. 
    Note that, for several MPGs, only representative SPGs are listed here; additional SPGs can be found in the SM \cite{Supplemental}.
    Furthermore, cubic MPGs are omitted because they do not have corresponding collinear SPGs.
    }
    %\vspace{1mm}
    \label{tab:SPG_MPG_relation_full}
    \centering
    \tabcolsep = 0.8mm
    \renewcommand\arraystretch{1.2}
    \begin{tabular}{c|c||c} \hline \hline
        MPG ($d$) & Collinear SPG & $M_{\mu,kl}$ \\ \hline 
        \multirow{3}{*}{$mmm$} & ${^{\text{-}1}}m{^{\text{-}1}}m{^1}m^{\infty_{[001]}m}1$ ($d$) & $M_{z,xy}$  \\ \cline{2-3}
        & ${^{\text{-}1}}6/{^{\text{-}1}}m{^{\text{-}1}}m{^1}m^{\infty_{[010]}m}1$ ($g$) & $-$ \\ \cline{2-3}
        & ${^{\text{-}1}}6/{^{\text{-}1}}m{^1}m{^{\text{-}1}}m^{\infty_{[100]}m}1$ ($g$) & $-$ \\ \hline
        \multirow{2}{*}{$4'/m$} & \multirow{2}{*}{${^{\text{-}1}}4/{^1}m^{\infty_{[001]}m}1$ ($d$)} & $M_{z,xx}=-M_{z,yy},$ \\
        & & $M_{z,xy}$ \\ \hline
        $4'/mmm'$ & ${^{\text{-}1}}4/{^1}m{^{\text{-}1}}m{^1}m^{\infty_{[001]}m}1$ ($d$) & $M_{z,xy}$ \\ \hline 
        \multirow{2}{*}{$4'/mm'm$} & ${^{\text{-}1}}4/{^1}m{^1}m{^{\text{-}1}}m^{\infty_{[001]}m}1$ ($d$) & $M_{z,xx}=-M_{z,yy}$ \\ \cline{2-3}
        & ${^{\text{-}1}}4/{^1}m{^1}\bar{3}{^{\text{-}1}}2/{^{\text{-}1}}m^{\infty_{[001]}m}1$ ($i$) & $-$ \\ \hline
        \multirow{2}{*}{$\bar{3}m$} & ${^1}\bar{3}{^{\text{-}1}}m^{\infty_{[001]}m}1$ ($g$) & $-$ \\ \cline{2-3}
        & ${^{\text{-}1}}4/{^1}m{^1}\bar{3}{^{\text{-}1}}2/{^{\text{-}1}}m^{\infty_{[111]}m}1$ ($i$) & $-$ \\ \hline
        $6'/m'$ & ${^{\text{-}1}}6/{^{\text{-}1}}m^{\infty_{[001]}m}1$ ($g$) & $-$ \\ \hline
        $6'/m'mm'$ & ${^{\text{-}1}}6/{^{\text{-}1}}m{^{\text{-}1}}m{^1}m^{\infty_{[001]}m}1$ ($g$) & $-$ \\ \hline
        $6'/m'm'm$ & ${^{\text{-}1}}6/{^{\text{-}1}}m{^1}m{^{\text{-}1}}m^{\infty_{[001]}m}1$ ($g$) & $-$ \\ \hline \hline
    \end{tabular}
\end{table}

The above analysis further reveals a limitation of the SPG-based classification of altermagnets when discussing physical responses.
Indeed, as shown in Table~\ref{tab:SPG_MPG_relation}, $\alpha$-\ce{MnTe} and \ce{CrSb} belong to the same nontrivial collinear SPG but correspond to different MPGs,
resulting in different sets of symmetry-allowed physical responses:
for example, $\alpha$-\ce{MnTe} can exhibit dipolar MOA \cite{Gray2024-us,Amin2024-ay,Yamamoto2025-kp,Yang2026-xr,Watanabe2026-ek,Yang2026-rw}, whereas \ce{CrSb} cannot.
This limitation arises because SOC is intrinsically present in real materials, and their physical symmetries and responses must be described by MPGs rather than by SPGs alone.
Conversely, although the MPG alone can determine which physical responses are symmetry-allowed,
it cannot determine their SOC dependence.
Indeed, as shown in Table~\ref{tab:SPG_MPG_relation_full}, octupolar MOA in $mmm$ and $4'/mm'm$ may or may not require SOC, depending on the corresponding SPG.
Therefore, a new framework is needed to simultaneously characterize the symmetry-allowed physical responses and their SOC dependence in collinear magnets.
Motivated by this need, we introduce below the strong/weak $X$-wave classification of collinear magnets based on the SPG--MPG correspondence.

%%%%%%%%%%%%%%%%%%%%%%%%%%%%%%%%%%%%%%%%%%%%%%%%%%%%%%%%%%%%%%%%%%

\subsection{Strong/weak $X$-wave classification ($X=s,d,g,i$)} \label{Subsec:strong_weak_altermagnets}

\tikzstyle{arrow} = [thick,->,>=stealth]
\tikzstyle{framework} = [rectangle, rounded corners, minimum width=1.4cm, minimum height=0.8cm, text centered, draw=black]

\begin{figure*}[t]
\centering
\begin{tikzpicture}[node distance=2cm,scale=1.0, transform shape]
    \node (SPG) [framework, align=center] {SPGs};
    \node (s-wave_SPG) [below of=SPG, yshift=0.9cm] {$s$-wave};
    \node (d-wave_SPG) [below of=s-wave_SPG, yshift=0.9cm] {$d$-wave};
    \node (g-wave_SPG) [below of=d-wave_SPG, yshift=0.9cm] {$g$-wave};
    \node (i-wave_SPG) [below of=g-wave_SPG, yshift=1.0cm] {$i$-wave};
    \node (PT_SPG) [below of=i-wave_SPG, yshift=1.0cm] {PT};

    \node (MPG) [framework, align=center, right of=SPG, xshift=0.5cm] {MPGs};
    \node (s-wave_MPG) [below of=MPG, yshift=0.9cm] {$s$-wave};
    \node (d-wave_MPG) [below of=s-wave_MPG, yshift=0.9cm] {$d$-wave};
    \node (g-wave_MPG) [below of=d-wave_MPG, yshift=1.0cm] {$g$-wave};
    \node (PT_MPG) [below of=g-wave_MPG, yshift=-0.1cm] {PT};

    \draw [arrow, color=blue] (s-wave_SPG.east) -- (s-wave_MPG.west) node[midway, above] {\textcolor{blue}{strong}};
    \draw [arrow, color=red] (d-wave_SPG.east) -- ([yshift=-1mm]s-wave_MPG.west) node[midway, above, xshift=-3mm] {\textcolor{red}{weak}};
    \draw [arrow, color=blue] (d-wave_SPG.east) -- (d-wave_MPG.west);
    \draw [arrow, color=red] (g-wave_SPG.east) -- ([yshift=-2mm]s-wave_MPG.west);
    \draw [arrow, color=red] (g-wave_SPG.east) -- ([yshift=-1mm]d-wave_MPG.west);
    \draw [arrow, color=red] (i-wave_SPG.east) -- ([yshift=-3mm]s-wave_MPG.west);
    \draw [arrow, color=red] (i-wave_SPG.east) -- ([yshift=-2mm]d-wave_MPG.west);
    \draw [arrow, color=blue] ([xshift=2mm]PT_SPG.east) -- ([xshift=-2mm]PT_MPG.west);

    \node[font=\scriptsize, right of=s-wave_MPG, xshift=3.9cm, yshift=0.0cm,  align=left] {\textcolor{blue}{\ce{Co3Sn2S2} (Strong ferromagnet): ${^1}\bar{3}{^1}m^{\infty_{[001]}m}1\ (s\text{-wave}) \to \bar{3}m'\ (s\text{-wave})$}
                                                                                            \\ [1mm] \textcolor{red}{$\alpha$-\ce{MnTe} (Weak ferromagnet): ${^{\text{-}1}}6/{^{\text{-}1}}m{^1}m{^{\text{-}1}}m^{\infty_{[010]}m}1\ (g\text{-wave}) \to m'm'm\ (s\text{-wave})$}};
    
    \node[font=\scriptsize, right of=d-wave_MPG, xshift=4.4cm, yshift=0.0cm,  align=left] {\textcolor{blue}{\ce{RuO2} (Strong $d$-wave altermagnet): ${^{\text{-}1}}4/{^1}m{^{\text{-}1}}m{^1}m^{\infty_{[001]}m}1\ (d\text{-wave}) \to 4'/mmm'\ (d\text{-wave})$}
                                                                                            \\ [1mm] \textcolor{red}{\ce{CrSb} (Weak $d$-wave altermagnet): ${^{\text{-}1}}6/{^{\text{-}1}}m{^1}m{^{\text{-}1}}m^{\infty_{[001]}m}1\ (g\text{-wave}) \to 6'/m'm'm\ (d\text{-wave})$}};
    
    \node[text=blue,font=\scriptsize, right of=PT_MPG, xshift=2.95cm, yshift=0.0cm,  align=left] {\ce{CuMnAs} (PT magnet): ${^1}4/{^{\text{-}1}}m{^1}m{^1}m^{\infty_{[010]}m}1\ (\mathrm{PT}) \to m'mm\ (\mathrm{PT})$ 
                                                                                            \\ [1mm] \ce{Cr2O3} (PT magnet): ${^{\text{-}1}}\bar{3}{^1}m^{\infty_{[001]}m}1\ (\mathrm{PT}) \to \bar{3}'m'\ (\mathrm{PT})$};
\end{tikzpicture}
\caption{Overview of the strong/weak $X$-wave classification together with the change in the PT class for all MPGs of collinear magnets, based on their SPG--MPG correspondences.
The blue arrows indicate pathways leading to strong $X$-wave magnets or PT magnets, whereas the red arrows indicate pathways leading to weak $X$-wave magnets.
Representative materials are shown for each category, together with their SPGs and MPGs obtained using \textit{FindSpinGroup} \cite{Yu2026-zs}.
Note that the $i$-wave class under SPGs has no counterpart under MPGs (see the SM \cite{Supplemental}).
}
\label{fig:Strong_Weak_magnets}
\end{figure*}
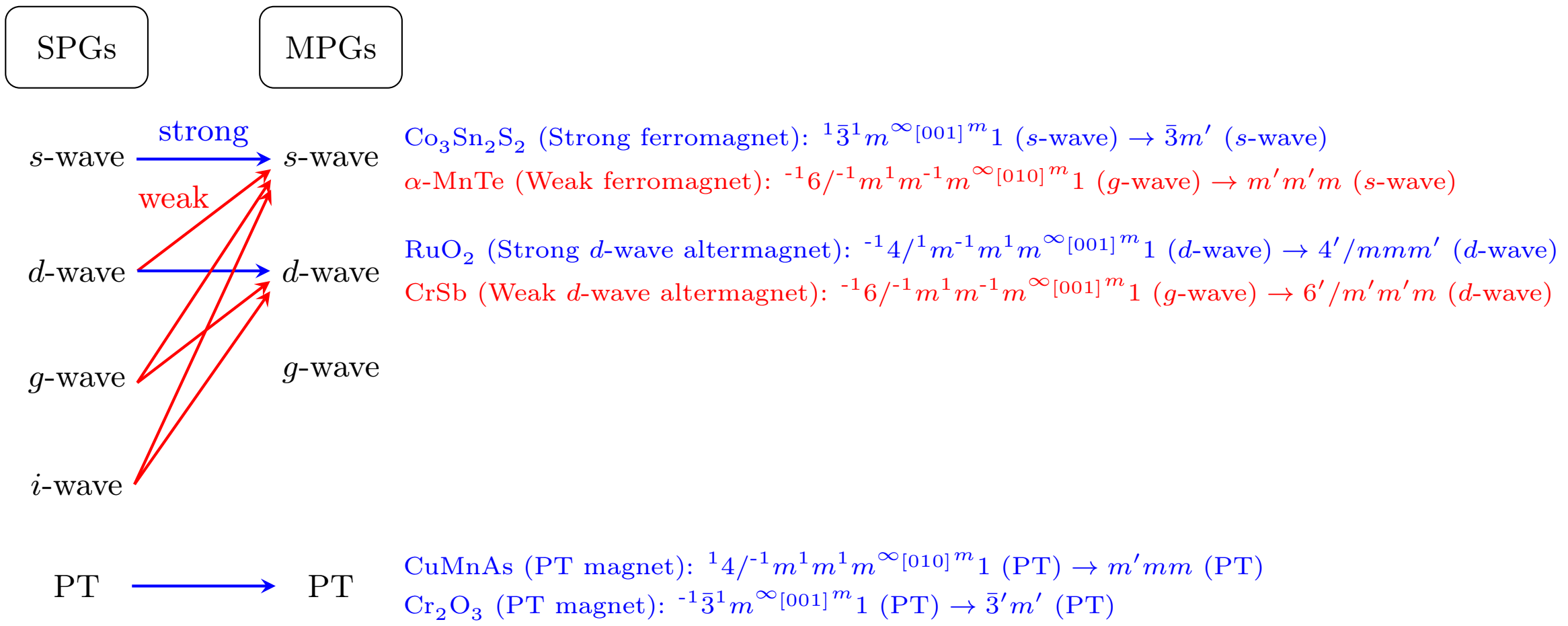

The strong/weak $X$-wave classification ($X=s,d,g,i$) categorizes collinear magnets belonging to an $X$-wave class under an MPG as strong or weak $X$-wave magnets,
with the strong/weak distinction determined by comparison with the corresponding SPG.
Specifically, for a strong $X$-wave magnet, the corresponding SPG belongs to the same $X$-wave class as its MPG,
whereas for a weak $X$-wave magnet, the corresponding SPG belongs to a higher-rank wave class than the $X$-wave class of its MPG (Fig.~\ref{fig:Strong_Weak_magnets}).
Thus, the $X$-wave character under the MPG persists in the absence of SOC in strong $X$-wave magnets,
whereas it emerges only in the presence of SOC in weak $X$-wave magnets.\footnote{
The terms ``strong'' and ``weak'' have previously been used in the classification of altermagnets under MPGs based on whether spin splitting persists in the absence of SOC \cite{Cheong2025-bt}.
Here, we use these terms with a different criterion: whether the $X$-wave character under the MPG persists in the absence of SOC.
}
Examples of strong and weak $X$-wave magnets can be found among the materials listed in Table~\ref{tab:SPG_MPG_relation}:
\ce{RuO2} is a strong $d$-wave altermagnet, \ce{CrSb} is a weak $d$-wave altermagnet, and $\alpha$-\ce{MnTe} is a weak $s$-wave magnet (i.e., a weak ferromagnet).

An important advantage of the strong/weak $X$-wave classification is that, in many cases, it provides a systematic framework for characterizing symmetry-allowed physical responses and their SOC dependence.
Specifically, the $X$-wave class determines which physical responses are symmetry-allowed based on its lowest-rank magnetic multipoles, while
the strong/weak distinction can characterize the SOC dependence of these responses.
As an example, we consider strong and weak $d$-wave altermagnets.
In both cases, their $d$-wave character implies that magnetic octupoles are the lowest-rank symmetry-allowed magnetic multipoles, thereby forbidding dipolar MOA while allowing octupolar MOA.
The strong/weak distinction further characterizes the SOC dependence of octupolar MOA:
it does not require SOC in strong $d$-wave altermagnets, whereas it requires SOC in weak $d$-wave altermagnets.
Recasting this result in terms of the SPG-based classification of altermagnets, we see that octupolar MOA in $d$-wave altermagnets does not require SOC, 
while SOC makes octupolar MOA symmetry-allowed in $g$- and $i$-wave altermagnets that are classified as weak $d$-wave altermagnets under their MPGs.
Thus, the strong/weak $X$-wave classification unifies collinear magnets that have different $X$-wave classes under SPGs but the same $X$-wave class under MPGs, thereby clarifying their response properties.

To fully realize the utility of this strong/weak $X$-wave classification, we now complete the classification by systematically screening all SPG--MPG correspondences.
The results are summarized in Fig.~\ref{fig:Strong_Weak_magnets},
together with the change in the PT class from SPGs to MPGs (see the SM \cite{Supplemental} for the full classification).
Figure~\ref{fig:Strong_Weak_magnets} reveals that strong $X$-wave magnets are not realized in all $X$-wave classes under SPGs.
Specifically, they occur only in the $s$- and $d$-wave classes and are absent in the $g$- and $i$-wave classes. 
In other words, $g$- and $i$-wave altermagnets under SPGs necessarily belong to lower-rank $s$- or $d$-wave classes under their MPGs,
thereby allowing lower-rank responses induced by SOC, as exemplified by dipolar MOA in $\alpha$-\ce{MnTe} \cite{Gray2024-us,Amin2024-ay,Yamamoto2025-kp,Yang2026-xr,Watanabe2026-ek,Yang2026-rw} and octupolar MOA in \ce{CrSb}.
In contrast, the PT class is preserved from the SPG to the MPG.
We refer to collinear magnets in this class as PT magnets, distinguishing them from strong and weak $X$-wave magnets.
This distinction completes the overall classification for all MPGs of collinear magnets based on their SPG--MPG correspondences (Fig.~\ref{fig:Strong_Weak_magnets}).

In light of the above discussion, we finally clarify our use of the term ``altermagnet''.
Conventionally, this term takes the SPG as its starting point:
the $X$-wave class is determined by the SPG itself, and the resulting $X$-wave designation is retained when SOC is included, irrespective of the corresponding MPG \cite{Smejkal2022-xr,Smejkal2022-xx}.
In contrast, the term ``altermagnet'' within the strong/weak $X$-wave classification takes the MPG as its starting point:
the $X$-wave class is determined by the MPG itself, and its strong/weak distinction is determined by comparison with the corresponding SPG.
Thus, the two usages emphasize different aspects of the same collinear magnet, and either can be adopted depending on the context.
In the following, we use the term ``altermagnet'' in its conventional SPG-based sense unless otherwise noted,
and explicitly use the term in its MPG-based sense within the strong/weak $X$-wave classification when discussing the role of SOC.

%%%%%%%%%%%%%%%%%%%%%%%%%%%%%%%%%%%%%%%%%%%%%%%%%%%%%%%%%%%%%%%%%%%%%%%%%%%%%%%%%%%%%%%%%%%%%%%%%%%%%%%%%%%%%%%%%%%%%%%%%%%%%%%%%%%%

\section{Complex Faraday angle for octupolar MOA} \label{Sec:Faraday_Rotation}

In this section, we derive analytical expressions for the complex Faraday angle to enable a quantitative evaluation of octupolar MOA.
We first derive a general expression for the complex Faraday angle in Sec.~\ref{Subsec:Faraday_angle_definition}.
We then apply this general expression to Type-I and Type-II optical activity in Secs.~\ref{Subsec:Faraday_Rotation_TypeIandII} and \ref{Subsec:Faraday_Rotation_explicit}, respectively.

%%%%%%%%%%%%%%%%%%%%%%%%%%%%%%%%%%%%%%%%%%%%%%%%%%%%%%%%%%%%%%%%%%

\subsection{General expression for the complex Faraday angle}  \label{Subsec:Faraday_angle_definition}

To derive the general expression, we first explain how to assess optical activity.
As a starting point, we consider linearly polarized light.
When interacting with an optically active medium, this light generally becomes elliptically polarized,
with its polarization state described by the rotation angle $\Theta$ and the ellipticity $\Delta$.
More precisely, $\Theta$ is the angle between the initial polarization direction and the major axis, 
while $\Delta$ is the ratio of the minor to major semiaxes.
Thus, determining $\Theta$ and $\Delta$ allows us to assess optical activity.

We next obtain explicit expressions for $\Theta$ and $\Delta$ from the electric-field components of elliptically polarized light,
\begin{align}
    E_{x/y}=\mathrm{Re} \bigl[ J_{x/y} e^{i\phi_{x/y}}e^{-i\omega t} \bigr]. \label{eq:ellipse_general_time}
\end{align}
Here, $J_{x/y}$ and $\phi_{x/y}$ are the corresponding real amplitudes and phases, respectively.
In terms of these amplitudes and phases, $\Theta$ and $\Delta$ are given by (see Appendix~\ref{Subsec:AppD1} for the derivation)
\begin{subequations}  \label{eq:Faraday_angle_general} 
    \begin{align}
        \tan 2\Theta &= \frac{-J^2_{x-y}\sin 2\theta + 2J_xJ_y \cos \delta \cos 2\theta}{J^2_{x-y}\cos 2\theta + 2J_xJ_y \cos \delta \sin 2\theta}, \label{eq:Faraday_angle_general_angle} \\
        \Delta^2 &= \frac{J^2_{x+y}-\sqrt{(J^2_{x-y})^2+(2J_xJ_y\cos \delta)^2}}{J^2_{x+y}+\sqrt{(J^2_{x-y})^2+(2J_xJ_y\cos \delta)^2}},  \label{eq:Faraday_angle_general_ellipticity}
    \end{align}
\end{subequations}
where $J^2_{x \pm y}=J^2_x \pm J^2_y$, $\delta=\phi_y-\phi_x$, and $\theta$ is the angle between the initial linear polarization direction and the $x$ axis.
These expressions can be applied to general elliptically polarized light.
In particular, when applied to elliptically polarized light transmitted through a medium, they yield the Faraday rotation angle $\Theta_{\mathrm{F}}$ and the Faraday ellipticity $\Delta_{\mathrm{F}}$,
which together define the complex Faraday angle $\Psi_{\mathrm{F}}=\Theta_{\mathrm{F}}+i \Delta_{\mathrm{F}}$.

We now apply Eq.~\eqref{eq:Faraday_angle_general} to transmitted light to derive a general expression for the complex Faraday angle $\Psi_{\mathrm{F}}$.
To this end, we consider light propagating along the $z$ axis and express $J_{x/y}e^{i\phi_{x/y}}$ in Eq.~\eqref{eq:ellipse_general_time} as a superposition of two eigenmodes:
\begin{align}
    J_{x/y} e^{i\phi_{x/y}}= a^+_{x/y} e^{i q_+ z} + a^-_{x/y} e^{iq_- z}, \label{eq:Faraday_angle_superposition}
\end{align}
where $a^{\pm}_{x/y}$ and $q_{\pm}$ are the complex electric-field amplitudes and the complex wave numbers of the two eigenmodes, respectively.
Using these amplitudes and writing $q_{\pm}=\alpha_{\pm}+i\beta_{\pm}$, we obtain the following expressions for the quantities appearing in Eq.~\eqref{eq:Faraday_angle_general}:
\begin{subequations} \label{eq:ellipse_Faraday_transmitted}
    \begin{align}
        &J_xJ_y\cos \delta = \sum_{n,m=\pm} e^{-(\beta_n+\beta_m)z} A^{nm}_{xy}, \label{eq:ellipse_Faraday_transmitted_offdiag} \\
        &J^2_{x\mp y} = \sum_{n,m=\pm} e^{-(\beta_n+\beta_m)z} (A^{nm}_{xx} \mp A^{nm}_{yy}), \label{eq:ellipse_Faraday_transmitted_diag}
    \end{align}
\end{subequations}
where $A^{nm}_{ij}$ is defined as
\begin{align}
    A^{nm}_{ij}=\mathrm{Re} \bigl[a^n_i (a^m_j)^*e^{i(\alpha_n-\alpha_m)z} \bigr]. \label{eq:ellipse_Faraday_transmitted_Anmij}
\end{align}
Thus, substituting the expressions in Eq.~\eqref{eq:ellipse_Faraday_transmitted} into Eq.~\eqref{eq:Faraday_angle_general} yields 
a general expression for $\Psi_{\mathrm{F}}$.

Finally, we explain how to evaluate the general expression for $\Psi_{\mathrm{F}}$.
This requires determining the wave numbers $q_{\pm}$ and the amplitudes $a^{\pm}_{x/y}$ of the two eigenmodes.
The wave numbers $q_{\pm}$ are obtained from the refractive indices $n_{\pm}$ through the relation $q_\pm=(\omega/c)n_\pm$, where $n_{\pm}$ are determined by the eigenvalue equation.
For the amplitudes $a^{\pm}_{x/y}$, we impose the initial condition for the incident linear polarization,
\begin{align}
    a_x^++a_x^-=E_{\mathrm{in}} \cos \theta, \quad a_y^++a_y^-=E_{\mathrm{in}} \sin \theta,
\end{align}
where $E_{\mathrm{in}}$ is the complex amplitude of the incident electric field.
Combining this initial condition with the polarization ratios $r_{\pm}=(E_{0y}/E_{0x})_{\pm}$ of the two eigenmodes yields
\begin{subequations}
    \begin{align}
        a^{\pm}_x&=\mp E_{\mathrm{in}}(r_\mp \cos\theta-\sin\theta)/(r_+-r_-), \\
        a^{\pm}_y&=r_{\pm}a^{\pm}_x.
    \end{align}
\end{subequations}
Thus, $a^{\pm}_{x/y}$ can be obtained once $r_{\pm}$ are determined from the eigenvalue equation.
Note that the polarization ratios $r_{\pm}$ are the same whether they are calculated from $\bm{E}_0$ or $\bm{D}_0$,
since $E_{0y}/E_{0x}=D_{0y}/D_{0x}$ holds for each eigenmode, as follows from Eqs.~\eqref{eq:eigenequation_general} and \eqref{eq:constitutive relation3}.
Therefore, the polarization ratios obtained in terms of $\bm{D}_0$ in Sec.~\ref{Sec:Macroscopic_theory_MO} can be directly used as $r_{\pm}$,
thereby allowing us to evaluate $\Psi_{\mathrm{F}}$ for octupolar MOA.

%%%%%%%%%%%%%%%%%%%%%%%%%%%%%%%%%%%%%%%%%%%%%%%%%%%%%%%%%%%%%%%%%%

\subsection{Complex Faraday angle for Type-I optical activity}  \label{Subsec:Faraday_Rotation_TypeIandII}

In this subsection, we derive the complex Faraday angle $\Psi_{\mathrm{F}}$ for Type-I optical activity following the procedure described in the previous subsection.
In doing so, we use the refractive indices and polarization ratios given in Eq.~\eqref{eq:eigenvalue_MO1_MO}.
We further consider a sample of thickness $\ell$ and evaluate $\Psi_{\mathrm{F}}$ at $z=\ell$, retaining terms up to first order in $\tilde{G}^{(\mathrm{o})}_z$.
The resulting expression for $\Psi_{\mathrm{F}}$ is independent of the incident linear polarization angle $\theta$ and is given by
\begin{align}
    \Psi_{\mathrm{F}}&=-\frac{\pi \ell}{\lambda}(n_+-n_-), \label{eq:Faraday_TypeI}
\end{align}
where $\lambda=2\pi c/\omega$ is the vacuum wavelength of the incident light.
The detailed derivation is presented in Appendix~\ref{Subsec:AppD2}.

A notable feature of Eq.~\eqref{eq:Faraday_TypeI} is that it has the same form as the well-known expression for dipolar MOA \cite{Sato2022-tu}.
Thus, Type-I optical activity shares the same mechanism for polarization rotation and ellipticity as dipolar MOA, but differs in its magnetic origin.
This difference enters the complex Faraday angle through the refractive-index difference $n_+-n_-$,
as can be seen from the explicit form of Eq.~\eqref{eq:Faraday_TypeI} obtained using Eq.~\eqref{eq:eigenvalue_MO1_MO},
\begin{align}
    \Psi_{\mathrm{F}}&=i\frac{\pi \ell}{\lambda}\sqrt{\varepsilon^{(0)}}(\varepsilon^{(0)})^2\tilde{G}^{(\mathrm{o})}_z. \label{eq:Faraday_TypeI_Gz}
\end{align}
Indeed, this expression contains the $z$ component of the octupolar MOA vector, $\tilde{G}^{(\mathrm{o})}_z$.
Furthermore, its linear dependence on $\tilde{G}^{(\mathrm{o})}_z$ implies that $\Psi_{\mathrm{F}}$ changes sign upon reversal of the magnetic octupole order, $\tilde{G}^{(\mathrm{o})}_z \to -\tilde{G}^{(\mathrm{o})}_z$,
thereby enabling imaging of opposite magnetic-octupole domains.

%%%%%%%%%%%%%%%%%%%%%%%%%%%%%%%%%%%%%%%%%%%%%%%%%%%%%%%%%%%%%%%%%%

\subsection{Complex Faraday angle for Type-II optical activity}  \label{Subsec:Faraday_Rotation_explicit}

We next derive the complex Faraday angle $\Psi_{\mathrm{F}}$ for Type-II optical activity.
The procedure and conditions are the same as for Type-I optical activity, but we use the refractive indices and polarization ratios given in Eq.~\eqref{eq:eigenvalue_MO2_MO} with $\tilde{G}^{(\mathrm{S})}_{xy}=0$.
The resulting expression for $\Psi_{\mathrm{F}}$ generally contains the exponential factor $e^{\pm i \Delta q \ell}$, where $\Delta q=(2\pi/\lambda)\Delta n$ and $\Delta n=n_x-n_y$ is the refractive-index anisotropy.
This factor, combined with the dependence on the incident linear polarization angle $\theta$, complicates the expression.
We therefore present two explicit expressions under different conditions.
The first expression is obtained by expanding the exponential factor to first order in $\Delta q \ell$ under the assumption $|\Delta q \ell| \ll 1 $.
Although this approximation is not always valid, the expression is useful for understanding the basic behavior of Type-II optical activity, including its dependence on $\theta$.
The second expression is obtained by imposing CAA ($\theta=0^\circ$, $90^\circ$, $180^\circ$, or $270^\circ$) without expanding the exponential factor.
Although restricted to CAA, the expression is suitable for quantitative evaluation 
because it does not require $|\Delta q \ell| \ll 1 $.
Following these considerations, we discuss the two expressions below, with their detailed derivations presented in Appendix~\ref{Subsec:AppD2}.

%%%%%%%%%%%%%%%%%%%%%%%%%%%%%%%%%

\subsubsection{Expression for small $|\Delta q \ell|$}

For small $|\Delta q \ell|$, the complex Faraday angle is given by 
\begin{align}
    \Psi_{\mathrm{F}}&=-i\frac{2\pi \ell}{\lambda} \Delta n 
    \bigl( \underbrace{\cos\theta \sin\theta \vphantom{A_x^{(\mathrm{o})}\cos^2\theta+A_y^{(\mathrm{o})}\sin^2\theta}}_{\text{NMOA}}
    + \underbrace{\tilde{a}^{(\mathrm{o})}_x \cos^2\theta +\tilde{a}^{(\mathrm{o})}_y \sin^2\theta}_{\text{Octupolar MOA}} \bigr), \label{eq:Faraday_TypeII}
\end{align}
where $\tilde{a}^{(\mathrm{o})}_x$ and $\tilde{a}^{(\mathrm{o})}_y$ are defined as
\begin{align}
    \tilde{a}^{(\mathrm{o})}_x=\frac{\tilde{G}^{(\mathrm{o})}_z}{\eta^{(\mathrm{S})}_{xx}(\eta^{(\mathrm{S})}_{xx}-\eta^{(\mathrm{S})}_{yy})}, \quad 
    \tilde{a}^{(\mathrm{o})}_y=\frac{\tilde{G}^{(\mathrm{o})}_z}{\eta^{(\mathrm{S})}_{yy}(\eta^{(\mathrm{S})}_{xx}-\eta^{(\mathrm{S})}_{yy})}. \label{eq:Faraday_TypeII_A}
\end{align}
Equation~\eqref{eq:Faraday_TypeII} reveals two notable features.
First, the first term gives rise to NMOA whenever the linear polarization direction deviates from CAA.
In this case, $\Theta_{\mathrm{F}}$ and $\Delta_{\mathrm{F}}$ arise from $\mathrm{Im}(\Delta n)$ and $\mathrm{Re}(\Delta n)$, respectively,
with $\Delta n$ contributing to both uniform NMOA and Lorentz NMOA (see Eq.~\eqref{eq:eigenvalue_MO2_MO}).
In particular, uniform NMOA has recently attracted attention for its connection to the quantum metric \cite{Li2025-qu,Li2025-tx,Wenhao2026-ix}, as discussed briefly in Sec.~\ref{Subsec:OCT_derivations}.
Although also predicted in altermagnets \cite{Li2025-tx}, uniform NMOA is not specific to altermagnetism
but arises from the refractive-index anisotropy $\Delta n$.

Second, the remaining two terms give rise to octupolar MOA in the presence of NMOA.
Under CAA, however, the NMOA contribution vanishes, leaving only octupolar MOA, which changes sign under $\tilde{G}^{(\mathrm{o})}_z \to -\tilde{G}^{(\mathrm{o})}_z$, as in the Type-I case.
This sign reversal also allows us to isolate octupolar MOA for an arbitrary polarization angle $\theta$ by taking half the difference between measurements before and after reversal of the magnetic octupole order:
\begin{align}
    \Delta \Psi_{\mathrm{F}}&=\frac{\Psi_{\mathrm{F}}(\tilde{G}^{(\mathrm{o})}_z)-\Psi_{\mathrm{F}}(-\tilde{G}^{(\mathrm{o})}_z)}{2} \notag \\
    &=-i\frac{2\pi \ell}{\lambda} \Delta n (\tilde{a}^{(\mathrm{o})}_x \cos^2\theta +\tilde{a}^{(\mathrm{o})}_y \sin^2\theta).
\end{align}
Moreover, this procedure can eliminate the contribution from $\tilde{G}^{(\mathrm{S})}_{xy}$ even when it is included in Eq.~\eqref{eq:Faraday_TypeII_A},
suggesting that the same procedure should also be effective for Type-III optical activity.

%%%%%%%%%%%%%%%%%%%%%%%%%%%%%%%%%

\subsubsection{Expression under CAA}

Under CAA, the complex Faraday angle is given by 
\begin{subequations} \label{eq:Faraday_TypeII_numerical}
    \begin{align}
        \Psi^x_{\mathrm{F}} &= -\tilde{a}^{(\mathrm{o})}_x \bigl[ 1-e^{-i(2\pi \ell/\lambda)\Delta n} \bigr], \\
        \Psi^y_{\mathrm{F}} &= \tilde{a}^{(\mathrm{o})}_y \bigl[ 1-e^{i(2\pi \ell/\lambda)\Delta n} \bigr], 
    \end{align}
\end{subequations}
where $\Psi^x_{\mathrm{F}}$ and $\Psi^y_{\mathrm{F}}$ correspond to incident linear polarization along the $x$ ($\theta=0^\circ$ or $180^\circ$)
and $y$ ($\theta=90^\circ$ or $270^\circ$) axes, respectively.
These expressions show that only octupolar MOA remains under CAA even without assuming $|\Delta q\ell|\ll 1$,
consistent with the conclusion drawn from Eq.~\eqref{eq:Faraday_TypeII}.
Moreover, $\Psi^x_{\mathrm{F}}$ and $\Psi^y_{\mathrm{F}}$ encompass the expression in Eq.~\eqref{eq:Faraday_TypeII} under CAA,
as can be seen from the following expressions obtained by Taylor-expanding the exponential factors:
\begin{subequations}  \label{eq:Faraday_TypeII_numerical_expansion}
    \begin{align}
        \Psi^x_{\mathrm{F}} &= \tilde{a}^{(\mathrm{o})}_x \sum_{m=1}^{\infty} \frac{1}{m!} \Bigl(-i\frac{2 \pi \ell}{\lambda}\Delta n \Bigr)^m, \\
        \Psi^y_{\mathrm{F}} &= -\tilde{a}^{(\mathrm{o})}_y \sum_{m=1}^{\infty} \frac{1}{m!} \Bigl(i\frac{2 \pi \ell}{\lambda}\Delta n \Bigr)^m.
    \end{align}
\end{subequations}
Indeed, retaining only terms up to first order in $\Delta q\ell$ reproduces Eq.~\eqref{eq:Faraday_TypeII} at $\theta=0^\circ$ and $90^\circ$.
Thus, under CAA, Eq.~\eqref{eq:Faraday_TypeII_numerical} generalizes Eq.~\eqref{eq:Faraday_TypeII} beyond the $|\Delta q\ell|\ll 1$ approximation,
allowing quantitative evaluation of octupolar MOA.

%%%%%%%%%%%%%%%%%%%%%%%%%%%%%%%%%%%%%%%%%%%%%%%%%%%%%%%%%%%%%%%%%%%%%%%%%%%%%%%%%%%%%%%%%%%%%%%%%%%%%%%%%%%%%%%%%%%%%%%%%%%%%%%%%%%%

\section{Microscopic theory of octupolar MOA} \label{Sec:Microscopic_theory_MO}

In this section, we derive a microscopic expression for octupolar MOA using the Kubo formula.
Specifically, our goal is to determine the second-order expansion coefficient $\varepsilon^{(\mathrm{A})}_{ij,kl}(\omega)$ of the permittivity tensor.
However, since the Kubo formula yields the current response rather than the polarization response, 
we instead begin with the optical conductivity tensor $\sigma_{ij}(\bm{q},\omega)$ and expand it in powers of $\bm{q}$ as 
\begin{align} 
    \sigma_{ij}(\bm{q},\omega)=\sigma_{ij}(\omega)+\sigma_{ij,k}(\omega)q_k+\sigma_{ij,kl}(\omega)q_kq_l + \cdots.
\end{align}
We then derive microscopic expressions for the expansion coefficients and relate them to the corresponding expansion coefficients of the permittivity tensor through
\begin{subequations}
    \begin{align}
        &\varepsilon_{ij}(\omega)=\delta_{ij}+\frac{i}{\varepsilon_0 \omega} \sigma_{ij}(\omega), 
    \end{align}
    for the zeroth-order expansion coefficient, and 
    \begin{align}
        &\varepsilon_{ij,k \cdots}(\omega)=\frac{i}{\varepsilon_0 \omega} \sigma_{ij,k \cdots}(\omega), \label{eq:epsilon_sigma_transformation2}
    \end{align}
    for all higher-order expansion coefficients.
\end{subequations}

Following this approach, we derive a microscopic expression for $\sigma^{(\mathrm{A})}_{ij,kl}(\omega)$ in Sec.~\ref{Subsec:OCT_derivations}.
We also derive microscopic expressions for the remaining expansion coefficients up to second order.
This enables a systematic comparison of their microscopic structures and
further provides the basis for a unified discussion of their physical significance in the following two subsections.
Specifically, Sec.~\ref{Subsec:Multipole_field_theory} develops a unified physical interpretation of the $\mathcal{T}$-odd expansion coefficients, including $\sigma^{(\mathrm{A})}_{ij,kl}(\omega)$, while
Sec.~\ref{Subsec:Sum_rule} derives universal $f$-sum rules satisfied by these coefficients, thereby establishing their relation to magnetic multipole moments formulated within the modern theory of magnetic multipoles.

%%%%%%%%%%%%%%%%%%%%%%%%%%%%%%%%%%%%%%%%%%%%%%%%%%%%%%%%%%%%%%%%%%

\subsection{Microscopic expressions for the expansion coefficients} \label{Subsec:OCT_derivations}

To derive the microscopic expressions, we first present a general expression for the optical conductivity tensor $\sigma_{ij}(\bm{q},\omega)$ in the Bloch representation.
In this representation, the Bloch Hamiltonian $\hat{H}_{\bm{k}}$ satisfies $\hat{H}_{\bm{k}}\ket{u_{n\bm{k}}}=\varepsilon_{n\bm{k}}\ket{u_{n\bm{k}}}$, where 
$\ket{u_{n\bm{k}}}$ is the cell-periodic part of the Bloch wave function in band $n$ with crystal momentum $\bm{k}$, and $\varepsilon_{n\bm{k}}$ is the corresponding band energy.
The optical conductivity tensor is then given by \cite{Malashevich2010-il,Zhong2016-kp,Pozo_Ocana2023-fk}
\begin{align}
    \sigma_{ij}(\bm{q},\omega)& =-ie^2 \hbar \sum_{n,m} \int_{\bm{k}} \frac{f_{n\bm{k}-\bm{q}/2}-f_{m\bm{k}+\bm{q}/2}}{\varepsilon_{n\bm{k}-\bm{q}/2}-\varepsilon_{m\bm{k}+\bm{q}/2}} \notag \\
    &\quad \times \frac{V^i_{nm}(\bm{k},\bm{q})V^j_{mn}(\bm{k},-\bm{q})}{\varepsilon_{n\bm{k}-\bm{q}/2}-\varepsilon_{m\bm{k}+\bm{q}/2}+\hbar\omega+i\delta}, \label{eq:sigma_kubo_general}
\end{align}
where $\int_{\bm{k}} \coloneqq \int_{\mathrm{BZ}} d^d k/(2\pi)^d$ denotes integration over the first Brillouin zone (BZ) in $d$-dimensional systems, 
and the other symbols are defined as follows.
The quantities $e$, $\hbar$, and $\delta$ are the electron charge, the reduced Planck constant, and a positive infinitesimal ensuring causality, respectively.
The function $f_{n\bm{k}}=[1+e^{(\varepsilon_{n\bm{k}}-\mu)/k_{\mathrm{B}}T}]^{-1}$ is the Fermi distribution function with chemical potential $\mu$, temperature $T$, and Boltzmann constant $k_\mathrm{B}$,
and $\bm{V}_{nm}(\bm{k},\bm{q})$ is the matrix element of the velocity operator $\hat{\bm{v}}_{\bm{k},\bm{q}}$, defined by
\begin{align}
    \bm{V}_{nm}(\bm{k},\bm{q})=\braket{u_{n\bm{k}-\bm{q}/2}|\hat{\bm{v}}_{\bm{k},\bm{q}}|u_{m\bm{k}+\bm{q}/2}}.
\end{align}
This velocity operator $\hat{\bm{v}}_{\bm{k},\bm{q}}$ is decomposed into orbital and spin contributions as \cite{Malashevich2010-il,Zhong2016-kp,Pozo_Ocana2023-fk}
\begin{align}
    \hat{\bm{v}}_{\bm{k},\bm{q}}=\hat{\bm{v}}_{\bm{k}}+\frac{i g_s}{2m} \bm{q} \times \hat{\bm{s}}, \label{eq:vq_decomposition}
\end{align}
where the first term corresponds to the orbital contribution, with the Bloch velocity operator $\hat{\bm{v}}_{\bm{k}}=\hbar^{-1} \partial \hat{H}_{\bm{k}}/\partial\bm{k}$, while 
the second term corresponds to the spin contribution, with the spin operator $\hat{\bm{s}}$,  the electron spin $g$ factor $g_s(\approx 2)$, and the electron mass $m$.
For later convenience, we define the Bloch velocity and spin matrix elements as $\bm{v}_{nm}(\bm{k})=\braket{u_{n\bm{k}}|\hat{\bm{v}}_{\bm{k}}|u_{m\bm{k}}}$ 
and $\bm{s}_{nm}(\bm{k})=\braket{u_{n\bm{k}}|\hat{\bm{s}}|u_{m\bm{k}}}$, respectively.

We next examine how the decomposition in Eq.~\eqref{eq:vq_decomposition} is reflected in the expansion coefficients $\sigma_{ij}(\omega)$, $\sigma_{ij,k}(\omega)$, and $\sigma_{ij,kl}(\omega)$.
The decomposition yields three operator contributions: 
the orbital-orbital contribution $\hat{\bm{v}}_{\bm{k}}\hat{\bm{v}}_{\bm{k}}$, the orbital-spin mixed contribution $\hat{\bm{v}}_{\bm{k}}(\bm{q} \times \hat{\bm{s}})+(\bm{q} \times \hat{\bm{s}})\hat{\bm{v}}_{\bm{k}}$,
and the spin-spin contribution $(\bm{q} \times \hat{\bm{s}})^2$.
Since these contributions appear at different orders in $\bm{q}$, they contribute differently to the expansion coefficients.
Specifically, $\sigma_{ij}(\omega)$ contains only the orbital-orbital contribution, whereas 
$\sigma_{ij,k}(\omega)$ and $\sigma_{ij,kl}(\omega)$ additionally contain the orbital-spin mixed contribution, with $\sigma_{ij,kl}(\omega)$ further containing the spin-spin contribution.
In the following, we consider the orbital-orbital contribution for $\sigma_{ij}(\omega)$ and the orbital-spin mixed contribution for $\sigma_{ij,k}(\omega)$ and $\sigma_{ij,kl}(\omega)$,
neglecting the other contributions to the latter two coefficients for simplicity.
With this simplification, it is convenient to introduce response tensors $\chi_{\mu,j}(\omega)$ and $\chi_{\mu,jk}(\omega)$,
through which $\sigma_{ij,k}(\omega)$ and $\sigma_{ij,kl}(\omega)$ are expressed as
\begin{subequations} \label{eq:1st_2nd_transformation} 
    \begin{align}
        \sigma^{(\mathrm{S/A})}_{ij,k}(\omega)&=ie_{ik\mu} \chi^{(\mathrm{o/e})}_{\mu,j}(\omega) \pm (i \leftrightarrow j), \label{eq:1st_transformation} \\
        \sigma^{(\mathrm{S/A})}_{ij,kl}(\omega)&=e_{il\mu} \chi^{(\mathrm{e/o})}_{\mu,jk}(\omega) \pm (i \leftrightarrow j), \label{eq:2nd_transformation}
    \end{align}
\end{subequations}
where the superscript (e/o) denotes the corresponding $\mathcal{T}$-even/$\mathcal{T}$-odd component, the $+$ ($-$) sign corresponds to the symmetric (antisymmetric) component, 
and $(i \leftrightarrow j)$ denotes the preceding term with $i$ and $j$ interchanged.

We now derive microscopic expressions for $\sigma_{ij}(\omega)$, $\chi_{\mu,j}(\omega)$, and $\chi_{\mu,jk}(\omega)$.
Their derivation, however, involves different treatments of the $\bm{q}$ expansion in Eq.~\eqref{eq:sigma_kubo_general}.
Specifically, $\sigma_{ij}(\omega)$ and $\chi_{\mu,j}(\omega)$ require no $\bm{q}$ expansion, 
whereas $\chi_{\mu,jk}(\omega)$ requires an explicit $\bm{q}$ expansion, resulting in a more intricate structure.
We therefore first derive the expressions for $\sigma_{ij}(\omega)$ and $\chi_{\mu,j}(\omega)$ together, and then turn to the expression for $\chi_{\mu,jk}(\omega)$.
Each expression is separated into intraband and interband contributions, with the detailed derivation presented in Appendix~\ref{Sec:AppE}.

%%%%%%%%%%%%%%%%%%%%%%%%%%%%%%%%%

\subsubsection{Zeroth- and first-order expansion coefficients}

The intraband contributions to $\sigma_{ij}(\omega)$ and $\chi_{\mu,j}(\omega)$ are given by 
$\sigma^{(\mathrm{A}),\mathrm{intra}}_{ij}(\omega)=\chi^{(\mathrm{o}),\mathrm{intra}}_{\mu,j}(\omega)=0$ and 
\begin{subequations} \label{eq:band_contribution_01}
    \begin{align}
        \! \! \! \sigma^{(\mathrm{S}),\mathrm{intra}}_{ij}(\omega) &= e^2 \frac{\tau}{1-i\omega \tau} \sum_{n} \int_{\bm{k}}  (-f'_n) \mathrm{Re}(v^i_n v^j_n), \label{eq:band_contribution_01_sigma} \\
        \! \! \! \chi^{(\mathrm{e}),\mathrm{intra}}_{\mu,j}(\omega) &= \frac{e^2 g_s}{2m} \frac{\tau}{1-i\omega \tau}  \sum_{n} \int_{\bm{k}}  (-f'_n) \mathrm{Re}(s^\mu_n v^j_n), \label{eq:band_contribution_01_chi} 
    \end{align}
\end{subequations}
where $f'_{n}=\partial f_{n}/\partial \varepsilon_{n}$, $v^i_n=v^i_{nn}$, $s^\mu_n=s^\mu_{nn}$, $\tau=\hbar/\delta$ is the effective relaxation time, 
and the explicit $\bm{k}$ dependence is omitted hereafter unless otherwise noted.
The nonzero contributions in Eq.~\eqref{eq:band_contribution_01} correspond to the $\mathcal{T}$-even components and exhibit characteristic Drude-type responses, 
making them particularly relevant in the DC limit of metals.

The interband contributions to $\sigma_{ij}(\omega)$ and $\chi_{\mu,j}(\omega)$ are given by 
\begin{subequations} \label{eq:geometric_contribution_01}
    \begin{align}
        \! \sigma^{\mathrm{inter}}_{ij}(\omega) &= -i\frac{e^2}{\hbar} \sum_{n,m (\neq n)} \int_{\bm{k}}  T^{ij}_{nm} f_{nm} \frac{\varepsilon_{nm}}{\varepsilon_{nm}+\hbar\tilde{\omega}}, \label{eq:geometric_contribution_01_sigma} \\
        \! \chi^{\mathrm{inter}}_{\mu,j}(\omega) &= -\frac{e^2 g_s}{2m} \sum_{n,m (\neq n)} \int_{\bm{k}} \tilde{T}^{\mu,j}_{nm} f_{nm} \frac{1}{\varepsilon_{nm}+\hbar\tilde{\omega}}, \label{eq:geometric_contribution_01_chi}
    \end{align}
\end{subequations}
where $\tilde{\omega}=\omega+i\delta/\hbar$, $\varepsilon_{nm}=\varepsilon_{n}-\varepsilon_{m}$, and $f_{nm}=f_{n}-f_{m}$.
These expressions are written in terms of the band-resolved quantum geometric tensor $T^{ij}_{nm}$ \cite{Provost1980-qb,Berry1984-fn} and 
the band-resolved Zeeman quantum geometric tensor $\tilde{T}^{\mu,j}_{nm}$ \cite{Xiang2025-cl,Xiang2025-xi}:
\begin{subequations}
    \begin{align}
        T^{ij}_{nm}&=\braket{D_i u_{n}|u_{m}} \braket{u_{m}|D_j u_{n}}, \\
        \tilde{T}^{\mu,j}_{nm}&=i\braket{u_{n}|\hat{s}^\mu|u_{m}} \braket{u_{m}|D_j u_{n}},
    \end{align}
\end{subequations}
where $\ket{D_i u_n}=(1-\ket{u_n}\bra{u_n})\ket{\partial_i u_n}$ is the covariant derivative of band $n$ with $\partial_i=\partial/\partial k_i$.
These tensors are further decomposed into their real and imaginary parts as
\begin{subequations}
\begin{align}
    T^{ij}_{nm}=g^{ij}_{nm} -\frac{i}{2} \Omega^{ij}_{nm}, \quad 
    \tilde{T}^{\mu,j}_{nm} =\tilde{g}^{\mu,j}_{nm} - \frac{i}{2} \tilde{\Omega}^{\mu,j}_{nm},
\end{align}
\end{subequations}
where $g^{ij}_{nm}$ ($\tilde{g}^{\mu,j}_{nm}$) and $\Omega^{ij}_{nm}$ ($\tilde{\Omega}^{\mu,j}_{nm}$) are the band-resolved (Zeeman) quantum metric and Berry curvature, respectively.
Substituting these decompositions into Eq.~\eqref{eq:geometric_contribution_01} establishes 
the following correspondence between the $\mathcal{T}$-even/$\mathcal{T}$-odd components and the band-resolved quantum geometric quantities:
\begin{subequations} \label{eq:correspondence_response_geometry_01}
\begin{align}
    \sigma^{(\mathrm{S}),\mathrm{inter}}_{ij} &\leftrightarrow g^{ij}_{nm}, & \sigma^{(\mathrm{A}),\mathrm{inter}}_{ij} &\leftrightarrow \Omega^{ij}_{nm}, \\
    \chi^{(\mathrm{e}),\mathrm{inter}}_{\mu,j}&\leftrightarrow \tilde{\Omega}^{\mu,j}_{nm}, & \chi^{(\mathrm{o}),\mathrm{inter}}_{\mu,j}&\leftrightarrow \tilde{g}^{\mu,j}_{nm}.
\end{align}  
\end{subequations}
In particular, relating $\sigma^{(\mathrm{S}),\mathrm{inter}}_{ij}$ and $\sigma^{(\mathrm{A}),\mathrm{inter}}_{ij}$ to their corresponding optical activities 
reveals that the quantum metric and Berry curvature govern the interband contributions to uniform NMOA and dipolar MOA, respectively.

%%%%%%%%%%%%%%%%%%%%%%%%%%%%%%%%%

\subsubsection{Second-order expansion coefficient}

The intraband contribution to $\chi_{\mu,jk}(\omega)$ is given by 
\begin{subequations} \label{eq:band_contribution_2}
    \begin{align}
        \chi^{(\mathrm{e}),\mathrm{intra}}_{\mu,jk}(\omega) &= \frac{e^2 g_s}{2m \hbar} \frac{\tau}{1-i\omega \tau} \sum_{n} \int_{\bm{k}}  \notag \\
        & \quad \times \Bigl( f_n \partial_j \tilde{g}^{\mu,k}_{n} + e^{-1} \hbar f'_n s^\mu_n m^{jk}_{n} \Bigr), \\
        \chi^{(\mathrm{o}),\mathrm{intra}}_{\mu,jk}(\omega) &= \frac{e^2 g_s}{2m} \Bigl( \frac{\tau}{1-i\omega \tau} \Bigr)^2\sum_{n} \int_{\bm{k}}  \notag \\
        & \quad \times (-f'_n)\mathrm{Re}(s^\mu_n v^j_n v^k_n). \label{eq:band_contribution_2_o}
    \end{align}
\end{subequations}
Unlike the intraband contributions to $\sigma_{ij}(\omega)$ and $\chi_{\mu,j}(\omega)$,
those to $\chi_{\mu,jk}(\omega)$ are present in both the $\mathcal{T}$-even and $\mathcal{T}$-odd components.
The $\mathcal{T}$-odd component scales as $\tau^2$ in the low-frequency limit, resembling the Drude contribution 
in nonlinear photocurrent responses \cite{Holder2020-qh,Watanabe2021-kw} and nonlinear nonreciprocal transport \cite{Ideue2017-ee,Watanabe2020-bs,Zhao2024-bu}.
In contrast, the $\mathcal{T}$-even component scales as $\tau$ in the low-frequency limit and involves two band-geometric quantities: 
the Zeeman quantum metric of band $n$, $\tilde{g}^{\mu,k}_n=\sum_{m} \tilde{g}^{\mu,k}_{nm}=\mathrm{Re}\bigl[i\braket{u_n|\hat{s}^\mu|D_k u_n} \bigr]$, and 
the orbital magnetic moment of band $n$, $m^{jk}_n=(e/\hbar)\mathrm{Im}\braket{D_j u_n|\hat{H}-\varepsilon_n|D_k u_n}$ \cite{Shi2007-xs}.
In particular, the component symmetric in $j$ and $k$ involves only the Zeeman quantum-metric-dipole contribution $\partial_j \tilde{g}^{\mu,k}_{n}$,
providing a counterpart to the Berry-curvature-dipole contribution 
in nonlinear photocurrent responses \cite{Moore2010-pm,Morimoto2016-dg,Holder2020-qh,Watanabe2021-kw} and the nonlinear Hall effect \cite{Sodemann2015-pu,Ma2019-oj,Watanabe2020-bs}.

The interband contribution to $\chi_{\mu,jk}(\omega)$ is given by 
\begin{subequations} \label{eq:geometric_contribution_2}
    \begin{align}
        \chi^{(\mathrm{e}),\mathrm{inter}}_{\mu,jk}(\omega) &= -i\frac{e^2 g_s}{4m}\sum_{n,m (\neq n)} \int_{\bm{k}} \Bigl[
        \mathrm{Re}\Bigl(\frac{\tilde{T}^{\mu,jk}_{nm}}{\varepsilon_{nm}} \Bigr) f_{nm} \notag \\
        & \quad +\tilde{g}^{\mu,j}_{nm}\hat{F}^k_{nm}(\omega) \Bigr] \frac{1}{\varepsilon_{nm}+\hbar \tilde{\omega}}, \\
        \chi^{(\mathrm{o}),\mathrm{inter}}_{\mu,jk}(\omega) &= -i\frac{e^2 g_s}{4m}\sum_{n,m (\neq n)} \int_{\bm{k}} \Bigl[
        i\mathrm{Im}\Bigl(\frac{\tilde{T}^{\mu,jk}_{nm}}{\varepsilon_{nm}} \Bigr) f_{nm} \notag \\
        & \quad -\frac{i}{2}\tilde{\Omega}^{\mu,j}_{nm}\hat{F}^k_{nm}(\omega) \Bigr] \frac{1}{\varepsilon_{nm}+\hbar \tilde{\omega}}, \label{eq:geometric_contribution_2_o}
    \end{align}
\end{subequations}
where $\hat{F}^k_{nm}(\omega)$ is defined by
\begin{align}
    \hat{F}^k_{nm}(\omega)=-\partial_k \tilde{f}_{nm}+\Bigl(2+ \omega \frac{d}{d \omega} \Bigr) \frac{f_{nm} \partial_k \tilde{\varepsilon}_{nm}}{\varepsilon_{nm}},
\end{align}
with $\tilde{\varepsilon}_{nm}=\varepsilon_{n}+\varepsilon_{m}$ and $\tilde{f}_{nm}=f_{n}+f_{m}$.
Here, $\tilde{T}^{\mu,jk}_{nm}$ is a gauge-invariant quantity defined by
\begin{align}
    \tilde{T}^{\mu,jk}_{nm} &= i\hbar \bigl[(-\braket{D_ku_n|\hat{s}^\mu|u_m}+\braket{u_n|\hat{s}^\mu|D_ku_m})v^j_{mn} \notag \\
    & \quad +s^\mu_{nm}(\braket{D_ku_m|\hat{v}^j|u_n}-\braket{u_m|\hat{v}^j|D_ku_n}) \bigr],
\end{align}
which has a band-geometric structure analogous to that arising from the $\bm{k}$-derivative of the Zeeman quantum geometric tensor $\tilde{T}^{\mu,j}_{nm}$.
In addition to $\tilde{T}^{\mu,jk}_{nm}$, Eq.~\eqref{eq:geometric_contribution_2} contains the Zeeman quantum-geometric quantities $\tilde{g}^{\mu,j}_{nm}$ and $\tilde{\Omega}^{\mu,j}_{nm}$.
These quantities establish the same correspondence between Zeeman quantum geometry and optical activity as in the zeroth-order case.
Specifically, the Zeeman quantum metric and Zeeman Berry curvature contribute to Lorentz NMOA and octupolar MOA, respectively.

%%%%%%%%%%%%%%%%%%%%%%%%%%%%%%%%%%%%%%%%%%%%%%%%%%%%%%%%%%%%%%%%%%

\subsection{Multipolar description of electromagnetic responses} \label{Subsec:Multipole_field_theory}

In this subsection, we develop a unified physical interpretation of the $\mathcal{T}$-odd expansion coefficients.
To this end, we first introduce two equivalent representations of the current response:
one in terms of the $\mathcal{T}$-odd optical conductivity tensor $\sigma^{(\mathrm{odd})}_{ij}(\bm{q},\omega)$, given by
\begin{subequations}
\begin{align}
    J_i (\bm{q},\omega) &=\sigma^{(\mathrm{odd})}_{ij}(\bm{q},\omega) E_j(\bm{q},\omega) \notag \\
    &= \bigl[\sigma^{(\mathrm{A})}_{ij}(\omega)+\sigma^{(\mathrm{S})}_{ij,k}(\omega)q_k + \sigma^{(\mathrm{A})}_{ij,kl}(\omega)q_kq_l + \cdots \bigr] \notag \\
    & \quad \times E_j(\bm{q},\omega), \label{eq:current_general_1}
\end{align}
and the other in terms of the induced electric and magnetic multipole moments, given by \cite{Raab2005}
\begin{align}
    J_i(\bm{q},\omega) &=(-i \omega) \bigl[P_i(\bm{q},\omega)+(-i q_k) P_{ik}(\bm{q},\omega) + \cdots \bigr] \notag \\
    & \quad + ie_{ik\mu} q_k \bigl[ M_\mu(\bm{q},\omega)+(-iq_l)M_{\mu,l}(\bm{q},\omega)+ \cdots \bigr]. \label{eq:current_general_2}
\end{align}
\end{subequations}
Here, $\bm{J}$ and $\bm{E}$ are the induced current density and the external electric field, respectively, while 
$P_i$ ($P_{ij}$) and $M_\mu$ ($M_{\mu,j}$) are the induced electric and magnetic dipole (quadrupole) moments, respectively.
Establishing the correspondence between these two representations at each expansion order reveals the physical interpretation of the $\mathcal{T}$-odd expansion coefficients.
In the following, we focus on the first- and second-order current responses and elucidate the physical interpretations of  
$\chi^{(\mathrm{o})}_{\mu,j}(\omega)$ and the symmetric part of $\chi^{(\mathrm{o})}_{\mu,jk}(\omega)$, respectively, where the symmetric part is defined as
\begin{align}
    \chi^{(\mathrm{o})}_{\mu,(jk)}(\omega) \coloneqq \bigl[\chi^{(\mathrm{o})}_{\mu,jk}(\omega)+\chi^{(\mathrm{o})}_{\mu,kj}(\omega)\bigr]/2.
\end{align}
The detailed discussion, including that of the remaining response coefficients, is presented in Appendix~\ref{Sec:AppF}; only the final results are given below.

These two response tensors can be physically interpreted as follows.
First, $\chi^{(\mathrm{o})}_{\mu,j}(\omega)$ describes the magnetoelectric response:
\begin{subequations}
    \begin{align}
        P_j(\bm{q},\omega)&=\chi^{(\mathrm{o})}_{\mu,j}(\omega)B_\mu(\bm{q},\omega), \\
        M_\mu(\bm{q},\omega)&=\chi^{(\mathrm{o})}_{\mu,j}(\omega)E_j(\bm{q},\omega), 
    \end{align}
\end{subequations}
where $\bm{B}$ is the external magnetic field.
Second, $\chi^{(\mathrm{o})}_{\mu,(jk)}(\omega)$ is found to describe two types of cross-correlated response: one is 
\begin{subequations} \label{eq:Octupolar_magnetoelectric_tensor_1}
    \begin{align}
        P_{jk}(\bm{q},\omega) &= - \chi^{(\mathrm{o})}_{\mu,(jk)}(\omega) B_\mu(\bm{q},\omega), \\
        M_\mu(\bm{q},\omega) &= - \chi^{(\mathrm{o})}_{\mu,(jk)}(\omega) E_{jk}(\bm{q},\omega),
    \end{align}
\end{subequations}
and the other is
\begin{subequations} 
    \begin{align} \label{eq:Octupolar_magnetoelectric_tensor_2}
        P_k(\bm{q},\omega) &= \chi^{(\mathrm{o})}_{\mu,(jk)} (\omega) B_{\mu,j}(\bm{q},\omega), \\
        M_{\mu,j}(\bm{q},\omega) &= \chi^{(\mathrm{o})}_{\mu,(jk)}(\omega)E_k(\bm{q},\omega).
    \end{align}
\end{subequations}
In the above expressions, we introduce the electric-field gradient $E_{ij}$ and the magnetic-field gradient $B_{\mu,j}$ as 
\begin{subequations} 
    \begin{align}
        E_{ij}(\bm{q},\omega) &\coloneqq i q_j E_i(\bm{q},\omega) \leftrightarrow \partial_{r_j}E_i(\bm r,\omega), \\
        B_{\mu,j}(\bm{q},\omega) &\coloneqq i q_j B_\mu(\bm{q},\omega) \leftrightarrow \partial_{r_j}B_\mu(\bm r,\omega),
    \end{align}
\end{subequations}
where the expressions on the right of $\leftrightarrow$ denote the corresponding real-space representations, with $\partial_{r_j}=\partial/\partial r_j$.

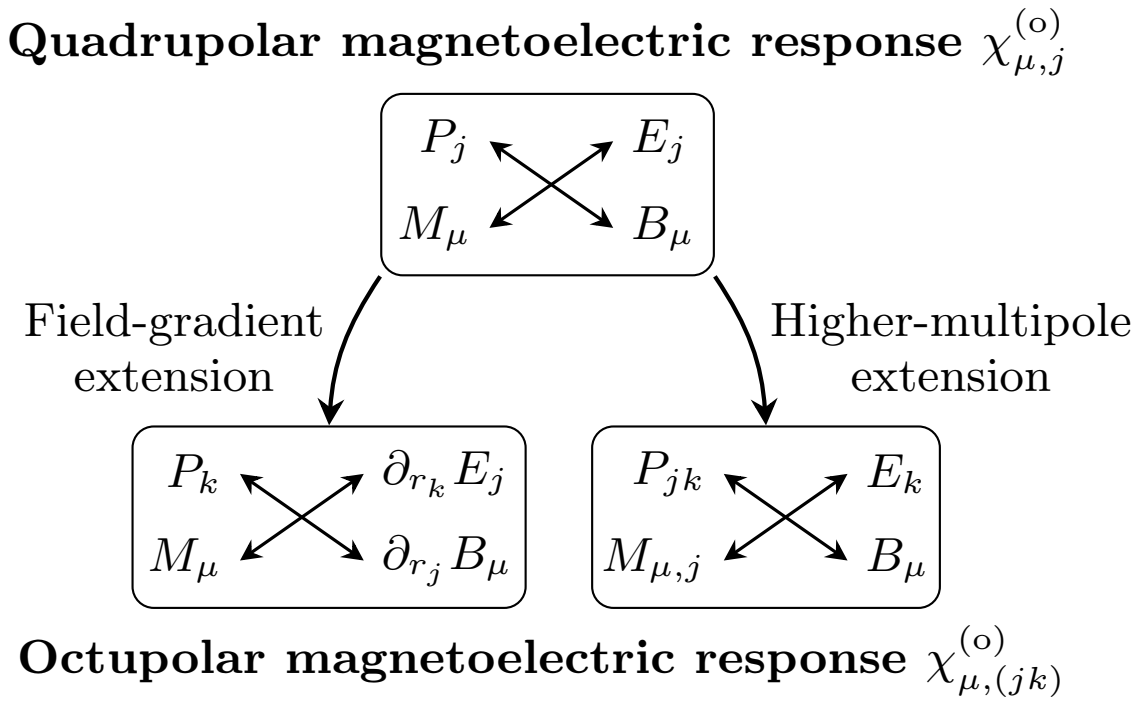
\begin{figure}[t]
    \centering
    \begin{tikzpicture}[
        >=Stealth,
        responsebox/.style={
            draw,
            rounded corners,
            minimum height=1.2cm,
            align=center
        },
        responsearrow/.style={
            -{Stealth[length=4pt,width=4pt]},
            line width=0.6pt
        },
        generalize/.style={
            -{Stealth[length=5pt,width=5pt]},
            line width=0.8pt
        }
    ]

    % ------------------------------------------------
    % Conventional magnetoelectric response
    % ------------------------------------------------
    \node[
        responsebox,
        minimum width=2.2cm
    ] (convbox) at (1.625,2.2) {};

    \node[
        above,
        align=center,
        font=\bfseries
    ] at (convbox.north) {
        Quadrupolar
        magnetoelectric response $\chi^{(\mathrm{o})}_{\mu,j}$
    };

    \coordinate (C0) at (1.65,2.2);

    \draw[responsearrow]
        (C0) -- ++(35:0.5)
        node[right] {$E_j$};

    \draw[responsearrow]
        (C0) -- ++(215:0.5)
        node[left] {$M_\mu$};

    \draw[responsearrow]
        (C0) -- ++(-35:0.5)
        node[right] {$B_\mu$};

    \draw[responsearrow]
        (C0) -- ++(145:0.5)
        node[left] {$P_j$};

    % ------------------------------------------------
    % Field-gradient responses
    % ------------------------------------------------
    \node[
        responsebox,
        minimum width=2.6cm
    ] (gradbox) at (0.18,0) {};

    \node[
        below,
        align=center,
        font=\bfseries
    ] at (gradbox.south) {
        % Field-gradient\\
        % responses
    };

    \coordinate (C1) at (0,0);

    \draw[responsearrow]
        (C1) -- ++(35:0.5)
        node[right] {$\partial_{r_k}E_j$};

    \draw[responsearrow]
        (C1) -- ++(215:0.5)
        node[left] {$M_\mu$};

    \draw[responsearrow]
        (C1) -- ++(-35:0.5)
        node[right] {$\partial_{r_j}B_\mu$};

    \draw[responsearrow]
        (C1) -- ++(145:0.5)
        node[left] {$P_k$};

    % ------------------------------------------------
    % Higher-multipole responses
    % ------------------------------------------------
    \node[
        responsebox,
        minimum width=2.3cm
    ] (multibox) at (3.07,0) {};

    \node[
        below,
        align=center,
        font=\bfseries
    ] at (multibox.south) {
        % Higher-multipole\\
        % responses
    };

    \coordinate (C2) at (3.2,0);

    \draw[responsearrow]
        (C2) -- ++(35:0.5)
        node[right] {$E_k$};

    \draw[responsearrow]
        (C2) -- ++(215:0.5)
        node[left] {$M_{\mu,j}$};

    \draw[responsearrow]
        (C2) -- ++(-35:0.5)
        node[right] {$B_\mu$};

    \draw[responsearrow]
        (C2) -- ++(145:0.5)
        node[left] {$P_{jk}$};

    % ------------------------------------------------
    % Generalization arrows
    % ------------------------------------------------
    \draw[generalize]
        (convbox.south west)
        to[bend right=15]
        node[left,align=center,font=\small] {
            Field-gradient\\
            extension
        }
        (gradbox.north);

    \draw[generalize]
        (convbox.south east)
        to[bend left=15]
        node[right,align=center,font=\small] {
            Higher-multipole \\
            extension
        }
        (multibox.north);

    % Overall interpretation
    \node[
        align=center,
        font=\small\bfseries
    ] at (1.65,-0.95) {
        Octupolar magnetoelectric response $\chi^{(\mathrm{o})}_{\mu,(jk)}$
    };

    % \node[
    %     align=center,
    %     font=\small\bfseries
    % ] at (1.65,-1.45) {
    %     Generalized magnetoelectric responses
    % };

    \end{tikzpicture}
    \caption{Schematic illustration of the higher-rank generalization from the magnetoelectric response $\chi^{(\mathrm{o})}_{\mu,j}$ to $\chi^{(\mathrm{o})}_{\mu,(jk)}$.
    The curved arrows indicate the field-gradient and higher-multipole extensions.
    }
    \label{fig:Multiferroics}
\end{figure}

The above results, summarized in Fig.~\ref{fig:Multiferroics}, reveal that $\chi^{(\mathrm{o})}_{\mu,(jk)}$ can be regarded as a higher-rank generalization of the magnetoelectric response $\chi^{(\mathrm{o})}_{\mu,j}$.
This generalization is achieved through two complementary extensions.
The first is a field-gradient extension, where the uniform electromagnetic fields are replaced by their spatial gradients.
The second is a higher-multipole extension, where the induced electric and magnetic dipole moments are replaced by the induced electric and magnetic quadrupole moments, respectively.
Accordingly, $\chi^{(\mathrm{o})}_{\mu,j}$ and $\chi^{(\mathrm{o})}_{\mu,(jk)}$ share the same fundamental role in describing magnetoelectric responses
but differ in the magnetic multipole orders that activate them.
Specifically, $\chi^{(\mathrm{o})}_{\mu,j}$ and $\chi^{(\mathrm{o})}_{\mu,(jk)}$ are activated by magnetic quadrupole and octupole orders, respectively (see Eq.~\eqref{eq:Multipole tensor_conductivity tensor}).
We therefore refer to these responses as the quadrupolar magnetoelectric response (QMER) and the octupolar magnetoelectric response (OMER), respectively (Fig.~\ref{fig:Multiferroics}).

Finally, we note two points regarding the above analysis.
First, the orbital-spin mixed contribution induces only spin magnetic multipole moments,
implying that the QMER and OMER derived here should be interpreted as the spin QMER and spin OMER, respectively.
Their orbital counterparts can be obtained from the orbital-orbital contribution, as already demonstrated for the orbital QMER \cite{Malashevich2010-il,Shinada2023-an,Pozo_Ocana2023-fk}.
Second, the multipolar description can also be applied to the $\mathcal{T}$-even optical conductivity tensor.
Accordingly, the $\mathcal{T}$-even response $\chi^{(\mathrm{e})}_{\mu,jk}$ is also expected to admit an analogous physical interpretation,
as already demonstrated for $\chi^{(\mathrm{e})}_{\mu,j}$ in terms of natural optical activity \cite{Landau-Lifshitz,Malashevich2010-il,Pozo_Ocana2023-fk,Shinada2023-uv,Shinada2024-pj} and the gyrotropic magnetic response \cite{Zhong2016-kp}.
A detailed analysis of the orbital OMER and the physical interpretation of $\chi^{(\mathrm{e})}_{\mu,jk}$ is left for future work.

%%%%%%%%%%%%%%%%%%%%%%%%%%%%%%%%%%%%%%%%%%%%%%%%%%%%%%%%%%%%%%%%%%

\subsection{$f$-sum rules} \label{Subsec:Sum_rule}

In this subsection, we derive universal $f$-sum rules satisfied by the $\mathcal{T}$-odd expansion coefficients for insulators at zero temperature.
These $f$-sum rules relate the $\mathcal{T}$-odd expansion coefficients to the corresponding magnetic multipole moments formulated within the modern theory of magnetic multipoles:
\begin{align}
    \braket{\chi} \coloneqq \int_{0}^{\infty} \mathrm{Im}\ \chi(\omega) d\omega \leftrightarrow M,
\end{align}
where $\chi(\omega)$ generically denotes a $\mathcal{T}$-odd expansion coefficient and $M$ denotes the corresponding magnetic multipole moment.
Specifically, $\sigma^{(\mathrm{A})}_{ij}(\omega)$, $\chi^{(\mathrm{o})}_{\mu,j}(\omega)$, and $\chi^{(\mathrm{o})}_{\mu,(jk)}(\omega)$ are related to
orbital magnetization $\bm{M}^{(\mathrm{orb})}$ \cite{Xiao2005-bu,Thonhauser2005-wp,Ceresoli2006-xy,Shi2007-xs}, 
the spin magnetic quadrupole moment $M^{(\mathrm{spin})}_{\mu,j}$ \cite{Gao2018-uk,Shitade2019-hq,Hua2026-rf}, and 
the spin magnetic octupole moment $M^{(\mathrm{spin})}_{\mu,(jk)}$ \cite{Oike2025-dk,Shitade2025-ue,Sato2026-vj,Hua2026-rf}, respectively.
In the following, we present their simplified expressions in the single-flat-band limit, where the occupied subspace consists of a single flat band with energy $\varepsilon_{\mathrm{occ}\bm{k}} \approx \varepsilon_{\mathrm{occ}}$.
The detailed derivations for the general case, together with St\v{r}eda formulas, are provided in Appendix~\ref{Sec:AppG}.

The resulting $f$-sum rules are given by
\begin{subequations} \label{eq:Sumrules_general}
\begin{align}
    M^{(\mathrm{orb})}_{ij} &= \frac{\hbar}{\pi e} \Bigl(\braket{\sigma^{(\mathrm{A})}_{ij}}+\frac{2 E_{\mathrm{occ}}}{\hbar}\braket{\omega^{-1}\sigma^{(\mathrm{A})}_{ij}} \Bigr), \\
    M^{(\mathrm{spin})}_{\mu,j} &= \frac{\hbar}{\pi e} \Bigl( \braket{\chi^{(\mathrm{o})}_{\mu,j}} + \frac{2 E_{\mathrm{occ}}}{\hbar} \braket{\omega^{-1}\chi^{(\mathrm{o})}_{\mu,j}} \Bigr), \\
    M^{(\mathrm{spin})}_{\mu,(jk)} &= \frac{\hbar}{\pi e} \Bigl( \braket{\chi^{(\mathrm{o})}_{\mu,(jk)}} + \frac{2 E_{\mathrm{occ}}}{\hbar} \braket{\omega^{-1}\chi^{(\mathrm{o})}_{\mu,(jk)}} \Bigr), \label{eq:Sumrules_general3}
\end{align}
\end{subequations}
where $M^{(\mathrm{orb})}_{ij}=e_{ijk}M^{(\mathrm{orb})}_k$, and $E_{\mathrm{occ}}=\varepsilon_{\mathrm{occ}}-\mu$ denotes the occupied-band energy measured from the chemical potential $\mu$.
Equation~\eqref{eq:Sumrules_general} establishes a universal relation between the $\mathcal{T}$-odd expansion coefficients and the corresponding magnetic multipole moments.
In particular, the first expression corresponds to the previously derived dichroic $f$-sum rule for the optical Hall conductivity $\sigma^{(\mathrm{A})}_{ij}(\omega)$ \cite{Souza2008-dl}.
The second and third expressions extend this sum rule to the spin QMER $\chi^{(\mathrm{o})}_{\mu,j}(\omega)$ and spin OMER $\chi^{(\mathrm{o})}_{\mu,(jk)}(\omega)$, respectively.
Similar sum rules are expected to hold for the orbital QMER and orbital OMER, which would establish their relation to the orbital magnetic quadrupole \cite{Shitade2018-rd,Gao2018-no} and octupole moments \cite{Sato2025-zd}, respectively.

An important advantage of the sum rules is that they allow the magnetic multipole moments to be estimated from the optical spectra.
In this regard, X-ray spectroscopy may provide a promising route because the core levels involved in X-ray transitions are typically highly localized,
making the single-flat-band limit a reasonable approximation.
Indeed, orbital magnetization can be estimated from X-ray magnetic circular dichroism spectra \cite{Thole1992-jl,Altarelli1993-mu}.
Similarly, the spin magnetic quadrupole and octupole moments may be estimated from the X-ray spectra of the spin QMER and spin OMER, respectively.

Another implication of the sum rules is that they help clarify the difference in SOC dependence between dipolar and octupolar MOA.
To see this, we recall that dipolar and octupolar MOA originate from magnetic dipoles and octupoles, respectively (see Eq.~\eqref{eq:Multipole tensor_conductivity tensor}). 
The sum rules in Eq.~\eqref{eq:Sumrules_general} further refine this correspondence by specifying the orbital and spin character of the magnetic multipoles.
Specifically, dipolar MOA described by the optical Hall conductivity is associated with the orbital magnetic dipole (orbital magnetization), 
whereas octupolar MOA described by the spin OMER is associated with the spin magnetic octupole.
This distinction explains why, in collinear magnets, dipolar MOA always requires SOC \cite{Watanabe2024-us}, whereas octupolar MOA does not necessarily require SOC.
Nevertheless, dipolar and octupolar MOA can share the same SOC dependence when they have the same microscopic origin.
Indeed, octupolar MOA described by the orbital OMER originates from the same orbital--orbital contribution as dipolar MOA and therefore always requires SOC in collinear magnets.
From this microscopic perspective, the discussion of the SOC dependence of octupolar MOA in Sec.~\ref{Sec:SOC_dependence_MOA} can now be understood as implicitly assuming the spin OMER.

%%%%%%%%%%%%%%%%%%%%%%%%%%%%%%%%%%%%%%%%%%%%%%%%%%%%%%%%%%%%%%%%%%%%%%%%%%%%%%%%%%%%%%%%%%%%%%%%%%%%%%%%%%%%%%%%%%%%%%%%%%%%%%%%%%%%

\section{Numerical results} \label{Sec:Numerical_results}

In this section, we numerically evaluate the complex Faraday angle for octupolar MOA using tight-binding models for three representative materials:
AIAO-ordered pyrochlore iridates \ce{\textit{R}2Ir2O7}, 
the strong $d$-wave altermagnet \ce{RuO2}, and the weak $d$-wave altermagnet \ce{CrSb}.
These materials allow octupolar MOA while dipolar MOA is symmetry-forbidden, offering suitable platforms for investigating octupolar MOA.
However, since the presence or absence of octupolar MOA depends on the light-propagation direction,
we choose a direction along which octupolar MOA is allowed for each material: $[111]$ for \ce{\textit{R}2Ir2O7}, $[111]$ for \ce{RuO2}, and $[010]$ for \ce{CrSb} [see Table~\ref{tab:Mijk_list}].
These configurations exhibit Type-I optical activity for \ce{\textit{R}2Ir2O7} and Type-II optical activity for \ce{RuO2} and \ce{CrSb},
allowing the complex Faraday angle to be evaluated using Eqs.~\eqref{eq:Faraday_TypeI_Gz} and \eqref{eq:Faraday_TypeII_numerical}, respectively.
The corresponding numerical results are presented in Secs.~\ref{Subsec:Numerical_results1} for \ce{\textit{R}2Ir2O7}
and \ref{Subsec:Numerical_results2} for \ce{RuO2} and \ce{CrSb}, 
while the details of the models and calculations are provided in Appendix~\ref{Sec:AppH}.

%%%%%%%%%%%%%%%%%%%%%%%%%%%%%%%%%%%%%%%%%%%%%%%%%%%%%%%%%%%%%%%%%%

\subsection{AIAO-ordered pyrochlore iridates: \ce{\textit{R}2Ir2O7}} \label{Subsec:Numerical_results1}

\begin{figure}[b]
    \centering
    \begin{subfigure}[t]{1.0\linewidth}
        \caption{}
        \includegraphics[width=1.0\linewidth,height=3.6cm]{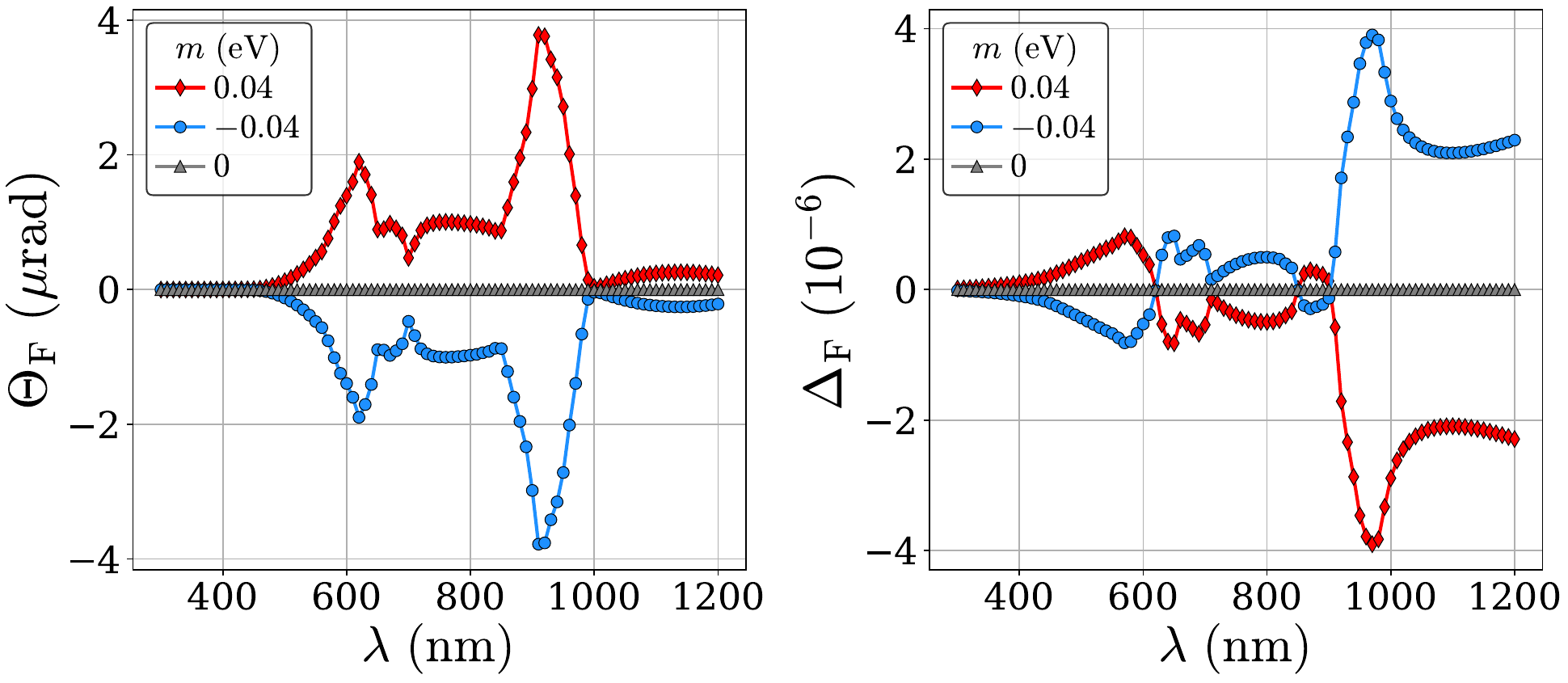}
        \label{fig:Pyrochlore1}
    \end{subfigure}
    \begin{subfigure}[t]{1.0\linewidth}
        \captionsetup{skip=-10pt}
        \caption{}
        \includegraphics[width=1.0\linewidth,height=3.6cm]{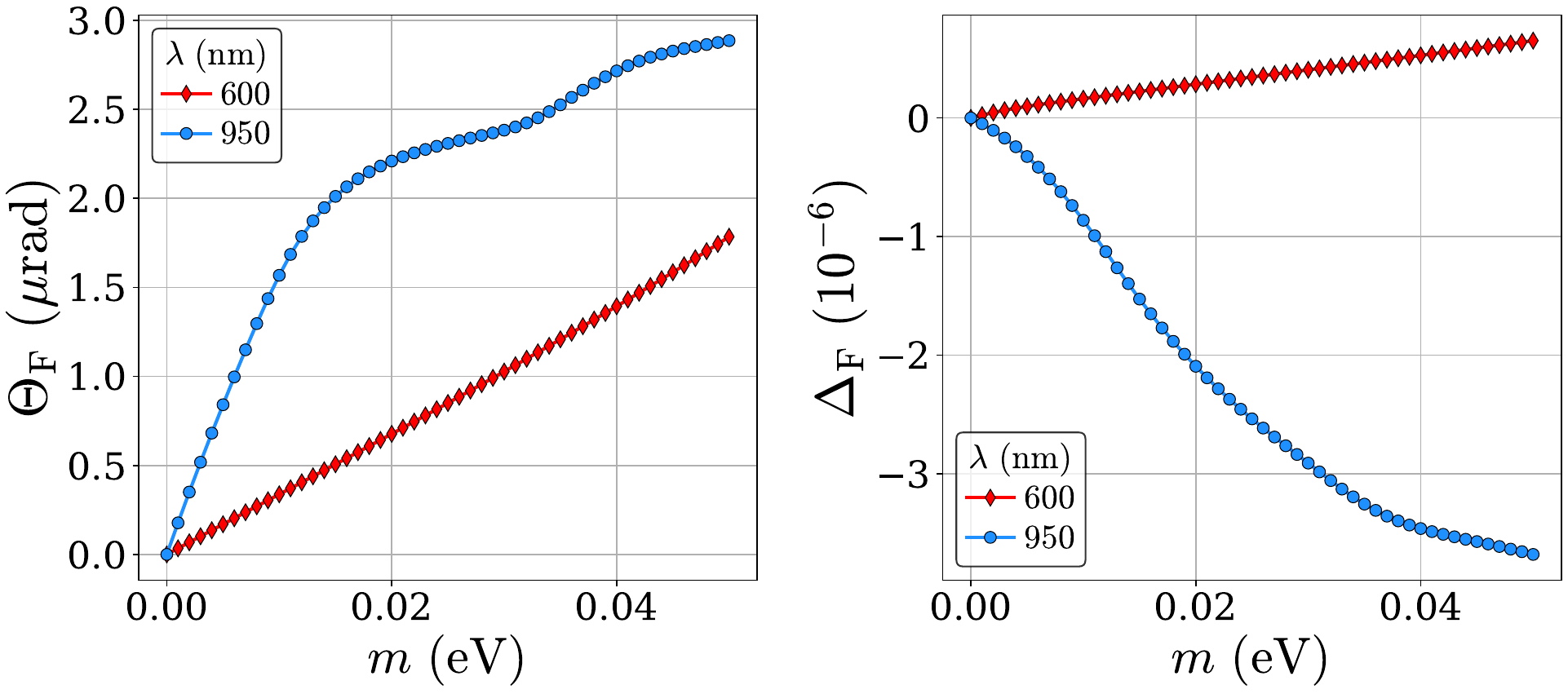}
        \label{fig:Pyrochlore2}
    \end{subfigure}
    \caption{Wavelength ($\lambda$) and $m$ dependences of the Faraday rotation angle $\Theta_{\mathrm{F}}$ and ellipticity $\Delta_{\mathrm{F}}$ in \ce{\textit{R}2Ir2O7}:
     (a) $\lambda$ dependence at $m=\pm 0.04$ and $0~\mathrm{eV}$ and (b) $m$ dependence at $\lambda=600$ and $950~\mathrm{nm}$. 
    }
    \label{fig:Pyrochlore}
\end{figure}

We first outline the model for AIAO-ordered pyrochlore iridates \ce{\textit{R}2Ir2O7} \cite{Witczak-Krempa2013-bv,Tian2015-zj}:
\begin{align}
    \hat{H}(\bm{k})=\hat{H}_t(\bm{k})+m\hat{O}_{\mathrm{AIAO}}.
\end{align}
Here, $\hat{H}_t(\bm{k})$ is the hopping Hamiltonian including SOC, and
$\hat{O}_{\mathrm{AIAO}}$ is the dimensionless operator describing the AIAO magnetic order with coupling strength $m$.
In particular, $m>0$ and $m<0$ represent the all-in and all-out configurations, respectively.

We next obtain an explicit form of the complex Faraday angle $\Psi_{\mathrm{F}}$ given in Eq.~\eqref{eq:Faraday_TypeI_Gz} for light propagating along the [111] direction.
Specifically, substituting Eq.~\eqref{eq:OAV_111_cubic} into Eq.~\eqref{eq:Faraday_TypeI_Gz} yields
\begin{align}
    \Psi_{\mathrm{F}}
    &= -i\frac{\ell}{\sqrt{3}} \Bigl(\frac{\omega}{c} \Bigr)^3 \sqrt{\varepsilon^{(\mathrm{S})}_{xx}}{\varepsilon}^{(\mathrm{A})}_{xy,xy}, \label{eq:Psi_TypeI}
\end{align}
where we use Eq.~\eqref{eq:tildegijk_eijkl} and set $\varepsilon^{(0)}=\varepsilon^{(\mathrm{S})}_{xx}$.
In the following, we numerically evaluate $\Psi_{\mathrm{F}}$ for a sample thickness of $\ell=100~\mathrm{nm}$,
using the expansion coefficients calculated from the microscopic expressions given in 
Eqs.~\eqref{eq:band_contribution_01_sigma}, \eqref{eq:geometric_contribution_01_sigma}, \eqref{eq:band_contribution_2_o}, and \eqref{eq:geometric_contribution_2_o}.

The numerical results are presented in Fig.~\ref{fig:Pyrochlore}.
Figure~\ref{fig:Pyrochlore1} shows the wavelength dependence of the Faraday rotation angle $\Theta_{\mathrm{F}}$ and ellipticity $\Delta_{\mathrm{F}}$ for different values of $m$,
revealing three notable features.
First, the response is of the order of $10^{-6}$ for a 100-nm-thick sample in the visible-light range, placing it well within an experimentally accessible regime.
Second, the response changes sign upon reversal of $m$, enabling imaging of AIAO domains with all-in and all-out configurations.
Third, the response vanishes at $m=0$ and becomes finite for $m \neq 0$, demonstrating that it provides an effective probe of AIAO magnetic order.
Moreover, Fig.~\ref{fig:Pyrochlore2} shows the $m$ dependence of the response at wavelengths near the two peaks in Fig.~\ref{fig:Pyrochlore1},
revealing an approximately linear dependence on $m$.
This behavior indicates that the response reflects not only the presence but also the strength of AIAO magnetic order.

Finally, we revisit the recently observed MOA in AIAO-ordered \ce{Eu2Ir2O7} \cite{Han2025-ma} from the perspective of octupolar MOA.
According to Ref.~\cite{Han2025-ma}, the observed MOA was interpreted as dipolar MOA, whereas we point out that it may instead be interpreted as octupolar MOA based on the following three considerations.
First, the observed MOA occurs at zero external magnetic field and in the absence of detectable magnetic dipole order, arguing against a dipolar-MOA origin.
Second, the observed MOA appears for light propagating along the [111] direction, for which octupolar MOA is allowed in \ce{\textit{R}2Ir2O7}.
Third, the magnitude of the observed MOA is roughly comparable to our numerical estimate in the same wavelength range, although the experimental value is somewhat larger.
These considerations suggest an octupolar-MOA origin for the observed MOA,
but this interpretation requires further experimental verification.

%%%%%%%%%%%%%%%%%%%%%%%%%%%%%%%%%%%%%%%%%%%%%%%%%%%%%%%%%%%%%%%%%%

\subsection{$d$-wave altermagnets: \ce{RuO2} and \ce{CrSb}} \label{Subsec:Numerical_results2}

We first outline the two-sublattice model for the strong $d$-wave altermagnet \ce{RuO2} and weak $d$-wave altermagnet \ce{CrSb}  \cite{Roig2024-zg}:
\begin{align}
    \hat{H}(\bm{k})&=\hat{H}_t(\bm{k})+\hat{H}_{\mathrm{intra}}(\bm{k}) 
    +\lambda_{\mathrm{SOC}} \hat{O}_{\mathrm{SOC}}(\bm{k})+J_z\hat{O}^z_{\mathrm{N}}. \label{eq:altermagnet_model_general}
\end{align}
Here, $\hat{H}_t(\bm{k})$ contains the sublattice-independent and intersublattice hopping terms,
$\hat{H}_{\mathrm{intra}}(\bm{k})$ is the intrasublattice hopping Hamiltonian,
$\hat{O}_{\mathrm{SOC}}(\bm{k})$ is the dimensionless operator describing SOC with coupling strength $\lambda_{\mathrm{SOC}}$, and 
$\hat{O}^z_{\mathrm{N}}$ is the dimensionless operator describing N\'{e}el order along the $z$ direction with coupling strength $J_z$.
In particular, $J_z>0$ and $J_z<0$ represent N\'{e}el order along the $+z$ and $-z$ directions, respectively.
Note that the Hamiltonian has the same structure for \ce{RuO2} and \ce{CrSb},
but the $\bm{k}$ dependences of its individual terms differ between the two materials.
This difference is particularly important in $\hat{H}_{\mathrm{intra}}(\bm{k})$, which contains terms characterizing the momentum dependence of altermagnetic spin splitting.
For example, in \ce{RuO2}, $\hat{H}_{\mathrm{intra}}(\bm{k})$ contains a term involving the hopping parameters $t_6$ and $t_7$, 
\begin{align}
    \sin k_x \sin k_y (t_6+t_7 \cos k_z),
\end{align}
which, combined with $J_z$, gives rise to $d$-wave spin splitting in the absence of SOC \cite{Roig2024-zg}.

We next obtain explicit forms of the complex Faraday angle $\Psi_{\mathrm{F}}$ given in Eq.~\eqref{eq:Faraday_TypeII_numerical} for light propagating along the [111] and [010] directions in \ce{RuO2} and \ce{CrSb}, respectively.
To this end, we satisfy CAA by choosing incident linear polarization along $[1\bar{1}0]$ for \ce{RuO2} and $[100]$ for \ce{CrSb}.
Under these conditions, using the expressions for $\tilde{G}^{(\mathrm{o})}_{[uvw]}$ given in Eqs.~\eqref{eq:OAV_111_tetragonal} and \eqref{eq:OAV_010} and 
the expressions for $\eta^{(\mathrm{S})}$ in coordinate system (ii) derived in Appendix~\ref{Sec:AppC}, we obtain
\begin{subequations}
    \begin{align}
        &\Psi_{\mathrm{F}} = -\tilde{a}^{(\mathrm{o})} \bigl[ 1-e^{-i(\omega \ell/c)\Delta n} \bigr], \label{eq:Numerical_RuO2_phi} \\
        &\left\{
        \begin{aligned}
        \tilde{a}^{(\mathrm{o})} &= \frac{1}{\sqrt{3}} \Bigl(\frac{\omega}{c} \Bigr)^2 \frac{2\varepsilon^{(\mathrm{S})}_{xx}\varepsilon^{(\mathrm{A})}_{yz,yz}+\varepsilon^{(\mathrm{S})}_{zz}\varepsilon^{(\mathrm{A})}_{xy,xy}}{\varepsilon^{(\mathrm{S})}_{xx}-\varepsilon^{(\mathrm{S})}_{zz}} \\
        \Delta n &= \frac{2(\varepsilon^{(\mathrm{S})}_{xx}-\varepsilon^{(\mathrm{S})}_{zz})\sqrt{\varepsilon^{(\mathrm{S})}_{xx}}}{\sqrt{2\varepsilon^{(\mathrm{S})}_{xx}+\varepsilon^{(\mathrm{S})}_{zz}}(\sqrt{2\varepsilon^{(\mathrm{S})}_{xx}+\varepsilon^{(\mathrm{S})}_{zz}}+\sqrt{3\varepsilon^{(\mathrm{S})}_{zz}})}
        \end{aligned}
        \right., \label{eq:Numerical_RuO2_coefficients}
    \end{align}
\end{subequations}
for \ce{RuO2} and 
\begin{subequations}
    \begin{align}
        &\Psi_{\mathrm{F}} = \tilde{a}^{(\mathrm{o})} \bigl[ 1-e^{i(\omega \ell/c)\Delta n} \bigr], \label{eq:Numerical_CrSb_phi} \\
        &\left\{
        \begin{aligned}
        \tilde{a}^{(\mathrm{o})} &= \Bigl(\frac{\omega}{c} \Bigr)^2 \frac{\varepsilon^{(\mathrm{S})}_{xx}\varepsilon^{(\mathrm{A})}_{zx,yy}}{\varepsilon^{(\mathrm{S})}_{zz}-\varepsilon^{(\mathrm{S})}_{xx}} \\
        \Delta n &= \sqrt{\varepsilon^{(\mathrm{S})}_{zz}}-\sqrt{\varepsilon^{(\mathrm{S})}_{xx}}
    \end{aligned}
    \right.,
    \end{align}
\end{subequations}
for \ce{CrSb}.
In the following, we numerically evaluate $\Psi_{\mathrm{F}}$ for a sample thickness of $\ell=100~\mathrm{nm}$,
using the same microscopic expressions for the expansion coefficients as in the case of \ce{\textit{R}2Ir2O7}.

\begin{figure}[b]
    \centering
    \includegraphics[width=1.0\linewidth,height=4.4cm]{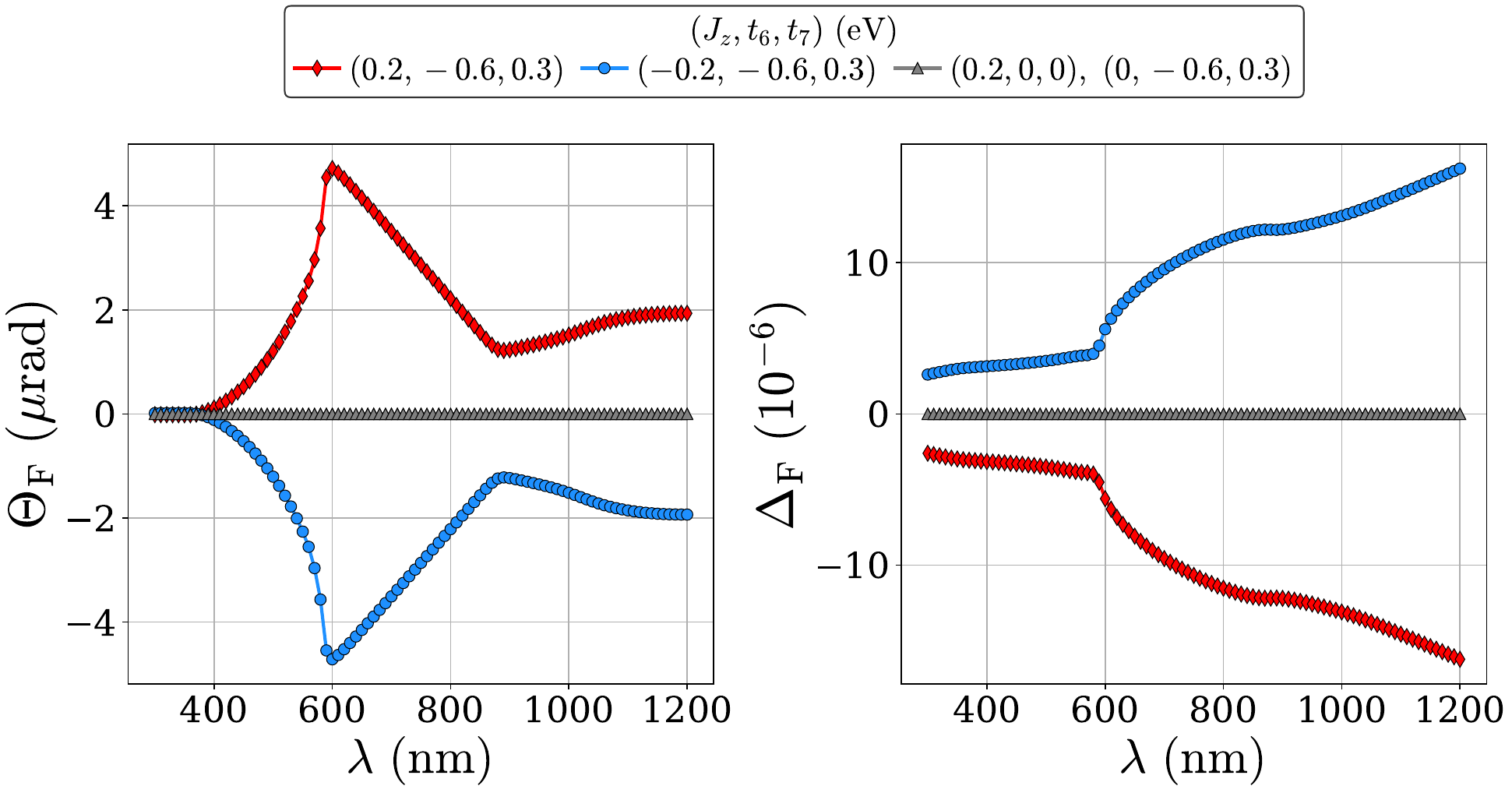}
    \caption{Wavelength ($\lambda$) dependence of the Faraday rotation angle $\Theta_{\mathrm{F}}$ and ellipticity $\Delta_{\mathrm{F}}$ in \ce{RuO2}
    for $(J_z,t_6,t_7)=(\pm 0.2,-0.6,0.3)$, $(0.2,0,0)$, and $(0,-0.6,0.3)~\mathrm{eV}$, with $\lambda_{\mathrm{SOC}}=0$.
    }\label{fig:dwave}
\end{figure}

%%%%%%%%%%%%%%%%%%%%%%%%%%%%%%%%%

\subsubsection{\ce{RuO2}}

The numerical results for \ce{RuO2} are presented in Fig.~\ref{fig:dwave},
which shows the wavelength dependence of the Faraday rotation angle $\Theta_{\mathrm{F}}$ and ellipticity $\Delta_{\mathrm{F}}$ 
for different values of $J_z$, $t_6$, and $t_7$ in the absence of SOC.
This figure reveals three notable features, as in the case of \ce{\textit{R}2Ir2O7}.
First, the response is of the order of $10^{-6}$ for a 100-nm-thick sample in the visible-light range, placing it well within an experimentally accessible regime.
Second, the response changes sign upon reversal of $J_z$, enabling imaging of altermagnetic domains with opposite N\'{e}el vectors.
Third, the response becomes finite only for $J_z \neq 0$ and $(t_6,t_7) \neq (0,0)$, demonstrating that it provides an effective probe of $d$-wave altermagnetism.
Moreover, the response is finite even in the absence of SOC, consistent with the fact that octupolar MOA does not require SOC in strong $d$-wave altermagnets (see Sec.~\ref{Sec:SOC_dependence_MOA}).

\begin{figure}[t]
    \centering
    \begin{subfigure}[t]{0.48\linewidth}
        \caption{}
        \includegraphics[width=1.0\linewidth,height=3.6cm]{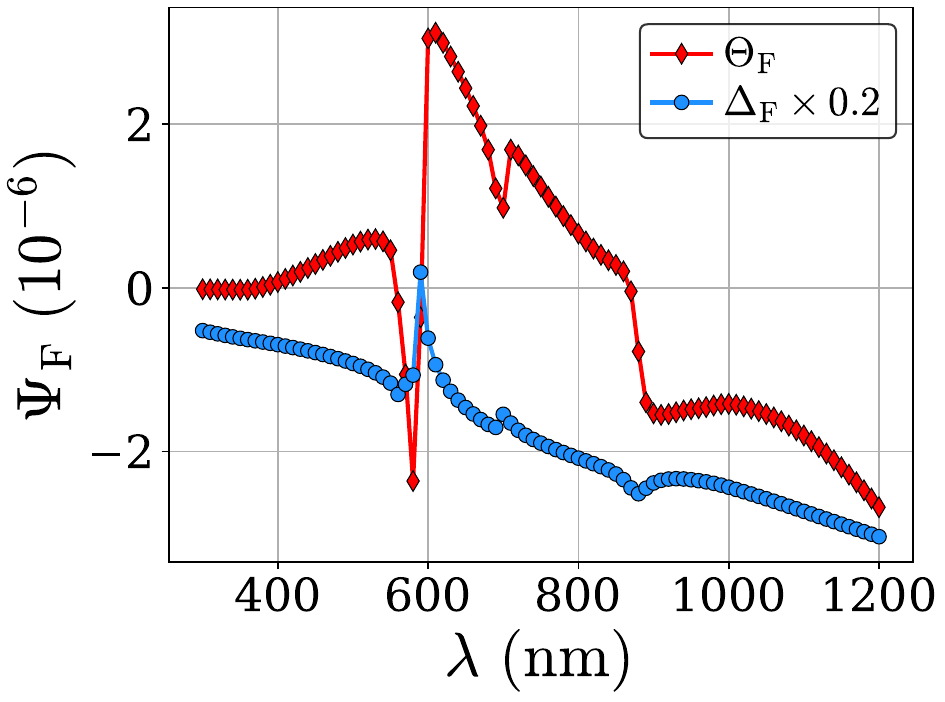}
        \label{fig:dwave_SOC1}
    \end{subfigure}
    \hfill
    \begin{subfigure}[t]{0.48\linewidth}
        % \captionsetup{skip=-10pt}
        \caption{}
        \includegraphics[width=1.0\linewidth,height=3.6cm]{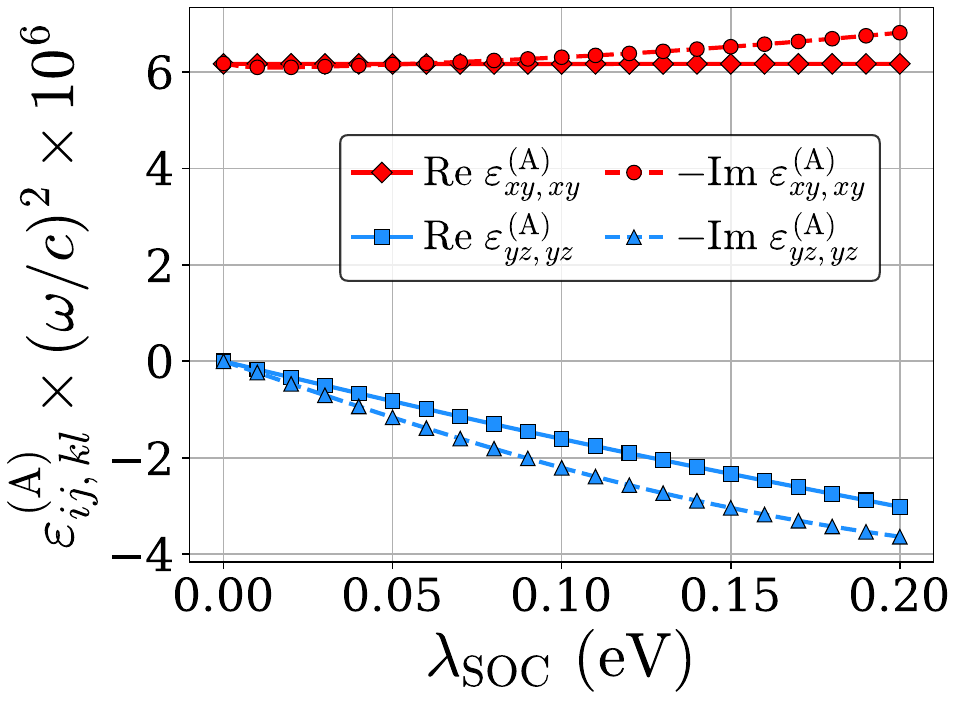}
        \label{fig:dwave_SOC2}
    \end{subfigure}
    \caption{(a) Wavelength ($\lambda$) dependence of the Faraday rotation angle $\Theta_{\mathrm{F}}$ and ellipticity $\Delta_{\mathrm{F}}$ in \ce{RuO2} at $\lambda_{\mathrm{SOC}}=0.1~\mathrm{eV}$.
    (b) SOC dependence of the second-order expansion coefficients $\varepsilon^{(\mathrm{A})}_{xy,xy}$ and $\varepsilon^{(\mathrm{A})}_{yz,yz}$ at $\lambda=600~\mathrm{nm}$.
    While $\varepsilon^{(\mathrm{A})}_{xy,xy}$ remains finite even without SOC and depends only weakly on SOC, 
    $\varepsilon^{(\mathrm{A})}_{yz,yz}$ vanishes in the absence of SOC and depends strongly on SOC.
    In both panels, $(J_z,t_6,t_7)=(0.2,-0.6,0.3)~\mathrm{eV}$.
    }
    \label{fig:dwave_SOC}
\end{figure}

We next examine the SOC dependence of the response by calculating its wavelength dependence at finite SOC, as shown in Fig.~\ref{fig:dwave_SOC1}.
A comparison with Fig.~\ref{fig:dwave} shows that including SOC changes the spectral shape, indicating that the effect of SOC is not negligible.
To understand this behavior, we recall that the second-order expansion coefficients $\varepsilon^{(\mathrm{A})}_{xy,xy}$ and $\varepsilon^{(\mathrm{A})}_{yz,yz}$ entering Eq.~\eqref{eq:Numerical_RuO2_coefficients}
are nonrelativistic and relativistic components, respectively [see Table~\ref{tab:SPG_MPG_relation_full}].
Based on this distinction, $\varepsilon^{(\mathrm{A})}_{yz,yz}$ is expected to acquire a sizable value in the presence of SOC.
We therefore examine the SOC dependence of these two expansion coefficients in Fig.~\ref{fig:dwave_SOC2}.
The results indeed confirm this expectation: $\varepsilon^{(\mathrm{A})}_{yz,yz}$ is small for weak SOC but increases with SOC and eventually becomes comparable to $\varepsilon^{(\mathrm{A})}_{xy,xy}$,
thereby accounting for the SOC-induced modification of the spectral shape.
These findings highlight that relativistic components should not be overlooked in altermagnets, even though their characteristic properties are often discussed in the nonrelativistic limit.

\begin{figure}[t]
    \centering
    \includegraphics[width=0.6\linewidth,height=4.0cm]{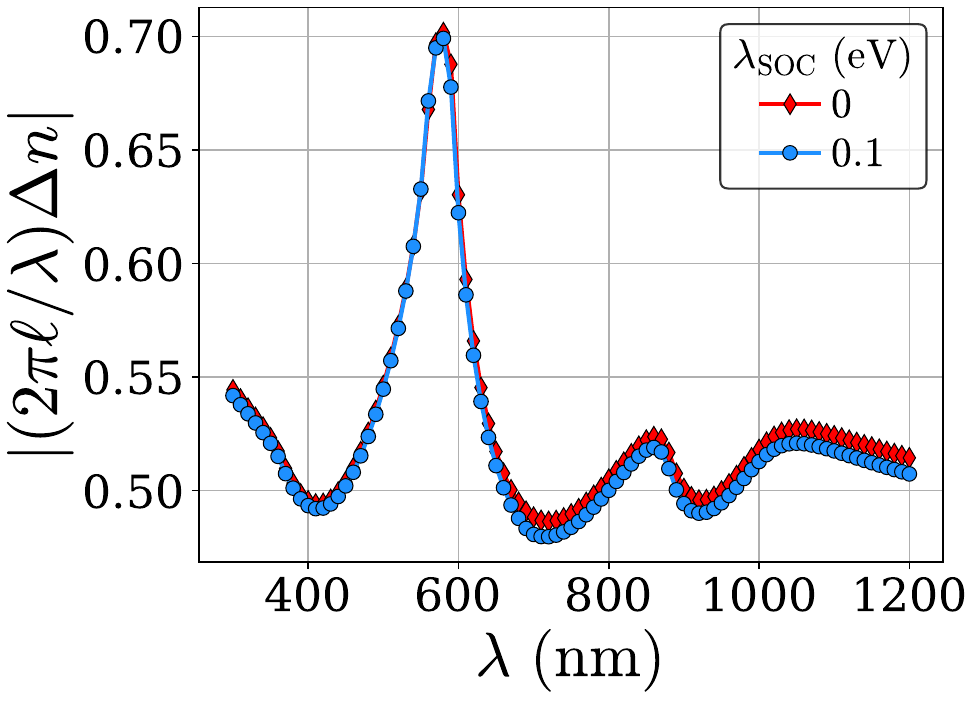}
    \caption{Wavelength ($\lambda$) dependence of $|(2 \pi \ell/\lambda)\Delta n|$ in \ce{RuO2} at $\lambda_{\mathrm{SOC}}=0$ and $0.1~\mathrm{eV}$ for $(J_z,t_6,t_7)=(0.2,-0.6,0.3)~\mathrm{eV}$.
    }\label{fig:dwave_phasedifference}
\end{figure}

Finally, we investigate the origin of the peak structure near $\lambda=600~\mathrm{nm}$ in Figs.~\ref{fig:dwave} and \ref{fig:dwave_SOC1}.
As a starting point, following the derivation of Eq.~\eqref{eq:Faraday_TypeII_numerical_expansion},
we expand the exponential factor $e^{-i(\omega \ell/c)\Delta n}=e^{-i(2 \pi \ell/\lambda)\Delta n}$ in Eq.~\eqref{eq:Numerical_RuO2_phi}
to obtain the following expression for $\Psi_{\mathrm{F}}$:
\begin{align}
    \Psi_{\mathrm{F}} = \tilde{a}^{(\mathrm{o})} \sum_{m=1}^{\infty} \frac{1}{m!} \Bigl(-i\frac{2 \pi \ell}{\lambda}\Delta n \Bigr)^m.
\end{align}
This expression suggests that $|\Psi_{\mathrm{F}}|$ may become larger as $|(2 \pi \ell/\lambda)\Delta n|$ increases.
We therefore examine the wavelength dependence of $|(2 \pi \ell/\lambda)\Delta n|$ in Fig.~\ref{fig:dwave_phasedifference}.
The result shows that $|(2 \pi \ell/\lambda)\Delta n|$ also exhibits a peak near $\lambda=600~\mathrm{nm}$,
suggesting that the enhancement of $|(2 \pi \ell/\lambda)\Delta n|$ may contribute to the peak structure.
In this regard, a large refractive-index anisotropy $\Delta n$ may be favorable for obtaining a large $|\Psi_{\mathrm{F}}|$.

%%%%%%%%%%%%%%%%%%%%%%%%%%%%%%%%%

\subsubsection{\ce{CrSb}}

The numerical results for \ce{CrSb} are presented in Fig.~\ref{fig:gwave}.
Figure~\ref{fig:gwave1} shows the wavelength dependence of the Faraday rotation angle $\Theta_{\mathrm{F}}$ and ellipticity $\Delta_{\mathrm{F}}$ for different values of $J_z$ and $\lambda_{\mathrm{SOC}}$.
As in the \ce{RuO2} case, the response is of the order of $10^{-6}$ for a 100-nm-thick sample in the visible-light range and changes sign upon reversal of $J_z$.
In contrast to the \ce{RuO2} case, the response becomes finite only for $\lambda_{\mathrm{SOC}} \neq 0$, 
consistent with the fact that octupolar MOA requires SOC in weak $d$-wave altermagnets (see Sec.~\ref{Sec:SOC_dependence_MOA}).
Nevertheless, the response remains within an experimentally accessible regime even at a very weak SOC strength of $\lambda_{\mathrm{SOC}}=0.004~\mathrm{eV}$,
indicating that SOC does not play a merely secondary role in determining the response.

We next examine the response in the wavelength range $\lambda=540$--$660~\mathrm{nm}$ omitted from Fig.~\ref{fig:gwave1}.
The results are shown in Fig.~\ref{fig:gwave2}, revealing two distinct peaks.
A possible origin of these peaks is a large $|(2 \pi \ell/\lambda)\Delta n|$, as in the \ce{RuO2} case.
We therefore compare the response shown in Fig.~\ref{fig:gwave2} with the wavelength dependence of $|(2 \pi \ell/\lambda)\Delta n|$ shown in Fig.~\ref{fig:gwave3}.
The comparison shows that only the peak near $\lambda=640~\mathrm{nm}$ coincides with a peak in $|(2 \pi \ell/\lambda)\Delta n|$,
suggesting that a large $|(2 \pi \ell/\lambda)\Delta n|$ may contribute to this peak.
In contrast, the peak near $\lambda=560~\mathrm{nm}$ is unlikely to be explained solely by a large $|(2 \pi \ell/\lambda)\Delta n|$.
As another possible origin of this peak, we consider resonant interband transitions and calculate the joint density of states (JDOS), defined by  
\begin{align}
    \mathrm{JDOS}=\sum_{n,m(\neq n)}\int_{\bm{k}}f_{nm} \delta(\varepsilon_{nm}+\hbar \omega),
\end{align}
which measures the number of interband transitions with an energy difference $\hbar\omega$.
The result is shown in Fig.~\ref{fig:gwave3}, revealing a peak near $\lambda=560~\mathrm{nm}$, which coincides with the peak in the response.
This correspondence suggests that resonant interband transitions may contribute to the enhanced response.

%%%%%%%%%%%%%%%%%%%%%%%%%%%%%%%%%%%%%%%%%%%%%%%%%%%%%%%%%%%%%%%%%%

\section{Summary and Outlook} \label{Sec:Conclusion}

In this work, we have developed a general and comprehensive theoretical framework for octupolar MOA.
We now summarize the main findings of each section in turn and discuss their broader implications and possible future directions.

\begin{figure}[t]
    \centering
    \begin{subfigure}[t]{1.0\linewidth}
        \caption{}
        \includegraphics[width=1.0\linewidth,height=4.4cm]{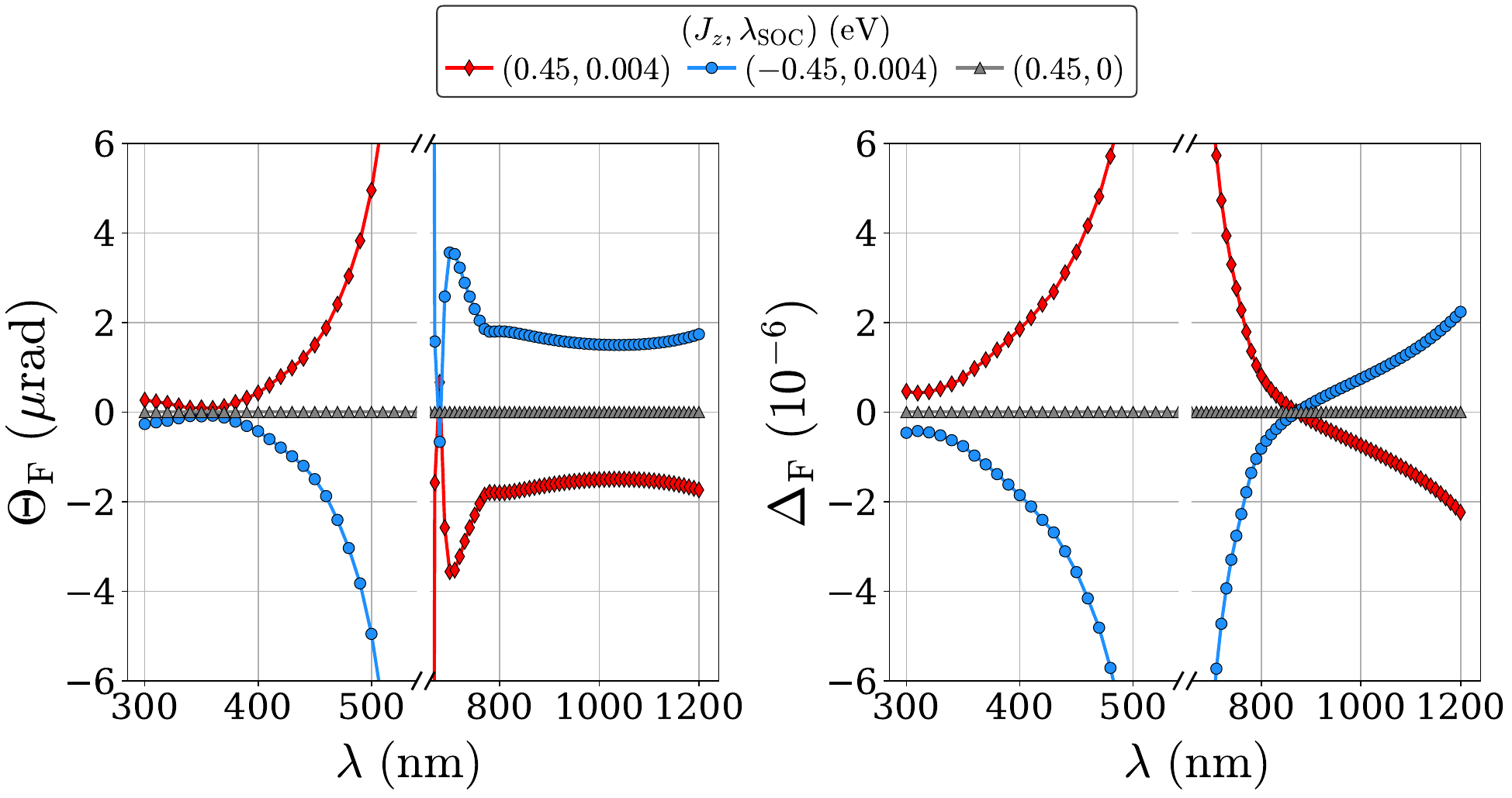}
        \label{fig:gwave1}
    \end{subfigure}
    \begin{subfigure}[t]{0.52\linewidth}
        \captionsetup{skip=-10pt}
        \caption{}
        \includegraphics[width=1.0\linewidth,height=3.5cm]{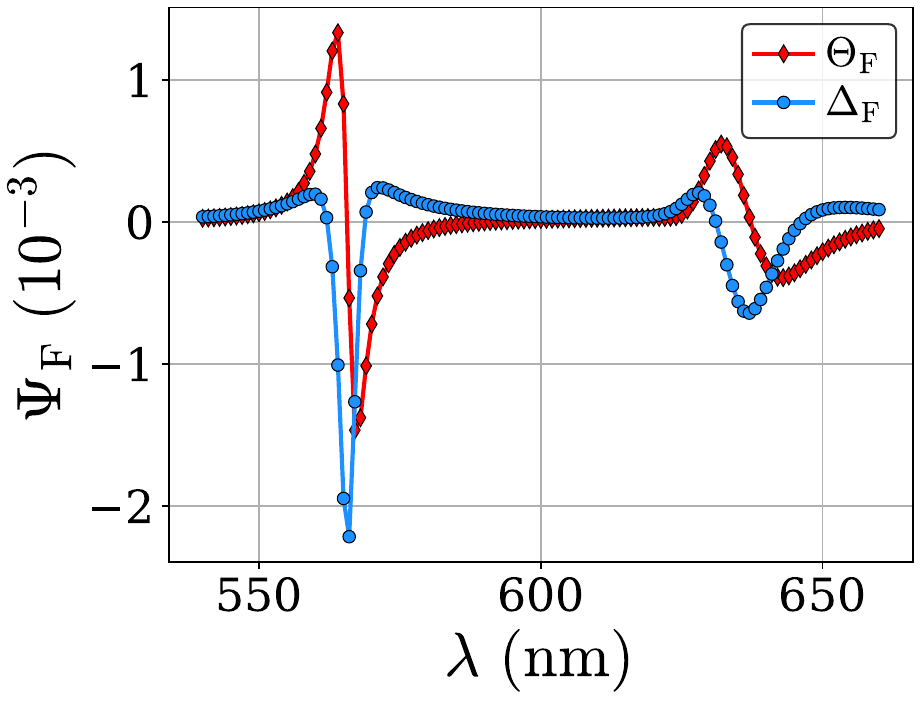}
        \label{fig:gwave2}
    \end{subfigure}
    \hfill
    \begin{subfigure}[t]{0.45\linewidth}
        \captionsetup{skip=-10pt}
        \caption{}
        \includegraphics[width=1.0\linewidth,height=3.5cm]{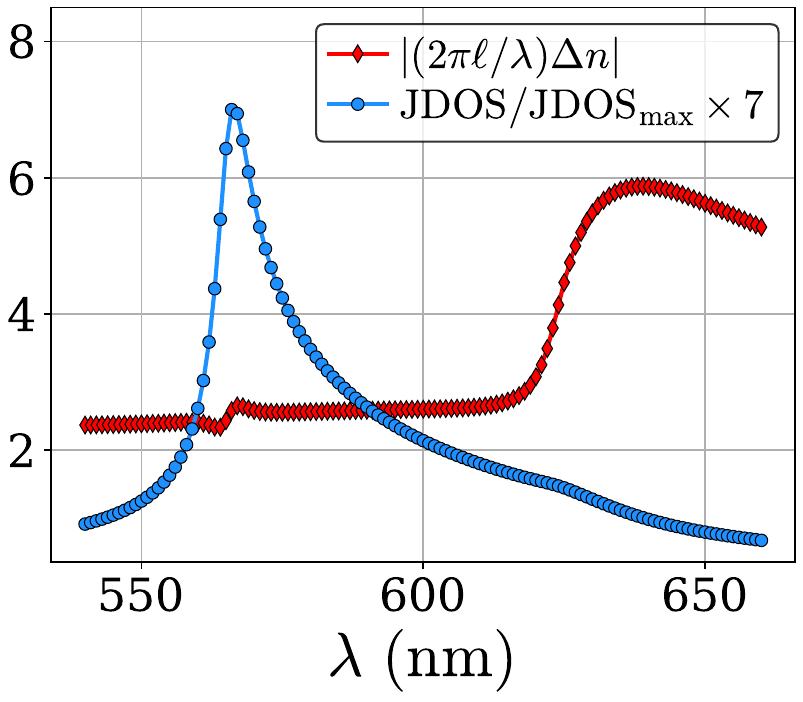}
        \label{fig:gwave3}
    \end{subfigure}
    \caption{(a) Wavelength ($\lambda$) dependence of the Faraday rotation angle $\Theta_{\mathrm{F}}$ and ellipticity $\Delta_{\mathrm{F}}$ in \ce{CrSb}
    for $(J_z,\lambda_{\mathrm{SOC}})=(\pm 0.45,0.004)$ and $(0.45,0)~\mathrm{eV}$.
    The $\lambda=540$--$660~\mathrm{nm}$ range is omitted and shown separately in panel (b). 
    (b) $\lambda$ dependence of $\Theta_{\mathrm{F}}$ and $\Delta_{\mathrm{F}}$ in the $\lambda=540$--$660~\mathrm{nm}$ range.
    (c) $\lambda$ dependence of $|(2 \pi \ell/\lambda)\Delta n|$ and the normalized JDOS, $\mathrm{JDOS}/\mathrm{JDOS}_{\max}$, in the $\lambda=540$--$660~\mathrm{nm}$ range.
    Here, $\mathrm{JDOS}_{\max}$ denotes the maximum value of the JDOS.
    In panels (b) and (c), $(J_z,\lambda_{\mathrm{SOC}})=(0.45,0.004)~\mathrm{eV}$.
    }
    \label{fig:gwave}
\end{figure}

Section~\ref{Sec:Macroscopic_theory_MO} shows that octupolar MOA arises from the second-order term $\varepsilon^{(\mathrm{A})}_{ij,kl}(\omega)q_kq_l$ in the $\bm{q}$ expansion of the $\mathcal{T}$-odd permittivity tensor $\varepsilon^{(\mathrm{odd})}_{ij}(\bm{q},\omega)$ [Eq.~\eqref{eq:qexpansion_odd}].
Although this spatially dispersive term vanishes in the $\bm{q} \to 0$ limit and is therefore usually neglected,
it becomes the leading nonvanishing response in centrosymmetric magnets where magnetic octupole order is present but magnetic dipole order is symmetry-forbidden.
However, octupolar MOA can coexist with uniform NMOA and Lorentz NMOA arising from the $\mathcal{T}$-even permittivity tensor $\varepsilon^{(\mathrm{even})}_{ij}(\bm{q},\omega)$ [Table~\ref{tab:classification_OA}],
complicating its experimental identification.
To address this issue, we first consider CAA, under which optical activity involving octupolar MOA can be classified into four distinct types [Fig.~\ref{fig:MO_Type}].
Among these, Types I and II enable octupolar MOA to be separated from uniform and Lorentz NMOA under CAA.
For Types III and IV, where such separation is not possible under CAA, Sagnac interferometry provides an alternative approach
by selectively isolating the $\mathcal{T}$-odd contribution to optical activity.

Section~\ref{Sec:MO_conditions} shows that the $\bm{q}$ dependence inherent in $\varepsilon^{(\mathrm{A})}_{ij,kl}(\omega)q_kq_l$ yields more complex conditions for observing octupolar MOA than for dipolar MOA [Eq.~\eqref{eq:MO_condition_MO}].
This complexity reflects the requirement that light couple to all three indices of the magnetic octupole [Fig.~\ref{fig:MO_condition}],
thereby restricting the light-propagation directions along which octupolar MOA is allowed.
Nevertheless, for many of these allowed directions,  
a broad range of materials, including \ce{RuO2}, \ce{KV2Se2O}, \ce{CrSb}, \ce{MnTe2}, and \ce{\textit{R}2Ir2O7}, exhibit Type-I or Type-II optical activity [Table~\ref{tab:Mijk_list}].
More broadly, although Table~\ref{tab:Mijk_list} lists the candidate centrosymmetric MPGs,
octupolar MOA can also emerge in some noncentrosymmetric MPGs.
For example, $6'$, $6'22'$, $6'mm'$, and $4'32'$ allow magnetic octupole order but neither magnetic dipole nor quadrupole order
and also satisfy the conditions for observing octupolar MOA (see SM \cite{Supplemental}).
These MPGs may therefore be additional candidates for observing octupolar MOA if NMOA arising from the first-order term $\varepsilon^{(\mathrm{A})}_{ij,k}(\omega)q_k$ can be avoided.

Section~\ref{Sec:SOC_dependence_MOA} shows that octupolar MOA in $d$-wave altermagnets does not require SOC, 
while SOC makes octupolar MOA symmetry-allowed in certain $g$- and $i$-wave altermagnets [Table~\ref{tab:SPG_MPG_relation_full}].
At first sight, these results lack a unified description in terms of $X$-wave classes
because the SPG-based classification of altermagnets cannot by itself capture the symmetry changes induced by SOC.
We therefore introduced the strong/weak $X$-wave classification under MPGs based on the SPG--MPG correspondence [Fig.~\ref{fig:Strong_Weak_magnets}].
Within this classification, octupolar MOA can arise in strong and weak $d$-wave altermagnets, while the strong/weak distinction characterizes its SOC dependence:
octupolar MOA does not require SOC in strong $d$-wave altermagnets, whereas it requires SOC in weak $d$-wave altermagnets.
In this classification, however, the $X$-wave character is defined by the lowest-rank symmetry-allowed even-parity magnetic multipole
rather than by the momentum dependence of spin splitting as in the SPG-based definition of altermagnetism.
This definition therefore does not directly establish how the $X$-wave character is related to spin splitting in the presence of SOC, a question that remains for future work.

Section~\ref{Sec:Faraday_Rotation} derives analytical expressions for the complex Faraday angle associated with octupolar MOA,
focusing on Type-I and Type-II optical activity.
The resulting expression for Type I takes the same form as the well-known expression for dipolar MOA [Eq.~\eqref{eq:Faraday_TypeI}].
For Type II, we present two expressions under different conditions.
The first expression is obtained under an approximation and explicitly shows that octupolar MOA generally coexists with NMOA, 
but that the NMOA contribution vanishes under CAA [Eq.~\eqref{eq:Faraday_TypeII}].
The second expression is obtained under CAA but without the approximation, allowing quantitative evaluation of octupolar MOA [Eq.~\eqref{eq:Faraday_TypeII_numerical}].
Overall, the expressions for Type-I and Type-II optical activity take different forms but share a common feature:
the octupolar MOA vector $\tilde{\bm{G}}^{(\mathrm{o})}$ governs the MOA contribution to the complex Faraday angle.
It would therefore be interesting to investigate how $\tilde{\bm{G}}^{(\mathrm{o})}$ contributes to octupolar MOA in other optical configurations.
In this regard, deriving analytical expressions for the complex Kerr angle of reflected light would be a natural direction for future work.

Section~\ref{Sec:Microscopic_theory_MO} derives a microscopic expression for $\varepsilon^{(\mathrm{A})}_{ij,kl}(\omega)$ using the Kubo formula,
focusing on the orbital-spin mixed contribution.
The resulting expression consists of an intraband contribution characterized by a Drude-like response [Eq.~\eqref{eq:band_contribution_2_o}] and 
an interband contribution involving the band-resolved Zeeman Berry curvature $\tilde{\Omega}^{\mu,j}_{nm}$ and a related band-geometric quantity $\mathrm{Im}(\tilde{T}^{\mu,jk}_{nm})$ [Eq.~\eqref{eq:geometric_contribution_2_o}].
This expression also enables a physical interpretation of $\varepsilon^{(\mathrm{A})}_{ij,kl}(\omega)$ within a multipolar description,
revealing that a component of $\varepsilon^{(\mathrm{A})}_{ij,kl}(\omega)$ describes the spin OMER, which generalizes the spin QMER to magnetic octupole order [Fig.~\ref{fig:Multiferroics}].
Beyond this physical interpretation, the microscopic expression demonstrates the utility of the spin OMER through the $f$-sum rule.
Specifically, the optical spectrum of the spin OMER can be used to estimate the spin magnetic octupole moment in the single-flat-band limit for insulators at zero temperature [Eq.~\eqref{eq:Sumrules_general3}].
This correspondence between the spin OMER and the spin magnetic octupole also provides a microscopic basis for the result in Sec.~\ref{Sec:SOC_dependence_MOA}
that octupolar MOA does not necessarily require SOC in collinear magnets.

Section~\ref{Sec:Numerical_results} numerically evaluates the complex Faraday angle for octupolar MOA in \ce{\textit{R}2Ir2O7}, \ce{RuO2}, and \ce{CrSb}
and reveals three experimentally promising features [Figs.~\ref{fig:Pyrochlore}, \ref{fig:dwave}, and \ref{fig:gwave1}].
First, the response is of the order of $10^{-6}$ for a 100-nm-thick sample in the visible-light range.
Second, the response changes sign upon reversal of the magnetic octupole order.
Third, the response becomes finite only in the presence of magnetic octupole order.
In addition to these three features, the numerical results provide two further insights.
First, the response can receive a sizable contribution even from weak SOC [Figs.~\ref{fig:dwave_SOC} and \ref{fig:gwave1}], 
suggesting that sizable responses can be expected even in weak $d$-wave altermagnets and also highlighting the importance of relativistic effects in altermagnets.
Second, the response may be further enhanced by a large refractive-index anisotropy $\Delta n$
or resonant interband transitions [Figs.~\ref{fig:dwave_phasedifference}, \ref{fig:gwave2}, and \ref{fig:gwave3}].

Finally, we discuss octupolar MOA from a broader experimental perspective.
One key advantage of octupolar MOA is that it can be observed without an external magnetic field or other additional manipulation.
Moreover, this zero-field response can reach experimentally accessible magnitudes, given that Faraday rotations down to the nrad scale have already been detected in the visible-light range \cite{Rowe2017-mt}.
Thus, octupolar MOA provides a field-free route to observing MOA in antiferromagnets where dipolar MOA is symmetry-forbidden, including \ce{RuO2}, \ce{KV2Se2O}, \ce{CrSb}, \ce{MnTe2}, and \ce{\textit{R}2Ir2O7} [Fig.~\ref{fig:General_concept}].
Octupolar MOA can also be used for magnetic-domain imaging in such antiferromagnets, including imaging of altermagnetic domains with opposite Néel vectors.
Beyond domain imaging, octupolar MOA can serve as a noncontact and bulk-sensitive optical probe of magnetic octupole order.
These characteristics offer distinct advantages over transport measurements and surface-sensitive techniques such as angle-resolved photoemission spectroscopy.
A promising application that exploits these advantages is to assess the presence of altermagnetism in candidate materials such as \ce{RuO2} \cite{Hiraishi2024-et,Kesler2024-sk,Liu2024-kv,Kiefer2025-gj} and \ce{KV2Se2O} \cite{Sun2025-mq,Bishal2026-mx}, where its presence remains under debate.

\textit{Note added.}
After completion of this work and shortly before submission, we became aware of a related study that also classifies multipoles under collinear SPGs \cite{Ogawa2026-rb}.

\section*{Acknowledgement}

J.O. was supported by JST SPRING (Grant Number JPMJSP2110), JSPS KAKENHI (Grant Number JP26KJ1450), and the Iwadare Scholarship Foundation.
R.P. was supported by JSPS KAKENHI (Grant Number JP23K03300).
K.S. acknowledges financial support from the RIKEN Special Postdoctoral Researcher (SPDR) Program.

\section*{Data availability}
The data that support the findings of this article are available from the authors upon reasonable request.

\appendix
\renewcommand{\appendixname}{APPENDIX}

%%%%%%%%%%%%%%%%%%%%%%%%%%%%%%%%%%%%%%%%%%%%%%%%%%%%%%%%%%%%%%%%%%%%%%%%%%%%%%%%%%%%%%%%%%%%%%%%%%%%%%%%%%%%%%%%%%%%%%%%%%%%%%%%%%%%

\section{FIELD-ASSISTED OPTICAL PROBING OF ALTERMAGNETISM} \label{Sec:AppA}

In this appendix, we briefly review how optical activity arising from the magnetic-order-dependent contribution to the $\mathcal{T}$-even component $\varepsilon^{(\mathrm{S})}_{ij}(\omega)$
can be used to probe altermagnetism \cite{Sunko2026-ty}.
We first illustrate such optical activity using the Voigt effect.
To this end, we expand the $\mathcal{T}$-even component $\varepsilon^{(\mathrm{S})}_{ij}(\bm{M})$ (with the argument $\omega$ omitted) in powers of the magnetic dipole $\bm{M}$ as 
\begin{align}
    \varepsilon^{(\mathrm{S})}_{ij}(\bm{M})=\varepsilon^{(\mathrm{S})}_{ij}(\bm{0})+c_{\mu,ijk}M_kM_\mu+\cdots, \label{eq:CME1}
\end{align}
where $c_{\mu,ijk}$ is the expansion coefficient, and the odd-order terms are forbidden by the Onsager reciprocal relation in Eq.~\eqref{eq:Onsager_0th}.
The first term is independent of $\bm{M}$ and can give rise to uniform NMOA discussed in the main text.
The second term, by contrast, is quadratic in $\bm{M}$ and gives rise to the Voigt (or Cotton--Mouton) effect \cite{GDZPPN002499460,CottonMouton1905,CottonMouton1905combined}.
Of these two types of optical activity, uniform NMOA is typically dominant.
However, when the first term is isotropic, $\varepsilon^{(\mathrm{S})}_{ij}(\bm{0})=\varepsilon^{(0)}\delta_{ij}$,
uniform NMOA vanishes, allowing the Voigt effect to become the dominant contribution.

We next consider applying Eq.~\eqref{eq:CME1} to altermagnets, assuming the isotropic condition $\varepsilon^{(\mathrm{S})}_{ij}(\bm{0})=\varepsilon^{(0)}\delta_{ij}$ for simplicity.
However, altermagnets generally have zero net magnetization, so the expansion in $\bm{M}$ is not directly applicable.
We therefore introduce an external magnetic field $\bm{B}$ and expand $\varepsilon^{(\mathrm{S})}_{ij}(\bm{B})$ in powers of $\bm{B}$ as 
\begin{align}
    \varepsilon^{(\mathrm{S})}_{ij}(\bm{B})=\varepsilon^{(0)}\delta_{ij}+\gamma_{\mu,ij}B_\mu+\cdots. \label{eq:CME2}
\end{align}
Unlike Eq.~\eqref{eq:CME1}, Eq.~\eqref{eq:CME2} includes a term linear in $\bm{B}$, with coefficient $\gamma_{\mu,ij}$.
Although this term is not always symmetry-allowed, it is allowed in $d$-wave altermagnets 
because $\gamma_{\mu,ij}$ is a $\mathcal{T}$-odd third-rank axial tensor that transforms in the same way as a magnetic octupole $M_{\mu,ij}$.
Thus, $d$-wave altermagnets can exhibit optical activity arising from the term linear in $\bm{B}$.
However, this approach necessarily relies on an external magnetic field.

%%%%%%%%%%%%%%%%%%%%%%%%%%%%%%%%%%%%%%%%%%%%%%%%%%%%%%%%%%%%%%%%%%%%%%%%%%%%%%%%%%%%%%%%%%%%%%%%%%%%%%%%%%%%%%%%%%%%%%%%%%%%%%%%%%%%

\begin{widetext}

\section{DETAILED DERIVATION OF EXPRESSIONS FOR OPTICAL ACTIVITY INDUCED BY THE SECOND-ORDER TERM $\varepsilon_{ij,kl}(\omega)q_kq_l$} \label{Sec:AppB}

In this appendix, we derive explicit expressions for optical activity induced by the second-order term $\varepsilon_{ij,kl}(\omega)q_kq_l$.
Specifically, we solve the eigenvalue equation in Eq.~\eqref{eq:eigenequation_MO_original}, which takes the explicit form
\begin{align}
    \begin{pmatrix}
    1/n^2-\eta^{(\mathrm{S})}_{xx}(1-n^2\tilde{\eta}^{(\mathrm{S})}_{xx}) & -\eta^{(\mathrm{S})}_{xy}+n^2(\eta^{(\mathrm{S})}_{xy}\tilde{\eta}_0-\tilde{G}^{(\mathrm{S})}_{xy}-\tilde{G}^{(\mathrm{o})}_z) \\
    -\eta^{(\mathrm{S})}_{xy}+n^2(\eta^{(\mathrm{S})}_{xy}\tilde{\eta}_0-\tilde{G}^{(\mathrm{S})}_{xy}+\tilde{G}^{(\mathrm{o})}_z) & 1/n^2-\eta^{(\mathrm{S})}_{yy}(1-n^2\tilde{\eta}^{(\mathrm{S})}_{yy})
    \end{pmatrix}
    \begin{pmatrix}
    D_{0x} \\ D_{0y}
    \end{pmatrix}=\bm{0}. \label{eq:eigenequation_MO_original_appendix}
\end{align}
However, since an exact solution would be cumbersome, we solve the equation perturbatively to first order in $\tilde{\eta}_0$, $\tilde{\eta}^{(\mathrm{S})}_{xx}$,
$\tilde{\eta}^{(\mathrm{S})}_{yy}$, $\tilde{G}^{(\mathrm{S})}_{xy}$, and $\tilde{G}^{(\mathrm{o})}_z$ as follows.
We first determine the eigenvalues $n^2$ from the following equation, obtained by setting the determinant of the coefficient matrix to zero: 
\begin{align}
    &\Bigl(\frac{1}{n^2}-\eta^{(\mathrm{S})}_{xx}\Bigr)\Bigl(\frac{1}{n^2}-\eta^{(\mathrm{S})}_{yy}\Bigr)-(\eta^{(\mathrm{S})}_{xy})^2 \notag \\
    &+n^2\Bigl[ \Bigl(\frac{1}{n^2}-\eta^{(\mathrm{S})}_{xx}\Bigr)\eta^{(\mathrm{S})}_{yy}\tilde{\eta}^{(\mathrm{S})}_{yy}
    +\Bigl(\frac{1}{n^2}-\eta^{(\mathrm{S})}_{yy}\Bigr)\eta^{(\mathrm{S})}_{xx}\tilde{\eta}^{(\mathrm{S})}_{xx}+2\eta^{(\mathrm{S})}_{xy}(\eta^{(\mathrm{S})}_{xy}\tilde{\eta}_0-\tilde{G}^{(\mathrm{S})}_{xy})\Bigr] \notag \\
    &+n^4\bigl[\eta^{(\mathrm{S})}_{xx}\eta^{(\mathrm{S})}_{yy}\tilde{\eta}^{(\mathrm{S})}_{xx}\tilde{\eta}^{(\mathrm{S})}_{yy}-(\eta^{(\mathrm{S})}_{xy}\tilde{\eta}_0-\tilde{G}^{(\mathrm{S})}_{xy})^2+(\tilde{G}^{(\mathrm{o})}_z)^2\bigr]=0. 
    \label{eq:eigenequation_MO}
\end{align}
Specifically, we obtain the unperturbed solutions $(n^{(0)})^2$ and calculate the first-order corrections $(n^{(1)})^2$, expressing the eigenvalues as $n^2 \simeq (n^{(0)})^2+(n^{(1)})^2$.
We then determine the corresponding eigenvectors $(D_{0x},D_{0y})$ by substituting the eigenvalues into Eq.~\eqref{eq:eigenequation_MO_original_appendix}.
Having outlined the procedure, we now turn to the analysis of the three cases introduced in the main text: 
(i) isotropic case: $\eta^{(\mathrm{S})}_{xx}=\eta^{(\mathrm{S})}_{yy}=1/\varepsilon^{(0)}$ and $\eta^{(\mathrm{S})}_{xy}=0$ [Appendix~\ref{Subsec:AppB1}],
(ii) anisotropic case (without off-diagonal components): $\eta^{(\mathrm{S})}_{xx} \neq \eta^{(\mathrm{S})}_{yy}$ and $\eta^{(\mathrm{S})}_{xy}=0$ [Appendix~\ref{Subsec:AppB2}], and 
(iii) anisotropic case (with off-diagonal components): $\eta^{(\mathrm{S})}_{xx} \neq \eta^{(\mathrm{S})}_{yy}$ and $\eta^{(\mathrm{S})}_{xy} \neq 0$ [Appendix~\ref{Subsec:AppB3}].

%%%%%%%%%%%%%%%%%%%%%%%%%%%%%%%%%%%%%%%%%%%%%%%%%%%%%%%%%%%%%%%%%%

\subsection{Isotropic case} \label{Subsec:AppB1}

First, we consider case (i) $\eta^{(\mathrm{S})}_{xx}=\eta^{(\mathrm{S})}_{yy}=1/\varepsilon^{(0)}$ and $\eta^{(\mathrm{S})}_{xy}=0$, for which Eq.~\eqref{eq:eigenequation_MO} reduces to
\begin{align}
    \Bigl(\frac{1}{n^2}-\frac{1}{\varepsilon^{(0)}}\Bigr)^2+n^2 \Bigl(\frac{1}{n^2}-\frac{1}{\varepsilon^{(0)}}\Bigr)\frac{\tilde{\eta}^{(\mathrm{S})}_{x+y}}{\varepsilon^{(0)}} 
    +n^4 \Bigl[ \frac{\tilde{\eta}^{(\mathrm{S})}_{xx} \tilde{\eta}^{(\mathrm{S})}_{yy}}{(\varepsilon^{(0)})^2}-(\tilde{G}^{(\mathrm{S})}_{xy})^2+(\tilde{G}^{(\mathrm{o})}_z)^2\Bigr]=0, \label{eq:eigenequation_MO1}
\end{align}
where $\tilde{\eta}^{(\mathrm{S})}_{x \pm y}=\tilde{\eta}^{(\mathrm{S})}_{xx}\pm \tilde{\eta}^{(\mathrm{S})}_{yy}$.
The unperturbed solution is $(n^{(0)})^2=\varepsilon^{(0)}$.
Substituting $n^2=\varepsilon^{(0)}+(n^{(1)})^2$ into Eq.~\eqref{eq:eigenequation_MO1}, we obtain the following equation for the first-order correction $(n^{(1)})^2$:
\begin{align}
    \Bigl(\frac{n^{(1)}}{\varepsilon^{(0)}}\Bigr)^4-\tilde{\eta}^{(\mathrm{S})}_{x+y}\Bigl(\frac{n^{(1)}}{\varepsilon^{(0)}}\Bigr)^2 
    +\tilde{\eta}^{(\mathrm{S})}_{xx}\tilde{\eta}^{(\mathrm{S})}_{yy}-\bigl[ (\varepsilon^{(0)}\tilde{G}^{(\mathrm{S})}_{xy})^2-(\varepsilon^{(0)}\tilde{G}^{(\mathrm{o})}_z)^2\bigr]=0,
\end{align}
where we use $1/n^2-1/\varepsilon^{(0)} \simeq -(n^{(1)}/\varepsilon^{(0)})^2$.
Solving this equation and combining the first-order corrections with the unperturbed solution yields the eigenvalues
\begin{subequations} \label{eq:eigenequation_solution_MO1}
\begin{align}
    n^2 \simeq \varepsilon^{(0)}\Bigl\{1+\frac{\varepsilon^{(0)}}{2}\Bigl[ \tilde{\eta}^{(\mathrm{S})}_{x+y} 
    \pm \sqrt{(\tilde{\eta}^{(\mathrm{S})}_{x-y})^2+(2\varepsilon^{(0)}\tilde{G}^{(\mathrm{S})}_{xy})^2-(2\varepsilon^{(0)}\tilde{G}^{(\mathrm{o})}_z)^2} \Bigr] \Bigr\}.
\end{align}
The corresponding eigenvectors $(D_{0x},D_{0y})$ are given by 
\begin{align}
    \frac{D_{0y}}{D_{0x}}&=\frac{1}{\tilde{G}^{(\mathrm{S})}_{xy}+\tilde{G}^{(\mathrm{o})}_z}\frac{1}{n^2}\Bigl(\frac{1}{n^2}-\frac{1}{\varepsilon^{(0)}}+n^2\frac{\tilde{\eta}^{(\mathrm{S})}_{xx}}{\varepsilon^{(0)}}\Bigr) \notag \\
    &\simeq \frac{1}{\tilde{G}^{(\mathrm{S})}_{xy}+\tilde{G}^{(\mathrm{o})}_z}\frac{1}{(n^{(0)})^2}\Bigl[\tilde{\eta}^{(\mathrm{S})}_{xx}-\Bigl(\frac{n^{(1)}}{\varepsilon^{(0)}}\Bigr)^2\Bigr] \notag \\
    &=\frac{1}{2\varepsilon^{(0)}(\tilde{G}^{(\mathrm{S})}_{xy}+\tilde{G}^{(\mathrm{o})}_z)} 
    \Bigl[ \tilde{\eta}^{(\mathrm{S})}_{x-y} \mp \sqrt{(\tilde{\eta}^{(\mathrm{S})}_{x-y})^2+(2\varepsilon^{(0)}\tilde{G}^{(\mathrm{S})}_{xy})^2-(2\varepsilon^{(0)}\tilde{G}^{(\mathrm{o})}_z)^2} \Bigr].
\end{align}
\end{subequations}
Equation~\eqref{eq:eigenequation_solution_MO1} corresponds to Eq.~\eqref{eq:eigenvalue_MO1} in the main text.

%%%%%%%%%%%%%%%%%%%%%%%%%%%%%%%%%%%%%%%%%%%%%%%%%%%%%%%%%%%%%%%%%%

\subsection{Anisotropic case (without off-diagonal components)} \label{Subsec:AppB2}  \label{Sec:AppB2}

Second, we turn to case (ii) $\eta^{(\mathrm{S})}_{xx} \neq \eta^{(\mathrm{S})}_{yy}$ and $\eta^{(\mathrm{S})}_{xy}=0$, for which Eq.~\eqref{eq:eigenequation_MO} reduces to
\begin{align}
    &\Bigl(\frac{1}{n^2}-\eta^{(\mathrm{S})}_{xx} \Bigr) \Bigl(\frac{1}{n^2}-\eta^{(\mathrm{S})}_{yy} \Bigr) \notag \\
    &+n^2 \Bigl[ \Bigl( \frac{1}{n^2}-\eta^{(\mathrm{S})}_{xx} \Bigr) \eta^{(\mathrm{S})}_{yy} \tilde{\eta}^{(\mathrm{S})}_{yy} 
    +\Bigl( \frac{1}{n^2}-\eta^{(\mathrm{S})}_{yy} \Bigr) \eta^{(\mathrm{S})}_{xx} \tilde{\eta}^{(\mathrm{S})}_{xx} \Bigr] 
    + n^4 \bigl[\eta^{(\mathrm{S})}_{xx}\eta^{(\mathrm{S})}_{yy}\tilde{\eta}^{(\mathrm{S})}_{xx} \tilde{\eta}^{(\mathrm{S})}_{yy} -(\tilde{G}^{(\mathrm{S})}_{xy})^2+(\tilde{G}^{(\mathrm{o})}_z)^2 \bigr]=0. \label{eq:eigenequation_MO2}
\end{align}
We now analyze Eq.~\eqref{eq:eigenequation_MO2} in two regimes: (a) away from degeneracy $(\eta^{(\mathrm{S})}_{xx} \not\simeq \eta^{(\mathrm{S})}_{yy})$
and (b) near degeneracy $(\eta^{(\mathrm{S})}_{xx} \simeq \eta^{(\mathrm{S})}_{yy})$.

%%%%%%%%%%%%%%%%%%%%%%%%%%%%%%%%%

\subsubsection{Away from degeneracy} 

For regime (a), the unperturbed solutions are $(n^{(0)})^2=1/\eta^{(\mathrm{S})}_{xx},1/\eta^{(\mathrm{S})}_{yy}$.
Of these two solutions, we focus on $(n^{(0)})^2=1/\eta^{(\mathrm{S})}_{xx}$ and substitute $n^2=1/\eta^{(\mathrm{S})}_{xx}+(n^{(1)})^2$ into Eq.~\eqref{eq:eigenequation_MO2}
to obtain the following equation for the first-order correction $(n^{(1)})^2$:
\begin{align}
    -( \eta^{(\mathrm{S})}_{xx}-\eta^{(\mathrm{S})}_{yy} ) \biggl[\Bigl(\frac{n^{(1)}}{1/\eta^{(\mathrm{S})}_{xx}} \Bigr)^2 - \tilde{\eta}^{(\mathrm{S})}_{xx}\biggr]=0,  \label{eq:eigenequation_MO2_firstorder_a}
\end{align}
where we use 
\begin{align}
    \frac{1}{n^2}-\eta^{(\mathrm{S})}_{xx} \simeq -\Bigl(\frac{n^{(1)}}{1/\eta^{(\mathrm{S})}_{xx}} \Bigr)^2, \quad
    \frac{1}{n^2}-\eta^{(\mathrm{S})}_{yy} \simeq \eta^{(\mathrm{S})}_{xx}-\eta^{(\mathrm{S})}_{yy}. 
\end{align}
Solving Eq.~\eqref{eq:eigenequation_MO2_firstorder_a} and combining the first-order correction with the unperturbed solution yields the eigenvalue
\begin{subequations} \label{eq:eigenequation_solution_MO2_ax}
\begin{align}
    n^2 \simeq \frac{1}{\eta^{(\mathrm{S})}_{xx}}\Bigl(1+\frac{\tilde{\eta}^{(\mathrm{S})}_{xx}}{\eta^{(\mathrm{S})}_{xx}}\Bigr).
\end{align}
The corresponding eigenvector $(D_{0x},D_{0y})$ is given by 
\begin{align}
    \frac{D_{0y}}{D_{0x}}&=\Bigl(\frac{1}{n^2}-\eta^{(\mathrm{S})}_{yy}+n^2 \eta^{(\mathrm{S})}_{yy} \tilde{\eta}^{(\mathrm{S})}_{yy} \Bigr)^{-1}n^2(\tilde{G}^{(\mathrm{S})}_{xy}-\tilde{G}^{(\mathrm{o})}_z) \notag \\
    &\simeq \Bigl[\frac{1}{(n^{(0)})^2}-\eta^{(\mathrm{S})}_{yy}\Bigr]^{-1} (n^{(0)})^2(\tilde{G}^{(\mathrm{S})}_{xy}-\tilde{G}^{(\mathrm{o})}_z) \notag \\
    &= \frac{\tilde{G}^{(\mathrm{S})}_{xy}-\tilde{G}^{(\mathrm{o})}_z}{\eta^{(\mathrm{S})}_{xx}(\eta^{(\mathrm{S})}_{xx}-\eta^{(\mathrm{S})}_{yy})}. \label{eq:eigenequation_MO2_firstorder_a_eigenvector1}
\end{align}
\end{subequations}
The same analysis can be applied to the other unperturbed solution $(n^{(0)})^2=1/\eta^{(\mathrm{S})}_{yy}$,
yielding the following eigenvalue and corresponding eigenvector:
\begin{align} 
    n^2 \simeq \frac{1}{\eta^{(\mathrm{S})}_{yy}}\Bigl(1+\frac{\tilde{\eta}^{(\mathrm{S})}_{yy}}{\eta^{(\mathrm{S})}_{yy}}\Bigr), \quad 
    \frac{D_{0x}}{D_{0y}}&\simeq -\frac{\tilde{G}^{(\mathrm{S})}_{xy}+\tilde{G}^{(\mathrm{o})}_z}{\eta^{(\mathrm{S})}_{yy}(\eta^{(\mathrm{S})}_{xx}-\eta^{(\mathrm{S})}_{yy})}. \label{eq:eigenequation_solution_MO2_ay}
\end{align}
Together, Eqs.~\eqref{eq:eigenequation_solution_MO2_ax} and \eqref{eq:eigenequation_solution_MO2_ay} correspond to Eq.~\eqref{eq:eigenvalue_MO2_MO} in the main text.

Here, we note that the above perturbative analysis is valid only when 
\begin{align}
    |\eta^{(\mathrm{S})}_{xx}-\eta^{(\mathrm{S})}_{yy}| \gg \left| \frac{\tilde{G}^{(\mathrm{S})}_{xy}-\tilde{G}^{(\mathrm{o})}_z}{\eta^{(\mathrm{S})}_{xx}} \right|,\ 
    \left|\frac{\tilde{G}^{(\mathrm{S})}_{xy}+\tilde{G}^{(\mathrm{o})}_z}{\eta^{(\mathrm{S})}_{yy}}\right|. \label{eq:eigenequation_MO2_validity_app}
\end{align}
Otherwise, at least one of the conditions $|D_{0y}/D_{0x}| \ll 1$ and $|D_{0x}/D_{0y}| \ll 1$ is no longer satisfied, so that the corresponding eigenvector cannot be treated perturbatively.
In this sense, ``away from degeneracy'' refers to the regime defined by the condition in Eq.~\eqref{eq:eigenequation_MO2_validity_app}.
When this condition is violated, we enter regime (b) near degeneracy, which requires a different treatment, as discussed below.

%%%%%%%%%%%%%%%%%%%%%%%%%%%%%%%%%

\subsubsection{Near degeneracy}

In regime (b), we treat $\eta^{(\mathrm{S})}_{xx}-\eta^{(\mathrm{S})}_{yy}$ as a small quantity of the same order as the other perturbative components and reformulate the perturbative analysis.
Specifically, we introduce the zeroth-order quantity $\eta^{(0)}$ and the difference $\Delta$ through $\eta^{(\mathrm{S})}_{xx}=\eta^{(0)}+\Delta/2$ and $\eta^{(\mathrm{S})}_{yy}=\eta^{(0)}-\Delta/2$,
and solve the eigenvalue equation to first order in $\Delta$ and the other perturbative components.
The resulting eigenvalues and eigenvectors are given by
\begin{subequations} \label{eq:eigenequation_solution_MO2_b}
\begin{align}
    n^2 &\simeq \frac{1}{\eta^{(0)}}\Bigl\{1+\frac{1}{2\eta^{(0)}}\Bigl[ \tilde{\eta}^{(\mathrm{S})}_{x+y} 
    \pm \sqrt{(\tilde{\eta}^{(\mathrm{S})}_{x-y}-\Delta)^2+(2\tilde{G}^{(\mathrm{S})}_{xy}/\eta^{(0)})^2-(2\tilde{G}^{(\mathrm{o})}_z/\eta^{(0)})^2} \Bigr] \Bigr\}, \\
    \frac{D_{0y}}{D_{0x}} &\simeq \frac{1}{2(\tilde{G}^{(\mathrm{S})}_{xy}+\tilde{G}^{(\mathrm{o})}_z)/\eta^{(0)}} 
    \Bigl[ (\tilde{\eta}^{(\mathrm{S})}_{x-y}-\Delta) \mp \sqrt{(\tilde{\eta}^{(\mathrm{S})}_{x-y}-\Delta)^2+(2\tilde{G}^{(\mathrm{S})}_{xy}/\eta^{(0)})^2-(2\tilde{G}^{(\mathrm{o})}_z/\eta^{(0)})^2} \Bigr].
\end{align}
\end{subequations}

The eigenvalues and eigenvectors in Eq.~\eqref{eq:eigenequation_solution_MO2_b} reproduce the results for case (i) and regime (a) in the appropriate limits of $\Delta$.
For case (i), taking the limit $\Delta \to 0$ and then setting $\eta^{(0)}=1/\varepsilon^{(0)}$ recovers Eq.~\eqref{eq:eigenequation_solution_MO1}.
For regime (a), taking $|\tilde{\eta}^{(\mathrm{S})}_{x-y}|, \ |(\tilde{G}^{(\mathrm{S})}_{xy}\mp\tilde{G}^{(\mathrm{o})}_{z})/\eta^{(0)}| \ll |\Delta| \ll |\eta^{(0)}|$
allows us to recover Eqs.~\eqref{eq:eigenequation_solution_MO2_ax} and \eqref{eq:eigenequation_solution_MO2_ay}.
To see this, we first expand the eigenvalues and eigenvectors in Eq.~\eqref{eq:eigenequation_solution_MO2_b}, denoting the upper and lower signs by the subscripts $+$ and $-$, respectively:
\begin{subequations} \label{eq:eigenequation_solution_MO2_b_limit}
\begin{alignat}{2}
    n_+^2 &\simeq \frac{1}{\eta^{(0)}}\Bigl[1+\frac{1}{\eta^{(0)}} \Bigl(\tilde{\eta}^{(\mathrm{S})}_{xx}-\frac{\Delta}{2}\Bigr) \Bigr], & \quad 
    \Bigl(\frac{D_{0y}}{D_{0x}}\Bigr)_+& \simeq -\frac{\tilde{G}^{(\mathrm{S})}_{xy}-\tilde{G}^{(\mathrm{o})}_z}{\eta^{(0)}(\tilde{\eta}^{(\mathrm{S})}_{x-y}-\Delta)} 
    \simeq \frac{\tilde{G}^{(\mathrm{S})}_{xy}-\tilde{G}^{(\mathrm{o})}_z}{\eta^{(0)}\Delta}, \\
    n_-^2 &\simeq \frac{1}{\eta^{(0)}}\Bigl[1+\frac{1}{\eta^{(0)}}\Bigl(\tilde{\eta}^{(\mathrm{S})}_{yy}+\frac{\Delta}{2}\Bigr) \Bigr], & \quad 
    \Bigl(\frac{D_{0x}}{D_{0y}}\Bigr)_-& \simeq \frac{\tilde{G}^{(\mathrm{S})}_{xy}+\tilde{G}^{(\mathrm{o})}_z}{\eta^{(0)}(\tilde{\eta}^{(\mathrm{S})}_{x-y}-\Delta)} 
    \simeq -\frac{\tilde{G}^{(\mathrm{S})}_{xy}+\tilde{G}^{(\mathrm{o})}_z}{\eta^{(0)}\Delta},
\end{alignat}
\end{subequations}
where we use 
\begin{align}
    \sqrt{(\tilde{\eta}^{(\mathrm{S})}_{x-y}-\Delta)^2+(2\tilde{G}^{(\mathrm{S})}_{xy}/\eta^{(0)})^2-(2\tilde{G}^{(\mathrm{o})}_z/\eta^{(0)})^2} 
    \simeq (\tilde{\eta}^{(\mathrm{S})}_{x-y}-\Delta)+\frac{(2\tilde{G}^{(\mathrm{S})}_{xy}/\eta^{(0)})^2-(2\tilde{G}^{(\mathrm{o})}_z/\eta^{(0)})^2}{2(\tilde{\eta}^{(\mathrm{S})}_{x-y}-\Delta)}.
\end{align}
We then substitute $\eta^{(\mathrm{S})}_{xx}=\eta^{(0)}+\Delta/2$ and $\eta^{(\mathrm{S})}_{yy}=\eta^{(0)}-\Delta/2$ into Eqs.~\eqref{eq:eigenequation_solution_MO2_ax} and \eqref{eq:eigenequation_solution_MO2_ay}
and expand the resulting expressions as 
\begin{subequations} \label{eq:eigenequation_solution_MO2_a_limit}
\begin{alignat}{2}
    n_x^2 \simeq \frac{1}{\eta^{(0)}}\Bigl[1+\frac{1}{\eta^{(0)}} \Bigl(\tilde{\eta}^{(\mathrm{S})}_{xx}-\frac{\Delta}{2}\Bigr) \Bigr], \quad
    \Bigl(\frac{D_{0y}}{D_{0x}}\Bigr)_x& \simeq \frac{\tilde{G}^{(\mathrm{S})}_{xy}-\tilde{G}^{(\mathrm{o})}_z}{\eta^{(0)}\Delta}, \\
    n_y^2 \simeq \frac{1}{\eta^{(0)}}\Bigl[1+\frac{1}{\eta^{(0)}}\Bigl(\tilde{\eta}^{(\mathrm{S})}_{yy}+\frac{\Delta}{2}\Bigr) \Bigr], \quad
    \Bigl(\frac{D_{0x}}{D_{0y}}\Bigr)_y& \simeq -\frac{\tilde{G}^{(\mathrm{S})}_{xy}+\tilde{G}^{(\mathrm{o})}_z}{\eta^{(0)}\Delta},
\end{alignat}
\end{subequations}
where the subscripts $x$ and $y$ label the solutions given in Eqs.~\eqref{eq:eigenequation_solution_MO2_ax} and \eqref{eq:eigenequation_solution_MO2_ay}, respectively.
Comparing Eqs.~\eqref{eq:eigenequation_solution_MO2_b_limit} and \eqref{eq:eigenequation_solution_MO2_a_limit},
we find that the $+$ and $-$ solutions agree with the $x$ and $y$ solutions, respectively.
Taken together, these results indicate that case (i) and regime (a) are continuously connected through regime (b).

%%%%%%%%%%%%%%%%%%%%%%%%%%%%%%%%%%%%%%%%%%%%%%%%%%%%%%%%%%%%%%%%%%

\subsection{Anisotropic case (with off-diagonal components)} \label{Subsec:AppB3}

Finally, we address case (iii) $\eta^{(\mathrm{S})}_{xx} \neq \eta^{(\mathrm{S})}_{yy}$ and $\eta^{(\mathrm{S})}_{xy} \neq 0$, for which Eq.~\eqref{eq:eigenequation_MO} reduces to
\begin{align}
    &\Bigl(\frac{1}{n^2}-\eta^{(\mathrm{S})}_{xx} \Bigr) \Bigl(\frac{1}{n^2}-\eta^{(\mathrm{S})}_{yy} \Bigr)-(\eta^{(\mathrm{S})}_{xy})^2 \notag \\
    &+n^2 \Bigl[\Bigl( \frac{1}{n^2}-\eta^{(\mathrm{S})}_{xx} \Bigr) \eta^{(\mathrm{S})}_{yy} \tilde{\eta}^{(\mathrm{S})}_{yy} +\Bigl( \frac{1}{n^2}-\eta^{(\mathrm{S})}_{yy} \Bigr) \eta^{(\mathrm{S})}_{xx} \tilde{\eta}^{(\mathrm{S})}_{xx}+2(\eta^{(\mathrm{S})}_{xy})^2\tilde{\eta}^{(\mathrm{S})}_{xy}  \Bigr] \notag \\
    &+n^4 \bigl[\eta^{(\mathrm{S})}_{xx}\eta^{(\mathrm{S})}_{yy}\tilde{\eta}^{(\mathrm{S})}_{xx} \tilde{\eta}^{(\mathrm{S})}_{yy}-(\eta^{(\mathrm{S})}_{xy})^2(\tilde{\eta}^{(\mathrm{S})}_{xy})^2+(\tilde{G}^{(\mathrm{o})}_z)^2 \bigr]=0, \label{eq:eigenequation_MO3}
\end{align}
where $\tilde{\eta}^{(\mathrm{S})}_{xy}=\tilde{\eta}_0-\tilde{G}^{(\mathrm{S})}_{xy}/\eta^{(\mathrm{S})}_{xy}$.
The unperturbed solutions $(n^{(0)})^2 \eqqcolon n^2_{xy}$ are given by 
\begin{align}
    \frac{1}{n^2_{xy}}=\frac{1}{2}\Bigl[ \eta^{(\mathrm{S})}_{x+y} \pm \sqrt{(\eta^{(\mathrm{S})}_{x-y})^2+(2\eta^{(\mathrm{S})}_{xy})^2}\Bigr],
\end{align}
where $\eta^{(\mathrm{S})}_{x \pm y}=\eta^{(\mathrm{S})}_{xx}\pm \eta^{(\mathrm{S})}_{yy}$.
Substituting $n^2=n^2_{xy}+(n^{(1)})^2$ into Eq.~\eqref{eq:eigenequation_MO3}, we obtain the following equation for the first-order correction $(n^{(1)})^2$:
\begin{align}
    \Bigl(\eta^{(\mathrm{S})}_{x+y}-\frac{2}{n^2_{xy}} \Bigr)\frac{(n^{(1)})^2}{n^4_{xy}}
    +\eta^{(\mathrm{S})}_{xx}\tilde{\eta}^{(\mathrm{S})}_{xx}+\eta^{(\mathrm{S})}_{yy}\tilde{\eta}^{(\mathrm{S})}_{yy} 
    -n^2_{xy} \bigl[\eta^{(\mathrm{S})}_{xx}\eta^{(\mathrm{S})}_{yy}\tilde{\eta}^{(\mathrm{S})}_{x+y}-2(\eta^{(\mathrm{S})}_{xy})^2\tilde{\eta}^{(\mathrm{S})}_{xy} \bigr]=0,  \label{eq:eigenequation_MO3_eigenvalues}
\end{align}
where we use 
\begin{align}
    \frac{1}{n^2}-\eta^{(\mathrm{S})}_{xx} \simeq \frac{1}{n^2_{xy}}-\eta^{(\mathrm{S})}_{xx}-\frac{(n^{(1)})^2}{n^4_{xy}}, \quad 
    \frac{1}{n^2}-\eta^{(\mathrm{S})}_{yy} \simeq \frac{1}{n^2_{xy}}-\eta^{(\mathrm{S})}_{yy}-\frac{(n^{(1)})^2}{n^4_{xy}}.
\end{align}
Solving Eq.~\eqref{eq:eigenequation_MO3_eigenvalues} and combining the first-order corrections with the unperturbed solutions yields the eigenvalues
\begin{align}  \label{eq:eigenequation_solution_MO3_eigenvalues}
    n^2 &\simeq n^2_{xy} \biggl\{ 1+\frac{n^4_{xy}}{2}\biggl[(\eta^{(\mathrm{S})}_{xx}\tilde{\eta}^{(\mathrm{S})}_{xx}+\eta^{(\mathrm{S})}_{yy}\tilde{\eta}^{(\mathrm{S})}_{yy}) 
    \pm \frac{\eta^{(\mathrm{S})}_{x-y}(\eta^{(\mathrm{S})}_{xx}\tilde{\eta}^{(\mathrm{S})}_{xx}-\eta^{(\mathrm{S})}_{yy}\tilde{\eta}^{(\mathrm{S})}_{yy})
    +4(\eta^{(\mathrm{S})}_{xy})^2\tilde{\eta}^{(\mathrm{S})}_{xy}}{\sqrt{(\eta^{(\mathrm{S})}_{x-y})^2+(2\eta^{(\mathrm{S})}_{xy})^2}}\biggr] \biggr\}, 
\end{align}     
where we use 
\begin{align}
    &\pm \frac{1}{\sqrt{(\eta^{(\mathrm{S})}_{x-y})^2+(2\eta^{(\mathrm{S})}_{xy})^2}}\Bigl[\frac{1}{n^2_{xy}} (\eta^{(\mathrm{S})}_{xx}\tilde{\eta}^{(\mathrm{S})}_{xx}+\eta^{(\mathrm{S})}_{yy}\tilde{\eta}^{(\mathrm{S})}_{yy})
    -\eta^{(\mathrm{S})}_{xx}\eta^{(\mathrm{S})}_{yy}\tilde{\eta}^{(\mathrm{S})}_{x+y}\Bigr] \notag \\
    &=\frac{1}{2}(\eta^{(\mathrm{S})}_{xx}\tilde{\eta}^{(\mathrm{S})}_{xx}+\eta^{(\mathrm{S})}_{yy}\tilde{\eta}^{(\mathrm{S})}_{yy}) 
    \pm \frac{1}{2\sqrt{(\eta^{(\mathrm{S})}_{x-y})^2+(2\eta^{(\mathrm{S})}_{xy})^2}} \underbrace{\bigl[ \eta^{(\mathrm{S})}_{x+y}(\eta^{(\mathrm{S})}_{xx}\tilde{\eta}^{(\mathrm{S})}_{xx}+\eta^{(\mathrm{S})}_{yy}\tilde{\eta}^{(\mathrm{S})}_{yy}) 
    -2\eta^{(\mathrm{S})}_{xx}\eta^{(\mathrm{S})}_{yy}\tilde{\eta}^{(\mathrm{S})}_{x+y} \bigr]}_{\eta^{(\mathrm{S})}_{x-y}(\eta^{(\mathrm{S})}_{xx}\tilde{\eta}^{(\mathrm{S})}_{xx}-\eta^{(\mathrm{S})}_{yy}\tilde{\eta}^{(\mathrm{S})}_{yy})}.
\end{align}
The corresponding eigenvectors $(D_{0x},D_{0y})$ are given by 
\begin{align} \label{eq:eigenequation_solution_MO3_eigenvectors}
    \frac{D_{0y}}{D_{0x}}&=\frac{1}{\eta^{(\mathrm{S})}_{xy}-n^2(\eta^{(\mathrm{S})}_{xy}\tilde{\eta}^{(\mathrm{S})}_{xy}-\tilde{G}^{(\mathrm{o})}_z)} 
    \Bigl(\frac{1}{n^2}-\eta^{(\mathrm{S})}_{xx}+n^2 \eta^{(\mathrm{S})}_{xx} \tilde{\eta}^{(\mathrm{S})}_{xx} \Bigr) \notag \\
    & \simeq \frac{1}{\eta^{(\mathrm{S})}_{xy}-(n^{(0)})^2(\eta^{(\mathrm{S})}_{xy}\tilde{\eta}^{(\mathrm{S})}_{xy}-\tilde{G}^{(\mathrm{o})}_z)} 
    \Bigl[\frac{1}{(n^{(0)})^2+(n^{(1)})^2}-\eta^{(\mathrm{S})}_{xx}+(n^{(0)})^2 \eta^{(\mathrm{S})}_{xx} \tilde{\eta}^{(\mathrm{S})}_{xx} \Bigr] \notag \\
    & \simeq \frac{-1}{2\eta^{(\mathrm{S})}_{xy}} \Bigl[\eta^{(\mathrm{S})}_{x-y} \mp \sqrt{(\eta^{(\mathrm{S})}_{x-y})^2+(2\eta^{(\mathrm{S})}_{xy})^2}\Bigr] \biggl\{ 1-n^2_{xy} 
    \biggl[\frac{\tilde{G}^{(\mathrm{o})}_z}{\eta^{(\mathrm{S})}_{xy}} 
    \mp \frac{\eta^{(\mathrm{S})}_{xx}\tilde{\eta}^{(\mathrm{S})}_{xx}-\eta^{(\mathrm{S})}_{yy}\tilde{\eta}^{(\mathrm{S})}_{yy}-\eta^{(\mathrm{S})}_{x-y}\tilde{\eta}^{(\mathrm{S})}_{xy}}{\sqrt{(\eta^{(\mathrm{S})}_{x-y})^2+(2\eta^{(\mathrm{S})}_{xy})^2}} \biggr] \biggr\},
\end{align}
where we use 
\begin{align}
    &-\frac{(n^{(1)})^2}{(n^{(0)})^4}+(n^{(0)})^2\eta^{(\mathrm{S})}_{xx}\tilde{\eta}^{(\mathrm{S})}_{xx}
    -\frac{1}{2}(n^{(0)})^2 \Bigl[\eta^{(\mathrm{S})}_{x-y} \mp \sqrt{(\eta^{(\mathrm{S})}_{x-y})^2+(2\eta^{(\mathrm{S})}_{xy})^2} \Bigr] \tilde{\eta}^{(\mathrm{S})}_{xy} \notag \\
    &= \frac{n_{xy}^2}{2}\biggl\{(\eta^{(\mathrm{S})}_{xx}\tilde{\eta}^{(\mathrm{S})}_{xx}-\eta^{(\mathrm{S})}_{yy}\tilde{\eta}^{(\mathrm{S})}_{yy}) 
    -\Bigl[\eta^{(\mathrm{S})}_{x-y} \mp \sqrt{(\eta^{(\mathrm{S})}_{x-y})^2+(2\eta^{(\mathrm{S})}_{xy})^2} \Bigr] \tilde{\eta}^{(\mathrm{S})}_{xy} 
    \mp \frac{\eta^{(\mathrm{S})}_{x-y}(\eta^{(\mathrm{S})}_{xx}\tilde{\eta}^{(\mathrm{S})}_{xx}-\eta^{(\mathrm{S})}_{yy}\tilde{\eta}^{(\mathrm{S})}_{yy})
    +4(\eta^{(\mathrm{S})}_{xy})^2\tilde{\eta}^{(\mathrm{S})}_{xy}}{\sqrt{(\eta^{(\mathrm{S})}_{x-y})^2+(2\eta^{(\mathrm{S})}_{xy})^2}}\biggr\} \notag \\
    &=\frac{n_{xy}^2}{2}\biggl( 1 \mp \frac{\eta^{(\mathrm{S})}_{x-y}}{\sqrt{(\eta^{(\mathrm{S})}_{x-y})^2+(2\eta^{(\mathrm{S})}_{xy})^2}} \biggr)
    (\eta^{(\mathrm{S})}_{xx}\tilde{\eta}^{(\mathrm{S})}_{xx}-\eta^{(\mathrm{S})}_{yy}\tilde{\eta}^{(\mathrm{S})}_{yy}-\eta^{(\mathrm{S})}_{x-y}\tilde{\eta}^{(\mathrm{S})}_{xy}).
\end{align}
Equations~\eqref{eq:eigenequation_solution_MO3_eigenvalues} and \eqref{eq:eigenequation_solution_MO3_eigenvectors} correspond to Eq.~\eqref{eq:eigenvalue_MO3} in the main text.

\end{widetext}

%%%%%%%%%%%%%%%%%%%%%%%%%%%%%%%%%%%%%%%%%%%%%%%%%%%%%%%%%%%%%%%%%%%%%%%%%%%%%%%%%%%%%%%%%%%%%%%%%%%%%%%%%%%%%%%%%%%%%%%%%%%%%%%%%%%%

\section{DETAILS OF THE ANALYSIS IN SEC.~\ref{Sec:MO_conditions}} \label{Sec:AppC}

In this appendix, we provide a detailed analysis for each light-propagation direction $[uvw]$ listed in Table~\ref{tab:Mijk_list},
deriving the expression for $\tilde{G}^{(\mathrm{o})}_{[uvw]}$ 
and determining which of the optical-activity types shown in Fig.~\ref{fig:MO_Type} is realized.
The analysis begins with the classification of optical activity, which proceeds as follows.
First, we introduce the rotation matrix $\mathcal{R}^{[uvw]}$ that transforms the coordinate components from coordinate system (i) ($x,y,z$) to (ii) ($X,Y,Z$) for the $[uvw]$ direction.
Second, using $\mathcal{R}^{[uvw]}$, we express the symmetry-allowed forms of $\varepsilon^{(\mathrm{S})}_{ij}$ and $\tilde{g}^{(\mathrm{S})}_{ij}=\varepsilon^{(\mathrm{S})}_{ij,kl}\tilde{q}_k\tilde{q}_l$ [Eq.~\eqref{eq:epsilon_normal_MO_coefficients_normalized_OA}] in coordinate system (ii),
with the factor $(\omega/c)^2$ in $\tilde{g}^{(\mathrm{S})}_{ij}$ omitted hereafter.
Finally, we consider CAA and determine the optical-activity type based on the resulting forms.
In doing so, we follow the flowchart shown in Fig.~\ref{fig:Flowchart}, which summarizes the discussion in Sec.~\ref{Sec:Macroscopic_theory_MO}.
We then turn to deriving the expression for $\tilde{G}^{(\mathrm{o})}_{[uvw]}$ from $\tilde{g}^{(\mathrm{o})}_{\mu,kl}$ [Eq.~\eqref{eq:tildegijk_eijkl}], for which we use the same rotation matrix $\mathcal{R}^{[uvw]}$.
Specifically, we express $\tilde{\bm{G}}^{(\mathrm{o})}$ in coordinate system (ii) and identify its component along the light-propagation direction as $\tilde{G}^{(\mathrm{o})}_{[uvw]}$.

For these analyses, we consider the symmetry-allowed forms of $\varepsilon^{(\mathrm{S})}_{ij}$, $\varepsilon^{(\mathrm{S})}_{ij,kl}$, and $\tilde{g}^{(\mathrm{o})}_{\mu,kl}$ in coordinate system (i).
Among these quantities, the forms of $\tilde{g}^{(\mathrm{o})}_{\mu,kl}$ have already been given in Table~\ref{tab:Mijk_list}.
We therefore summarize those of $\varepsilon^{(\mathrm{S})}_{ij}$ and $\varepsilon^{(\mathrm{S})}_{ij,kl}$ in Table~\ref{tab:eijkl_list} \cite{Agranovich1984}, where Voigt notation is used for $\varepsilon^{(\mathrm{S})}_{ij,kl}$: $xx \to 1$, $yy \to 2$, $zz \to 3$, $yz \to 4$, $zx \to 5$, and $xy \to 6$.
Their features can be briefly summarized as follows.
For $\varepsilon^{(\mathrm{S})}_{ij}$, there are three possible forms, with all other components vanishing:
\begin{subequations} \label{eq:eij_bui}
    \begin{align}
        &\varepsilon^{(\mathrm{S})}_{xx} \neq \varepsilon^{(\mathrm{S})}_{yy} \neq \varepsilon^{(\mathrm{S})}_{zz} \quad (\mathrm{Biaxial}), \label{eq:eij_biaxial} \\
        &\varepsilon^{(\mathrm{S})}_{xx} = \varepsilon^{(\mathrm{S})}_{yy} \neq \varepsilon^{(\mathrm{S})}_{zz} \quad (\mathrm{Uniaxial}), \label{eq:eij_uniaxial} \\
        &\varepsilon^{(\mathrm{S})}_{xx} = \varepsilon^{(\mathrm{S})}_{yy} = \varepsilon^{(\mathrm{S})}_{zz} \quad (\mathrm{Isotropic}). \label{eq:eij_isotropic}
    \end{align}
\end{subequations}
These forms are common within each crystal system, leading to the same conditions for uniform NMOA.
For $\varepsilon^{(\mathrm{S})}_{ij,kl}$, in contrast, the forms can differ within the same crystal system, leading to different conditions for Lorentz NMOA.
With these preliminaries in place, we now carry out the analyses for each direction.
Specifically, we treat the [111], [101]/[011], and [110]/[100]/[010] directions in Appendices~\ref{Subsec:111_direction}, \ref{Subsec:101_011_direction}, and \ref{Subsec:110_100_010_direction}, respectively.

\tikzstyle{startstop} = [rectangle, rounded corners, minimum width=1mm, minimum height=7mm, text centered, draw=black, fill=red!20]
\tikzstyle{process} = [rectangle, minimum width=15mm, minimum height=8mm, text centered, draw=black, fill=blue!20]
\tikzstyle{decision} = [diamond, minimum width=0.5mm, minimum height=0.5mm, text centered, draw=black, fill=green!20, aspect=1.5, inner sep=0.5pt]

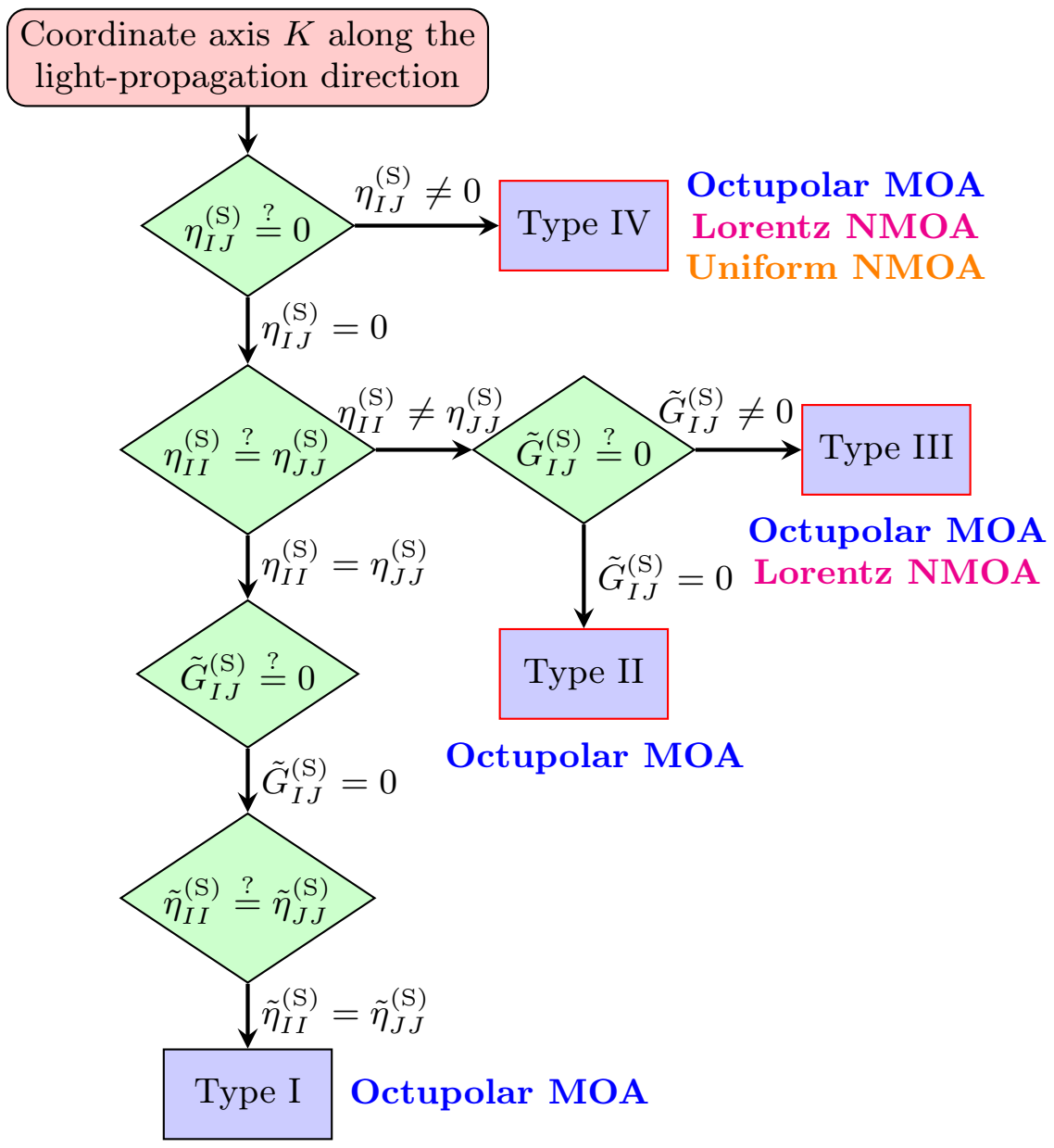
\begin{figure}[t]
\centering
\begin{tikzpicture}[scale=0.8, transform shape]
    \node (start) [startstop, align=center] {Coordinate axis $K$ along the \\ light-propagation direction};
    \node (dec1) [decision, below of=start, yshift=-0.5cm] {$\eta^{(\mathrm{S})}_{IJ} \overset{?}{=} 0$};
    \node (proc1) [process, right of=dec1, xshift=2.0cm, draw=red] {Type IV};
    \node[right of=proc1, xshift=1.25cm, yshift=0cm, align=center] {\textcolor{blue}{\textbf{Octupolar MOA}} \\ \textcolor{magenta}{\textbf{Lorentz NMOA}} \\ \textcolor{orange}{\textbf{Uniform NMOA}}};
    \node (dec2) [decision, below of=dec1, yshift=-1.0cm] {$\eta^{(\mathrm{S})}_{II} \overset{?}{=} \eta^{(\mathrm{S})}_{JJ}$};
    \node (dec3) [decision, right of=dec2, xshift=2.0cm] {$\tilde{G}^{(\mathrm{S})}_{IJ} \overset{?}{=} 0$};
    \node (proc2) [process, right of=dec3, xshift=1.7cm, draw=red] {Type III};
    \node[below of=proc2, xshift=0.1cm, yshift=0.1cm, align=center] {\textcolor{blue}{\textbf{Octupolar MOA}} \\ \textcolor{magenta}{\textbf{Lorentz NMOA}}};
    \node (proc3) [process, below of=dec3, yshift=-1.0cm, draw=red] {Type II};
    \node[below of=proc3, xshift=0.1cm, yshift=0.25cm, align=center] {\textcolor{blue}{\textbf{Octupolar MOA}}};
    \node (dec4) [decision, below of=dec2, yshift=-1.0cm] {$\tilde{G}^{(\mathrm{S})}_{IJ} \overset{?}{=} 0$};
    \node (dec5) [decision, below of=dec4, yshift=-1.0cm] {$\tilde{\eta}^{(\mathrm{S})}_{II} \overset{?}{=}{\tilde{\eta}^{(\mathrm{S})}_{JJ}}$};
    \node (proc4) [process, below of=dec5, yshift=-0.75cm] {Type I};
    \node[right of=proc4, xshift=1.25cm] {\textcolor{blue}{\textbf{Octupolar MOA}}};

    \draw [arrow] (start) -- (dec1);
    \draw [arrow] (dec1.east) -- (proc1) node[midway, above, xshift=-0.8mm] {$\eta^{(\mathrm{S})}_{IJ} \neq 0$};
    \draw [arrow] (dec1.south) -- (dec2) node[midway, right, right, yshift=0.5mm] {$\eta^{(\mathrm{S})}_{IJ}=0$};
    \draw [arrow] (dec2.east) -- (dec3) node[midway, above, xshift=-0.2mm, yshift=0.3mm] {$\eta^{(\mathrm{S})}_{II} \neq \eta^{(\mathrm{S})}_{JJ}$};
    \draw [arrow] (dec3.east) -- (proc2.west) node[midway, above, xshift=-2.0mm, yshift=0.3mm] {$\tilde{G}^{(\mathrm{S})}_{IJ} \neq 0$};
    \draw [arrow] (dec3.south) -- (proc3) node[midway, right] {$\tilde{G}^{(\mathrm{S})}_{IJ} = 0$};
    \draw [arrow] (dec2.south) -- (dec4) node[midway, right, right, yshift=0.5mm] {$\eta^{(\mathrm{S})}_{II}=\eta^{(\mathrm{S})}_{JJ}$};
    % \draw [arrow,dashed] (dec4.east) -- ++(4.7,0) -- (proc2.south) node[midway, right] {$\tilde{G}^{(\mathrm{S})}_{ij} \neq 0$};
    \draw [arrow] (dec4.south) -- (dec5) node[midway, right] {$\tilde{G}^{(\mathrm{S})}_{IJ} = 0$};
    % \draw [arrow,dashed] (dec5.east) -- ++(1.85,0) -- (proc3.south) node[midway, right] {$\tilde{\eta}^{(\mathrm{S})}_{ii} \neq \tilde{\eta}^{(\mathrm{S})}_{jj}$};
    \draw [arrow] (dec5.south) -- (proc4) node[midway, right, yshift=0.5mm] {$\tilde{\eta}^{(\mathrm{S})}_{II}=\tilde{\eta}^{(\mathrm{S})}_{JJ}$};
\end{tikzpicture}
\caption{Flowchart illustrating the procedure for determining the optical-activity type under CAA.
Once the axis $K$ in coordinate system (ii) is chosen along the light-propagation direction, the tensor components in the $I$-$J$ plane perpendicular to $K$ are examined.}
\label{fig:Flowchart}
\end{figure}

\begin{table}[t]
    \caption{Symmetry-allowed forms of $\varepsilon^{(\mathrm{S})}_{ij}$ and $\varepsilon^{(\mathrm{S})}_{ij,kl}$ for octupolar MPGs
    (the superscript (S) is omitted below).
    The coefficient $\varepsilon_{ij}$ takes one of the three forms defined in Eq.~\eqref{eq:eij_bui}: biaxial (B), uniaxial (U), and isotropic (I).
    The coefficient $\varepsilon_{ij,kl}$ is expressed in Voigt notation: $xx \to 1$, $yy \to 2$, $zz \to 3$, $yz \to 4$, $zx \to 5$, and $xy \to 6$.
    }
    %\vspace{1mm}
    \label{tab:eijkl_list}
    \centering
    \tabcolsep = 0.8mm
    \renewcommand\arraystretch{1.2}
    \begin{tabular}{c||c|c} \hline \hline
        MPG & $\varepsilon_{ij}$ & $\varepsilon_{ij,kl}$ \\ \hline 
        \multirow{2}{*}{$mmm$} & \multirow{2}{*}{B} & $\varepsilon_{11},\varepsilon_{22},\varepsilon_{33},\varepsilon_{44},\varepsilon_{55},\varepsilon_{66},$  \\ 
        & & $\varepsilon_{12},\varepsilon_{13},\varepsilon_{21},\varepsilon_{23},\varepsilon_{31},\varepsilon_{32},$  \\ \hline
        \multirow{3}{*}{$4'/m$} & \multirow{5}{*}{U} & $\varepsilon_{11}=\varepsilon_{22},\varepsilon_{33},\varepsilon_{44}=\varepsilon_{55},\varepsilon_{66},$ \\
        & & $\varepsilon_{12}=\varepsilon_{21},\varepsilon_{13}=\varepsilon_{23},\varepsilon_{31}=\varepsilon_{32},$  \\
        & & $\varepsilon_{16}=-\varepsilon_{26},\varepsilon_{61}=-\varepsilon_{62},\varepsilon_{45}=-\varepsilon_{54}$  \\ \cline{1-1} \cline{3-3}
        $4'/mmm'$ & & $\varepsilon_{11}=\varepsilon_{22},\varepsilon_{33},\varepsilon_{44}=\varepsilon_{55},\varepsilon_{66},$ \\ 
        $4'/mm'm$ & & $\varepsilon_{12}=\varepsilon_{21},\varepsilon_{13}=\varepsilon_{23},\varepsilon_{31}=\varepsilon_{32}$ \\ \hline
        \multirow{3}{*}{$\bar{3}m$} & \multirow{3}{*}{U} & $\varepsilon_{11}=\varepsilon_{22},\varepsilon_{33},\varepsilon_{44}=\varepsilon_{55},\varepsilon_{66}=(\varepsilon_{11}-\varepsilon_{12})/2,$ \\ 
        & & $\varepsilon_{12}=\varepsilon_{21},\varepsilon_{13}=\varepsilon_{23},\varepsilon_{31}=\varepsilon_{32},$ \\ 
        & & $\varepsilon_{14}=-\varepsilon_{24}=\varepsilon_{65}, \varepsilon_{41}=-\varepsilon_{42}=\varepsilon_{56}$ \\ \hline
        \multirow{3}{*}{$6'/m'$} & \multirow{5}{*}{U} & $\varepsilon_{11}=\varepsilon_{22},\varepsilon_{33},\varepsilon_{44}=\varepsilon_{55},\varepsilon_{66}=(\varepsilon_{11}-\varepsilon_{12})/2,$ \\ 
        & & $\varepsilon_{12}=\varepsilon_{21},\varepsilon_{13}=\varepsilon_{23},\varepsilon_{31}=\varepsilon_{32},$ \\ 
        & & $\varepsilon_{16}=-\varepsilon_{26}=-\varepsilon_{61}=\varepsilon_{62},\varepsilon_{45}=-\varepsilon_{54}$ \\ \cline{1-1} \cline{3-3}
        $6'/m'mm'$ & & $\varepsilon_{11}=\varepsilon_{22},\varepsilon_{33},\varepsilon_{44}=\varepsilon_{55},\varepsilon_{66}=(\varepsilon_{11}-\varepsilon_{12})/2,$ \\ 
        $6'/m'm'm$ & & $\varepsilon_{12}=\varepsilon_{21},\varepsilon_{13}=\varepsilon_{23},\varepsilon_{31}=\varepsilon_{32}$ \\ \hline
        \multirow{2}{*}{$m\bar{3}$} & \multirow{4}{*}{I} & $\varepsilon_{11}=\varepsilon_{22}=\varepsilon_{33},\varepsilon_{44}=\varepsilon_{55}=\varepsilon_{66},$ \\ 
        & & $\varepsilon_{12}=\varepsilon_{23}=\varepsilon_{31},\varepsilon_{13}=\varepsilon_{21}=\varepsilon_{32}$ \\ \cline{1-1} \cline{3-3}
        \multirow{2}{*}{$m\bar{3}m'$} & & $\varepsilon_{11}=\varepsilon_{22}=\varepsilon_{33},\varepsilon_{44}=\varepsilon_{55}=\varepsilon_{66},$ \\ 
        & & $\varepsilon_{12}=\varepsilon_{23}=\varepsilon_{31}=\varepsilon_{13}=\varepsilon_{21}=\varepsilon_{32}$ \\ \hline \hline
    \end{tabular}
\end{table}

%%%%%%%%%%%%%%%%%%%%%%%%%%%%%%%%%%%%%%%%%%%%%%%%%%%%%%%%%%%%%%%%%%

\subsection{[111] direction} \label{Subsec:111_direction}

First, we consider the [111] direction, where coordinate system (ii) is defined by $X \parallel [1\bar{1}0]$, $Y \parallel [11\bar{2}]$, and $Z \parallel [111]$, and 
the corresponding rotation matrix $\mathcal{R}^{[111]}$ is given by
\begin{align}
    \mathcal{R}^{[111]}=
    \begin{pmatrix}
        1/\sqrt{2} & -1/\sqrt{2} & 0 \\
        1/\sqrt{6} & 1/\sqrt{6} & -2/\sqrt{6} \\
        1/\sqrt{3} & 1/\sqrt{3} & 1/\sqrt{3}
    \end{pmatrix}. \label{eq:transformation_rotation_111}
\end{align}
With this rotation matrix, $\varepsilon^{(\mathrm{S})}_{ij}$ and $\tilde{g}^{(\mathrm{S})}_{ij}=\varepsilon^{(\mathrm{S})}_{ij,kl}\tilde{q}_k\tilde{q}_l$ are expressed in coordinate system (ii) as 
\begin{subequations} \label{eq:transformation_OA_111}
    \begin{align}
        \varepsilon^{(\mathrm{S})}_{IJ}&=\mathcal{R}^{[111]}_{Ii}\mathcal{R}^{[111]}_{Jj}\varepsilon^{(\mathrm{S})}_{ij}, \label{eq:transformation_OA_111_eij} \\
        \tilde{g}^{(\mathrm{S})}_{IJ}&=\frac{1}{3} \mathcal{R}^{[111]}_{Ii}\mathcal{R}^{[111]}_{Jj}(D^{(\mathrm{S})}_{ij}+2O^{(\mathrm{S})}_{ij}), \label{eq:transformation_OA_111_eijkl}
    \end{align}
\end{subequations}
where we use $\tilde{q}_x=\tilde{q}_y=\tilde{q}_z=1/\sqrt{3}$ and introduce
\begin{subequations}
    \begin{align}
        D^{(\mathrm{S})}_{ij}&=\varepsilon^{(\mathrm{S})}_{ij,xx}+\varepsilon^{(\mathrm{S})}_{ij,yy}+\varepsilon^{(\mathrm{S})}_{ij,zz}, \\
        O^{(\mathrm{S})}_{ij}&=\varepsilon^{(\mathrm{S})}_{ij,yz}+\varepsilon^{(\mathrm{S})}_{ij,zx}+\varepsilon^{(\mathrm{S})}_{ij,xy}.
    \end{align}
\end{subequations}
Here, uppercase Latin indices denote the components $X$, $Y$, and $Z$.
Combining Eq.~\eqref{eq:transformation_OA_111} with the symmetry-allowed forms listed in Table~\ref{tab:eijkl_list},
we obtain the tensor components appearing in the flowchart and thereby determine the optical-activity type.
The results are Type I for $m\bar{3}$ and $m\bar{3}m'$, Type II for $4'/mmm'$, Type III for $4'/m$, and Type IV for $mmm$.
The detailed derivation of these results is given below.

We first consider $mmm$, for which we use the explicit form of $\varepsilon^{(\mathrm{S})}_{IJ}$ together with the general expression for $\eta^{(\mathrm{S})}_{IJ}$.
The former is obtained from Eq.~\eqref{eq:transformation_OA_111_eij} as 
\begin{align}
    \left\{
    \begin{aligned}
        \varepsilon^{(\mathrm{S})}_{XX} &= \frac{1}{2}\varepsilon^{(\mathrm{S})}_{x+y}, 
        & \varepsilon^{(\mathrm{S})}_{YZ} &= \frac{1}{3\sqrt{2}}(\varepsilon^{(\mathrm{S})}_{x+y}-2\varepsilon^{(\mathrm{S})}_{zz}), \\
        \varepsilon^{(\mathrm{S})}_{YY} &= \frac{1}{6}(\varepsilon^{(\mathrm{S})}_{x+y}+4\varepsilon^{(\mathrm{S})}_{zz}), 
        & \varepsilon^{(\mathrm{S})}_{ZX} &= \frac{1}{\sqrt{6}}\varepsilon^{(\mathrm{S})}_{x-y}, \\
        \varepsilon^{(\mathrm{S})}_{ZZ} &= \frac{1}{3}(\varepsilon^{(\mathrm{S})}_{x+y}+\varepsilon^{(\mathrm{S})}_{zz}), 
        & \varepsilon^{(\mathrm{S})}_{XY} &= \frac{1}{2\sqrt{3}}\varepsilon^{(\mathrm{S})}_{x-y},
    \end{aligned}
    \right. \label{eq:etaIJ_111}
\end{align}
where $\varepsilon^{(\mathrm{S})}_{x\pm y}=\varepsilon^{(\mathrm{S})}_{xx} \pm \varepsilon^{(\mathrm{S})}_{yy}$, and 
the off-diagonal components $\varepsilon^{(\mathrm{S})}_{i \neq j}$ vanish by symmetry.
The latter is given separately for the diagonal and off-diagonal components as [Eq.~\eqref{eq:eta_G_relation_MD_approx}]
\begin{subequations} \label{eq:eta_G_relation_MD_approx_app}
    \begin{align}
        \eta^{(\mathrm{S})}_{II}&=\frac{1}{|\varepsilon^{(\mathrm{S})}|}\bigl[\varepsilon^{(\mathrm{S})}_{JJ}\varepsilon^{(\mathrm{S})}_{KK}-(\varepsilon^{(\mathrm{S})}_{JK})^2\bigr], \label{eq:eta_G_relation_MD_approx_diag} \\ 
        \eta^{(\mathrm{S})}_{IJ}&=\frac{-1}{|\varepsilon^{(\mathrm{S})}|}\bigl[\varepsilon^{(\mathrm{S})}_{IJ}\varepsilon^{(\mathrm{S})}_{KK}-\varepsilon^{(\mathrm{S})}_{JK}\varepsilon^{(\mathrm{S})}_{KI}\bigr], \quad (I \neq J) \label{eq:eta_G_relation_MD_approx_offdiag}
    \end{align}
\end{subequations}
where $(I,J,K)$ runs over the cyclic permutations of $(X,Y,Z)$, with no summation over $I$, $J$, or $K$.
Combining these two expressions, we obtain the tensor component $\eta^{(\mathrm{S})}_{XY}$ examined in the first step of the flowchart as 
\begin{align}
    \eta^{(\mathrm{S})}_{XY}=-\frac{1}{2\sqrt{3}}\frac{\varepsilon^{(\mathrm{S})}_{x-y}}{\varepsilon^{(\mathrm{S})}_{xx}\varepsilon^{(\mathrm{S})}_{yy}}.
\end{align}
This component remains finite for $mmm$ because $\varepsilon^{(\mathrm{S})}_{xx}\neq \varepsilon^{(\mathrm{S})}_{yy}$, thereby classifying its optical activity as Type IV.

We next turn to the remaining MPGs, which satisfy $\varepsilon^{(\mathrm{S})}_{xx}=\varepsilon^{(\mathrm{S})}_{yy}$ and hence $\eta^{(\mathrm{S})}_{XY}=0$.
We therefore proceed to the next step of the flowchart, first considering $4'/m$ and $4'/mmm'$.
For these MPGs, the tensor components relevant to this step are given by
\begin{align}
    &\eta^{(\mathrm{S})}_{XX}=\frac{1}{\varepsilon^{(\mathrm{S})}_{xx}}, \quad \eta^{(\mathrm{S})}_{YY}=\frac{1}{3}\frac{2\varepsilon^{(\mathrm{S})}_{xx}+\varepsilon^{(\mathrm{S})}_{zz}}{\varepsilon^{(\mathrm{S})}_{xx}\varepsilon^{(\mathrm{S})}_{zz}},
    \label{eq:eta_XX_eta_YY_111} 
\end{align}
which are obtained from Eq.~\eqref{eq:eta_G_relation_MD_approx_diag} using Eq.~\eqref{eq:etaIJ_111} under $\varepsilon^{(\mathrm{S})}_{xx}=\varepsilon^{(\mathrm{S})}_{yy}$.
These expressions give $\eta^{(\mathrm{S})}_{XX} \neq \eta^{(\mathrm{S})}_{YY}$ for both $4'/m$ and $4'/mmm'$ because $\varepsilon^{(\mathrm{S})}_{xx} \neq \varepsilon^{(\mathrm{S})}_{zz}$,
leaving $\tilde{G}^{(\mathrm{S})}_{XY}$ to be examined in the final step of the flowchart.
For this purpose, we start from the general expression for $\tilde{G}^{(\mathrm{S})}_{IJ}$ [Eq.~\eqref{eq:Gij_definition}],
\begin{align}
    \tilde{G}^{(\mathrm{S})}_{IJ} &= \frac{-1}{|\varepsilon^{(\mathrm{S})}|}(\varepsilon^{(\mathrm{S})}_{KK}\tilde{g}^{(\mathrm{S})}_{IJ}+\varepsilon^{(\mathrm{S})}_{IJ}\tilde{g}^{(\mathrm{S})}_{KK} \notag \\
    & \qquad \qquad -\varepsilon^{(\mathrm{S})}_{JK}\tilde{g}^{(\mathrm{S})}_{KI}-\varepsilon^{(\mathrm{S})}_{KI}\tilde{g}^{(\mathrm{S})}_{JK}), \quad (I \neq J), \label{eq:GIJ_general_app}
\end{align}
where $(I,J,K)$ runs over the cyclic permutations of $(X,Y,Z)$, with no summation over $I$, $J$, or $K$.
Taking the $XY$ component and applying Eq.~\eqref{eq:etaIJ_111} under $\varepsilon^{(\mathrm{S})}_{xx}=\varepsilon^{(\mathrm{S})}_{yy}$, we obtain
\begin{align}
    \tilde{G}^{(\mathrm{S})}_{XY} = \frac{-1}{|\varepsilon^{(\mathrm{S})}|}(\varepsilon^{(\mathrm{S})}_{ZZ}\tilde{g}^{(\mathrm{S})}_{XY}-\varepsilon^{(\mathrm{S})}_{YZ}\tilde{g}^{(\mathrm{S})}_{ZX}), \label{eq:GXY_111}
\end{align}
where $\tilde{g}^{(\mathrm{S})}_{ZX}$ and $\tilde{g}^{(\mathrm{S})}_{XY}$ are obtained from Eq.~\eqref{eq:transformation_OA_111_eijkl} as
\begin{subequations} \label{eq:gZXandgXY_111}
    \begin{align}
        \tilde{g}^{(\mathrm{S})}_{ZX}&=\frac{1}{3\sqrt{6}}\Bigl[D^{(\mathrm{S})}_1-D^{(\mathrm{S})}_2-2(\varepsilon^{(\mathrm{S})}_{44}-\varepsilon^{(\mathrm{S})}_{55}) \notag \\
        & \quad +2(\varepsilon^{(\mathrm{S})}_{16}-\varepsilon^{(\mathrm{S})}_{26})-2(\varepsilon^{(\mathrm{S})}_{45}-\varepsilon^{(\mathrm{S})}_{54}) \Bigr], \label{eq:gZXandgXY_ZX_111} \\
        \tilde{g}^{(\mathrm{S})}_{XY}&=\frac{1}{6\sqrt{3}}\Bigl[D^{(\mathrm{S})}_1-D^{(\mathrm{S})}_2+4(\varepsilon^{(\mathrm{S})}_{44}-\varepsilon^{(\mathrm{S})}_{55}) \notag \\
        & \quad +2(\varepsilon^{(\mathrm{S})}_{16}-\varepsilon^{(\mathrm{S})}_{26})+4(\varepsilon^{(\mathrm{S})}_{45}-\varepsilon^{(\mathrm{S})}_{54}) \Bigr].
    \end{align}
\end{subequations}
Here, only the tensor components that can be nonzero in either $4'/m$ or $4'/mmm'$ are retained, with additional constraints imposed for each MPG.
Specifically, $D^{(\mathrm{S})}_1-D^{(\mathrm{S})}_2$ and $\varepsilon^{(\mathrm{S})}_{44}-\varepsilon^{(\mathrm{S})}_{55}$ vanish for both $4'/m$ and $4'/mmm'$,
whereas $\varepsilon^{(\mathrm{S})}_{16}-\varepsilon^{(\mathrm{S})}_{26}$ and $\varepsilon^{(\mathrm{S})}_{45}-\varepsilon^{(\mathrm{S})}_{54}$ remain finite for $4'/m$ but are absent for $4'/mmm'$ (see Table~\ref{tab:eijkl_list}),
resulting in a finite $\tilde{G}^{(\mathrm{S})}_{XY}$ only for $4'/m$:
\begin{align}
    \tilde{G}^{(\mathrm{S})}_{XY} = -\frac{2}{3\sqrt{3}} \frac{1}{\varepsilon^{(\mathrm{S})}_{xx}} \biggl(\frac{\varepsilon^{(\mathrm{S})}_{16}}{\varepsilon^{(\mathrm{S})}_{xx}}+\frac{2\varepsilon^{(\mathrm{S})}_{45}}{\varepsilon^{(\mathrm{S})}_{zz}} \biggr).
    \label{eq:GXY_tetragonal_111}
\end{align}
Thus, $4'/mmm'$ and $4'/m$ exhibit Type-II and Type-III optical activity, respectively.

We then consider $m\bar{3}$ and $m\bar{3}m'$, which satisfy $\eta^{(\mathrm{S})}_{XX}=\eta^{(\mathrm{S})}_{YY}$ because $\varepsilon^{(\mathrm{S})}_{xx}=\varepsilon^{(\mathrm{S})}_{zz}$
and $\tilde{G}^{(\mathrm{S})}_{XY}=0$ because $\varepsilon^{(\mathrm{S})}_{16}=\varepsilon^{(\mathrm{S})}_{45}=0$,
leaving only $\tilde{\eta}^{(\mathrm{S})}_{XX}$ and $\tilde{\eta}^{(\mathrm{S})}_{YY}$ to be examined in the final step of the flowchart.
These components are given by $\tilde{\eta}^{(\mathrm{S})}_{XX}=\eta^{(\mathrm{S})}_{XX}\tilde{g}^{(\mathrm{S})}_{XX}$ and $\tilde{\eta}^{(\mathrm{S})}_{YY}=\eta^{(\mathrm{S})}_{YY}\tilde{g}^{(\mathrm{S})}_{YY}$,
which are obtained from Eqs.~\eqref{eq:tilde_eta0_definition} and \eqref{eq:tilde_etaii_definition} using Eqs.~\eqref{eq:etaIJ_111} and \eqref{eq:eta_G_relation_MD_approx_app} under $\varepsilon^{(\mathrm{S})}_{xx}=\varepsilon^{(\mathrm{S})}_{yy}=\varepsilon^{(\mathrm{S})}_{zz}$.
We thus evaluate $\tilde{g}^{(\mathrm{S})}_{XX}$ and $\tilde{g}^{(\mathrm{S})}_{YY}$, which are obtained from Eq.~\eqref{eq:transformation_OA_111_eijkl} as
\begin{subequations} 
    \begin{align}
        \tilde{g}^{(\mathrm{S})}_{XX}&=\frac{1}{6} (D^{(\mathrm{S})}_1+D^{(\mathrm{S})}_2-4\varepsilon^{(\mathrm{S})}_{66}), \\
        \tilde{g}^{(\mathrm{S})}_{YY}&=\frac{1}{18}\Bigl[D^{(\mathrm{S})}_1+D^{(\mathrm{S})}_2+4D^{(\mathrm{S})}_3 \notag \\
        & \quad -4(2\varepsilon^{(\mathrm{S})}_{44}+2\varepsilon^{(\mathrm{S})}_{55}-\varepsilon^{(\mathrm{S})}_{66}) \Bigr],
    \end{align}
\end{subequations}
where only the tensor components that can be nonzero in both $m\bar{3}$ and $m\bar{3}m'$ are retained.
For these components, $D^{(\mathrm{S})}_1=D^{(\mathrm{S})}_2=D^{(\mathrm{S})}_3$ and $\varepsilon^{(\mathrm{S})}_{44}=\varepsilon^{(\mathrm{S})}_{55}=\varepsilon^{(\mathrm{S})}_{66}$ in both MPGs (see Table~\ref{tab:eijkl_list}),
resulting in $\tilde{g}^{(\mathrm{S})}_{XX}=\tilde{g}^{(\mathrm{S})}_{YY}$.
This result, together with $\eta^{(\mathrm{S})}_{XX}=\eta^{(\mathrm{S})}_{YY}$, yields $\tilde{\eta}^{(\mathrm{S})}_{XX}=\tilde{\eta}^{(\mathrm{S})}_{YY}$,
thereby classifying the optical activity of $m\bar{3}$ and $m\bar{3}m'$ as Type I.

We finally derive explicit expressions for $\tilde{G}^{(\mathrm{o})}_{[111]}$ in the five MPGs considered here.
To this end, we start from the general expression for $\tilde{\bm{G}}^{(\mathrm{o})}$ in coordinate system (i) [Eq.~\eqref{eq:tildeGijk_definition}],
\begin{align}
    \tilde{G}^{(\mathrm{o})}_\mu=\tilde{G}^{(\mathrm{o})}_{\mu,kl}\tilde{q}_k\tilde{q}_l, \quad 
    \tilde{G}^{(\mathrm{o})}_{\mu,kl}=\frac{-1}{|\varepsilon^{(\mathrm{S})}|}\varepsilon^{(\mathrm{S})}_{\mu \nu}\tilde{g}^{(\mathrm{o})}_{\nu,kl}. \label{eq:tildeGijk_definition_app}
\end{align}
Applying $\mathcal{R}^{[111]}$ to this expression, we obtain $\tilde{\bm{G}}^{(\mathrm{o})}$ in coordinate system (ii) as
\begin{align}
    \tilde{G}^{(\mathrm{o})}_U&=\mathcal{R}^{[111]}_{U\mu}\mathcal{R}^{[111]}_{Kk}\mathcal{R}^{[111]}_{Ll}\tilde{G}^{(\mathrm{o})}_{\mu,kl}\tilde{q}_K\tilde{q}_L \notag \\
    &=\mathcal{R}^{[111]}_{U\mu}\mathcal{R}^{[111]}_{Zk}\mathcal{R}^{[111]}_{Zl} \notag \\
    & \quad \times \frac{-1}{|\varepsilon^{(\mathrm{S})}|}(\varepsilon^{(\mathrm{S})}_{\mu x}\tilde{g}^{(\mathrm{o})}_{x,kl}+\varepsilon^{(\mathrm{S})}_{\mu y}\tilde{g}^{(\mathrm{o})}_{y,kl}+\varepsilon^{(\mathrm{S})}_{\mu z}\tilde{g}^{(\mathrm{o})}_{z,kl}), \label{eq:GZ_111}
\end{align}
where we use $\tilde{q}_X=\tilde{q}_Y=0$ and $\tilde{q}_Z=1$.
In particular, setting $U=Z$ gives $\tilde{G}^{(\mathrm{o})}_Z=\tilde{G}^{(\mathrm{o})}_{[111]}$ as 
\begin{align}
    \tilde{G}^{(\mathrm{o})}_Z&=-\frac{1}{3\sqrt{3}}
    \biggl[\frac{2(\tilde{g}^{(\mathrm{o})}_{x,zx}+\tilde{g}^{(\mathrm{o})}_{x,yz})}{\varepsilon^{(\mathrm{S})}_{yy}\varepsilon^{(\mathrm{S})}_{zz}} 
    +\frac{2(\tilde{g}^{(\mathrm{o})}_{y,yz}+\tilde{g}^{(\mathrm{o})}_{y,zx})}{\varepsilon^{(\mathrm{S})}_{zz}\varepsilon^{(\mathrm{S})}_{xx}} \notag \\
    & \quad +\frac{\tilde{g}^{(\mathrm{o})}_{z,xx}+\tilde{g}^{(\mathrm{o})}_{z,yy}+2\tilde{g}^{(\mathrm{o})}_{z,xy}}{\varepsilon^{(\mathrm{S})}_{xx}\varepsilon^{(\mathrm{S})}_{yy}}\biggr], \label{eq:OAV_111_general_App}
\end{align}
where only the tensor components that can be nonzero in at least one of the five MPGs are retained.
Combining this expression with the symmetry-allowed forms listed in Tables~\ref{tab:Mijk_list} and \ref{tab:eijkl_list},
we obtain the expressions given in Eq.~\eqref{eq:OAV_111} of the main text for each MPG:
\begin{subequations} \label{eq:OAV_111_App}
    \begin{align}
        \tilde{G}^{(\mathrm{o})}_Z&=-\frac{2}{3\sqrt{3}}\biggl(\frac{\tilde{g}^{(\mathrm{o})}_{x,yz}}{\varepsilon^{(\mathrm{S})}_{yy}\varepsilon^{(\mathrm{S})}_{zz}}
        +\frac{\tilde{g}^{(\mathrm{o})}_{y,zx}}{\varepsilon^{(\mathrm{S})}_{zz}\varepsilon^{(\mathrm{S})}_{xx}}+\frac{\tilde{g}^{(\mathrm{o})}_{z,xy}}{\varepsilon^{(\mathrm{S})}_{xx}\varepsilon^{(\mathrm{S})}_{yy}} \biggr), \label{eq:OAV_111_orthorombic_App} \\
        \tilde{G}^{(\mathrm{o})}_Z&=-\frac{2}{3\sqrt{3}}\frac{1}{\varepsilon^{(\mathrm{S})}_{xx}}\biggl(\frac{2\tilde{g}^{(\mathrm{o})}_{x,yz}}{\varepsilon^{(\mathrm{S})}_{zz}}
        +\frac{\tilde{g}^{(\mathrm{o})}_{z,xy}}{\varepsilon^{(\mathrm{S})}_{xx}} \biggr), \label{eq:OAV_111_tetragonal_App} \\
        \tilde{G}^{(\mathrm{o})}_Z&=-\frac{2}{\sqrt{3}}\frac{\tilde{g}^{(\mathrm{o})}_{z,xy}}{(\varepsilon^{(\mathrm{S})}_{xx})^2}, \label{eq:OAV_111_cubic_App}
    \end{align}
\end{subequations}
where the first line applies to $mmm$, the second line to $4'/m$ and $4'/mmm'$, and the third line to $m\bar{3}$ and $m\bar{3}m'$.

%%%%%%%%%%%%%%%%%%%%%%%%%%%%%%%%%%%%%%%%%%%%%%%%%%%%%%%%%%%%%%%%%%

\subsection{[101] and [011] directions} \label{Subsec:101_011_direction}

Second, we turn to the [101] and [011] directions for $4'/m$ and $4'/mm'm$.
Here, we consider the [101] direction as representative, since the analyses for the two directions are analogous.
For the [101] direction, coordinate system (ii) is defined by $X \parallel [10\bar{1}]$, $Y \parallel [010]$, and $Z \parallel [101]$, and 
the corresponding rotation matrix $\mathcal{R}^{[101]}$ is given by
\begin{align}
    \mathcal{R}^{[101]}=
    \begin{pmatrix}
        1/\sqrt{2} & 0 & -1/\sqrt{2} \\
        0 & 1 & 0 \\
        1/\sqrt{2} & 0 & 1/\sqrt{2}
    \end{pmatrix}. \label{eq:transformation_rotation_101}
\end{align}
With this rotation matrix, $\varepsilon^{(\mathrm{S})}_{ij}$ is expressed in coordinate system (ii) as
\begin{align}
    \varepsilon^{(\mathrm{S})}_{XX}&=\varepsilon^{(\mathrm{S})}_{ZZ}=(\varepsilon^{(\mathrm{S})}_{xx}+\varepsilon^{(\mathrm{S})}_{zz})/2, \quad \varepsilon^{(\mathrm{S})}_{YY}=\varepsilon^{(\mathrm{S})}_{yy}, \notag \\
    \varepsilon^{(\mathrm{S})}_{YZ}&=\varepsilon^{(\mathrm{S})}_{XY}=0, \quad \varepsilon^{(\mathrm{S})}_{ZX}=(\varepsilon^{(\mathrm{S})}_{xx}-\varepsilon^{(\mathrm{S})}_{zz})/2, \label{eq:transformation_OA_101_eij}
\end{align}
where the off-diagonal components $\varepsilon^{(\mathrm{S})}_{i \neq j}$ vanish by symmetry.
Combining these expressions with Eq.~\eqref{eq:eta_G_relation_MD_approx_app} under $\varepsilon^{(\mathrm{S})}_{xx}=\varepsilon^{(\mathrm{S})}_{yy} \neq \varepsilon^{(\mathrm{S})}_{zz}$ immediately gives 
\begin{align}
    \eta^{(\mathrm{S})}_{XY}=0, \quad 
    \eta^{(\mathrm{S})}_{XX}=\frac{1}{2}\frac{\varepsilon^{(\mathrm{S})}_{xx}+\varepsilon^{(\mathrm{S})}_{zz}}{\varepsilon^{(\mathrm{S})}_{xx}\varepsilon^{(\mathrm{S})}_{zz}} \neq
    \eta^{(\mathrm{S})}_{YY}=\frac{1}{\varepsilon^{(\mathrm{S})}_{xx}},
\end{align}
leaving $\tilde{G}^{(\mathrm{S})}_{XY}$ to be examined in the final step of the flowchart.
For this purpose, we use Eq.~\eqref{eq:GIJ_general_app} together with Eq.~\eqref{eq:transformation_OA_101_eij} to obtain $\tilde{G}^{(\mathrm{S})}_{XY}$:
\begin{align}
    \tilde{G}^{(\mathrm{S})}_{XY} = \frac{-1}{|\varepsilon^{(\mathrm{S})}|}(\varepsilon^{(\mathrm{S})}_{ZZ}\tilde{g}^{(\mathrm{S})}_{XY}-\varepsilon^{(\mathrm{S})}_{ZX}\tilde{g}^{(\mathrm{S})}_{YZ}), \label{eq:GXY_101}
\end{align}
where $\tilde{g}^{(\mathrm{S})}_{YZ}$ and $\tilde{g}^{(\mathrm{S})}_{XY}$ are obtained from $\tilde{g}^{(\mathrm{S})}_{ij}=\varepsilon^{(\mathrm{S})}_{ij,kl}\tilde{q}_k\tilde{q}_l$ by 
using $\tilde{q}_x=\tilde{q}_z=1/\sqrt{2}$ and $\tilde{q}_y=0$ and applying $\mathcal{R}^{[101]}$:
\begin{subequations}
    \begin{align}
        \tilde{g}^{(\mathrm{S})}_{YZ}&=\frac{1}{2\sqrt{2}}(\varepsilon^{(\mathrm{S})}_{61}+2\varepsilon^{(\mathrm{S})}_{45}), \\
        \tilde{g}^{(\mathrm{S})}_{XY}&=\frac{1}{2\sqrt{2}}(\varepsilon^{(\mathrm{S})}_{61}-2\varepsilon^{(\mathrm{S})}_{45}). \label{eq:gZXandgXY_101}
    \end{align}
\end{subequations}
Here, only the tensor components that can be nonzero in either $4'/m$ or $4'/mm'm$ are retained, with additional constraints imposed for each MPG.
Specifically, $\varepsilon^{(\mathrm{S})}_{16}$ and $\varepsilon^{(\mathrm{S})}_{45}$ are nonzero for $4'/m$ but are absent for $4'/mm'm$ (see Table~\ref{tab:eijkl_list}),
resulting in a finite $\tilde{G}^{(\mathrm{S})}_{XY}$ only for $4'/m$:
\begin{align}
    \tilde{G}^{(\mathrm{S})}_{XY} = -\frac{1}{2\sqrt{2}} \frac{1}{\varepsilon^{(\mathrm{S})}_{xx}} \biggl(\frac{\varepsilon^{(\mathrm{S})}_{61}}{\varepsilon^{(\mathrm{S})}_{xx}}-\frac{2\varepsilon^{(\mathrm{S})}_{45}}{\varepsilon^{(\mathrm{S})}_{zz}} \biggr).
    \label{eq:GXY_tetragonal_101}
\end{align}
Thus, $4'/mm'm$ and $4'/m$ exhibit Type-II and Type-III optical activity, respectively.

We finally derive explicit expressions for $\tilde{G}^{(\mathrm{o})}_{[101]}$ in $4'/m$ and $4'/mm'm$.
To this end, we apply $\mathcal{R}^{[101]}$ to Eq.~\eqref{eq:tildeGijk_definition_app} and express $\tilde{\bm{G}}^{(\mathrm{o})}$ in coordinate system (ii) as
\begin{align}
    \tilde{G}^{(\mathrm{o})}_U&=\mathcal{R}^{[101]}_{U\mu}\mathcal{R}^{[101]}_{Zk}\mathcal{R}^{[101]}_{Zl} \notag \\
    & \quad \times \frac{-1}{|\varepsilon^{(\mathrm{S})}|}(\varepsilon^{(\mathrm{S})}_{\mu x}\tilde{g}^{(\mathrm{o})}_{x,kl}+\varepsilon^{(\mathrm{S})}_{\mu y}\tilde{g}^{(\mathrm{o})}_{y,kl}+\varepsilon^{(\mathrm{S})}_{\mu z}\tilde{g}^{(\mathrm{o})}_{z,kl}), \label{eq:GZ_101}
\end{align}
where we use $\tilde{q}_X=\tilde{q}_Y=0$ and $\tilde{q}_Z=1$.
In particular, setting $U=Z$ and combining the resulting expression with the symmetry-allowed forms listed in Tables~\ref{tab:Mijk_list} and \ref{tab:eijkl_list},
we obtain the expression for $\tilde{G}^{(\mathrm{o})}_Z=\tilde{G}^{(\mathrm{o})}_{[101]}$ in both MPGs:
\begin{align}
    \tilde{G}^{(\mathrm{o})}_Z&=-\frac{1}{2\sqrt{2}} \frac{1}{\varepsilon^{(\mathrm{S})}_{xx}} \biggl( \frac{2\tilde{g}^{(\mathrm{o})}_{x,zx}}{\varepsilon^{(\mathrm{S})}_{zz}} + \frac{\tilde{g}^{(\mathrm{o})}_{z,xx}}{\varepsilon^{(\mathrm{S})}_{xx}} \biggr), \label{eq:OAV_101_App} 
\end{align}
which reproduces Eq.~\eqref{eq:OAV_101} of the main text.

%%%%%%%%%%%%%%%%%%%%%%%%%%%%%%%%%%%%%%%%%%%%%%%%%%%%%%%%%%%%%%%%%%

\subsection{[110], [100], and [010] directions} \label{Subsec:110_100_010_direction}

Finally, we address the [110], [100], and [010] directions
for $\bar{3}m$, $6'/m'$, $6'/m'mm'$, and $6'/m'm'm$, with the latter two directions absent for some of these MPGs.
Here, we consider the [110] direction as representative,
since it is common to all four MPGs and the analyses for the three directions are analogous.\footnote{
For the $[100]$ and $[010]$ directions, coordinate system (i) can be used directly as coordinate system (ii),
and the subsequent analysis follows the same procedure.}
For the [110] direction, coordinate system (ii) is defined by $X \parallel [110]$, $Y \parallel [\bar{1}10]$, and $Z \parallel [001]$, and 
the corresponding rotation matrix $\mathcal{R}^{[110]}$ is given by
\begin{align}
    \mathcal{R}^{[110]}=
    \begin{pmatrix}
        1/\sqrt{2} & 1/\sqrt{2} & 0 \\
        -1/\sqrt{2} & 1/\sqrt{2} & 0 \\
        0 & 0 & 1
    \end{pmatrix}. \label{eq:transformation_rotation_110}
\end{align}
With this rotation matrix, $\varepsilon^{(\mathrm{S})}_{ij}$ is expressed in coordinate system (ii) as
\begin{align}
    \varepsilon^{(\mathrm{S})}_{XX}&=\varepsilon^{(\mathrm{S})}_{YY}=\varepsilon^{(\mathrm{S})}_{x+y}/2, \quad \varepsilon^{(\mathrm{S})}_{ZZ}=\varepsilon^{(\mathrm{S})}_{zz}, \notag \\
    \varepsilon^{(\mathrm{S})}_{YZ}&=\varepsilon^{(\mathrm{S})}_{ZX}=0, \quad \varepsilon^{(\mathrm{S})}_{XY}=-\varepsilon^{(\mathrm{S})}_{x-y}/2, \label{eq:transformation_OA_110_eij}
\end{align}
where the off-diagonal components $\varepsilon^{(\mathrm{S})}_{i \neq j}$ vanish by symmetry.
Combining these expressions with Eq.~\eqref{eq:eta_G_relation_MD_approx_app} under $\varepsilon^{(\mathrm{S})}_{xx}=\varepsilon^{(\mathrm{S})}_{yy} \neq \varepsilon^{(\mathrm{S})}_{zz}$ immediately gives 
\begin{align}
    \eta^{(\mathrm{S})}_{YZ}=0, \quad \eta^{(\mathrm{S})}_{YY}=\frac{1}{\varepsilon^{(\mathrm{S})}_{xx}} \neq 
    \eta^{(\mathrm{S})}_{ZZ}=\frac{1}{\varepsilon^{(\mathrm{S})}_{zz}}, 
\end{align}
leaving $\tilde{G}^{(\mathrm{S})}_{YZ}$ to be examined in the final step of the flowchart.
For this purpose, we use Eq.~\eqref{eq:GIJ_general_app} together with Eq.~\eqref{eq:transformation_OA_110_eij} under $\varepsilon^{(\mathrm{S})}_{xx}=\varepsilon^{(\mathrm{S})}_{yy}$ to obtain $\tilde{G}^{(\mathrm{S})}_{YZ}$:
\begin{align}
    \tilde{G}^{(\mathrm{S})}_{YZ} = \frac{-1}{|\varepsilon^{(\mathrm{S})}|}\varepsilon^{(\mathrm{S})}_{XX}\tilde{g}^{(\mathrm{S})}_{YZ}, \label{eq:GYZ_110}
\end{align}
where $\tilde{g}^{(\mathrm{S})}_{YZ}$ is obtained from $\tilde{g}^{(\mathrm{S})}_{ij}=\varepsilon^{(\mathrm{S})}_{ij,kl}\tilde{q}_k\tilde{q}_l$ by using $\tilde{q}_x=\tilde{q}_y=1/\sqrt{2}$ and $\tilde{q}_z=0$ and applying $\mathcal{R}^{[110]}$:
\begin{align}
    \tilde{g}^{(\mathrm{S})}_{YZ}=\frac{1}{2\sqrt{2}}(\varepsilon^{(\mathrm{S})}_{41}+\varepsilon^{(\mathrm{S})}_{42}-2\varepsilon^{(\mathrm{S})}_{56}). \label{eq:gYZ_110}
\end{align}
Here, only the tensor components that can be nonzero in at least one of the four MPGs are retained, with additional constraints imposed for each MPG.
Specifically, $\varepsilon^{(\mathrm{S})}_{41}$, $\varepsilon^{(\mathrm{S})}_{42}$, and $\varepsilon^{(\mathrm{S})}_{56}$ are nonzero for $\bar{3}m$ 
but are absent for $6'/m'$, $6'/m'mm'$, and $6'/m'm'm$ (see Table~\ref{tab:eijkl_list}),
resulting in a finite $\tilde{G}^{(\mathrm{S})}_{YZ}$ only for $\bar{3}m$:
\begin{align}
    \tilde{G}^{(\mathrm{S})}_{YZ} = \frac{1}{\sqrt{2}} \frac{\varepsilon^{(\mathrm{S})}_{41}}{\varepsilon^{(\mathrm{S})}_{xx}\varepsilon^{(\mathrm{S})}_{zz}}.
    \label{eq:GYZ_trigonal_110}
\end{align}
Thus, $6'/m'$, $6'/m'mm'$, and $6'/m'm'm$ exhibit Type-II optical activity, whereas $\bar{3}m$ exhibits Type-III optical activity.

We finally derive explicit expressions for $\tilde{G}^{(\mathrm{o})}_{[110]}$ in the four MPGs considered here.
To this end, we apply $\mathcal{R}^{[110]}$ to Eq.~\eqref{eq:tildeGijk_definition_app} and express $\tilde{\bm{G}}^{(\mathrm{o})}$ in coordinate system (ii) as
\begin{align}
    \tilde{G}^{(\mathrm{o})}_U&=\mathcal{R}^{[110]}_{U\mu}\mathcal{R}^{[110]}_{Xk}\mathcal{R}^{[110]}_{Xl} \notag \\
    & \quad \times \frac{-1}{|\varepsilon^{(\mathrm{S})}|}(\varepsilon^{(\mathrm{S})}_{\mu x}\tilde{g}^{(\mathrm{o})}_{x,kl}+\varepsilon^{(\mathrm{S})}_{\mu y}\tilde{g}^{(\mathrm{o})}_{y,kl}+\varepsilon^{(\mathrm{S})}_{\mu z}\tilde{g}^{(\mathrm{o})}_{z,kl}), \label{eq:GX_110}
\end{align}
where we use $\tilde{q}_X=1$ and $\tilde{q}_Y=\tilde{q}_Z=0$.
In particular, setting $U=X$ gives $\tilde{G}^{(\mathrm{o})}_X=\tilde{G}^{(\mathrm{o})}_{[110]}$ as 
\begin{align}
    \tilde{G}^{(\mathrm{o})}_X&=-\frac{1}{2\sqrt{2}} \frac{1}{\varepsilon^{(\mathrm{S})}_{xx}\varepsilon^{(\mathrm{S})}_{zz}}
    \bigl[(\tilde{g}^{(\mathrm{o})}_{x,xx}+\tilde{g}^{(\mathrm{o})}_{x,yy}+2\tilde{g}^{(\mathrm{o})}_{x,xy}) \notag \\
    & \quad + (\tilde{g}^{(\mathrm{o})}_{y,xx}+\tilde{g}^{(\mathrm{o})}_{y,yy}+2\tilde{g}^{(\mathrm{o})}_{y,xy})\bigr], \label{eq:OAV_110_general_App}
\end{align}
where only the tensor components that can be nonzero in at least one of the four MPGs are retained.
Combining this expression with the symmetry-allowed forms listed in Tables~\ref{tab:Mijk_list} and \ref{tab:eijkl_list},
we obtain the expressions corresponding to Eq.~\eqref{eq:OAV_110} of the main text for each MPG:
\begin{subequations} \label{eq:OAV_110_App}
    \begin{align}
        \tilde{G}^{(\mathrm{o})}_X&=-\frac{1}{\sqrt{2}} \frac{\tilde{g}^{(\mathrm{o})}_{x,xy}}{\varepsilon^{(\mathrm{S})}_{xx}\varepsilon^{(\mathrm{S})}_{zz}}, \\
        \tilde{G}^{(\mathrm{o})}_X&=-\frac{1}{\sqrt{2}} \frac{\tilde{g}^{(\mathrm{o})}_{y,xy}}{\varepsilon^{(\mathrm{S})}_{xx}\varepsilon^{(\mathrm{S})}_{zz}}, \\
        \tilde{G}^{(\mathrm{o})}_X&=-\frac{1}{\sqrt{2}} \frac{\tilde{g}^{(\mathrm{o})}_{x,xy}+\tilde{g}^{(\mathrm{o})}_{y,xy}}{\varepsilon^{(\mathrm{S})}_{xx}\varepsilon^{(\mathrm{S})}_{zz}}, 
    \end{align}
\end{subequations}
where the first line applies to $6'/m'm'm$, the second line to $\bar{3}m$ and $6'/m'mm'$, and the third line to $6'/m'$.

%%%%%%%%%%%%%%%%%%%%%%%%%%%%%%%%%%%%%%%%%%%%%%%%%%%%%%%%%%%%%%%%%%%%%%%%%%%%%%%%%%%%%%%%%%%%%%%%%%%%%%%%%%%%%%%%%%%%%%%%%%%%%%%%%%%%

\section{DETAILED DERIVATION OF THE COMPLEX FARADAY ANGLE} \label{Sec:AppD}

In this appendix, we present the detailed derivation of the complex Faraday angle $\Psi_{\mathrm{F}}=\Theta_{\mathrm{F}}+i \Delta_{\mathrm{F}}$ for octupolar MOA.
We first derive general expressions for the Faraday rotation angle $\Theta_{\mathrm{F}}$ and the Faraday ellipticity $\Delta_{\mathrm{F}}$ in Appendix~\ref{Subsec:AppD1}.
We then apply these general expressions to Type-I and Type-II optical activity in Appendix~\ref{Subsec:AppD2}.

%%%%%%%%%%%%%%%%%%%%%%%%%%%%%%%%%%%%%%%%%%%%%%%%%%%%%%%%%%%%%%%%%%

\subsection{General expressions for the Faraday rotation angle and ellipticity} \label{Subsec:AppD1}

The outline is as follows.
We first derive the rotation angle $\Theta$ and the ellipticity $\Delta$ for a general ellipse in the $x$-$y$ plane defined by the equation
\begin{align}
    Ax^2+By^2+2Cxy=1, \label{eq:ellipse_general}
\end{align}
where $A$, $B$, and $C$ are coefficients.
Here, $\Theta$ is defined as the angle between the $x'$ axis of the rotated coordinate system ($x',y'$) and the major axis of the ellipse,
where ($x',y'$) is obtained by rotating the original coordinate system ($x,y$) through an angle $\theta$.
Meanwhile, $\Delta$ is defined as the ratio of the minor to major semiaxes of the ellipse.
We next obtain $\Theta$ and $\Delta$ for elliptically polarized light by applying the resulting expressions to its electric-field components,
\begin{align}
    E_i=\mathrm{Re} \bigl[ J_i e^{i\phi_i}e^{-i\omega t} \bigr], \label{eq:ellipse_general_t}
\end{align}
where $i=x,y$, and $J_i$ and $\phi_i$ are the corresponding real amplitudes and phases, respectively. 
We then focus on transmitted light to derive general expressions for the Faraday rotation angle $\Theta_{\mathrm{F}}$ and the Faraday ellipticity $\Delta_{\mathrm{F}}$.
To this end, we consider light propagating along the $z$ axis and express $J_ie^{i\phi_i}$ in Eq.~\eqref{eq:ellipse_general_t} as a superposition of two eigenmodes:
\begin{align}
    J_i e^{i\phi_i}= a^+_i e^{i q_+ z} + a^-_i e^{iq_- z}, \label{eq:Faraday_angle_z_superposition}
\end{align}
where $a^{\pm}_i$ and $q_{\pm}$ are the complex electric-field amplitudes and the complex wave numbers of the two eigenmodes, respectively.
We finally enable the explicit evaluation of $\Theta_{\mathrm{F}}$ and $\Delta_{\mathrm{F}}$ by imposing the initial condition for the incident linear polarization,
\begin{align}
    a_x^++a_x^-=E_{\mathrm{in},x}, \quad a_y^++a_y^-=E_{\mathrm{in},y},  \label{eq:Faraday_angle_initial}
\end{align}
where $\bm{E}_{\mathrm{in}}=(E_{\mathrm{in},x},E_{\mathrm{in},y})$ is the complex electric-field amplitude of the incident light.

Following this outline, we first rewrite Eq.~\eqref{eq:ellipse_general} in the rotated coordinate system ($x',y'$) by applying the coordinate transformation,
\begin{align}
    \begin{pmatrix} x \\ y \end{pmatrix}
    =\begin{pmatrix} \cos\theta & -\sin\theta \\ \sin\theta & \cos \theta \end{pmatrix}
    \begin{pmatrix} x' \\ y' \end{pmatrix}.
\end{align}
Specifically, substituting this expression into Eq.~\eqref{eq:ellipse_general}, we obtain
\begin{align}
    A'x'^2+B'y'^2+2C'x'y'=1, \label{eq:ellipse_general_transformation}
\end{align}
where the coefficients $A'$, $B'$, and $C'$ are given by
\begin{subequations}
    \begin{align}
        A'&=A\cos^2\theta+B\sin^2\theta+C\sin2\theta, \\
        B'&=A\sin^2\theta+B\cos^2\theta-C\sin2\theta, \\
        C'&=-\frac{A-B}{2}\sin 2\theta+C\cos2\theta. 
    \end{align}
\end{subequations}
Here, if we choose the $x'$ axis to coincide with the major axis of the ellipse,
the angle $\theta$ can be determined by requiring the cross term $x'y'$ to vanish:
\begin{align}
    C'=0 \ \Rightarrow \ \tan 2\theta=\frac{2C}{A-B}. \label{eq:ellipse_angle1_general}
\end{align}
In this case, $\theta$ is the angle between the $x$ axis and the major axis of the ellipse.
Thus, the angle $\Theta$ between the $x'$ axis and the major axis can also be obtained by following the same procedure, 
starting from Eq.~\eqref{eq:ellipse_general_transformation} instead of Eq.~\eqref{eq:ellipse_general}:
\begin{align}
    \tan2\Theta=\frac{2C'}{A'-B'}=\frac{-(A-B)\sin2\theta+2C\cos2\theta}{(A-B)\cos2\theta+2C\sin2\theta}. \label{eq:ellipse_angle2_general}
\end{align}
Meanwhile, the major and minor semiaxes, $a$ and $b$, are given by
\begin{subequations} \label{eq:major_minor_length}
    \begin{align}
        a^{-2}&=A'=\frac{1}{2}\Bigl[(A+B)-\frac{(A-B)^2+4C^2}{\sqrt{(A-B)^2+4C^2}}\Bigr], \\
        b^{-2}&=B'=\frac{1}{2}\Bigl[(A+B)+\frac{(A-B)^2+4C^2}{\sqrt{(A-B)^2+4C^2}}\Bigr], 
    \end{align}
\end{subequations}
where we use Eq.~\eqref{eq:ellipse_angle1_general} to eliminate $\theta$.
Using these expressions, we obtain the ellipticity $\Delta$ as
\begin{align}
    \Delta^2=\Bigl( \frac{b}{a}\Bigr)^2=\frac{(A+B)-\sqrt{(A-B)^2+4C^2}}{(A+B)+\sqrt{(A-B)^2+4C^2}}. \label{eq:ellipse_ellipticity_general}
\end{align}

We next eliminate the time variable $t$ from the two electric-field components in Eq.~\eqref{eq:ellipse_general_t} to obtain the ellipse equation,
\begin{align}
    \frac{E^2_x}{J^2_x}+\frac{E^2_y}{J^2_y}-\frac{2E_xE_y}{J_xJ_y}\cos\delta=\sin^2\delta,
\end{align}
where $\delta=\phi_y-\phi_x$.
This ellipse equation has the same form as Eq.~\eqref{eq:ellipse_general}.
We can therefore apply Eqs.~\eqref{eq:ellipse_angle2_general} and \eqref{eq:ellipse_ellipticity_general}
to obtain $\Theta$ and $\Delta$ for elliptically polarized light as 
\begin{subequations} \label{eq:ellipse_angle_ellipticity_general}
    \begin{align}
        \tan 2\Theta &= \frac{-J^2_{x-y}\sin 2\theta + 2J_xJ_y \cos \delta \cos 2\theta}{J^2_{x-y}\cos 2\theta + 2J_xJ_y \cos \delta \sin 2\theta}, \label{eq:ellipse_angle_ellipticity_general_angle} \\
        \Delta^2 &= \frac{J^2_{x+y}-\sqrt{(J^2_{x-y})^2+(2J_xJ_y\cos \delta)^2}}{J^2_{x+y}+\sqrt{(J^2_{x-y})^2+(2J_xJ_y\cos \delta)^2}},  \label{eq:ellipse_angle_ellipticity_general_ellipticity}
    \end{align}
\end{subequations}
where $J^2_{x \pm y}=J^2_x \pm J^2_y$ and $\theta$ is the angle between the $x'$ and $x$ axes.
These expressions correspond to Eq.~\eqref{eq:Faraday_angle_general} in the main text when the $x'$ axis is aligned with the initial linear polarization direction.
We adopt this choice in the following analysis.

We then use Eq.~\eqref{eq:Faraday_angle_z_superposition} to express $J_x J_y\cos \delta$ and $J^2_{x\mp y}$ for transmitted light as 
\begin{subequations} \label{eq:ellipse_Faraday_general}
    \begin{align}
        &J_x J_y\cos \delta = \mathrm{Re}\bigl[(J_x e^{i\phi_x}) (J_y e^{i\phi_y})^* \bigr] \notag \\
        &\hspace{42.5pt} = \mathrm{Re} \Bigl[ \sum_{n,m=\pm} a^n_x (a^m_y)^*e^{i(q_n-q^*_m)z} \Bigr] \notag \\
        &\hspace{42.5pt} =\sum_{n,m=\pm} e^{-(\beta_n+\beta_m)z} A^{nm}_{xy}, \label{eq:ellipse_Faraday_general_offdiag} \\
        &J^2_{x\mp y} = \mathrm{Re}\bigl[(J_x e^{i\phi_x}) (J_x e^{i\phi_x})^* \mp (J_y e^{i\phi_y}) (J_y e^{i\phi_y})^* \bigr] \notag \\
        &\hspace{21pt} = \mathrm{Re} \Bigl\{ \sum_{n,m=\pm} \bigl[ a^n_x (a^m_x)^* \mp a^n_y (a^m_y)^* \bigr]e^{i(q_n-q^*_m)z} \Bigr\} \notag \\
        &\hspace{21pt} = \sum_{n,m=\pm} e^{-(\beta_n+\beta_m)z} (A^{nm}_{xx} \mp A^{nm}_{yy}), \label{eq:ellipse_Faraday_general_diag}
    \end{align}
\end{subequations}
where $q_{\pm}=\alpha_{\pm}+i\beta_{\pm}$ and $A^{nm}_{ij}$ is defined as
\begin{align}
    A^{nm}_{ij}=\mathrm{Re} \bigl[a^n_i (a^m_j)^*e^{i(\alpha_n-\alpha_m)z} \bigr]. \label{eq:ellipse_Anm_ij}
\end{align}
Substituting the expressions in Eq.~\eqref{eq:ellipse_Faraday_general} into Eq.~\eqref{eq:ellipse_angle_ellipticity_general},
we can obtain general expressions for the Faraday rotation angle $\Theta_{\mathrm{F}}$ and the Faraday ellipticity $\Delta_{\mathrm{F}}$.

We finally solve Eq.~\eqref{eq:Faraday_angle_initial} using the polarization ratios $r_{\pm}=(E_{0y}/E_{0x})_{\pm}$ of the two eigenmodes
to determine the amplitudes $a^{\pm}_i$:
\begin{subequations}
    \begin{align}
        a^{\pm}_x&=\mp E_{\mathrm{in}}(r_\mp \cos\theta-\sin\theta)/(r_+-r_-), \\
        a^{\pm}_y&=r_{\pm}a^{\pm}_x,
    \end{align}
\end{subequations}
where $E_{\mathrm{in},x}=E_{\mathrm{in}} \cos \theta$ and $E_{\mathrm{in},y}=E_{\mathrm{in}} \sin \theta$, 
with $E_{\mathrm{in}}$ denoting the complex amplitude of the incident electric field.
Thus, once $r_{\pm}$ are determined from the eigenvalue equation, we can evaluate $\Theta_{\mathrm{F}}$ and $\Delta_{\mathrm{F}}$ 
through the amplitudes $a^{\pm}_i$ and the resulting quantities $A^{nm}_{ij}$ defined in Eq.~\eqref{eq:ellipse_Anm_ij}.

%%%%%%%%%%%%%%%%%%%%%%%%%%%%%%%%%%%%%%%%%%%%%%%%%%%%%%%%%%%%%%%%%%

\subsection{Faraday rotation angle and ellipticity for Type-I and Type-II optical activity} \label{Subsec:AppD2}

In this subsection, we derive the Faraday rotation angle $\Theta_{\mathrm{F}}$ and the Faraday ellipticity $\Delta_{\mathrm{F}}$ 
for Type-I and Type-II optical activity following the procedure described in the previous subsection.
In doing so, recalling that $E_{0y}/E_{0x}=D_{0y}/D_{0x}$,
we use the polarization ratios $r_{\pm}=\pm i$ for Type-I optical activity [Eq.~\eqref{eq:eigenvalue_MO1_MO}]
and $r_+=-\tilde{G}^{(\mathrm{o})}_z/(\eta^{(\mathrm{S})}_{xx}\eta^{(\mathrm{S})}_{x-y})$ and $r_-=-\eta^{(\mathrm{S})}_{yy}\eta^{(\mathrm{S})}_{x-y}/\tilde{G}^{(\mathrm{o})}_z$ for Type-II optical activity [Eq.~\eqref{eq:eigenvalue_MO2_MO}].
We further consider a sample of thickness $\ell$ and evaluate $\Theta_{\mathrm{F}}$ and $\Delta_{\mathrm{F}}$ at $z=\ell$, retaining terms up to first order in $\tilde{G}^{(\mathrm{o})}_z$.
Note that, in the following derivations, we use the approximation $\tan 2\Theta_{\mathrm{F}} \simeq 2\Theta_{\mathrm{F}}$
and choose the sign of $\Delta_{\mathrm{F}}$ to be consistent with the expression for the complex Faraday angle $\Psi_{\mathrm{F}}$ in the main text.

%%%%%%%%%%%%%%%%%%%%%%%%%%%%%%%%%

\subsubsection{Type-I optical activity} 

First, we consider Type-I optical activity, for which the amplitudes $a^{\pm}_i$ are given by 
\begin{subequations}
\begin{align}
    a^{\pm}_x&=\frac{E_{\mathrm{in}}}{2}(\cos\theta \mp i \sin\theta)=\frac{E_{\mathrm{in}}}{2}e^{\mp i \theta}, \\
    a^{\pm}_y&=\frac{E_{\mathrm{in}}}{2}(\sin\theta \pm i \cos\theta)=\frac{E_{\mathrm{in}}}{2}e^{\pm i(\pi/2-\theta)}.
\end{align}
\end{subequations}
Substituting these expressions into the definition of $A^{nm}_{ij}$ in Eq.~\eqref{eq:ellipse_Anm_ij} yields
\begin{align}
    \left\{
    \begin{aligned}
    A^{\pm \pm}_{xy}&=0, &A^{\pm \mp}_{xy}&=\frac{|E_{\mathrm{in}}|^2}{4}\mathrm{Re} \bigl[e^{\pm i(\Delta \alpha \ell+\pi/2-2\theta)}\bigr], \\
    A^{\pm \pm}_{xx}&=\frac{|E_{\mathrm{in}}|^2}{4}, &A^{\pm \mp}_{xx}&=\frac{|E_{\mathrm{in}}|^2}{4} \mathrm{Re} \bigl[e^{\pm i (\Delta \alpha \ell-2\theta)}\bigr], \\
    A^{\pm \pm}_{yy}&=\frac{|E_{\mathrm{in}}|^2}{4}, &A^{\pm \mp}_{yy}&=\frac{|E_{\mathrm{in}}|^2}{4} \mathrm{Re} \bigl[e^{\pm i (\Delta \alpha \ell +\pi-2\theta)}\bigr],
    \end{aligned}
    \right.
\end{align}
from which Eq.~\eqref{eq:ellipse_Faraday_general} reduces to 
\begin{subequations}
    \begin{align}
        J_xJ_y \cos\delta
        &=-\frac{|E_{\mathrm{in}}|^2}{2}e^{-(\beta_++\beta_-)\ell} \sin(\Delta \alpha \ell-2\theta), \\
        J^2_{x-y}&=|E_{\mathrm{in}}|^2e^{-(\beta_++\beta_-)\ell} \cos(\Delta \alpha \ell-2\theta), \\
        J^2_{x+y}&=\frac{|E_{\mathrm{in}}|^2}{2}(e^{-2\beta_+ \ell}+e^{-2\beta_- \ell}),
    \end{align}
\end{subequations}
where $\Delta \alpha=\alpha_+-\alpha_-$.
Using these expressions in Eq.~\eqref{eq:ellipse_angle_ellipticity_general}, we obtain $\Theta_{\mathrm{F}}$ and $\Delta_{\mathrm{F}}$ as
\begin{subequations} \label{eq:Faraday1_theta_ellipticity}
    \begin{align}
        &\tan 2\Theta_{\mathrm{F}} = -\frac{\sin(\Delta \alpha \ell)}{\cos(\Delta \alpha \ell)}=-\tan(\Delta \alpha \ell) \notag \\
        &\Rightarrow \Theta_{\mathrm{F}} \simeq -\tan(\Delta \alpha \ell)/2, \\
        &\Delta^2_{\mathrm{F}}=\Bigl(\frac{e^{-\beta_+ \ell}-e^{-\beta_- \ell}}{e^{-\beta_+ \ell}+e^{-\beta_- \ell}} \Bigr)^2 \notag \\
        &\Rightarrow \Delta_{\mathrm{F}} = \frac{e^{-\beta_+ \ell}-e^{-\beta_- \ell}}{e^{-\beta_+ \ell}+e^{-\beta_- \ell}}
        = \frac{1-e^{\Delta \beta \ell}}{1+e^{\Delta \beta \ell}},
    \end{align}
\end{subequations}
where $\Delta \beta=\beta_+ - \beta_-$.

The expressions in Eq.~\eqref{eq:Faraday1_theta_ellipticity} can be further simplified by considering the properties of the wave numbers $q_\pm$. 
Specifically, we note that the wave-number difference $\Delta q=q_+-q_-=(2\pi/\lambda)(n_+-n_-)$ is proportional to $\tilde{G}^{(\mathrm{o})}_z$ (see Eq.~\eqref{eq:eigenvalue_MO1_MO}),
where $\lambda$ is the vacuum wavelength of the incident light and $n_{\pm}$ are the refractive indices of the two eigenmodes.
Using this property, we can expand Eq.~\eqref{eq:Faraday1_theta_ellipticity} to first order in $\Delta q \ell=\Delta \alpha \ell + i \Delta \beta \ell$ 
to obtain the well-known expressions for $\Theta_{\mathrm{F}}$ and $\Delta_{\mathrm{F}}$:
\begin{subequations}
    \begin{align}
        \Theta_{\mathrm{F}}&\simeq -\frac{\Delta \alpha \ell}{2} =-\frac{\pi \ell}{\lambda}\mathrm{Re}(n_+-n_-), \\
        \Delta_{\mathrm{F}}&\simeq -\frac{\Delta \beta \ell}{2} =-\frac{\pi \ell}{\lambda}\mathrm{Im}(n_+-n_-).
    \end{align}
\end{subequations}
Combining these two expressions immediately yields $\Psi_{\mathrm{F}}$ given in Eq.~\eqref{eq:Faraday_TypeI} of the main text.

%%%%%%%%%%%%%%%%%%%%%%%%%%%%%%%%%

\subsubsection{Type-II optical activity}

Second, we consider Type-II optical activity, for which the amplitudes $a^{\pm}_i$ are given to first order in $\tilde{G}^{(\mathrm{o})}_z$ by 
\begin{subequations} \label{eq:ai_typeII}
    \begin{align}
        a^+_x& \simeq E_{\mathrm{in}}(c_1+a_y s_1), \quad a^-_x \simeq -E_{\mathrm{in}}a_y s_1, \\
        a^+_y& \simeq -E_{\mathrm{in}}a_x c_1, \quad a^-_y \simeq E_{\mathrm{in}}(s_1+a_x c_1),
    \end{align}
\end{subequations}
where $c_n=\cos n\theta$, $s_n=\sin n\theta$, and $a_i$ are defined by
\begin{align}
    a_x=\frac{\tilde{G}^{(\mathrm{o})}_z}{\eta^{(\mathrm{S})}_{xx}\eta^{(\mathrm{S})}_{x-y}}, \quad 
    a_y=\frac{\tilde{G}^{(\mathrm{o})}_z}{\eta^{(\mathrm{S})}_{yy}\eta^{(\mathrm{S})}_{x-y}}. \label{eq:Faraday_TypeII_A_app}
\end{align}
Substituting the expressions in Eq.~\eqref{eq:ai_typeII} into the definition of $A^{nm}_{ij}$ 
and retaining terms up to second order in $a_i$ yields
\begin{widetext}
\begin{subequations}
\begin{align}
    \left\{
    \begin{aligned}
        A^{++}_{xy} &=-|E_{\mathrm{in}}|^2\mathrm{Re} (a_x c_1^2 +a^*_x a_y c_1s_1), \quad
        A^{--}_{xy} =-|E_{\mathrm{in}}|^2\mathrm{Re} (a_y s_1^2 +a^*_x a_y c_1s_1), \\
        %%%%%%%%%%%%%%%%%%%%%%%%%%%%%%%%%
        A^{+-}_{xy} &=|E_{\mathrm{in}}|^2 \bigl[ c_1s_1 \cos(\Delta \alpha \ell) 
        +\mathrm{Re} (a_x c_1^2 +a_y s_1^2 +a^*_x a_y c_1s_1)\cos(\Delta \alpha \ell) 
        +\mathrm{Im} (a_x c_1^2 -a_y s_1^2 -a^*_x a_y c_1s_1)\sin(\Delta \alpha \ell) \bigr], \\
        A^{-+}_{xy} &=|E_{\mathrm{in}}|^2 \bigl[ \mathrm{Re} (a^*_x a_y c_1s_1)\cos(\Delta \alpha \ell) 
        +\mathrm{Im} (a^*_x a_y c_1s_1)\sin(\Delta \alpha \ell) \bigr], 
    \end{aligned}
    \right.
\end{align}
for Eq.~\eqref{eq:ellipse_Faraday_general_offdiag}, and 
\begin{align}
    \left\{
    \begin{aligned}
        A^{++}_{xx} &=|E_{\mathrm{in}}|^2 \bigl[c_1^2 +2\mathrm{Re}(a_y)c_1s_1+|a_y|^2 s_1^2\bigr], \quad
        A^{--}_{xx} =|E_{\mathrm{in}}|^2|a_y|^2 s_1^2, \\ 
        A^{--}_{yy} &=|E_{\mathrm{in}}|^2 \bigl[s_1^2 +2\mathrm{Re}(a_x)c_1s_1+|a_x|^2 c_1^2 \bigr], \quad 
        A^{++}_{yy} =|E_{\mathrm{in}}|^2|a_x|^2 c_1^2, \\
        %%%%%%%%%%%%%%%%%%%%%%%%%%%%%%%%%
        A^{\pm \mp}_{xx} &=-|E_{\mathrm{in}}|^2 \bigl\{ \bigl[\mathrm{Re}(a_y)c_1s_1+|a_y|^2 s_1^2 \bigr] \cos(\Delta \alpha \ell) 
        + \mathrm{Im}(a_y)c_1s_1 \sin(\Delta \alpha \ell) \bigr\}, \\
        A^{\pm \mp}_{yy} &=-|E_{\mathrm{in}}|^2 \bigl\{ \bigl[ \mathrm{Re}(a_x)c_1s_1+|a_x|^2 c_1^2 \bigr] \cos(\Delta \alpha \ell) 
        - \mathrm{Im}(a_x)c_1s_1 \sin(\Delta \alpha \ell) \bigr\},
    \end{aligned}
    \right.
\end{align}
\end{subequations}
for Eq.~\eqref{eq:ellipse_Faraday_general_diag}.
Using these expressions in Eq.~\eqref{eq:ellipse_Faraday_general}, we obtain
\begin{subequations} \label{eq:JxJy_JxJx_JyJy_typeII}
    \begin{align}
        J_xJ_y\cos \delta &=|E_{\mathrm{in}}|^2 e^{-(\beta_+ + \beta_-)\ell}\bigl[c_1s_1 \cos(\Delta \alpha \ell) 
        -(s_1^2 A_y G^+_{\varphi_y}+c_1^2 A_x G^-_{\varphi_x}) 
        - c_1s_1 A_x A_y \cos (\varphi_y -\varphi_x) H \bigr], \label{eq:JxJy_typeII} \\
        J^2_{x \mp y}&=|E_{\mathrm{in}}|^2 e^{-(\beta_+ + \beta_-)\ell}\bigl[(c_1^2e^{-\Delta \beta \ell} \mp s_1^2e^{\Delta \beta \ell}) 
        +2c_1s_1 (A_yG^-_{\varphi_y}\mp A_xG^+_{\varphi_x}) 
        +(s_1^2 A^2_y \mp c_1^2 A^2_x )H \bigr], \label{eq:JxJx_JyJy_typeII}
    \end{align}
\end{subequations}
where we express $a_i$ in polar form as $a_i=A_i e^{i\varphi_i}$ and introduce $G^{\pm}_\varphi$ and $H$, defined by
\begin{align}
    G^{\pm}_\varphi = e^{\pm\Delta \beta \ell}\cos\varphi-\cos(\Delta \alpha \ell \pm \varphi), \quad
    H=e^{\Delta \beta \ell}+e^{-\Delta \beta \ell}-2\cos(\Delta \alpha \ell).
\end{align}
We now derive $\Theta_{\mathrm{F}}$ and $\Delta_{\mathrm{F}}$ by substituting the expressions in Eq.~\eqref{eq:JxJy_JxJx_JyJy_typeII} into Eq.~\eqref{eq:ellipse_angle_ellipticity_general}
and retaining terms up to first order in $A_i$.

%%%%%%%%%%%%%%%%%

\subsubsubsection{Faraday rotation angle}

We first consider the Faraday rotation angle $\Theta_{\mathrm{F}}$, for which Eq.~\eqref{eq:ellipse_angle_ellipticity_general_angle} reduces to
\begin{align}
    \tan 2\Theta_{\mathrm{F}}=\frac{F'_0+2F'_1}{F_0-2F_1}, \label{eq:Faraday_rotation_typeII_general}
\end{align}
where $F_0$ and $F'_0$ are the zeroth-order terms in $A_i$, and $F_1$ and $F'_1$ are the first-order terms:
\begin{subequations}  \label{eq:Faraday_rotation_typeII_F}
\begin{align}
    F_0 &= (c_1^2e^{-\Delta \beta \ell}-s_1^2e^{\Delta \beta \ell})c_2+s_2^2 \cos(\Delta \alpha \ell), \\
    F'_0 &= -(c_1^2e^{-\Delta \beta \ell}-s_1^2e^{\Delta \beta \ell})s_2+c_2s_2 \cos(\Delta \alpha \ell), \\
    F_1 &= c_1 A_x (s_1 c_2G^+_{\varphi_x}+c_1s_2G^-_{\varphi_x}) 
    + s_1 A_y (s_1s_2G^+_{\varphi_y}-c_1c_2G^-_{\varphi_y}), \\
    F'_1 &= c_1 A_x (s_1 s_2G^+_{\varphi_x}-c_1c_2G^-_{\varphi_x}) 
    - s_1 A_y (s_1c_2G^+_{\varphi_y}+c_1s_2G^-_{\varphi_y}).
\end{align}
\end{subequations}
Expanding Eq.~\eqref{eq:Faraday_rotation_typeII_general} to first order in $A_i$ yields
\begin{align}
    \tan 2\Theta_{\mathrm{F}} \simeq \frac{F_0'}{F_0}\Bigl(1+\frac{2F_1}{F_0} \Bigr)+\frac{2F_1'}{F_0}. \label{eq:Faraday_rotation_typeII_firstorderA}
\end{align}
Here, following the discussion in the main text, we analyze this expression in two cases: (i) for small $|\Delta q \ell|$ and (ii) under CAA ($\theta=0^\circ$, $90^\circ$, $180^\circ$, or $270^\circ$).

We begin with case (i), where Eq.~\eqref{eq:Faraday_rotation_typeII_firstorderA} can be expanded to first order in $\Delta q \ell=\Delta \alpha \ell + i \Delta \beta \ell$ to obtain $\Theta_{\mathrm{F}}$ as 
\begin{align}
    & \tan 2\Theta_{\mathrm{F}} \simeq 2c_1 s_1 \Delta \beta \ell + 2c_1^2A_x (\Delta \alpha \ell \sin \varphi_x+ \Delta \beta \ell \cos \varphi_x) 
    +2s_1^2A_y (\Delta \alpha \ell \sin \varphi_y+ \Delta \beta \ell \cos \varphi_y), \notag \\
    & \Rightarrow \Theta_{\mathrm{F}} \simeq \mathrm{Re}\bigl[- i\Delta q \ell (c_1s_1+a_x c_1^2+a_ys_1^2) \bigr]. \label{eq:ThetaF_typeII_final}
\end{align}
In the derivation, we use the following expansions:
\begin{subequations}
    \begin{align}
        F_0 &\simeq 1- c_2\Delta \beta \ell, \quad F'_0 \simeq s_2 \Delta \beta \ell, \\
        G^{\pm}_{\varphi} & \simeq \pm (\Delta \alpha \ell \sin \varphi + \Delta \beta \ell \cos \varphi) 
        +\frac{1}{2} \bigl[ (\Delta \alpha \ell)^2+(\Delta \beta \ell)^2 \bigr] \cos \varphi, \quad 
        H \simeq (\Delta \alpha \ell)^2+(\Delta \beta \ell)^2 \eqqcolon H^{(2)}, \label{eq:expansion_G_H}
    \end{align}
\end{subequations}
where the second-order contributions to $G^{\pm}_{\varphi}$ and $H$ are retained for later use.

We next turn to case (ii), where Eq.~\eqref{eq:Faraday_rotation_typeII_F} reduces to 
\begin{subequations}
\begin{align}
    F_0 &=e^{-\Delta \beta \ell}, \quad F_0' = F_1 = 0, \quad 
    F_1' = - A_x G^-_{\varphi_x} = -e^{-\Delta \beta \ell} A_x \bigl[ \cos \varphi_x - e^{\Delta \beta \ell} \cos(\Delta \alpha \ell - \varphi_x)\bigr],
\end{align}
for incident linear polarization along the $x$ axis ($\theta=0^\circ$ or $180^\circ$), and  
\begin{align}
    F_0 &=e^{\Delta \beta \ell}, \quad F_0' = F_1 = 0, \quad 
    F_1' = A_y G^+_{\varphi_y} = e^{\Delta \beta \ell} A_y \bigl[ \cos \varphi_y - e^{-\Delta \beta \ell} \cos(\Delta \alpha \ell + \varphi_y)\bigr], 
\end{align}
\end{subequations}
for incident linear polarization along the $y$ axis ($\theta=90^\circ$ or $270^\circ$).
Substituting these expressions into Eq.~\eqref{eq:Faraday_rotation_typeII_firstorderA} without expanding in $\Delta q \ell$,
we obtain $\Theta_{\mathrm{F}}^x$ and $\Theta_{\mathrm{F}}^y$ for incident linear polarization along the $x$ and $y$ axes, respectively:
\begin{subequations} \label{eq:Faraday_TypeII_numerical_rotation}
\begin{align}
    \Theta^x_{\mathrm{F}} &\simeq -A_x \bigl[ \cos \varphi_x - e^{\Delta \beta \ell} \cos(\Delta \alpha \ell - \varphi_x)\bigr] 
    =-\mathrm{Re} \bigl[ a_x(1-e^{-i \Delta q \ell})\bigr],  \\
    \Theta^y_{\mathrm{F}} &\simeq A_y \bigl[ \cos \varphi_y - e^{-\Delta \beta \ell} \cos(\Delta \alpha \ell + \varphi_y)\bigr]
    =\mathrm{Re} \bigl[ a_y(1-e^{i \Delta q \ell})\bigr].
\end{align}
\end{subequations}

%%%%%%%%%%%%%%%%%

\subsubsubsection{Faraday ellipticity}

We next consider the Faraday ellipticity $\Delta_{\mathrm{F}}$, beginning by expanding $\Delta^2_{\mathrm{F}}$ in Eq.~\eqref{eq:ellipse_angle_ellipticity_general_ellipticity} to second order in $A_i$.
To this end, we expand the square-root term to second order in $A_i$ as 
\begin{align}
    \sqrt{(J^2_{x-y})^2+(2J_xJ_y\cos \delta)^2} 
    &=|E_{\mathrm{in}}|^2e^{-(\beta_+ + \beta_-)\ell}\sqrt{f_0+f_1+f_2} \notag \\
    &\simeq |E_{\mathrm{in}}|^2e^{-(\beta_+ + \beta_-)\ell} \sqrt{f_0} \Bigl[1+\frac{f_1+f_2}{2f_0}-\frac{f_1^2}{8f_0^2} \Bigr], \label{eq:Faraday_ellipticity_typeII_square_root}
\end{align}
where $f_0$, $f_1$, and $f_2$ are the zeroth-, first-, and second-order terms in $A_i$, respectively:
\begin{subequations} \label{eq:Faraday_ellipticity_coefficients_f}
    \begin{align}
        f_0&=(c_1^2e^{-\Delta \beta \ell}-s_1^2e^{\Delta \beta \ell})^2+4c_1^2s_1^2\cos^2(\Delta \alpha \ell), \\
        f_1&=-4c_1s_1(c_1^2e^{-\Delta \beta \ell}-s_1^2e^{\Delta \beta \ell})(A_x G^+_{\varphi_x}-A_y G^-_{\varphi_y}) 
        -8c_1s_1(c_1^2A_xG^-_{\varphi_x}+s_1^2A_yG^+_{\varphi_y})\cos(\Delta \alpha \ell), \\
        f_2&=-2(c_1^2e^{-\Delta \beta \ell}-s_1^2e^{\Delta \beta \ell})(c_1^2A^2_x-s_1^2A^2_y)H 
        +4c_1^2s_1^2(A_x G^+_{\varphi_x}-A_y G^-_{\varphi_y})^2 \notag \\
        & \quad -8c_1^2 s_1^2 A_xA_y \cos(\varphi_y -\varphi_x) \cos (\Delta \alpha \ell) H 
        +4(c_1^2A_xG^-_{\varphi_x}+s_1^2A_yG^+_{\varphi_y})^2.
    \end{align}
\end{subequations}
Substituting the expanded square-root term in Eq.~\eqref{eq:Faraday_ellipticity_typeII_square_root} into Eq.~\eqref{eq:ellipse_angle_ellipticity_general_ellipticity}, we obtain
\begin{align}
    \Delta^2_{\mathrm{F}}=\frac{F^-_0+F^-_1+F^-_2}{F^+_0+F^+_1+F^+_2}, \label{eq:Faraday_ellipticity_typeII_general}
\end{align}
where $F^{\mp}_0$, $F^{\mp}_1$, and $F^{\mp}_2$ are the zeroth-, first-, and second-order terms in $A_i$, respectively:
\begin{subequations} \label{eq:Faraday_ellipticity_coefficients_F}
    \begin{align}
        F^\mp_0&=(s_1^2e^{\Delta \beta \ell}+c_1^2e^{-\Delta \beta \ell}) \mp \sqrt{f_0}, \\
        F^\mp_1&=2c_1s_1(A_x G^+_{\varphi_x}+A_y G^-_{\varphi_y}) \mp \frac{f_1}{2\sqrt{f_0}}, \\
        F^\mp_2&=(c_1^2A^2_x +s_1^2A^2_y)H \mp \frac{1}{2\sqrt{f_0}}\Bigl(f_2-\frac{f_1^2}{4f_0} \Bigr).
    \end{align}
\end{subequations}
Expanding Eq.~\eqref{eq:Faraday_ellipticity_typeII_general} to second order in $A_i$ yields
\begin{align}
    \Delta^2_{\mathrm{F}} &\simeq \frac{F^-_0}{F^+_0} \Bigl[1-\frac{F^+_1+F^+_2}{F^+_0}+\Bigl(\frac{F^+_1}{F^+_0}\Bigr)^2 \Bigr] 
    +\frac{F^-_1}{F^+_0} \Bigl(1-\frac{F^+_1}{F^+_0} \Bigr)+\frac{F^-_2}{F^+_0}. \label{eq:Faraday_ellipticity_typeII_Delta2}
\end{align}
Here, as in the case of the Faraday rotation angle, we analyze this expression in two cases: (i) for small $|\Delta q \ell|$ and (ii) under CAA ($\theta=0^\circ$, $90^\circ$, $180^\circ$, or $270^\circ$).

We begin with case (i), where we first derive the expressions for $F^{\mp}_0$, $F^{\mp}_1$, and $F^{\mp}_2$ to second order in $\Delta q \ell=\Delta \alpha \ell + i \Delta \beta \ell$.
To this end, we use the expansions in Eq.~\eqref{eq:expansion_G_H} and define $A^{(1)}_i$ and $\tilde{A}^{(1)}_i$ as
\begin{align}
    A^{(1)}_i&=A_i(\Delta \alpha \ell \sin \varphi_i + \Delta \beta \ell \cos \varphi_i), \quad
    \tilde{A}^{(1)}_i=A_i (\Delta \alpha \ell \cos \varphi_i - \Delta \beta \ell \sin \varphi_i).
\end{align}
The resulting expressions are obtained as follows.
For $F^{\mp}_0$, we expand $f_0$ as 
\begin{align}
    f_0 & \simeq \Bigl[c_2-\Delta \beta \ell+\frac{1}{2}c_2 (\Delta \beta \ell)^2\Bigr]^2+s^2_2 \Bigl[1-\frac{1}{2}(\Delta \alpha \ell)^2 \Bigr]^2 \notag \\
    & \simeq 1-2c_2 \Delta \beta \ell +(1+c^2_2)(\Delta \beta \ell)^2-s^2_2(\Delta \alpha \ell)^2 \notag \\
    \Rightarrow & \sqrt{f_0} \simeq 1-c_2 \Delta \beta \ell+\frac{1}{2}(\Delta \beta \ell)^2-\frac{1}{2}s^2_2(\Delta \alpha \ell)^2,
\end{align}
which yields $F^{\mp}_0$ as 
\begin{align}
    F_0^{\mp} &\simeq \Bigl[ 1-c_2 \Delta \beta \ell+\frac{1}{2}(\Delta \beta \ell)^2\Bigr] \mp \sqrt{f_0} 
    = \begin{cases}
        2c^2_1s^2_1 (\Delta \alpha \ell)^2, & \mathrm{for}\ F^-_0, \\
        2+\mathcal{O}(|\Delta q \ell|), & \mathrm{for}\ F^+_0. \\
    \end{cases}
\end{align}
For $F^{\mp}_1$, we expand $f_1$ as 
\begin{align}
    f_1 &\simeq -4c_1s_1(c^2_1-s^2_1-\Delta \beta \ell)\bigl[ A^{(1)}_x+A^{(1)}_y+(A_x \cos \varphi_x-A_y \cos \varphi_y)H^{(2)}/2 \bigr] \notag \\
    &\quad +8c_1s_1 \bigl[ c_1^2A^{(1)}_x-s_1^2A^{(1)}_y - (c^2_1A_x\cos \varphi_x+s^2_1A_y \cos \varphi_y)H^{(2)}/2 \bigr] \notag \\
    &=4c_1s_1 \Bigl\{ A^{(1)}_x(1+\Delta \beta \ell)-A^{(1)}_y(1-\Delta \beta \ell) 
    +\bigl[(1-4c^2_1)A_x \cos\varphi_x + (1-4s^2_1)A_y \cos\varphi_y\bigr] H^{(2)}/2\Bigr\},
\end{align}
which yields $F^{\mp}_1$ as $F^+_1=\mathcal{O}(|\Delta q \ell|)$ and 
\begin{align}
    F^-_1 &\simeq 2c_1 s_1 \bigl[ A^{(1)}_x+A^{(1)}_y+(A_x \cos \varphi_x+A_y \cos \varphi_y)H^{(2)}/2 \bigr] - (1+c_2 \Delta \beta \ell)f_1/2 \notag \\
    &=-2c_1s_1 \Bigl\{\bigl[(1+c_2)A^{(1)}_x+(1-c_2)A^{(1)}_y \bigr]\Delta \beta \ell-2(c^2_1 A_x \cos \varphi_x+s^2_1 A_y \cos \varphi_y)H^{(2)}\Bigr\} \notag \\
    &=4c_1s_1(c^2_1\tilde{A}^{(1)}_x+s^2_1\tilde{A}^{(1)}_y)\Delta \alpha \ell.
\end{align}
For $F^{\mp}_2$, we expand $f_2$ as 
\begin{align}
    f_2 &\simeq -2(c^2_1-s^2_1)(c_1^2A^2_x-s_1^2A^2_y)H^{(2)}+4c^2_1s^2_1(A^{(1)}_x+A^{(1)}_y)^2-8c_1^2 s_1^2 A_xA_y \cos(\varphi_y -\varphi_x) H^{(2)}+4(c_1^2A^{(1)}_x-s_1^2A^{(1)}_y)^2 \notag \\
    &=-2(c^2_1-s^2_1)(c_1^2A^2_x-s_1^2A^2_y)H^{(2)}+4 \bigl[(c_1A^{(1)}_x)^2+(s_1A^{(1)}_y)^2 \bigr]-8c_1^2 s_1^2 A_xA_y \cos(\varphi_y -\varphi_x) H^{(2)},
\end{align}
which yields $F^{\mp}_2$ as $F^+_2=\mathcal{O}(|\Delta q \ell|^2)$ and 
\begin{align}
    F^-_2 &\simeq (c_1^2A^2_x +s_1^2A^2_y)H^{(2)} -(f_2/2-f^2_1/8) \notag \\
    &=2(c^4_1A^2_x+s^4_1A^2_y)H^{(2)}-2\bigl[(c_1A^{(1)}_x)^2+(s_1A^{(1)}_y)^2 \bigr]+4c_1^2 s_1^2 A_xA_y \cos(\varphi_y -\varphi_x)H^{(2)}+2c_1^2s_1^2 (A^{(1)}_x-A^{(1)}_y)^2 \notag \\
    &=2(c^4_1A^2_x+s^4_1A^2_y)H^{(2)}-2\bigl[(c^2_1A^{(1)}_x)^2+(s^2_1A^{(1)}_y)^2 \bigr]+4c_1^2 s_1^2 \bigl[A_xA_y \cos(\varphi_y -\varphi_x)H^{(2)}-A^{(1)}_xA^{(1)}_y \bigr]\notag \\
    &=2(c^2_1 \tilde{A}^{(1)}_x+ s^2_1 \tilde{A}^{(1)}_y)^2.
\end{align}
In deriving $F^-_2$, we use the following relations:
\begin{subequations}
    \begin{align}
        &(c^2_1A^{(1)}_x)^2+(s^2_1A^{(1)}_y)^2=-(c^2_1\tilde{A}^{(1)}_x)^2-(s^2_1\tilde{A}^{(1)}_y)^2 + (c^4_1A^2_x+s^4_1A^2_y)H^{(2)}, \\
        &A_xA_y\cos(\varphi_y-\varphi_x)H^{(2)}-A^{(1)}_xA^{(1)}_y=\tilde{A}^{(1)}_x \tilde{A}^{(1)}_y.
    \end{align}
\end{subequations}

Having obtained the expressions for $F^{\mp}_0$, $F^{\mp}_1$, and $F^{\mp}_2$, we substitute them into Eq.~\eqref{eq:Faraday_ellipticity_typeII_Delta2} to evaluate $\Delta^2_{\mathrm{F}}$:
\begin{align}
    \Delta^2_{\mathrm{F}} &\simeq c^2_1 s^2_1(\Delta \alpha \ell)^2 +2c_1s_1(c^2_1\tilde{A}^{(1)}_x+s^2_1\tilde{A}^{(1)}_y)\Delta \alpha \ell 
    +(c^2_1 \tilde{A}^{(1)}_x+ s^2_1 \tilde{A}^{(1)}_y)^2 \notag \\
    &=(c_1s_1\Delta \alpha \ell+c^2_1 \tilde{A}^{(1)}_x+ s^2_1 \tilde{A}^{(1)}_y)^2. \label{eq:Faraday_ellipticity_typeII_expansion}
\end{align}
Taking the square root, we obtain $\Delta_{\mathrm{F}}$ as  
\begin{align}
    \Delta_{\mathrm{F}} & \simeq  -c_1 s_1 \Delta \alpha \ell - c_1^2A_x (\Delta \alpha \ell \cos \varphi_x- \Delta \beta \ell \sin \varphi_x) 
    -s_1^2A_y (\Delta \alpha \ell \cos \varphi_y-\Delta \beta \ell \sin \varphi_y) \notag \\
    & = \mathrm{Im}\bigl[- i\Delta q \ell (c_1s_1+a_x c_1^2+a_ys_1^2) \bigr].  \label{eq:DeltaF_typeII_final}
\end{align}
Combining Eqs.~\eqref{eq:ThetaF_typeII_final} and \eqref{eq:DeltaF_typeII_final} and using $\Delta q=(2 \pi/\lambda)\Delta n$
immediately yields $\Psi_{\mathrm{F}}$ given in Eq.~\eqref{eq:Faraday_TypeII} of the main text.

We next turn to case (ii), where Eqs.~\eqref{eq:Faraday_ellipticity_coefficients_f} and \eqref{eq:Faraday_ellipticity_coefficients_F} reduce to
\begin{subequations}
\begin{align}
    f_0 &= e^{-2 \Delta \beta \ell}, \quad f_1=0, \quad f_2=-2 A^2_x \bigl[ e^{-\Delta \beta \ell} H -2(G^-_{\varphi_x})^2\bigr], \\
    F_0^+&=2 e^{-\Delta \beta \ell}, \quad F_0^-=F_1^{\mp}=0, \quad F_2^-=2 e^{-\Delta \beta \ell} A^2_x \bigl[ \sin \varphi_x + e^{\Delta \beta \ell} \sin(\Delta \alpha \ell - \varphi_x)\bigr]^2,
\end{align}
for incident linear polarization along the $x$ axis ($\theta=0^\circ$ or $180^\circ$), and  
\begin{align}
    f_0 &= e^{2 \Delta \beta \ell}, \quad f_1=0, \quad f_2=-2 A^2_y \bigl[ e^{\Delta \beta \ell} H -2(G^+_{\varphi_y})^2\bigr], \\
    F_0^+&=2 e^{\Delta \beta \ell}, \quad F_0^-=F_1^{\mp}=0, \quad F_2^-=2 e^{\Delta \beta \ell} A^2_y \bigl[ \sin \varphi_y - e^{-\Delta \beta \ell} \sin(\Delta \alpha \ell + \varphi_y)\bigr]^2, 
\end{align}
\end{subequations}
for incident linear polarization along the $y$ axis ($\theta=90^\circ$ or $270^\circ$).
Substituting the expressions for $F^{\mp}_0$, $F^{\mp}_1$, and $F^{\mp}_2$ into Eq.~\eqref{eq:Faraday_ellipticity_typeII_Delta2} without expanding in $\Delta q \ell$ and then taking the square root,
we obtain $\Delta_{\mathrm{F}}^x$ and $\Delta_{\mathrm{F}}^y$ for incident linear polarization along the $x$ and $y$ axes, respectively:
\begin{subequations} \label{eq:Faraday_TypeII_numerical_ellipticity}
\begin{align}
    \Delta^x_{\mathrm{F}} &\simeq -A_x \bigl[ \sin \varphi_x + e^{\Delta \beta \ell} \sin(\Delta \alpha \ell - \varphi_x)\bigr]
    =-\mathrm{Im} \bigl[ a_x(1-e^{-i \Delta q \ell})\bigr],  \\
    \Delta^y_{\mathrm{F}} &\simeq A_y \bigl[ \sin \varphi_y - e^{-\Delta \beta \ell} \sin(\Delta \alpha \ell + \varphi_y)\bigr]
    =\mathrm{Im} \bigl[ a_y(1-e^{i \Delta q \ell})\bigr].
\end{align}
\end{subequations}
Combining Eqs.~\eqref{eq:Faraday_TypeII_numerical_rotation} and \eqref{eq:Faraday_TypeII_numerical_ellipticity} and using $\Delta q=(2 \pi/\lambda)\Delta n$
immediately yields $\Psi^{x}_{\mathrm{F}}$ and $\Psi^{y}_{\mathrm{F}}$ given in Eq.~\eqref{eq:Faraday_TypeII_numerical} of the main text.

%%%%%%%%%%%%%%%%%%%%%%%%%%%%%%%%%%%%%%%%%%%%%%%%%%%%%%%%%%%%%%%%%%%%%%%%%%%%%%%%%%%%%%%%%%%%%%%%%%%%%%%%%%%%%%%%%%%%%%%%%%%%%%%%%%%%

\section{DETAILED DERIVATION OF MICROSCOPIC EXPRESSIONS FOR THE EXPANSION COEFFICIENTS} \label{Sec:AppE}

In this appendix, we derive microscopic expressions for the expansion coefficients $\sigma_{ij}(\omega)$, $\sigma_{ij,k}(\omega)$, and $\sigma_{ij,kl}(\omega)$ in units with $\hbar=1$.
The derivation begins by decomposing the optical conductivity tensor $\sigma_{ij}(\bm{q},\omega)$ in Eq.~\eqref{eq:sigma_kubo_general} into the orbital-orbital, orbital-spin mixed, and spin-spin contributions using Eq.~\eqref{eq:vq_decomposition}:
\begin{subequations} \label{eq:sigma_oo_os_ss}
    \begin{align}
        \sigma^{(\mathrm{orb\text{-}orb})}_{ij}(\bm{q},\omega)& =-ie^2 \sum_{n,m} \int_{\bm{k}} \mathcal{K}_{nm}(\bm{k},\bm{q},\omega)
        \braket{u_{n \bm{k}-\bm{q}/2}| \hat{v}^i_{\bm{k}}|u_{m\bm{k}+\bm{q}/2}}\braket{u_{m \bm{k}+\bm{q}/2}| \hat{v}^j_{\bm{k}}|u_{n\bm{k}-\bm{q}/2}}, \label{eq:sigma_oo} \\
        \sigma^{(\mathrm{orb\text{-}spin})}_{ij}(\bm{q},\omega)& =\frac{e^2g_s}{2m} e_{ik\mu}q_k \sum_{n,m} \int_{\bm{k}} \mathcal{K}_{nm}(\bm{k},\bm{q},\omega)
        \bigl[\braket{u_{n \bm{k}-\bm{q}/2}| \hat{s}^\mu|u_{m\bm{k}+\bm{q}/2}}\braket{u_{m \bm{k}+\bm{q}/2}| \hat{v}^j_{\bm{k}}|u_{n\bm{k}-\bm{q}/2}} -(i\leftrightarrow j)^*\bigr], \label{eq:sigma_os} \\
        \sigma^{(\mathrm{spin\text{-}spin})}_{ij}(\bm{q},\omega)& =-i \Bigl(\frac{e g_s}{2m} \Bigr)^2 e_{ik\mu} e_{jl\nu}q_k q_l \sum_{n,m} \int_{\bm{k}} \mathcal{K}_{nm}(\bm{k},\bm{q},\omega)
        \braket{u_{n \bm{k}-\bm{q}/2}| \hat{s}^\mu|u_{m\bm{k}+\bm{q}/2}}\braket{u_{m \bm{k}+\bm{q}/2}| \hat{s}^\nu|u_{n\bm{k}-\bm{q}/2}}, 
    \end{align}
\end{subequations}
where $(i \leftrightarrow j)^*$ denotes the complex conjugate of the preceding term with $i$ and $j$ interchanged, and the spectral kernel $\mathcal{K}_{nm}(\bm{k},\bm{q},\omega)$ is given by
\begin{align}
    \mathcal{K}_{nm}(\bm{k},\bm{q},\omega)= \frac{f_{n\bm{k}-\bm{q}/2}-f_{m\bm{k}+\bm{q}/2}}{\varepsilon_{n\bm{k}-\bm{q}/2}-\varepsilon_{m\bm{k}+\bm{q}/2}} 
    \frac{1}{\varepsilon_{n\bm{k}-\bm{q}/2}-\varepsilon_{m\bm{k}+\bm{q}/2}+\omega+i\delta}.
\end{align}
Expanding each expression in Eq.~\eqref{eq:sigma_oo_os_ss} in powers of $\bm{q}$, we can derive microscopic expressions for the expansion coefficients.
In particular, we focus on the orbital-orbital contribution in Eq.~\eqref{eq:sigma_oo} for $\sigma_{ij}(\omega)$ and 
the orbital-spin mixed contribution in Eq.~\eqref{eq:sigma_os} for $\sigma_{ij,k}(\omega)$ and $\sigma_{ij,kl}(\omega)$.

Before proceeding to the detailed derivation of each expansion coefficient, we first expand the spectral kernel $\mathcal{K}_{nm}(\bm{k},\bm{q},\omega)$ to first order in $\bm{q}$:
\begin{subequations} \label{eq:Spectral_kernel_expansion}
    \begin{align}
        \mathcal{K}_{n=m}(\bm{k},\bm{q},\omega) &\simeq \frac{1}{\tilde{\omega}}f'_{n}+q_k\frac{\partial_k \varepsilon_{n}}{\tilde{\omega}^2}f'_{n}
        =-i \frac{\tau}{1-i \omega \tau} f'_{n}+q_k \Bigl( \frac{\tau}{1-i \omega \tau} \Bigr)^2 (-f'_{n}) (\partial_k \varepsilon_{n}), \\
        \mathcal{K}_{n \neq m}(\bm{k},\bm{q},\omega) &\simeq \Bigl[f_{nm}+q_k \Bigl(-\frac{1}{2}\partial_k \tilde{f}_{nm}+\frac{f_{nm}\partial_k\tilde{\varepsilon}_{nm}}{\varepsilon_{nm}}
        +\frac{f_{nm}\partial_k\tilde{\varepsilon}_{nm}}{2\varepsilon_{nm}} \omega\frac{d}{d\omega}\Bigr)\Bigr]\frac{1}{\varepsilon_{nm}(\varepsilon_{nm}+\tilde{\omega})},
    \end{align}
\end{subequations}
where $\tilde{\omega}=\omega+i\delta$, $\tau=1/\delta$, $\partial_k=\partial/\partial k_k$, $f'_{n}=\partial f_{n}/\partial \varepsilon_{n}$,
$\varepsilon_{nm}=\varepsilon_{n}-\varepsilon_{m}$, $\tilde{\varepsilon}_{nm}=\varepsilon_{n}+\varepsilon_{m}$,
$f_{nm}=f_{n}-f_{m}$, and $\tilde{f}_{nm}=f_{n}+f_{m}$, and we henceforth omit the explicit $\bm{k}$ dependence in the expanded expressions for notational simplicity.
In deriving Eq.~\eqref{eq:Spectral_kernel_expansion}, we separately consider the intraband ($n=m$) and interband ($n\neq m$) cases and use the following expansion formulas: 
\begin{subequations}
\begin{align}
    \frac{f_{n\bm{k}-\bm{q}/2}-f_{n\bm{k}+\bm{q}/2}}{\varepsilon_{n\bm{k}-\bm{q}/2}-\varepsilon_{n\bm{k}+\bm{q}/2}}
    &=\frac{f_{n\bm{k}-\bm{q}/2+\bm{q}}-f_{n\bm{k}-\bm{q}/2}}{\varepsilon_{n\bm{k}-\bm{q}/2+\bm{q}}-\varepsilon_{n\bm{k}-\bm{q}/2}}
    \simeq f'_{n\bm{k}-\bm{q}/2}+\frac{1}{2}f''_{n\bm{k}-\bm{q}/2}(\varepsilon_{n\bm{k}+\bm{q}/2}-\varepsilon_{n\bm{k}-\bm{q}/2}) \simeq f'_{n}, \\
    %%%%%%%%%%%%%%%%%%%%%%%%%%%%%%%%%%%%%%%%%%%%%%%%%%%%%%%%%%%%%%%%%%
    \frac{1}{\varepsilon_{n\bm{k}-\bm{q}/2}-\varepsilon_{n\bm{k}+\bm{q}/2}+\tilde{\omega}}
    &\simeq \frac{1}{\tilde{\omega}-q_k \partial_k \varepsilon_{n\bm{k}}}
    \simeq \frac{1}{\tilde{\omega}}+q_k\frac{\partial_k \varepsilon_{n}}{\tilde{\omega}^2},
\end{align}
\end{subequations}
for the intraband case and 
\begin{subequations}
    \begin{align}
        \frac{f_{n\bm{k}-\bm{q}/2}-f_{m\bm{k}+\bm{q}/2}}{\varepsilon_{n\bm{k}-\bm{q}/2}-\varepsilon_{m\bm{k}+\bm{q}/2}}
        &\simeq \frac{f_{nm\bm{k}}-q_k \partial_k \tilde{f}_{nm\bm{k}}/2}{\varepsilon_{nm\bm{k}}-q_k \partial_k \tilde{\varepsilon}_{nm\bm{k}}/2}
        \simeq \Bigl(\frac{1}{\varepsilon_{nm}}+q_k\frac{\partial_k \tilde{\varepsilon}_{nm}}{2\varepsilon^2_{nm}} \Bigr) f_{nm} -q_k \frac{\partial_k \tilde{f}_{nm}}{2\varepsilon_{nm}}, \\
        %%%%%%%%%%%%%%%%%%%%%%%%%%%%%%%%%%%%%%%%%%%%%%%%%%%%%%%%%%%%%%%%%%
        \frac{1}{\varepsilon_{n\bm{k}-\bm{q}/2}-\varepsilon_{m\bm{k}+\bm{q}/2}+\tilde{\omega}}
        &\simeq \frac{1}{\varepsilon_{nm}+\tilde{\omega}}+q_k \Bigl[ \frac{\partial_k \tilde{\varepsilon}_{nm}}{2\varepsilon_{nm}(\varepsilon_{nm}+\tilde{\omega})}
        +\frac{\partial_k \tilde{\varepsilon}_{nm}}{2\varepsilon_{nm}} \omega \frac{d}{d\omega}\Bigl( \frac{1}{\varepsilon_{nm}+\tilde{\omega}}\Bigr)\Bigr], \label{eq:Spectral_kernel_expansion_inter2}
    \end{align}
\end{subequations}
for the interband case, where we use the following algebraic manipulation to obtain Eq.~\eqref{eq:Spectral_kernel_expansion_inter2}, assuming $\varepsilon_{nm}+i\delta \approx \varepsilon_{nm}$:
\begin{align}
    \frac{\partial_k \tilde{\varepsilon}_{nm}}{2(\varepsilon_{nm}+\tilde{\omega})^2}
    =\frac{\partial_k \tilde{\varepsilon}_{nm}}{2 \varepsilon_{nm}(\varepsilon_{nm}+\tilde{\omega})}\Bigl( 1-\frac{\omega}{\varepsilon_{nm}+\tilde{\omega}}\Bigr).
\end{align}
We then use the expansion of the spectral kernel in Eq.~\eqref{eq:Spectral_kernel_expansion} 
to derive the expressions for $\sigma_{ij}(\omega)$, $\sigma_{ij,k}(\omega)$, and $\sigma_{ij,kl}(\omega)$ in Appendices~\ref{Subsec:AppE1}, \ref{Subsec:AppE2}, and \ref{Subsec:AppE3}, respectively.

%%%%%%%%%%%%%%%%%%%%%%%%%%%%%%%%%%%%%%%%%%%%%%%%%%%%%%%%%%%%%%%%%%

\subsection{Zeroth-order expansion coefficient $\sigma_{ij}(\omega)$} \label{Subsec:AppE1}

The zeroth-order expansion coefficient $\sigma_{ij}(\omega)$ is obtained by taking the $\bm{q} \to 0 $ limit in Eqs.~\eqref{eq:sigma_oo} and \eqref{eq:Spectral_kernel_expansion}:
\begin{align}
    \sigma_{ij}(\omega)&=e^2 \sum_{n} \int_{\bm{k}} \Bigl[ \Bigl( \frac{\tau}{1-i \omega \tau}\Bigr) (-f'_n) v^i_n v^j_n 
    -i \sum_{m (\neq n)}  v^i_{nm} v^j_{mn} \frac{f_{nm}}{\varepsilon_{nm}(\varepsilon_{nm}+\tilde{\omega})} \Bigr],
\end{align}
where, for an arbitrary operator $\hat{O}$, we define $O_{nm}=\braket{u_n|\hat{O}|u_m}$ and $O_{n}=O_{nn}$.
The coefficient $\sigma_{ij}(\omega)$ is further decomposed into its symmetric and antisymmetric parts with respect to $i$ and $j$ as 
\begin{subequations}
\begin{align}
    \sigma^{(\mathrm{S})}_{ij}(\omega)&=e^2 \sum_{n} \int_{\bm{k}} \Bigl[ \Bigl( \frac{\tau}{1-i \omega \tau}\Bigr) (-f'_n) \mathrm{Re}(v^i_n v^j_n) 
    -i \sum_{m (\neq n)} \mathrm{Re}(v^i_{nm}v^j_{mn})  \frac{f_{nm}}{\varepsilon_{nm}(\varepsilon_{nm}+\tilde{\omega})} \Bigr], \\
    \sigma^{(\mathrm{A})}_{ij}(\omega)&=-i e^2 \sum_{n,m (\neq n)} \int_{\bm{k}} i \mathrm{Im}(v^i_{nm}v^j_{mn}) \frac{f_{nm}}{\varepsilon_{nm}(\varepsilon_{nm}+\tilde{\omega})}.
\end{align}
\end{subequations}
These expressions can be further simplified in terms of the band-resolved quantum metric $g^{ij}_{nm}$ and Berry curvature $\Omega^{ij}_{nm}$, defined as \cite{Provost1980-qb,Berry1984-fn} 
\begin{subequations}
    \begin{align}
        g^{ij}_{nm} &= \mathrm{Re} \Bigl( \frac{v^i_{nm}v^j_{mn}}{\varepsilon^2_{nm}}\Bigr) = \mathrm{Re} \braket{D_i u_n|u_m} \braket{u_m|D_j u_n}, \\
        \Omega^{ij}_{nm} &=-2\mathrm{Im} \Bigl( \frac{v^i_{nm}v^j_{mn}}{\varepsilon^2_{nm}}\Bigr) = -2\mathrm{Im} \braket{D_i u_n|u_m} \braket{u_m|D_j u_n}, 
    \end{align}
\end{subequations}
where we use the Hellmann-Feynman theorem, $v^i_{nm}=-\varepsilon_{nm}\braket{u_n|D_i u_m}\ (n \neq m)$, and $\ket{D_i u_n}=(1-\ket{u_n}\bra{u_n}) \ket{\partial_i u_n}$ is the covariant derivative of band $n$.
The resulting expressions yield Eqs.~\eqref{eq:band_contribution_01_sigma} and \eqref{eq:geometric_contribution_01_sigma} of the main text.

%%%%%%%%%%%%%%%%%%%%%%%%%%%%%%%%%%%%%%%%%%%%%%%%%%%%%%%%%%%%%%%%%%

\subsection{First-order expansion coefficient $\sigma_{ij,k}(\omega)$} \label{Subsec:AppE2}

The first-order expansion coefficient $\sigma_{ij,k}(\omega)$ is obtained by retaining the explicit prefactor $q_k$ in Eq.~\eqref{eq:sigma_os} and taking the $\bm{q} \to 0 $ limit of the remaining factors:
\begin{align}
    \sigma_{ij,k}(\omega)& =\frac{e^2g_s}{2m} \lim_{\bm{q} \to 0} \sum_{n,m} \int_{\bm{k}} \mathcal{K}_{nm}(\bm{k},\bm{q},\omega)
    \bigl[e_{ik\mu}\braket{u_{n \bm{k}-\bm{q}/2}| \hat{s}^\mu|u_{m\bm{k}+\bm{q}/2}}\braket{u_{m \bm{k}+\bm{q}/2}| \hat{v}^j_{\bm{k}}|u_{n\bm{k}-\bm{q}/2}} -(i\leftrightarrow j)^*\bigr].
\end{align}  
The coefficient $\sigma_{ij,k}(\omega)$ is further decomposed into its symmetric and antisymmetric parts with respect to $i$ and $j$ as 
\begin{subequations}
    \begin{align}
        \sigma^{(\mathrm{S})}_{ij,k}(\omega)& =i\frac{e^2g_s}{2m} \lim_{\bm{q} \to 0} \sum_{n,m} \int_{\bm{k}} \mathcal{K}_{nm}(\bm{k},\bm{q},\omega)
        \mathrm{Im}\bigl[e_{ik\mu}\braket{u_{n \bm{k}-\bm{q}/2}| \hat{s}^\mu|u_{m\bm{k}+\bm{q}/2}}\braket{u_{m \bm{k}+\bm{q}/2}| \hat{v}^j_{\bm{k}}|u_{n\bm{k}-\bm{q}/2}} +(i\leftrightarrow j)\bigr], \\
        \sigma^{(\mathrm{A})}_{ij,k}(\omega)& =\frac{e^2g_s}{2m} \lim_{\bm{q} \to 0} \sum_{n,m} \int_{\bm{k}} \mathcal{K}_{nm}(\bm{k},\bm{q},\omega)
        \mathrm{Re}\bigl[e_{ik\mu}\braket{u_{n \bm{k}-\bm{q}/2}| \hat{s}^\mu|u_{m\bm{k}+\bm{q}/2}}\braket{u_{m \bm{k}+\bm{q}/2}| \hat{v}^j_{\bm{k}}|u_{n\bm{k}-\bm{q}/2}} -(i\leftrightarrow j)\bigr],
    \end{align}  
\end{subequations}
which can be evaluated as
\begin{subequations}
    \begin{align}
        \sigma^{(\mathrm{S})}_{ij,k}(\omega)&=-i\frac{e^2g_s}{2m}e_{ik\mu} \sum_{n,m (\neq n)} \int_{\bm{k}} \mathrm{Re}(is^\mu_{nm} v^j_{mn})\frac{f_{nm}}{\varepsilon_{nm}(\varepsilon_{nm}+\tilde{\omega})} +(i\leftrightarrow j), \\
        \sigma^{(\mathrm{A})}_{ij,k}(\omega)&=i\frac{e^2g_s}{2m}e_{ik\mu} \sum_{n} \int_{\bm{k}} \Bigl[ \Bigl( \frac{\tau}{1-i \omega \tau}\Bigr) (-f'_n) \mathrm{Re}(s^\mu_n v^j_n) 
        -i \sum_{m (\neq n)} \mathrm{Im}(is^\mu_{nm}v^j_{mn})  \frac{f_{nm}}{\varepsilon_{nm}(\varepsilon_{nm}+\tilde{\omega})} \Bigr] -(i\leftrightarrow j).
    \end{align}
\end{subequations}
These expressions can be further simplified in terms of the band-resolved Zeeman quantum metric $\tilde{g}^{\mu,j}_{nm}$ and Berry curvature $\tilde{\Omega}^{\mu,j}_{nm}$, defined as \cite{Xiang2025-cl,Xiang2025-xi}
\begin{subequations} \label{eq:Zeeman_quantum_geometry_app}
    \begin{align}
        \tilde{g}^{\mu,j}_{nm} &= \mathrm{Re} \Bigl( i\frac{s^\mu_{nm}v^j_{mn}}{\varepsilon_{nm}}\Bigr) = \mathrm{Re} \bigl[ i\braket{u_n|\hat{s}^\mu|u_m} \braket{u_m|D_j u_n} \bigr], \\
        \tilde{\Omega}^{\mu,j}_{nm} &=-2\mathrm{Im} \Bigl( i\frac{s^\mu_{nm}v^j_{mn}}{\varepsilon_{nm}}\Bigr) = -2\mathrm{Im} \bigl[ i\braket{u_n|\hat{s}^\mu|u_m} \braket{u_m|D_j u_n} \bigr].
    \end{align}
\end{subequations}
The resulting expressions yield Eqs.~\eqref{eq:band_contribution_01_chi} and \eqref{eq:geometric_contribution_01_chi} of the main text after introducing the response tensors $\chi^{(\mathrm{o/e})}_{\mu,j}(\omega)$, defined through
\begin{align}
    \sigma^{(\mathrm{S})}_{ij,k}(\omega)=ie_{ik\mu} \chi^{(\mathrm{o})}_{\mu,j}(\omega) + (i \leftrightarrow j), \quad 
    \sigma^{(\mathrm{A})}_{ij,k}(\omega)=ie_{ik\mu} \chi^{(\mathrm{e})}_{\mu,j}(\omega) - (i \leftrightarrow j).
\end{align}

%%%%%%%%%%%%%%%%%%%%%%%%%%%%%%%%%%%%%%%%%%%%%%%%%%%%%%%%%%%%%%%%%%

\subsection{Second-order expansion coefficient $\sigma_{ij,kl}(\omega)$} \label{Subsec:AppE3}

The second-order expansion coefficient $\sigma_{ij,kl}(\omega)$ is obtained by retaining the explicit prefactor $q_k$ in Eq.~\eqref{eq:sigma_os},
expanding the remaining factors to first order in $q_l$, and then taking the $\bm{q} \to 0 $ limit:
\begin{subequations}
    \begin{align}
        \sigma^{(\mathrm{S})}_{ij,kl}(\omega)& =i\frac{e^2g_s}{2m} \lim_{\bm{q} \to 0} \partial_l \Bigl\{ \sum_{n,m} \int_{\bm{k}} \mathcal{K}_{nm}(\bm{k},\bm{q},\omega)
        \mathrm{Im}\bigl[e_{ik\mu}\braket{u_{n \bm{k}-\bm{q}/2}| \hat{s}^\mu|u_{m\bm{k}+\bm{q}/2}}\braket{u_{m \bm{k}+\bm{q}/2}| \hat{v}^j_{\bm{k}}|u_{n\bm{k}-\bm{q}/2}} +(i\leftrightarrow j)\bigr] \Bigr\}, \\
        \sigma^{(\mathrm{A})}_{ij,kl}(\omega)& =\frac{e^2g_s}{2m} \lim_{\bm{q} \to 0} \partial_l \Bigl\{ \sum_{n,m} \int_{\bm{k}} \mathcal{K}_{nm}(\bm{k},\bm{q},\omega)
        \mathrm{Re}\bigl[e_{ik\mu}\braket{u_{n \bm{k}-\bm{q}/2}| \hat{s}^\mu|u_{m\bm{k}+\bm{q}/2}}\braket{u_{m \bm{k}+\bm{q}/2}| \hat{v}^j_{\bm{k}}|u_{n\bm{k}-\bm{q}/2}} -(i\leftrightarrow j)\bigr] \Bigr\},
    \end{align}  
\end{subequations}
where $\sigma^{(\mathrm{S})}_{ij,kl}(\omega)$ and $\sigma^{(\mathrm{A})}_{ij,kl}(\omega)$ are the symmetric and antisymmetric parts with respect to $i$ and $j$, respectively.
Deriving explicit expressions for these coefficients requires expanding the operator matrix elements to first order in $\bm{q}$:
\begin{align}
    &\braket{u_{n \bm{k}-\bm{q}/2}| \hat{s}^\mu|u_{m\bm{k}+\bm{q}/2}}\braket{u_{m \bm{k}+\bm{q}/2}| \hat{v}^j_{\bm{k}}|u_{n\bm{k}-\bm{q}/2}} \notag \\
    &\simeq s^\mu_{nm} v^j_{mn} +\frac{q_l}{2} \Bigl[ (-\braket{\partial_l u_n|\hat{s}^\mu|u_m}+\braket{u_n|\hat{s}^\mu|\partial_l  u_m}) v^j_{mn}
    +s^\mu_{nm}(\braket{\partial_l u_m|\hat{v}^j|u_n}-\braket{u_m|\hat{v}^j|\partial_l u_n}) \Bigr] \notag \\
    &= s^\mu_{nm} v^j_{mn} +\frac{q_l}{2} \Bigl[ (-\braket{D_l u_n|\hat{s}^\mu|u_m}+\braket{u_n|\hat{s}^\mu|D_l u_m}) v^j_{mn}
    +s^\mu_{nm}(\braket{D_l u_m|\hat{v}^j|u_n}-\braket{u_m|\hat{v}^j|D_l u_n}) \Bigr] \notag \\
    &\quad +\frac{q_l}{2} s^\mu_{nm}v^j_{mn} \underbrace{\bigl[ (-\braket{\partial_l u_n|u_n}+\braket{u_m|\partial_l u_m})+(\braket{\partial_l u_m|u_m}-\braket{u_n|\partial_l u_n}) \bigr]}_{=0}.
\end{align}
Here, we introduce the covariant derivative $D_l$, thereby eliminating the terms involving the diagonal Berry connection $\mathcal{A}^l_n=\braket{u_n|i\partial_l u_n}$ and  
rendering the resulting expression manifestly gauge invariant.
Using this expansion together with the expansion of the spectral kernel in Eq.~\eqref{eq:Spectral_kernel_expansion}, we obtain the following explicit expressions:
\begin{subequations} \label{eq:sigma2S}
    \begin{align}
        \sigma^{(\mathrm{S})}_{ij,kl}(\omega) &=-\frac{e^2g_s}{2m} e_{i k\mu}\sum_{n} \int_{\bm{k}} \Bigl\{ \frac{1}{2}\Bigl( \frac{\tau}{1-i\omega \tau} \Bigr) \mathrm{Re}(\tilde{T}^{\mu,jl}_{nn})f'_n \notag \\
        & \qquad \qquad \qquad \qquad  +\frac{i}{2} \sum_{m (\neq n)} \mathrm{Re} \bigl[\tilde{T}^{\mu,jl}_{nm} f_{nm}+is^\mu_{nm}v^j_{mn}\hat{F}^l_{nm}(\omega) \bigr] \frac{1}{\varepsilon_{nm}(\varepsilon_{nm}+\tilde{\omega})}\Bigr\}+(i \leftrightarrow j), \label{eq:sigma2S_os} \\
        \sigma^{(\mathrm{A})}_{ij,kl}(\omega) &=\frac{e^2g_s}{2m} e_{i k\mu}\sum_{n} \int_{\bm{k}} \Bigl\{\Bigl( \frac{\tau}{1-i\omega \tau} \Bigr)^2 (-f'_n)\mathrm{Re}(s^\mu_n v^j_n v^l_n) \notag \\
        & \qquad \qquad \qquad \qquad  +\frac{1}{2} \sum_{m (\neq n)} \mathrm{Im} \bigl[\tilde{T}^{\mu,jl}_{nm} f_{nm}+is^\mu_{nm}v^j_{mn}\hat{F}^l_{nm}(\omega) \bigr] \frac{1}{\varepsilon_{nm}(\varepsilon_{nm}+\tilde{\omega})}\Bigr\}-(i \leftrightarrow j), 
    \end{align}
\end{subequations}
where we define $\hat{F}^l_{nm}(\omega)$ and the gauge-invariant quantity $\tilde{T}^{\mu,jl}_{nm}$ as
\begin{subequations}
    \begin{align}
        \hat{F}^l_{nm}(\omega) &= -\partial_l \tilde{f}_{nm}+\Bigl(2+\omega \frac{d}{d\omega} \Bigr) \frac{f_{nm}\partial_l\tilde{\varepsilon}_{nm}}{\varepsilon_{nm}}, \\
        \tilde{T}^{\mu,jl}_{nm} &= i\bigl[ (-\braket{D_l u_n|\hat{s}^\mu|u_m}+\braket{u_n|\hat{s}^\mu|D_l u_m}) v^j_{mn}+s^\mu_{nm}(\braket{D_l u_m|\hat{v}^j|u_n}-\braket{u_m|\hat{v}^j|D_l u_n}) \bigr].
    \end{align}
\end{subequations}

The expressions in Eq.~\eqref{eq:sigma2S} can be further simplified in terms of band-geometric quantities.
To this end, we first transform $\tilde{T}^{\mu,jl}_{nm}$ into the form
\begin{align}
    \tilde{T}^{\mu,jl}_{nm} &= i\bigl[(-\braket{D_l u_n|\hat{s}^\mu|u_m}+\braket{u_n|\hat{s}^\mu|D_l u_m}) v^j_{mn} \notag \\
    & \quad -s^\mu_{nm}(\partial_j \tilde{\varepsilon}_{nm}\braket{u_m|D_l u_n}+\braket{D_l u_m|\hat{H}-\varepsilon_n|D_j u_n}-\braket{D_j u_m|\hat{H}-\varepsilon_m|D_l u_n})\bigr], \label{eq:tildeT_transform}
\end{align}
where we use the following relation obtained by differentiating $\hat{H}\ket{u_n}=\varepsilon_n \ket{u_n}$ with respect to $k_j$:
\begin{align}
    \hat{v}^j \ket{u_n}=\partial_j \varepsilon_n \ket{u_n}-(\hat{H}-\varepsilon_n)\ket{\partial_j u_n}=\partial_j \varepsilon_n \ket{u_n}-(\hat{H}-\varepsilon_n)\ket{D_j u_n}.
\end{align}
From Eq.~\eqref{eq:tildeT_transform}, $\mathrm{Re}(\tilde{T}^{\mu,jl}_{nn})$ in the intraband term of Eq.~\eqref{eq:sigma2S_os} reduces to 
\begin{align}
    \mathrm{Re}(\tilde{T}^{\mu,jl}_{nn}) &= 2\mathrm{Re} \bigl[i\braket{u_n|\hat{s}^\mu|D_l u_n} \bigr] v^j_n - 2s^\mu_n\ \mathrm{Im} \braket{D_ju_n|\hat{H}-\varepsilon_n|D_l u_n}
    =2(\tilde{g}^{\mu,l}_{n} v^j_n - e^{-1} s^\mu_n m^{jl}_{n}),
\end{align}
where $\tilde{g}^{\mu,l}_n=\sum_m \tilde{g}^{\mu,l}_{nm}=\mathrm{Re}\bigl[i\braket{u_n|\hat{s}^\mu|D_l u_n}\bigr]$
and $m^{jl}_{n}=e \mathrm{Im}\braket{D_ju_n|\hat{H}-\varepsilon_n|D_l u_n}$ are the Zeeman quantum metric of band $n$ and the orbital magnetic moment of band $n$ \cite{Shi2007-xs}, respectively.
The interband terms in Eq.~\eqref{eq:sigma2S} can likewise be simplified in terms of the band-resolved Zeeman quantum metric $\tilde{g}^{\mu,j}_{nm}$ and Berry curvature $\tilde{\Omega}^{\mu,j}_{nm}$, defined in Eq.~\eqref{eq:Zeeman_quantum_geometry_app}.
With these simplifications, the expressions in Eq.~\eqref{eq:sigma2S} yield Eqs.~\eqref{eq:band_contribution_2} and \eqref{eq:geometric_contribution_2} of the main text after interchanging the indices $k$ and $l$ and introducing the response tensors $\chi^{(\mathrm{e/o})}_{\mu,jk}(\omega)$, defined through
\begin{align}
    \sigma^{(\mathrm{S})}_{ij,kl}(\omega)=e_{il\mu} \chi^{(\mathrm{e})}_{\mu,jk}(\omega) + (i \leftrightarrow j), \quad 
    \sigma^{(\mathrm{A})}_{ij,kl}(\omega)=e_{il\mu} \chi^{(\mathrm{o})}_{\mu,jk}(\omega) - (i \leftrightarrow j).
\end{align}

%%%%%%%%%%%%%%%%%%%%%%%%%%%%%%%%%%%%%%%%%%%%%%%%%%%%%%%%%%%%%%%%%%%%%%%%%%%%%%%%%%%%%%%%%%%%%%%%%%%%%%%%%%%%%%%%%%%%%%%%%%%%%%%%%%%%

\section{DETAILED DISCUSSION OF THE MULTIPOLAR DESCRIPTION} \label{Sec:AppF}

In this appendix, we develop a unified physical interpretation of the $\mathcal{T}$-odd expansion coefficients 
by comparing the following two equivalent expressions for the current response order by order:
\begin{subequations}
\begin{align}
    J_i (\bm{q},\omega) &= \bigl\{\sigma^{(\mathrm{A})}_{ij}(\omega)+\bigl[ie_{ik\mu}\chi^{(\mathrm{o})}_{\mu,j}(\omega)+(i \leftrightarrow j)\bigr]q_k 
    + \bigl[e_{il\mu}\chi^{(\mathrm{o})}_{\mu,jk}(\omega)-(i \leftrightarrow j)\bigr]q_kq_l + \cdots \bigr\} E_j(\bm{q},\omega), \\
    J_i(\bm{q},\omega) &=(-i \omega) \bigl[P_i(\bm{q},\omega)+(-i q_k) P_{ik}(\bm{q},\omega) + \cdots \bigr] + ie_{ik\mu} q_k \bigl[ M_\mu(\bm{q},\omega)+(-iq_l)M_{\mu,l}(\bm{q},\omega)+ \cdots \bigr],
\end{align}
\end{subequations}
\end{widetext}
where, in the first expression, $\sigma^{(\mathrm{S})}_{ij,k}(\omega)$ and $\sigma^{(\mathrm{A})}_{ij,kl}(\omega)$ are written in terms of 
$\chi^{(\mathrm{o})}_{\mu,j}(\omega)$ and $\chi^{(\mathrm{o})}_{\mu,jk}(\omega)$, respectively, according to Eq.~\eqref{eq:1st_2nd_transformation}.
The zeroth-, first-, and second-order current responses are discussed separately in Appendices~\ref{Subsec:AppF1}, \ref{Subsec:AppF2}, and \ref{Subsec:AppF3}, respectively.

%%%%%%%%%%%%%%%%%%%%%%%%%%%%%%%%%%%%%%%%%%%%%%%%%%%%%%%%%%%%%%%%%%

\subsection{Zeroth-order current response} \label{Subsec:AppF1}

The zeroth-order current response can be rewritten as 
\begin{align}
    J_i (\bm{q},\omega) &=(-i \omega) i\sigma^{(\mathrm{A})}_{ij}(\omega)/\omega E_j(\bm{q},\omega),
\end{align}
from which the corresponding induced multipole field is identified as 
\begin{align}
    P_i(\bm{q},\omega)&=\frac{i}{\omega}\sigma^{(\mathrm{A})}_{ij}(\omega)E_j(\bm{q},\omega) \notag \\
    &=\varepsilon_0 \varepsilon^{(\mathrm{A})}_{ij}(\omega)E_j(\bm{q},\omega).
\end{align}
Thus, $\sigma^{(\mathrm{A})}_{ij}$, or equivalently $\varepsilon^{(\mathrm{A})}_{ij}$, describes an electric polarization response.

%%%%%%%%%%%%%%%%%%%%%%%%%%%%%%%%%%%%%%%%%%%%%%%%%%%%%%%%%%%%%%%%%%

\subsection{First-order current response} \label{Subsec:AppF2}

The first-order current response can be rewritten as 
\begin{align}
    J_i(\bm{q},\omega) &= \bigl[ ie_{ik\mu} \chi^{(\mathrm{o})}_{\mu,j}(\omega) + ie_{jk\mu} \chi^{(\mathrm{o})}_{\mu,i}(\omega)\bigr] q_k E_j(\bm{q},\omega) \notag \\
    &=(-i\omega) \chi^{(\mathrm{o})}_{\mu,i}(\omega) e_{\mu kj}q_k E_j(\bm{q},\omega)/\omega \notag \\
    &\quad +i e_{ik\mu} q_k \chi^{(\mathrm{o})}_{\mu,j}(\omega)E_j(\bm{q},\omega), 
\end{align}
from which the corresponding induced multipole fields are identified as 
\begin{subequations}
    \begin{align}
        P_j(\bm{q},\omega)&=\chi^{(\mathrm{o})}_{\mu,j}(\omega)B_\mu(\bm{q},\omega), \\
        M_\mu(\bm{q},\omega)&=\chi^{(\mathrm{o})}_{\mu,j}(\omega)E_j(\bm{q},\omega), 
    \end{align}
\end{subequations}
where the first expression is rewritten in terms of the external magnetic field $\bm{B}$ using Faraday's law $\bm{B}=(\bm{q} \times \bm{E} )/ \omega$ in Eq.~\eqref{eq:Maxwell1}.

%%%%%%%%%%%%%%%%%%%%%%%%%%%%%%%%%%%%%%%%%%%%%%%%%%%%%%%%%%%%%%%%%%

\subsection{Second-order current response} \label{Subsec:AppF3}

The second-order current response can be rewritten as 
\begin{align}
    J_i(\bm{q},\omega) &= \bigl[ e_{il\mu}  \chi^{(\mathrm{o})}_{\mu,jk}(\omega) - e_{jl\mu} \chi^{(\mathrm{o})}_{\mu,ik}(\omega)\bigr] q_k q_l E_j(\bm{q},\omega).
\end{align}
Here, we decompose $\chi^{(\mathrm{o})}_{\mu,jk}$ into its symmetric component $\chi^{(\mathrm{o})}_{\mu,(jk)}$ 
and antisymmetric component $\chi^{(\mathrm{o})}_{\mu,[jk]}$ with respect to $j$ and $k$.
In the following, we separately consider the current response arising from each component.

We first consider the current response arising from the symmetric component:
\begin{align}
    J_i(\bm{q},\omega) 
    &=-(-i\omega) (-i q_k) \chi^{(\mathrm{o})}_{\mu,(ik)}(\omega) e_{\mu lj}q_l E_j(\bm{q},\omega)/\omega \notag \\
    &\quad -i e_{ik\mu} q_k \chi^{(\mathrm{o})}_{\mu,(jl)}(\omega) iq_l E_j(\bm{q},\omega), 
\end{align}
from which the corresponding induced multipole fields are identified as
\begin{subequations}
    \begin{align}
        P_{jk}(\bm{q},\omega) = - \chi^{(\mathrm{o})}_{\mu,(jk)}(\omega) B_\mu(\bm{q},\omega), \\
        M_\mu(\bm{q},\omega) = - \chi^{(\mathrm{o})}_{\mu,(jk)}(\omega) E_{jk}(\bm{q},\omega).
    \end{align}
\end{subequations}
Alternatively, the same current response can also be rewritten as 
\begin{align}
    J_i(\bm{q},\omega) 
    &= (-i\omega) \chi^{(\mathrm{o})}_{\mu,(ki)}(\omega)(iq_k)e_{\mu lj}q_l E_j(\bm{q},\omega)/\omega \notag \\
    &\quad +ie_{ik\mu}q_k(-i q_l) \chi^{(\mathrm{o})}_{\mu,(lj)}(\omega)E_j(\bm{q},\omega), 
\end{align}
from which another set of induced multipole fields is identified as 
\begin{subequations}
    \begin{align} 
        P_k(\bm{q},\omega) &= \chi^{(\mathrm{o})}_{\mu,(jk)} (\omega) B_{\mu,j}(\bm{q},\omega), \\
        M_{\mu,j}(\bm{q},\omega) &= \chi^{(\mathrm{o})}_{\mu,(jk)}(\omega)E_k(\bm{q},\omega).
    \end{align}
\end{subequations}
In the above expressions, we introduce the electric-field gradient $E_{ij}$ and the magnetic-field gradient $B_{\mu,j}$ as 
\begin{subequations}
    \begin{align}
        E_{ij}(\bm{q},\omega) &\coloneqq i q_j E_i(\bm{q},\omega) \ \left(=\partial_{r_j}E_i(\bm r,\omega)\right), \\
        B_{\mu,j}(\bm{q},\omega) &\coloneqq i q_j B_\mu(\bm{q},\omega) \ \left(=\partial_{r_j}B_\mu(\bm r,\omega)\right),
    \end{align}
\end{subequations}
where the expressions in parentheses denote the corresponding real-space representations, with $\partial_{r_j}=\partial/\partial r_j$.

We next consider the current response arising from the antisymmetric component.
To this end, we introduce a new polar tensor $\gamma^{(\mathrm{o})}_{\mu\nu}(\omega)$, defined as $\chi^{(\mathrm{o})}_{\mu,[jk]}(\omega)=e_{jk\nu} \gamma^{(\mathrm{o})}_{\mu\nu}(\omega)$.
The current response can then be rewritten as 
\begin{align}
    J_i(\bm{q},\omega) 
    &=i e_{ik\mu} q_k \bigl[ i\omega \gamma^{(\mathrm{o})}_{\mu\nu}(\omega) e_{\nu lj} q_l E_j(\bm{q},\omega)/\omega \notag \\
    &\quad -i\omega \gamma^{(\mathrm{o})}_{\nu\mu}(\omega) e_{\nu lj} q_l E_j(\bm{q},\omega)/\omega \bigr],
\end{align}
from which the corresponding induced multipole field is identified as 
\begin{align}
    M_\mu(\bm{q},\omega) &=i \omega \bigl[\gamma^{(\mathrm{o})}_{\mu\nu}(\omega) -\gamma^{(\mathrm{o})}_{\nu\mu}(\omega) \bigr] B_\nu(\bm{q},\omega) \notag \\
    &= -2 \gamma^{(\mathrm{o})}_{[\mu\nu]}(\omega) \dot{B}_\nu(\bm{q},\omega).
\end{align}
Here, $\gamma^{(\mathrm{o})}_{[\mu\nu]}$ denotes the antisymmetric part of $\gamma^{(\mathrm{o})}_{\mu\nu}$, and 
we define the time derivative of the magnetic field as 
\begin{align}
    \dot{\bm{B}}(\bm{q},\omega)=-i\omega \bm{B}(\bm{q},\omega) \ \left(=\partial_t \bm{B}(\bm{q},t)\right),
\end{align}
where the expression in parentheses denotes the corresponding time-domain representation, with $\partial_t=\partial/\partial t$.
Thus, $\gamma^{(\mathrm{o})}_{\mu\nu}$, especially $\gamma^{(\mathrm{o})}_{[\mu\nu]}$, describes a dynamic magnetization response that vanishes in the DC limit,
while its polar character implies a direct response rather than a cross response, in contrast to $\chi^{(\mathrm{o})}_{\mu,(jk)}$.

%%%%%%%%%%%%%%%%%%%%%%%%%%%%%%%%%%%%%%%%%%%%%%%%%%%%%%%%%%%%%%%%%%%%%%%%%%%%%%%%%%%%%%%%%%%%%%%%%%%%%%%%%%%%%%%%%%%%%%%%%%%%%%%%%%%%

\section{DETAILED DERIVATION OF THE ST\v{R}EDA FORMULAS AND $f$-SUM RULES} \label{Sec:AppG}

In this appendix, we derive St\v{r}eda formulas and $f$-sum rules relating the $\mathcal{T}$-odd expansion coefficients to magnetic multipole moments formulated within the modern theory of magnetic multipoles.
These relations can be schematically summarized as 
\begin{subequations}
    \begin{align}
        \text{St\v{r}eda formula}&: \chi \equiv \chi(\omega \to 0) \leftrightarrow \frac{\partial M}{\partial \mu}, \\
        f\text{-sum rule}&: \braket{\chi} \equiv \int_{0}^{\infty} \mathrm{Im}\ \chi(\omega) d\omega \leftrightarrow M, 
    \end{align}
\end{subequations}
where $\chi(\omega)$ generically denotes a $\mathcal{T}$-odd expansion coefficient and $M$ denotes the corresponding magnetic multipole moment.
Specifically, we relate the optical Hall conductivity $\sigma^{(\mathrm{A})}_{ij}(\omega)$ [Eq.~\eqref{eq:geometric_contribution_01_sigma}] to orbital magnetization $M^{(\mathrm{orb})}_{ij}$ in Appendix~\ref{Subsec:AppG1},
the QMER $\chi^{(\mathrm{o})}_{\mu,j}(\omega)$ [Eq.~\eqref{eq:geometric_contribution_01_chi}] to the spin magnetic quadrupole moment $M^{(\mathrm{spin})}_{\mu,j}$ in Appendix~\ref{Subsec:AppG2}, and 
the OMER $\chi^{(\mathrm{o})}_{\mu,(jk)}(\omega)$ [Eq.~\eqref{eq:geometric_contribution_2_o}] to the spin magnetic octupole moment $M^{(\mathrm{spin})}_{\mu,(jk)}$ in Appendix~\ref{Subsec:AppG3}.
These results are summarized in Appendix~\ref{Subsec:AppG4}.

In each subsection, we first derive the St\v{r}eda formula and then obtain the $f$-sum rule.
For the St\v{r}eda formula, we use $\partial \mathcal{G}_n/\partial \mu=- f_n$ and 
impose the following condition: (i) the system is a zero-temperature insulator with the chemical potential $\mu$ in the band gap, 
for which $\partial f_n/\partial \mu=\delta(\varepsilon_n-\mu)=0$.
Here, $\mathcal{G}_n=-\int^\mu_{-\infty}f_n(\mu')d\mu'$ is the grand-potential density of band $n$ and
$\delta(x)$ is the Dirac delta function.
For the $f$-sum rule, we again use condition (i) together with the following relations:
\begin{subequations} \label{eq:preliminary_sum_rule}
\begin{align}
    & \frac{1}{\varepsilon_{nm}+\omega+i\delta}=\mathcal{P}\frac{1}{\varepsilon_{nm}+\omega}-i\pi \delta(\varepsilon_{nm}+\omega), \label{eq:preliminary_sum_rule1} \\
    & \sum_{n,m (\neq n)} f_{nm} \int_0^\infty \omega^{-l}\delta(\varepsilon_{nm}+\omega) d\omega \notag \\
    & \quad = \sum_n^{\mathrm{occ}}\sum_m^{\mathrm{unocc}}(-\varepsilon_{nm})^{-l}, \quad (l=0,1), \label{eq:preliminary_sum_rule2} \\
    & f_n=\Theta(\mu-\varepsilon_n), \quad \mathcal{G}_n= (\varepsilon_n-\mu) \Theta(\mu-\varepsilon_{n}), \label{eq:preliminary_sum_rule3}
\end{align}
\end{subequations}
where $\mathcal{P}$ denotes the Cauchy principal value, `occ' and `unocc' denote occupied and unoccupied bands, respectively,
$\Theta(x)$ is the Heaviside step function, and we set $\hbar=1$ in this appendix.

%%%%%%%%%%%%%%%%%%%%%%%%%%%%%%%%%%%%%%%%%%%%%%%%%%%%%%%%%%%%%%%%%%

\subsection{Optical Hall conductivity $\sigma^{(\mathrm{A})}_{ij}(\omega)$} \label{Subsec:AppG1}

We first derive the St\v{r}eda formula from the expressions for orbital magnetization $M^{(\mathrm{orb})}_{ij}$ \cite{Xiao2005-bu,Thonhauser2005-wp,Ceresoli2006-xy,Shi2007-xs} 
and the DC limit of $\sigma^{(\mathrm{A})}_{ij}(\omega)$,
\begin{subequations}
    \begin{align}
        M^{(\mathrm{orb})}_{ij} &= e \mathrm{Im} \sum_{n}  \int_{\bm{k}} \Bigl[f_n \braket{D_i u_n|\hat{H}-\varepsilon_n|D_j u_n} \notag \\
        & \quad +2\mathcal{G}_n \braket{D_i u_n|D_j u_n} \Bigr], \label{eq:OM_modern} \\
        \sigma^{(\mathrm{A})}_{ij} &=2e^2 \mathrm{Im} \sum_{n}\int_{\bm{k}} f_n \braket{D_i u_n|D_j u_n}.
    \end{align}
\end{subequations}
Comparing these expressions and imposing condition (i) yield the following St\v{r}eda formula:
\begin{align}
    e\frac{\partial M^{(\mathrm{orb})}_{ij}}{\partial \mu}=-\sigma^{(\mathrm{A})}_{ij},
\end{align}
which was originally derived in Ref.~\cite{Streda1982-xz}.

We next derive the $f$-sum rule from the following two spectral integrals under condition (i):
\begin{subequations} \label{eq:OHC_sum_rule}
\begin{align}
    &\braket{\sigma^{(\mathrm{A})}_{ij}} \notag \\
    &= -\pi e^2 \mathrm{Im}\sum_n^{\mathrm{occ}}\sum_m^{\mathrm{unocc}} \int_{\bm{k}}\braket{D_i u_n|u_m}\varepsilon_{nm}\braket{u_m|D_j u_n} \notag \\
    &= \pi e^2 \mathrm{Im} \sum_n^{\mathrm{occ}} \int_{\bm{k}} \braket{D_i u_n|(\hat{H}-\varepsilon_n)\hat{Q}|D_j u_n}, \\
    &\braket{\omega^{-1}\sigma^{(\mathrm{A})}_{ij}} =\pi e^2 \mathrm{Im}\sum_n^{\mathrm{occ}} \int_{\bm{k}}\braket{D_i u_n|\hat{Q}|D_j u_n}, \label{eq:OHC_sum_rule2}
\end{align}
\end{subequations}
where $\hat{Q}=\sum_m^{\mathrm{unocc}}\ket{u_m}\bra{u_m}$ is the projection operator onto the unoccupied subspace.
These spectral integrals appear explicitly in $M^{(\mathrm{orb})}_{ij}$ when Eq.~\eqref{eq:OM_modern} is rewritten under condition (i) as
\begin{align}
    M^{(\mathrm{orb})}_{ij} &= e \mathrm{Im} \sum_{n}^{\mathrm{occ}}  \int_{\bm{k}} \Bigl[\braket{D_i u_n|(\hat{H}-\varepsilon_n)\hat{Q}|D_j u_n} \notag \\
    & \quad +2(\varepsilon_n -\mu) \braket{D_i u_n|\hat{Q}|D_j u_n} \Bigr], \label{eq:OM_modern_zerotemperature}
\end{align}
which yields the following $f$-sum rule:
\begin{align}
    M^{(\mathrm{orb})}_{ij}=\frac{1}{\pi e}\Bigl[ \braket{\sigma^{(\mathrm{A})}_{ij}} +2 \sum_n^{\mathrm{occ}}\int_{\bm{k}}(\varepsilon_{n \bm{k}}-\mu) \braket{\omega^{-1}\sigma^{(\mathrm{A})}_{ij}}_{n\bm{k}}\Bigr],
\end{align}
where $\braket{\omega^{-1}\sigma^{(\mathrm{A})}_{ij}}=\sum_n^{\mathrm{occ}}\int_{\bm{k}}\braket{\omega^{-1}\sigma^{(\mathrm{A})}_{ij}}_{n\bm{k}}$.
In deriving Eq.~\eqref{eq:OM_modern_zerotemperature}, we use the completeness relation $\hat{1}=\hat{P}+\hat{Q}$, where $\hat{P}=\sum_n^{\mathrm{occ}}\ket{u_n}\bra{u_n}$ is the projection operator onto the occupied subspace, 
and the fact that the following additional term vanishes due to cancellation under the interchange $n \leftrightarrow n'$:
\begin{align}
    &\mathrm{Im}\sum_n^{\mathrm{occ}}\braket{D_i u_n|(\hat{H}+\varepsilon_n-2\mu)\hat{P}|D_j u_n} \notag \\
    &=\mathrm{Im}\sum_{n,n'}^{\mathrm{occ}}(\varepsilon_n+\varepsilon_{n'}-2\mu)\braket{D_i u_n|u_{n'}}\braket{u_{n'}|D_j u_n} \notag \\
    &=0.
\end{align}

%%%%%%%%%%%%%%%%%%%%%%%%%%%%%%%%%%%%%%%%%%%%%%%%%%%%%%%%%%%%%%%%%%

\subsection{Quadrupolar magnetoelectric response $\chi^{(\mathrm{o})}_{\mu,j}(\omega)$} \label{Subsec:AppG2}

We first derive the St\v{r}eda formula from the expressions for the spin magnetic quadrupole moment $M^{(\mathrm{spin})}_{\mu,j}$ \cite{Gao2018-uk,Shitade2019-hq,Hua2026-rf} 
and the DC limit of $\chi^{(\mathrm{o})}_{\mu,j}(\omega)$,
\begin{subequations}
    \begin{align}
        M^{(\mathrm{spin})}_{\mu,j} &=-\frac{e g_s}{2m} \mathrm{Im}\sum_{n} \int_{\bm{k}} \Bigl[f_{n} \braket{u_n|\hat{s}^\mu|D_j u_n} \notag \\
        & \quad -\sum_{m (\neq n)} \mathcal{G}_{nm} \frac{s^\mu_{nm}\braket{u_m|D_j u_n}}{\varepsilon_{nm}}\Bigr], \label{eq:SQM_modern} \\
        \chi^{(\mathrm{o})}_{\mu,j} & = \frac{e^2 g_s}{2m} \mathrm{Im} \sum_{n,m(\neq n)} \int_{\bm{k}} f_{nm} \frac{s^\mu_{nm}\braket{u_m|D_j u_n}}{\varepsilon_{nm}},
    \end{align}
\end{subequations}
where $\mathcal{G}_{nm}=\mathcal{G}_n-\mathcal{G}_m$ and $s^\mu_{nm}=\braket{u_n|\hat{s}^\mu|u_m}$.
Comparing these expressions and imposing condition (i) yield the following St\v{r}eda formula:
\begin{align}
    e\frac{\partial M^{(\mathrm{spin})}_{\mu,j}}{\partial \mu}=-\chi^{(\mathrm{o})}_{\mu,j},
\end{align}
which was previously derived in Refs.~\cite{Gao2018-uk,Shitade2019-hq}.

We next derive the $f$-sum rule from the following two spectral integrals under condition (i):
\begin{subequations} \label{eq:QMER_sum_rule}
    \begin{align} 
        \braket{\chi^{(\mathrm{o})}_{\mu,j}} &= -\pi \frac{e^2 g_s}{2m}\mathrm{Im}\sum_n^{\mathrm{occ}} \int_{\bm{k}}\braket{u_n|\hat{s}^\mu \hat{Q}|D_j u_n}, \\
        \braket{\omega^{-1}\chi^{(\mathrm{o})}_{\mu,j}} &= \pi \frac{e^2 g_s}{2m}\mathrm{Im}\sum_n^{\mathrm{occ}} \sum_m^{\mathrm{unocc}} \int_{\bm{k}} \frac{s^\mu_{nm}\braket{u_m|D_j u_n}}{\varepsilon_{nm}}.
    \end{align}
\end{subequations}
These spectral integrals appear explicitly in $M^{(\mathrm{spin})}_{\mu,j}$ when Eq.~\eqref{eq:SQM_modern} is rewritten under condition (i) as 
\begin{widetext}
\begin{align}
    M^{(\mathrm{spin})}_{\mu,j} &=-\frac{e g_s}{2m} \mathrm{Im} \int_{\bm{k}} \Bigl\{\sum_{n}^{\mathrm{occ}} \braket{u_n|\hat{s}^\mu|D_j u_n} 
    -\sum_{n}^{\mathrm{occ}}\sum_{n' (\neq n)}^{\mathrm{occ}} \braket{u_n|\hat{s}^\mu|u_{n'}}\braket{u_{n'}|D_j u_n} \notag \\
    &\quad -\Bigl[\sum_{n}^{\mathrm{occ}}\sum_{m}^{\mathrm{unocc}} (\varepsilon_n -\mu) \frac{s^\mu_{nm}\braket{u_m|D_j u_n}}{\varepsilon_{nm}} - \sum_{n}^{\mathrm{unocc}}\sum_{m}^{\mathrm{occ}} (\varepsilon_m -\mu) \frac{s^\mu_{nm}\braket{u_m|D_j u_n}}{\varepsilon_{nm}} \Bigr]\Bigr\} \notag \\
    &=-\frac{e g_s}{2m} \mathrm{Im}\sum_{n}^{\mathrm{occ}} \int_{\bm{k}} \Bigl[\braket{u_n|\hat{s}^\mu \hat{Q}|D_j u_n} 
    -\sum_{m}^{\mathrm{unocc}} 2(\varepsilon_n -\mu) \frac{s^\mu_{nm}\braket{u_m|D_j u_n}}{\varepsilon_{nm}}\Bigr], \label{eq:SQM_modern_zerotemperature} 
\end{align}
which yields the following $f$-sum rule:
\begin{align}
    M^{(\mathrm{spin})}_{\mu,j} = \frac{1}{\pi e}\Bigl[ \braket{\chi^{(\mathrm{o})}_{\mu,j}}+2 \sum_n^{\mathrm{occ}} \int_{\bm{k}}(\varepsilon_{n\bm{k}} -\mu)\braket{\omega^{-1}\chi^{(\mathrm{o})}_{\mu,j}}_{n\bm{k}}\Bigr],
\end{align}
where $\braket{\omega^{-1}\chi^{(\mathrm{o})}_{\mu,j}}=\sum_n^{\mathrm{occ}}\int_{\bm{k}}\braket{\omega^{-1}\chi^{(\mathrm{o})}_{\mu,j}}_{n\bm{k}}$.
In deriving Eq.~\eqref{eq:SQM_modern_zerotemperature}, we use $\braket{u_{n}|D_j u_n}=0$, $\hat{1}-\hat{P}=\hat{Q}$, and the following identity obtained by interchanging $n$ and $m$:
\begin{align}
    -\mathrm{Im}\sum_{n}^{\mathrm{unocc}}\sum_{m}^{\mathrm{occ}}(\varepsilon_m-\mu)\frac{s^\mu_{nm}\braket{u_m|D_j u_n}}{\varepsilon_{nm}}
    =\mathrm{Im}\sum_{n}^{\mathrm{occ}}\sum_{m}^{\mathrm{unocc}}(\varepsilon_n-\mu)\frac{s^\mu_{nm}\braket{u_m|D_j u_n}}{\varepsilon_{nm}}.
\end{align}

%%%%%%%%%%%%%%%%%%%%%%%%%%%%%%%%%%%%%%%%%%%%%%%%%%%%%%%%%%%%%%%%%%

\subsection{Octupolar magnetoelectric response $\chi^{(\mathrm{o})}_{\mu,(jk)}(\omega)$} \label{Subsec:AppG3}

We first derive the St\v{r}eda formula from the expressions for the spin magnetic octupole moment $M^{(\mathrm{spin})}_{\mu,(jk)}$ \cite{Oike2025-dk,Shitade2025-ue,Sato2026-vj,Hua2026-rf}
and the DC limit of $\chi^{(\mathrm{o})}_{\mu,(jk)}(\omega)$,
\begin{subequations} \label{eq:OMER_Streda}
    \begin{align}
        M^{(\mathrm{spin})}_{\mu,(jk)} &= \frac{e g_s}{8m} \mathrm{Re}\sum_{n} \int_{\bm{k}} \Bigl\{ f_{n} \braket{D_j u_{n}|\hat{s}^\mu+s^\mu_{n}|D_k u_{n}} \notag \\
        & \quad +\sum_{m (\neq n)} \Bigl[ \mathcal{G}_{nm}\frac{R^{\mu,jk}_{nm}}{\varepsilon_{nm}} - \Bigl( \partial_k \tilde{\mathcal{G}}_{nm}-\frac{\mathcal{G}_{nm}\partial_k \tilde{\varepsilon}_{nm}}{\varepsilon_{nm}} \Bigr) \frac{s^\mu_{nm}\braket{u_m|D_j u_n}}{\varepsilon_{nm}} \Bigr] \Bigr\}+(j \leftrightarrow k), \label{eq:OMER_Streda1} \\
        \chi^{(\mathrm{o})}_{\mu,(jk)} &= \frac{e^2 g_s}{8m} \mathrm{Re} \sum_{n,m(\neq n)} \int_{\bm{k}} \Bigl(f_{nm}\frac{R^{\mu,jk}_{nm}}{\varepsilon_{nm}}+\frac{f_{nm}\partial_k \tilde{\varepsilon}_{nm}}{\varepsilon_{nm}} \frac{s^\mu_{nm}\braket{u_m|D_j u_n}}{\varepsilon_{nm}} \Bigr)+(j \leftrightarrow k), \label{eq:OMER_Streda2}
    \end{align}
\end{subequations}
where $\tilde{\mathcal{G}}_{nm}=\mathcal{G}_n+\mathcal{G}_m$, $s^\mu_n=\braket{u_n|\hat{s}^\mu|u_n}$, and 
\begin{align}
    R^{\mu,jk}_{nm} &= (-\braket{D_j u_n|\hat{s}^\mu|u_m}+\braket{u_n|\hat{s}^\mu|D_j u_m}) \braket{u_m|D_k u_n}+ s^\mu_{nm} \braket{D_j u_m|D_k u_n}. \label{eq:Rujk_general}
\end{align}
Note that the terms proportional to $\partial f_n/\partial \mu$ are omitted in Eq.~\eqref{eq:OMER_Streda}, 
and that the expression for $\tilde{T}^{\mu,jk}_{nm}$ given in Eq.~\eqref{eq:tildeT_transform} is used in deriving Eq.~\eqref{eq:OMER_Streda2}.
Comparing the two expressions in Eq.~\eqref{eq:OMER_Streda} and imposing condition (i) yield the following St\v{r}eda formula:
\begin{align}
    e\frac{\partial M^{(\mathrm{spin})}_{\mu,(jk)}}{\partial \mu}=-\chi^{(\mathrm{o})}_{\mu,(jk)},
\end{align} 
which was previously derived microscopically in Ref.~\cite{Shitade2025-ue}.

We next derive the $f$-sum rule from the following two spectral integrals under condition (i):
\begin{subequations} \label{eq:OMER_sum_rule}
    \begin{align} 
        \braket{\chi^{(\mathrm{o})}_{\mu,(jk)}} &= -\pi \frac{e^2 g_s}{8m}\mathrm{Re}\sum_n^{\mathrm{occ}} \sum_m^{\mathrm{unocc}}\int_{\bm{k}}R^{\mu,jk}_{nm}+(j \leftrightarrow k) \notag \\
        &= \pi \frac{e^2 g_s}{8m}\mathrm{Re}\sum_n^{\mathrm{occ}} \int_{\bm{k}} \bigl(\braket{D_j u_n|\hat{s}^\mu\hat{Q}|D_k u_n}-\braket{u_n|\hat{s}^\mu \partial_j \hat{Q}|D_k u_n}\bigr)+(j \leftrightarrow k), \label{eq:OMER_sum_rule1} \\
        \braket{\omega^{-1}\chi^{(\mathrm{o})}_{\mu,(jk)}} &= \pi \frac{e^2 g_s}{8m}\mathrm{Re}\sum_n^{\mathrm{occ}} \sum_m^{\mathrm{unocc}} \int_{\bm{k}} \Bigl( \frac{R^{\mu,jk}_{nm}}{\varepsilon_{nm}}+\frac{\partial_k \tilde{\varepsilon}_{nm}}{\varepsilon_{nm}}\frac{s^\mu_{nm}\braket{u_m|D_j u_n}}{\varepsilon_{nm}} \Bigr)+(j \leftrightarrow k), \label{eq:OMER_sum_rule2}
    \end{align}
\end{subequations}
where we use the expression for $\tilde{T}^{\mu,jk}_{nm}$ given in Eq.~\eqref{eq:tildeT_transform} and the following relations:
\begin{subequations}
    \begin{align}
        &\int^{\infty}_0 \omega \frac{d}{d\omega}\delta(\varepsilon_{nm}+\omega)d\omega=-\int^{\infty}_0 \delta(\varepsilon_{nm}+\omega)d\omega, \quad 
        \int^{\infty}_0 \frac{d}{d \omega}\delta(\varepsilon_{nm}+\omega)d\omega=0, \\
        &\sum_{m}^{\mathrm{unocc}}R^{\mu,jk}_{nm}
        =-\braket{D_j u_n|\hat{s}^\mu\hat{Q}|D_k u_n}+\sum_{m}^{\mathrm{unocc}}\bra{u_n}\hat{s}^\mu(\ket{D_j u_m}\bra{u_m}+\ket{u_m}\bra{D_j u_m})\ket{D_k u_n}, \\
        &\partial_j \hat{Q}=\sum_{m}^{\mathrm{unocc}}(\ket{\partial_j u_m}\bra{u_m}+\ket{u_m}\bra{\partial_j u_m})=\sum_{m}^{\mathrm{unocc}}(\ket{D_j u_m}\bra{u_m}+\ket{u_m}\bra{D_j u_m}).
    \end{align}
\end{subequations}
These spectral integrals appear explicitly in $M^{(\mathrm{spin})}_{\mu,(jk)}$ when Eq.~\eqref{eq:OMER_Streda1} is rewritten under condition (i) as
\begin{align}
    M^{(\mathrm{spin})}_{\mu,(jk)} &= \frac{e g_s}{8m} \mathrm{Re} \int_{\bm{k}} \Bigl\{ \sum_{n}^{\mathrm{occ}} \braket{D_j u_{n}|\hat{s}^\mu+s^\mu_{n}|D_k u_{n}} + \sum_{n}^{\mathrm{occ}}\sum_{n' (\neq n)}^{\mathrm{occ}} R^{\mu,jk}_{nn'} \notag \\
    & \quad + \sum_{n}^{\mathrm{occ}}\sum_m^{\mathrm{unocc}} \Bigl[ (\varepsilon_n - \mu) \frac{R^{\mu,jk}_{nm}}{\varepsilon_{nm}}- \Bigl(\partial_k (\varepsilon_n-\mu)-\frac{(\varepsilon_n-\mu)\partial_k \tilde{\varepsilon}_{nm}}{\varepsilon_{nm}} \Bigr)\frac{s^\mu_{nm}\braket{u_m|D_j u_n}}{\varepsilon_{nm}} \Bigr] \notag \\
    & \quad + \sum_{n}^{\mathrm{unocc}}\sum_m^{\mathrm{occ}} \Bigl[ -(\varepsilon_m - \mu) \frac{R^{\mu,jk}_{nm}}{\varepsilon_{nm}}- \Bigl(\partial_k (\varepsilon_m-\mu)+\frac{(\varepsilon_m-\mu)\partial_k \tilde{\varepsilon}_{nm}}{\varepsilon_{nm}} \Bigr)\frac{s^\mu_{nm}\braket{u_m|D_j u_n}}{\varepsilon_{nm}} \Bigr] \Bigr\} +(j \leftrightarrow k), \notag \\
    &=\frac{e g_s}{8m} \mathrm{Re}\sum_{n}^{\mathrm{occ}} \int_{\bm{k}} \Bigl[ \braket{D_j u_n|\hat{s}^\mu\hat{Q}|D_k u_n}-\braket{u_n|\hat{s}^\mu \partial_j\hat{Q}|D_k u_n} \notag \\
    & \quad + \sum_m^{\mathrm{unocc}} 2 (\varepsilon_n - \mu) \Bigl( \frac{R^{\mu,jk}_{nm}}{\varepsilon_{nm}}+\frac{\partial_k \tilde{\varepsilon}_{nm}}{\varepsilon_{nm}}\frac{s^\mu_{nm}\braket{u_m|D_j u_n}}{\varepsilon_{nm}} \Bigr) -\sum_{m}^{\mathrm{unocc}} 2\partial_k \varepsilon_n \frac{s^\mu_{nm}\braket{u_m|D_ju_n}}{\varepsilon_{nm}} \Bigr]+(j \leftrightarrow k), \label{eq:SOM_modern_zerotemperature} 
\end{align}
which yields the following $f$-sum rule:
\begin{align}
    M^{(\mathrm{spin})}_{\mu,(jk)}=\frac{1}{\pi e}\Bigl[ \braket{\chi^{(\mathrm{o})}_{\mu,(jk)}}+2 \sum_n^{\mathrm{occ}}\int_{\bm{k}}(\varepsilon_{n\bm{k}}-\mu)\braket{\omega^{-1}\chi^{(\mathrm{o})}_{\mu,(jk)}}_{n\bm{k}}\Bigr]+(\text{band-velocity terms}),
\end{align}
where $\braket{\omega^{-1}\chi^{(\mathrm{o})}_{\mu,(jk)}}=\sum_n^{\mathrm{occ}}\int_{\bm{k}}\braket{\omega^{-1}\chi^{(\mathrm{o})}_{\mu,(jk)}}_{n\bm{k}}$,
and the band-velocity terms denote contributions involving the band velocity $\partial_k \varepsilon_n$.
In deriving Eq.~\eqref{eq:SOM_modern_zerotemperature}, we use the following two relations.
The first is 
\begin{align}
    &\braket{D_j u_{n}|\hat{s}^\mu+s^\mu_{n}|D_k u_{n}} +\sum_{n' (\neq n)}^{\mathrm{occ}} R^{\mu,jk}_{nn'} \notag \\
    &=\braket{D_j u_n|\hat{s}^\mu\hat{Q}|D_k u_n}+\underbrace{\sum_{n'}^{\mathrm{occ}}\braket{D_j u_n|\hat{s}^\mu|u_{n'}}\braket{u_{n'}|D_k u_n} -\sum_{n'(\neq n)}^{\mathrm{occ}}\braket{D_j u_n|\hat{s}^\mu|u_{n'}}\braket{u_{n'}|D_k u_n}}_{=0} \notag \\
    & \quad + \underbrace{\sum_{n'(\neq n)}^{\mathrm{occ}}\braket{u_n|\hat{s}^\mu|D_ju_{n'}}\braket{u_{n'}|D_k u_n}+ \sum_{n'}^{\mathrm{occ}}\braket{u_n|\hat{s}^\mu|u_{n'}}\braket{D_j u_{n'}|D_k u_n}}_{\braket{u_n|\hat{s}^\mu\partial_j \hat{P}|D_k u_n}=-\braket{u_n|\hat{s}^\mu \partial_j \hat{Q}|D_k u_n}},
\end{align}
where we use $\hat{1}=\hat{P}+\hat{Q}$, $\braket{u_{n}|D_j u_n}=0$, and 
\begin{align}
    \partial_j \hat{P}=\sum_{n}^{\mathrm{occ}}(\ket{\partial_j u_n}\bra{u_n}+\ket{u_n}\bra{\partial_j u_n})=\sum_{n}^{\mathrm{occ}}(\ket{D_j u_n}\bra{u_n}+\ket{u_n}\bra{D_j u_n})=-\partial_j \hat{Q}.
\end{align}
The second is 
\begin{align}
    &\mathrm{Re}\sum_{n}^{\mathrm{unocc}}\sum_{m}^{\mathrm{occ}} \Bigl[ -(\varepsilon_m - \mu) \frac{R^{\mu,jk}_{nm}}{\varepsilon_{nm}}-\Bigl(\partial_k (\varepsilon_m-\mu)+\frac{(\varepsilon_m-\mu)\partial_k \tilde{\varepsilon}_{nm}}{\varepsilon_{nm}} \Bigr)\frac{s^\mu_{nm}\braket{u_m|D_j u_n}}{\varepsilon_{nm}} \Bigr]+(j \leftrightarrow k) \notag \\
    &=\mathrm{Re} \sum_n^{\mathrm{occ}}\sum_m^{\mathrm{unocc}} \Bigl[ (\varepsilon_n - \mu) \frac{R^{\mu,jk}_{nm}}{\varepsilon_{nm}}- \Bigl(\partial_k (\varepsilon_n-\mu)-\frac{(\varepsilon_n-\mu)\partial_k \tilde{\varepsilon}_{nm}}{\varepsilon_{nm}} \Bigr)\frac{s^\mu_{nm}\braket{u_m|D_j u_n}}{\varepsilon_{nm}} \Bigr]+(j \leftrightarrow k),
\end{align}
which is obtained by interchanging $n$ and $m$.

%%%%%%%%%%%%%%%%%%%%%%%%%%%%%%%%%%%%%%%%%%%%%%%%%%%%%%%%%%%%%%%%%%

\subsection{Short summary} \label{Subsec:AppG4}

In summary, the St\v{r}eda formulas and $f$-sum rules under condition (i) can be unified as 
\begin{subequations} \label{eq:streda_fsumrule_general}
    \begin{align}
        e\frac{\partial M^{(\mathrm{orb})}_{ij}}{\partial \mu}&=-\sigma^{(\mathrm{A})}_{ij}, \quad 
        &M^{(\mathrm{orb})}_{ij} &=\frac{1}{\pi e}\Bigl[ \braket{\sigma^{(\mathrm{A})}_{ij}} +2 \sum_n^{\mathrm{occ}}\int_{\bm{k}}(\varepsilon_{n \bm{k}}-\mu) \braket{\omega^{-1}\sigma^{(\mathrm{A})}_{ij}}_{n\bm{k}}\Bigr], \\
        e\frac{\partial M^{(\mathrm{spin})}_{\mu,j}}{\partial \mu}&=-\chi^{(\mathrm{o})}_{\mu,j}, \quad 
        &M^{(\mathrm{spin})}_{\mu,j} &= \frac{1}{\pi e}\Bigl[ \braket{\chi^{(\mathrm{o})}_{\mu,j}}+2 \sum_n^{\mathrm{occ}} \int_{\bm{k}}(\varepsilon_{n\bm{k}} -\mu)\braket{\omega^{-1}\chi^{(\mathrm{o})}_{\mu,j}}_{n\bm{k}}\Bigr], \\
        e\frac{\partial M^{(\mathrm{spin})}_{\mu,(jk)}}{\partial \mu}&=-\chi^{(\mathrm{o})}_{\mu,(jk)}, \quad 
        &M^{(\mathrm{spin})}_{\mu,(jk)}&=\frac{1}{\pi e}\Bigl[ \braket{\chi^{(\mathrm{o})}_{\mu,(jk)}}+2 \sum_n^{\mathrm{occ}}\int_{\bm{k}}(\varepsilon_{n\bm{k}}-\mu)\braket{\omega^{-1}\chi^{(\mathrm{o})}_{\mu,(jk)}}_{n\bm{k}}\Bigr]+(\text{band-velocity terms}), 
    \end{align}
\end{subequations}
\end{widetext}
revealing universal relations between the $\mathcal{T}$-odd expansion coefficients and the corresponding magnetic multipole moments.
Here, we further impose the following condition: (ii) the occupied subspace consists of a single flat band with Bloch state $\ket{u_{\mathrm{occ}\bm{k}}}$ and energy $\varepsilon_{\mathrm{occ}\bm{k}} \approx \varepsilon_{\mathrm{occ}}$.
Under this condition, the $f$-sum rules in Eq.~\eqref{eq:streda_fsumrule_general} reduce to 
\begin{subequations} 
    \begin{align}
        M^{(\mathrm{orb})}_{ij} &=\frac{1}{\pi e}\Bigl[ \braket{\sigma^{(\mathrm{A})}_{ij}} +2 (\varepsilon_{\mathrm{occ}}-\mu) \braket{\omega^{-1}\sigma^{(\mathrm{A})}_{ij}}\Bigr], \\
        M^{(\mathrm{spin})}_{\mu,j} &= \frac{1}{\pi e}\Bigl[ \braket{\chi^{(\mathrm{o})}_{\mu,j}}+2 (\varepsilon_{\mathrm{occ}} -\mu)\braket{\omega^{-1}\chi^{(\mathrm{o})}_{\mu,j}}\Bigr], \\
        M^{(\mathrm{spin})}_{\mu,(jk)}&=\frac{1}{\pi e}\Bigl[ \braket{\chi^{(\mathrm{o})}_{\mu,(jk)}}+2 (\varepsilon_{\mathrm{occ}}-\mu)\braket{\omega^{-1}\chi^{(\mathrm{o})}_{\mu,(jk)}}\Bigr],
    \end{align}
\end{subequations}
which correspond to Eq.~\eqref{eq:Sumrules_general} of the main text.

We finally discuss two implications of the above results.
First, $\braket{\omega^{-1}\sigma^{(\mathrm{A})}_{ij}}$ is directly related to the Chern number of the occupied band in two-dimensional systems.
To see this, we consider a system in the $x$-$y$ plane, for which $\braket{\omega^{-1}\sigma^{(\mathrm{A})}_{xy}}$ can be expressed as [Eq.~\eqref{eq:OHC_sum_rule2}]
\begin{align}
    \braket{\omega^{-1}\sigma^{(\mathrm{A})}_{xy}} &= \pi e^2 \mathrm{Im}\int_{\mathrm{BZ}}\frac{d^2k}{(2\pi)^2} \notag \\
    &\quad \times \bra{D_x u_{\mathrm{occ}}}(1-\ket{u_{\mathrm{occ}}}\bra{u_{\mathrm{occ}}})\ket{D_y u_{\mathrm{occ}}} \notag \\
    &=-\frac{\pi e^2}{2}\int_{\mathrm{BZ}}\frac{d^2k}{(2\pi)^2} \Omega^{xy}_{\mathrm{occ}}
    =-\frac{e^2}{4}\mathrm{Ch}_{\mathrm{occ}},
\end{align}
where we use $\hat{Q}=1-\hat{P}=1-\ket{u_{\mathrm{occ}}}\bra{u_{\mathrm{occ}}}$ and $\braket{u_{\mathrm{occ}}|D_i u_{\mathrm{occ}}}=0$, $\Omega_{\mathrm{occ}}^{xy}=-2\mathrm{Im}\braket{D_x u_{\mathrm{occ}}|D_y u_{\mathrm{occ}}}$ is the Berry curvature of the occupied band, and 
$\mathrm{Ch}_{\mathrm{occ}}$ is the corresponding Chern number, defined by 
\begin{align}
    \mathrm{Ch}_{\mathrm{occ}}=\frac{1}{2\pi}\int_{\mathrm{BZ}}d^2k \Omega^{xy}_{\mathrm{occ}}.
\end{align}
Thus, $\braket{\omega^{-1}\sigma^{(\mathrm{A})}_{xy}}$ vanishes for a topologically trivial insulator with $\mathrm{Ch}_{\mathrm{occ}}=0$,
whereas no analogous relation generally holds for $\braket{\omega^{-1}\chi^{(\mathrm{o})}_{\mu,j}}$ and $\braket{\omega^{-1}\chi^{(\mathrm{o})}_{\mu,(jk)}}$.

Second, the St\v{r}eda formulas and $f$-sum rules have different requirements for estimating magnetic multipole moments from the corresponding $\mathcal{T}$-odd expansion coefficients.
The St\v{r}eda formulas require only condition (i), but involve integration with respect to the chemical potential $\mu$.
In contrast, the $f$-sum rules allow direct estimation from the optical spectra, but require the additional condition (ii).
Thus, these two approaches have complementary advantages and limitations.

%%%%%%%%%%%%%%%%%%%%%%%%%%%%%%%%%%%%%%%%%%%%%%%%%%%%%%%%%%%%%%%%%%%%%%%%%%%%%%%%%%%%%%%%%%%%%%%%%%%%%%%%%%%%%%%%%%%%%%%%%%%%%%%%%%%%

\section{DETAILS OF THE NUMERICAL CALCULATIONS} \label{Sec:AppH}

\begin{figure*}[t]
    \centering
    \begin{subfigure}[t]{0.32\linewidth}
        \caption{\ce{\textit{R}2Ir2O7}}
        \includegraphics[width=1.0\linewidth,height=4.0cm]{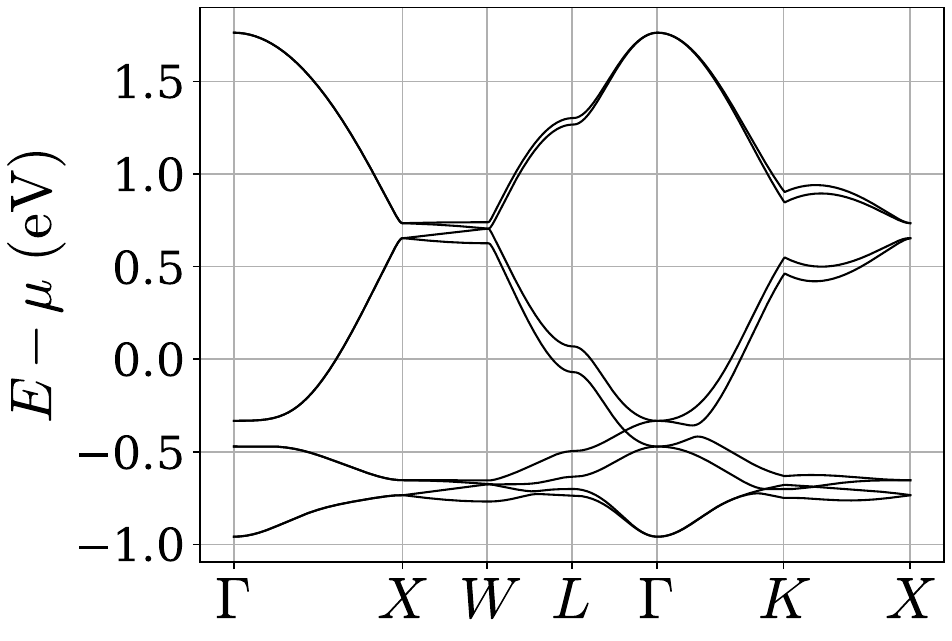}
        \label{fig:Band_structure1}
    \end{subfigure}
    \hfill
    \begin{subfigure}[t]{0.32\linewidth}
        \caption{\ce{RuO2}}
        \includegraphics[width=1.0\linewidth,height=4.0cm]{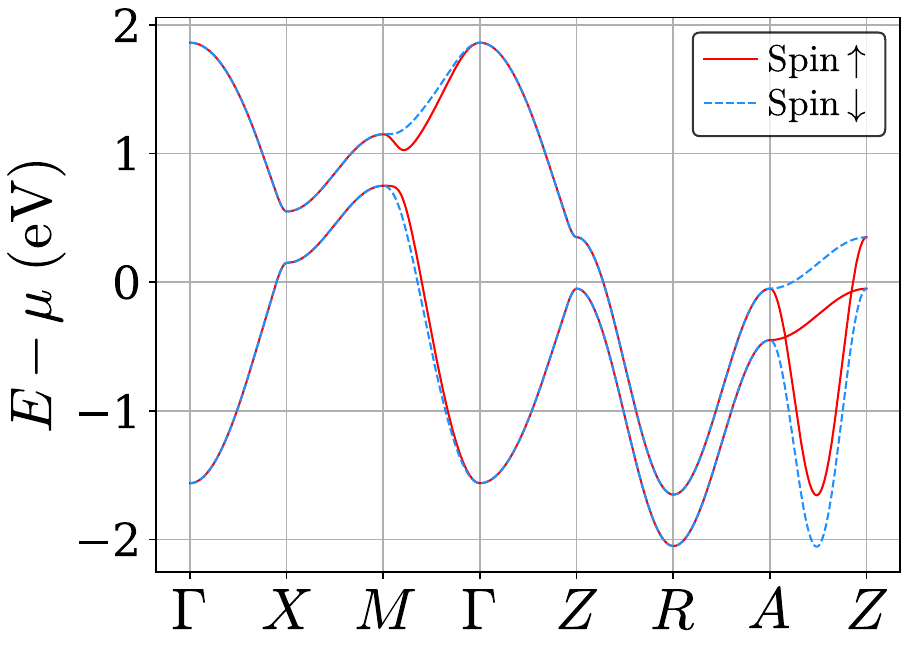}
        \label{fig:Band_structure2}
    \end{subfigure}
    \hfill
    \begin{subfigure}[t]{0.32\linewidth}
        \caption{\ce{CrSb}}
        \includegraphics[width=1.0\linewidth,height=4.0cm]{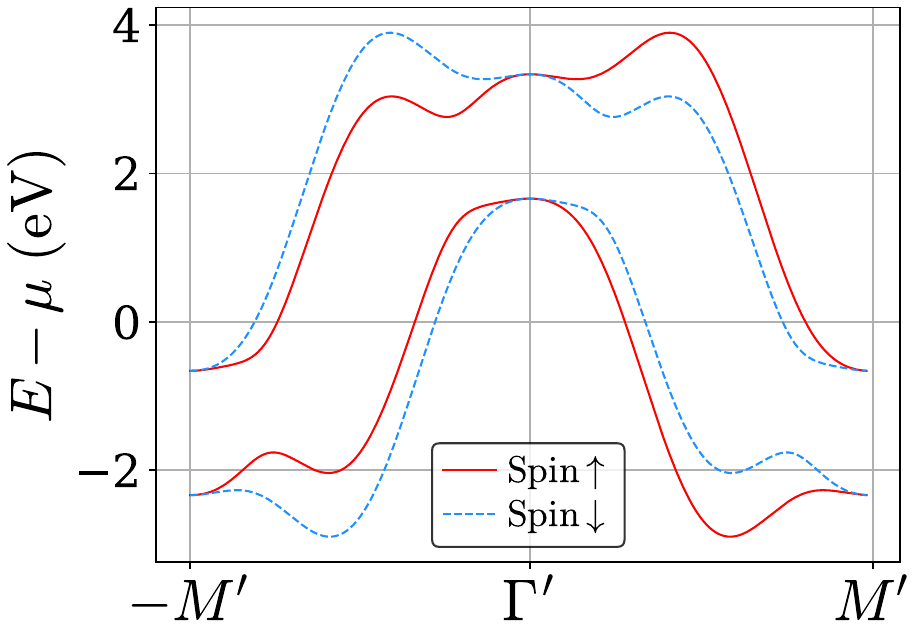}
        \label{fig:Band_structure3}
    \end{subfigure}
    \caption{Band structures along selected $k$ paths for (a) \ce{\textit{R}2Ir2O7}, (b) \ce{RuO2}, and (c) \ce{CrSb}, with
    the points $-M'$, $\Gamma'$, and $M'$ in panel (c) given by $(0,-2\pi/\sqrt{3},\pi/2)$, $(0,0,\pi/2)$, and $(0,2\pi/\sqrt{3},\pi/2)$, respectively.
    Panels (b) and (c) exhibit spin splitting in the absence of SOC. }
    \label{fig:Band_structure}
\end{figure*}

In this appendix, we present the details of the models and calculations used to obtain the numerical results.
We first discuss two technical points concerning the use of Eqs.~\eqref{eq:band_contribution_01_sigma}, \eqref{eq:geometric_contribution_01_sigma}, \eqref{eq:band_contribution_2_o}, and \eqref{eq:geometric_contribution_2_o} in the numerical calculations.
One is that these microscopic expressions contain the covariant derivative $\ket{D_iu_n}$, which is not well suited for direct numerical evaluation.
We therefore use the Hellmann--Feynman theorem $\braket{u_n|D_i u_m}=-\hbar v^i_{nm}/\varepsilon_{nm}$ ($n \neq m$) to rewrite the covariant derivative in terms of the Bloch velocity matrix element $v^i_{nm}$.
The other is that the microscopic expressions contain $1/\varepsilon_{nm}$, which can hinder convergence near band degeneracies.
To address this issue, we regularize $1/\varepsilon_{nm}$ by introducing a smearing parameter $\gamma$: $1/\varepsilon_{nm} \to \varepsilon_{nm}/(\varepsilon^2_{nm}+\gamma^2)$.
However, this regularization is not applied to the frequency-dependent part $1/(\varepsilon_{nm}+\hbar \omega+i \delta)$ due to the finite broadening $\delta$.
These parameters $\gamma$ and $\delta$ are both set to 0.01~eV in the numerical calculations.

We next choose the calculation temperature for each material to be below its antiferromagnetic ordering temperature.
Specifically, we set $T=10$~K for \ce{\textit{R}2Ir2O7}, $T=300$~K for \ce{RuO2}, and $T=300$~K for \ce{CrSb}.
Indeed, these temperatures are below the experimentally reported ordering temperatures of 
33~K for AIAO-ordered \ce{Nd2Ir2O7} \cite{Guo2013-mn,Guo2016-is}, above 300~K for \ce{RuO2} \cite{Berlijn2017-yg,Zhu2019-fx}, and 705~K for \ce{CrSb} \cite{Takei1963-nv},
where the $R=\mathrm{Nd}$ value is used for \ce{\textit{R}2Ir2O7}.

We then describe the model for each material in the following subsections, with all energy parameters given in eV.
Specifically, Appendix~\ref{Subsec:AppH1} is devoted to \ce{\textit{R}2Ir2O7}, while Appendix~\ref{Subsec:AppH2} covers \ce{RuO2} and \ce{CrSb}.

%%%%%%%%%%%%%%%%%%%%%%%%%%%%%%%%%%%%%%%%%%%%%%%%%%%%%%%%%%%%%%%%%%

\subsection{\ce{\textit{R}2Ir2O7}} \label{Subsec:AppH1}

The model for \ce{\textit{R}2Ir2O7} consists of four sublattices and takes the form 
\begin{align}
    \hat{H}(\bm{k})=\hat{H}_t(\bm{k})+m\hat{O}_{\mathrm{AIAO}}. \label{eq:Hamiltonian_Pyrochlore}
\end{align}
Here, $\hat{H}_t(\bm{k})$ is the hopping Hamiltonian including SOC, and
$\hat{O}_{\mathrm{AIAO}}$ is the dimensionless operator describing the AIAO magnetic order with coupling strength $m$.
The explicit forms of these two terms are given below.

First, the hopping Hamiltonian $\hat{H}_t(\bm{k})$ consists of the nearest-neighbor (NN) and next-nearest-neighbor (NNN) contributions for the cubic pyrochlore structure:
$\hat{H}_t(\bm{k})=\hat{H}^{\mathrm{NN}}(\bm{k})+\hat{H}^{\mathrm{NNN}}(\bm{k})$.
In the sublattice basis, these contributions are represented by the matrix blocks \cite{Witczak-Krempa2013-bv,Tian2015-zj}
\begin{subequations}
    \begin{align}
        H^{\mathrm{NN}}_{ij}(\bm{k})&=2\bigl[ t_1 (1-\delta_{ij}) + it_2(\bm{d}_{ij} \cdot \bm{\sigma})\bigr] \cos (\bm{b}_{ij} \cdot \bm{k}), \\
        H^{\mathrm{NNN}}_{ij}(\bm{k})&=\sum_{k (\neq i,j)}2 \bigl\{ t'_1(1-\delta_{ij})+i\bigl[t'_2(\bm{b}_{ik}\times\bm{b}_{kj}) \notag \\
        & \quad +t'_3(\bm{d}_{ik}\times\bm{d}_{kj}) \bigr] \cdot \bm{\sigma} \bigr\} \cos \bigl[(-\bm{b}_{ik}+\bm{b}_{kj}) \cdot \bm{k} \bigr],
    \end{align}
\end{subequations}
where $i$ and $j$ label the four sublattices and each block is a $2\times2$ matrix in spin space, with $\bm{\sigma}=(\sigma_x,\sigma_y,\sigma_z)$ denoting the Pauli matrices.
The vectors entering these expressions are given by $\bm{d}_{ij}=2\bm{a}_{ij}\times\bm{b}_{ij}$, $\bm{a}_{ij}=(\bm{x}_i+\bm{x}_j)/2-\bm{x}_{\mathrm{G}}$, and $\bm{b}_{ij}=\bm{x}_j-\bm{x}_i$,
where $\bm{x}_i$ is the position vector of sublattice $i$; we choose $\bm{x}_1=(0,0,0)$, $\bm{x}_2=(1,1,0)$, $\bm{x}_3=(1,0,1)$, and $\bm{x}_4=(0,1,1)$, with $\bm{x}_{\mathrm{G}}=(1,1,1)/2$.
The hopping parameters are expressed in terms of $t_{\mathrm{oxy}}$, $t_\sigma$, $t_\pi$, $t'_\sigma$, and $t'_\pi$ as follows:
\begin{subequations}
\begin{align}
    \left\{
    \begin{aligned}
    t_1&=\frac{130}{243}t_{\mathrm{oxy}}+\frac{17}{324}t_\sigma-\frac{79}{243}t_\pi, \\
    t_2&=\frac{28}{243}t_{\mathrm{oxy}}+\frac{15}{243}t_\sigma-\frac{40}{243}t_\pi, \\
    \end{aligned}
    \right.
\end{align}
for the NN contribution and 
\begin{align}
    \left\{
    \begin{aligned}
    t_1'&=\frac{233}{2916}t'_\sigma-\frac{407}{2187}t'_\pi, \\
    t_2'&=\frac{1}{1458}t'_\sigma+\frac{220}{2187}t'_\pi, \\
    t_3'&=\frac{25}{1458}t'_\sigma+\frac{460}{2187}t'_\pi, 
    \end{aligned}
    \right.
\end{align}
\end{subequations}
for the NNN contribution.
These underlying parameters are adopted from Ref.~\cite{Witczak-Krempa2013-bv} and set to $t_{\mathrm{oxy}}=1.0$, $t_\sigma=-0.8$, $t_\pi=(-2/3)t_\sigma$, and $t'_{\sigma,\pi}=0.08 t_{\sigma,\pi}$ 
in the numerical calculations, with the chemical potential set to $\mu=0$.

Second, the AIAO-order operator $\hat{O}_{\mathrm{AIAO}}$ is given by 
\begin{align}
    \hat{O}_{\mathrm{AIAO}}&=\mathrm{diag}(1,-1,-1,1)\sigma_x+\mathrm{diag}(1,-1,1,-1)\sigma_y \notag \\
    &\quad +\mathrm{diag}(1,1,-1,-1)\sigma_z,
\end{align}
where $\mathrm{diag}(\cdots)$ denotes a $4\times4$ diagonal matrix in the sublattice space.
For the coupling strength $m$, we adopt the value from Ref.~\cite{Ueda2018-wq} and set $m=0.04$ in the numerical calculations unless otherwise noted.
This choice, together with the hopping parameters and chemical potential specified above, yields the band structure of \ce{\textit{R}2Ir2O7} shown in Fig.~\ref{fig:Band_structure1}.

%%%%%%%%%%%%%%%%%%%%%%%%%%%%%%%%%%%%%%%%%%%%%%%%%%%%%%%%%%%%%%%%%%

\subsection{\ce{RuO2} and \ce{CrSb}} \label{Subsec:AppH2}

The models for \ce{RuO2} and \ce{CrSb} both consist of two sublattices and take the form 
\begin{align}
    \hat{H}(\bm{k})&=\hat{H}_t(\bm{k})+\hat{H}_{\mathrm{intra}}(\bm{k}) 
    +\lambda_{\mathrm{SOC}} \hat{O}_{\mathrm{SOC}}(\bm{k})+J_z\hat{O}^z_{\mathrm{N}}. \label{eq:Hamiltonian_Altermagnet}
\end{align}
Here, $\hat{H}_t(\bm{k})$ consists of the sublattice-independent hopping Hamiltonian $\hat{H}_0(\bm{k})$ and the intersublattice hopping Hamiltonian $\hat{H}_{\mathrm{inter}}(\bm{k})$,
$\hat{H}_{\mathrm{intra}}(\bm{k})$ is the intrasublattice hopping Hamiltonian,
$\hat{O}_{\mathrm{SOC}}(\bm{k})$ is the dimensionless operator describing SOC with coupling strength $\lambda_{\mathrm{SOC}}$, and 
$\hat{O}^z_{\mathrm{N}}$ is the dimensionless operator describing N\'{e}el order along the $z$ direction with coupling strength $J_z$.
These terms are explicitly written in terms of the Pauli matrices $\bm{\tau}=(\tau_x,\tau_y,\tau_z)$ and $\bm{\sigma}=(\sigma_x,\sigma_y,\sigma_z)$ in the sublattice and spin spaces, respectively,
and the corresponding identity matrices $\tau_0$ and $\sigma_0$:
$\hat{H}_0(\bm{k})=\varepsilon_0(\bm{k})\tau_0 \sigma_0$, $\hat{H}_{\mathrm{inter}}(\bm{k})=t_x(\bm{k})\tau_x \sigma_0$,
$\hat{H}_{\mathrm{intra}}(\bm{k})=t_z(\bm{k})\tau_z \sigma_0$, $\hat{O}_{\mathrm{SOC}}(\bm{k})=\tau_y \bm{g}(\bm{k})\cdot\bm{\sigma}$, and $\hat{O}^z_{\mathrm{N}}=\tau_z \sigma_z$ \cite{Roig2024-zg}.
While these operator forms are common to both models, the $\bm{k}$ dependences of the individual terms differ between \ce{RuO2} and \ce{CrSb}.

For \ce{RuO2}, the $\bm{k}$ dependences reflect its tetragonal crystal structure and are given by 
\begin{subequations}
    \begin{align}
        \varepsilon_0(\bm{k})&=t_1(\cos k_x+\cos k_y)+t_2 \cos k_z \notag \\
        & \quad + t_3 \cos k_x \cos k_y + t_4(\cos k_x+\cos k_y)\cos k_z \notag \\
        & \quad +t_5\cos k_x \cos k_y \cos k_z, \\
        t_x(\bm{k})&=t_8 \cos\frac{k_x}{2} \cos\frac{k_y}{2} \cos\frac{k_z}{2}, \\
        t_z(\bm{k})&=\sin k_x \sin k_y (t_6+t_7 \cos k_z), \\
        g_x(\bm{k})&=\sin\frac{k_x}{2}\cos\frac{k_y}{2}\sin\frac{k_z}{2}, \\
        g_y(\bm{k})&=-\cos\frac{k_x}{2}\sin\frac{k_y}{2}\sin\frac{k_z}{2},
    \end{align}
\end{subequations}
with $g_z(\bm{k})=0$.
The hopping parameters entering these expressions, together with the coupling strengths and chemical potential, are adopted from Ref.~\cite{Roig2024-zg} and set to
$t_1=-0.05$, $t_2=0.7$, $t_3=0.5$, $t_4=-0.15$, $t_5=-0.4$, $t_6=-0.6$, $t_7=0.3$, $t_8=1.7$, $\lambda_{\mathrm{SOC}}=0.1$, $J_z=0.2$, and $\mu=0.25$ in the numerical calculations unless otherwise noted.
For these parameter values, but with $\lambda_{\mathrm{SOC}}=0$, the band structure of \ce{RuO2} is shown in Fig.~\ref{fig:Band_structure2}.

For \ce{CrSb}, the $\bm{k}$ dependences reflect its hexagonal crystal structure and are given by 
\begin{subequations}
    \begin{align}
        \varepsilon_0(\bm{k})&=t_1 \Bigl(\cos k_x + 2 \cos\frac{k_x}{2}\cos\frac{\sqrt{3}k_y}{2} \Bigr) \notag \\
        & \quad +t_2 \cos k_z, \\
        t_x(\bm{k})&=t_3 \cos\frac{k_z}{2}, \\
        t_z(\bm{k})&=t_4 f_y(f^2_y-3f^2_x) \sin k_z, \\
        g_x(\bm{k})&=(f^2_x-f^2_y)\cos\frac{k_z}{2}, \\
        g_y(\bm{k})&=-2f_x f_y \cos\frac{k_z}{2}, \\
        g_z(\bm{k})&=f_x(f^2_x-3f^2_y)\sin\frac{k_z}{2}, 
    \end{align}
\end{subequations}
where $f_x$ and $f_y$ are defined as
\begin{subequations}
    \begin{align}
        f_x&=\sin k_x+\sin\frac{k_x}{2}\cos\frac{\sqrt{3}k_y}{2}, \\
        f_y&=\sqrt{3}\cos\frac{k_x}{2}\sin\frac{\sqrt{3}k_y}{2}.
    \end{align}
\end{subequations}
The hopping parameters entering these expressions, together with the coupling strengths and chemical potential, are adopted from Ref.~\cite{Zheng2026-wu} and set to
$t_1=1.0$, $t_2=1.0$, $t_3=1.0$, $t_4=0.5$, $\lambda_{\mathrm{SOC}}=0.004$, $J_z=0.45$, and $\mu=0.5$ in the numerical calculations unless otherwise noted.
For these parameter values, but with $\lambda_{\mathrm{SOC}}=0$, the band structure of \ce{CrSb} is shown in Fig.~\ref{fig:Band_structure3}.

\bibliography{main.bib}

%%%%%%%%%%%%%%%%%%%%%%%%%%%%%%%%%%%%%%%%%%%%%%%%%%%%%%%%%%%%%%%%%%%%%%%%%%%%%%%%%%%%%%%%%%%%%%%%%%%%%%%%%%%%%%%%%%%%%%%%%%%%%%%%%%%%

\onecolumngrid
\clearpage

\foreach \x in {1,...,\the\numbersupplementpages}{%
    \includepdf[pages={\x}]{\supplementfilename}%
}

\end{document}